\documentclass[a4paper,11pt]{article}
\pdfoutput=1
\usepackage[utf8]{inputenc}
\usepackage[T1]{fontenc}
\usepackage{youngtab}
\usepackage[english]{babel}
\usepackage[colorlinks=true
,urlcolor=blue
,anchorcolor=blue
,citecolor=blue
,filecolor=blue
,linkcolor=blue
,menucolor=blue
,linktocpage=true
,pdfproducer=medialab
,pdfa=true
]{hyperref}
\usepackage{amsmath,amsthm,amsbsy,amssymb,amsfonts,graphicx,mathrsfs,bm,cite,color,accents}
\allowdisplaybreaks
\newtheorem{theorem}{Theorem}
\usepackage{empheq}
\usepackage[customcolors]{hf-tikz}
\hfsetfillcolor{blue!5}
\hfsetbordercolor{white}
\hyphenchar\font=\string"7F
\usepackage{tikz}
\usetikzlibrary{calc,matrix}
\DeclareMathAlphabet{\mathdsl}{U}{bbm}{m}{sl}

\usepackage[all]{xy}
\usepackage{enumitem}
\usepackage{refcount}

\usepackage[utf8]{inputenc}
\renewcommand{\thefootnote}{\arabic{footnote}}

\usepackage{ulem}

\makeatletter
\catcode`\@=11
\@addtoreset{equation}{section}

\makeatother

\newcommand{\Ah}{\widehat{A}}
\newcommand{\Bh}{\widehat{B}}
\newcommand{\Ch}{\widehat{C}}
\newcommand{\Dh}{\widehat{D}}
\newcommand{\Eh}{\widehat{E}}
\newcommand{\Fh}{\widehat{F}}
\newcommand{\Ih}{\widehat{I}}
\newcommand{\Jh}{\widehat{J}}
\newcommand{\Kh}{\widehat{K}}
\newcommand{\Lh}{\widehat{L}}
\newcommand{\Mh}{\widehat{M}}
\newcommand{\Nh}{\widehat{N}}
\newcommand{\Vh}{\widehat{V}}

\newcommand{\rmd}{\mathrm{d}}

\newcommand{\cA}{\mathcal A}\newcommand{\cB}{\mathcal B}
\newcommand{\cC}{\mathcal C}\newcommand{\cD}{\mathcal D}
\newcommand{\cE}{\mathcal E}\newcommand{\cF}{\mathcal F}
\newcommand{\cG}{\mathcal G}\newcommand{\cH}{\mathcal H}
\newcommand{\cI}{\mathcal I}
\newcommand{\cL}{\mathcal L}
\newcommand{\cM}{\mathcal M}\newcommand{\cN}{\mathcal N}

\newcommand{\cR}{\mathcal R}
\newcommand{\cT}{\mathcal T}

\newcommand{\cW}{\mathcal W}

\newcommand{\nn}{\nonumber}
\newcommand{\OO}{\text{O}}
\newcommand{\SL}{\text{SL}}
\newcommand{\GL}{\text{GL}}
\newcommand{\SO}{\text{SO}}

\makeatletter
\newcommand{\ubar}[1]{\underaccent{\bar}{#1}}
\makeatother

\makeatletter
\newcommand{\Exp}[1]{\mathrm{e}^{#1}}
\makeatother

\newcommand{\Lie}{\cL}
\newcommand{\gLie}{\hat{\cL}}
\newcommand{\GLie}{\hat{\mathbb{L}}}
\newcommand{\tr}{\text{tr}}
\newcommand{\diag}{\text{diag}}

\newcommand{\sfa}{\mathsf{a}}
\newcommand{\sfb}{\mathsf{b}}
\newcommand{\sfc}{\mathsf{c}}
\newcommand{\sfd}{\mathsf{d}}
\newcommand{\sfe}{\mathsf{e}}
\newcommand{\sff}{\mathsf{f}}
\newcommand{\sfg}{\mathsf{g}}

\newcommand{\sfi}{\mathsf{i}}
\newcommand{\sfj}{\mathsf{j}}

\newcommand{\adja}{\bm{a}}
\newcommand{\adjb}{\bm{b}}
\newcommand{\adjc}{\bm{c}}

\newcommand{\adji}{\dot{\bm{a}}}
\newcommand{\adjj}{\dot{\bm{b}}}
\newcommand{\adjk}{\dot{\bm{c}}}

\newcommand{\adjah}{\bm{\widehat{a}}}

\newcommand{\GS}{G_{\mathrm{S}}}

\newcommand{\GD}{G_{\mathrm{D}}}
\newcommand{\GM}{G_{\mathrm{M}}}
\newcommand{\GDM}{\hat{G}_{\mathrm{D}}}

\newcommand{\Edd}[1][d]{\mathrm{E}_{#1(#1)}}
\newcommand{\edd}[1][d]{\mathfrak{e}_{#1(#1)}}
\newcommand{\Odd}[1][d]{\mathrm{O}(#1,#1+\mathfrak{n})}

\newcommand{\irrep}[1]{\mathbf{#1}}
\newcommand{\irrepb}[1]{\mathbf{\overline{#1}}}

\newcommand{\bra}[1]{\langle{#1}\rvert}
\newcommand{\ket}[1]{\lvert{#1}\rangle}
\newcommand{\braket}[2]{\langle{#1}\vert{#2}\rangle}

\newcommand{\Calpha}{\boldsymbol{\alpha}}
\newcommand{\Cbeta}{\boldsymbol{\beta}}

\renewcommand{\AA}{\overline{A}}
\newcommand{\BB}{\overline{B}}
\newcommand{\CC}{\overline{C}}
\newcommand{\DD}{\overline{D}}
\newcommand{\EE}{\overline{E}}
\newcommand{\FF}{\overline{F}}
\newcommand{\GG}{\overline{G}}
\newcommand{\II}{\overline{I}}
\newcommand{\JJ}{\overline{J}}
\newcommand{\KK}{\overline{K}}
\newcommand{\LL}{\overline{L}}

\newcommand{\AAA}{\overline{\AA}}
\newcommand{\BBB}{\overline{\BB}}
\newcommand{\CCC}{\overline{\CC}}
\newcommand{\DDD}{\overline{\DD}}
\newcommand{\EEE}{\overline{\EE}}

\newcommand{\GGG}{\overline{\GG}}
\newcommand{\III}{\overline{\II}}
\newcommand{\JJJ}{\overline{\JJ}}

\newcommand{\EXm}{\bm{\mu}}
\newcommand{\EXn}{\bm{\nu}}
\newcommand{\EXp}{\bm{\rho}}
\newcommand{\EXq}{\bm{\sigma}}
\newcommand{\EXa}{\bm{\alpha}}
\newcommand{\EXb}{\bm{\beta}}

\newcommand{\tnabla}{\nabla}
\newcommand{\bnabla}{\bar{\nabla}}

\newcommand{\exA}{\alpha}
\newcommand{\exB}{\beta}
\newcommand{\exC}{\gamma}
\newcommand{\exD}{\delta}
\newcommand{\exE}{\epsilon}

\newcommand{\lA}{\mathbb{A}}
\newcommand{\lB}{\mathbb{B}}
\newcommand{\lC}{\mathbb{C}}
\newcommand{\lD}{\mathbb{D}}
\newcommand{\lE}{\mathbb{E}}

\newcommand{\tp}{\mathrm{t}}

\newcommand{\phiS}{\phi}
\newcommand{\phiD}{\varphi}

\newcommand{\gD}{\bm{g}}
\newcommand{\eD}{\bm{e}}

\newcommand{\SSS}{K}

\newcommand{\pagediff}[2]{%
  \the\numexpr\getpagerefnumber{#2} - \getpagerefnumber{#1} + 1\relax
}

\allowdisplaybreaks[3]

\begin{document}

\hypersetup{pageanchor=false}
\begin{titlepage}
\renewcommand{\thefootnote}{\fnsymbol{footnote}}

\vspace*{1.0cm}

\centerline{\LARGE\textbf{Consistent truncation and generalized duality}}
\vspace{2mm}
\centerline{\LARGE\textbf{based on exceptional generalized cosets}}

\vspace{1.0cm}

\centerline{
{Falk Hassler}%
\footnote{E-mail address: falk.hassler@uwr.edu.pl} 
and
{Yuho Sakatani}%
\footnote{E-mail address: yuho@koto.kpu-m.ac.jp}%
}

\begin{center}
${}^\ast${\it University of Wroc\l{}aw, Faculty of Physics and Astronomy,}\\
{\it Maksa Borna 9, 50-204 Wroclaw, Poland}

${}^\dagger${\it Department of Physics, Kyoto Prefectural University of Medicine,}\\
{\it 1-5 Shimogamohangi-cho, Sakyo-ku, Kyoto 606-0823, Japan\\}
\end{center}

\begin{abstract}
We present a systematic framework for constructing consistent truncations of supergravity based on exceptional generalized cosets of the form $\GS \backslash G/H$. This approach generalizes the well-established generalized Scherk--Schwarz reductions on generalized parallelizable spaces $G/H$, which preserve maximal supersymmetry, to scenarios with reduced supersymmetry by introducing a non-trivial generalized structure group $\GS$. The double coset structure plays two distinct roles: for a given $G$, the choice of subgroup $\GS$ determines the (constant) generalized torsion/curvature and the pattern of supersymmetry breaking, while $H$ parameterizes inequivalent supergravity backgrounds that share the same truncated theory. The entire construction proceeds algebraically, systematically building $\GS$-invariant tensors from generalized frame fields, with the intrinsic torsion automatically constant and a $\GS$-singlet. Different choices of $H$ lead to distinct higher-dimensional backgrounds that truncate to the same lower-dimensional theory, thereby realizing U-duality. We illustrate the framework through explicit examples in double field theory and exceptional field theory.
\end{abstract}

\thispagestyle{empty}
\end{titlepage}
\hypersetup{pageanchor=true}

\setcounter{footnote}{0}

\newpage
\hrule
\tableofcontents
\vspace{2em}
\hrule
\newpage

\section{Introduction}\label{start}
\nocite{Hassler:2023axp}%
A fundamental challenge in supergravity is identifying independent subsectors of the full theory. A practical framework for this purpose is provided by consistent truncations, where a restricted set of fields can be retained without sourcing the discarded modes. These truncations enable the systematic construction of new solutions and exploration of supergravity vacua. However, the non-linear structure of the field equations makes ensuring consistency highly non-trivial and often requires detailed case-by-case analysis.

Generalized geometry \cite{Hitchin:2003cxu,Gualtieri:2003dx,Hull:2007zu,Berman:2010is,Coimbra:2011ky,Coimbra:2012af} and extended field theories \cite{Siegel:1993xq,Siegel:1993th,Siegel:1993bj,Hull:2009mi,Hohm:2010pp,Jeon:2011cn,Hohm:2011si,Geissbuhler:2013uka,West:2000ga,West:2001as,Hillmann:2009ci,Berman:2011jh,Berman:2012vc,West:2012qz,Hohm:2013pua,Hohm:2013vpa,Hohm:2013uia,Hohm:2014fxa,Hohm:2015xna,Abzalov:2015ega,Musaev:2015ces,Berman:2015rcc} have introduced systematic methods for constructing and validating such reductions. These frameworks unify the various gauge symmetries of ten- and eleven-dimensional supergravities with diffeomorphisms, and reveal the hidden geometric structures that govern consistent truncations. While they have significantly advanced the study of supergravity reductions, a fundamental tension between generality and tractability remains. Within these frameworks, two complementary approaches have emerged:
\begin{enumerate}[label=\alph*)]
\item\label{item:maximal} explicit constructions preserving maximal supersymmetry, formulated entirely in terms of algebraic constraints, and
\item\label{item:recast} a systematic reformulation of the conditions for preserving reduced supersymmetry in terms of generalized $\GS$-structures \cite{Coimbra:2014uxa,Coimbra:2016ydd,Coimbra:2017fqv}, which requires solving differential equations to determine the $\GS$-invariant tensors.
\end{enumerate}

This article presents a new framework built upon \textit{exceptional generalized cosets}, synthesizing the advantages of both approaches. It extends the algebraic techniques of the first approach to scenarios with partial or no supersymmetry, significantly broadening their scope, while preserving the underlying algebraic structure that eliminates the need to solve differential equations. Simultaneously, it systematically constructs the objects required in the second approach, thereby resolving one of its main technical obstacles. In this way, exceptional generalized cosets provide a unifying framework that combines the algebraic simplicity of \ref{item:maximal} with the flexibility of \ref{item:recast}. A wide range of physically relevant scenarios with reduced supersymmetry, previously beyond the reach of existing methods, now becomes accessible.

To motivate and define the framework of exceptional generalized cosets, we begin by reviewing generalized Scherk--Schwarz reductions, the prototypical realization of approach~\ref{item:maximal}, as a concrete example of how purely algebraic constraints can ensure the consistency of a truncation. We then show how this algebraic structure generalizes to incorporate non-trivial structure groups $\GS$, leading to the notion of exceptional generalized cosets.

\paragraph{(Generalized) Scherk--Schwarz reductions:}
Generalized Scherk--Schwarz reductions \cite{Aldazabal:2011nj,Geissbuhler:2011mx,Grana:2012rr,Geissbuhler:2013uka,Lee:2014mla,Hohm:2014qga} are inspired by the original Scherk--Schwarz reductions \cite{Scherk:1978ta,Scherk:1979zr} on group manifolds, where only left- or right-invariant tensor fields are considered. As a consequence, at least locally, all differential structure is captured by a Lie algebra. While this approach yields a variety of reductions, it was initially developed for theories with only diffeomorphism invariance and neglects the fact that maximal supergravities in ten and eleven dimensions also possess additional gauge symmetries (such as the gauge transformations of $p$-form fields). This is exactly where generalized geometry and extended field theories become essential: they unify these additional gauge symmetries with diffeomorphisms via a single generalized Lie derivative \cite{Hitchin:2003cxu,Gualtieri:2003dx,Hull:2007zu,Berman:2012vc}.

In the definition of left- or right-invariant tensor fields on group manifolds, the ordinary Lie derivative plays a key role, as it defines invariance under the group action. This naturally raises the question of whether a similar notion exists for generalized Lie derivatives. Recall that all right-invariant tensors on a Lie group $G$ can be constructed from right-invariant vector fields $e_a$\footnote{They are dual to the right-invariant Maurer--Cartan form on $G$, satisfying $\rmd g\, g^{-1} = e^a\,t_a$ with $g\in G$ and $[t_a, t_b] = f_{ab}{}^c\, t_c$, where $t_a$ are the generators of the Lie algebra.} with $a=1,\dotsc,\dim G$. Under the Lie derivative, they satisfy
\begin{align}\label{eq:LieLeftInv}
  \Lie_{e_a} e_b = - f_{ab}{}^c\, e_c\,,
\end{align}
where $f_{ab}{}^c$ are the structure constants of the Lie algebra of $G$.

When lifting the structure from ordinary geometry to generalized geometry, the closest analogue of \eqref{eq:LieLeftInv} is
\begin{align}\label{eq:genLieGenPara}
  \gLie_{E_A} E_B = - X_{AB}{}^C\, E_C
\end{align}
where $\gLie$ is the generalized Lie derivative and $E_A$ are generalized vector fields, also referred to as generalized frame fields. A key difference in the generalized setting is that the generalized Lie derivative is generally not antisymmetric in its arguments. As a consequence, the constants $X_{AB}{}^C$ define a Leibniz algebra rather than a Lie algebra. For consistency with the algebraic properties of the generalized Lie derivative, the symmetric part $\gLie_{E_A} E_B + \gLie_{E_B} E_A = - Z_{AB}{}^C\, E_C$ should correspond to generalized vector fields that generate the identity transformation, and is therefore physically trivial. Modding out these trivial generators then gives the physically relevant Lie algebra $\mathrm{Lie}(G)$.

Another key difference from the group manifold case is that the dimension of $\mathrm{Lie}(G)$ will exceed that of the physical space $M$ itself. Extended field theories initially address this by enlarging the coordinate space (introducing both momentum and winding-type coordinates), but the section condition must then be imposed for consistency, and upon solving this condition, the same dimensional mismatch reappears. Consequently, at each point $p\in M$, there exists a subset of generalized vector fields $E_A$ whose projections onto the ordinary tangent space $T_pM$ vanish. These vector fields leave each point unchanged, forming an isotropy subalgebra $\mathrm{Lie}(H)$ under the generalized Lie derivative. By contrast, the complementary set of generalized frame fields has non-vanishing components along $T_pM$ and generates genuine displacements on $M$. In this way, the physical space has the structure of a coset space $M=G/H$.

For such coset spaces $G/H$ (where $H$ is required to be a suitable subgroup, technically a ``minimally coisotropic'' subgroup \cite{Bugden:2021wxg}), there exists, up to constant transformations in the duality group, a unique generalized frame field $E_A$ satisfying the above algebra \eqref{eq:genLieGenPara}. This frame is globally defined, endowing the space with a generalized parallelizable structure \cite{Lee:2014mla}. The triviality of the generalized structure group for such spaces ensures that generalized Scherk--Schwarz reductions on coset spaces $G/H$ preserve maximal supersymmetry.

\paragraph{Generalized dualities:}
A novel and crucial feature in generalized geometry is that the choice of the subgroup $H$ is not unique. The constants $X_{AB}{}^C$ in \eqref{eq:genLieGenPara} encode only the Lie algebra of $G$, while the subgroup $H$ can be chosen freely provided it is minimally coisotropic. Consequently, several different cosets $G/H$, $G/H'$, \dots, with different subgroups can admit generalized frame fields satisfying the same algebraic relations \eqref{eq:genLieGenPara}. Because the reduced or truncated theory depends only on the constants $X_{AB}{}^C$ (known as generalized fluxes), distinct choices of $H$ nevertheless yield the same effective lower-dimensional theory.

Put differently, different higher-dimensional backgrounds that realize the same generalized fluxes are physically equivalent upon reduction. This equivalence is precisely the manifestation of the duality symmetries of string and M-theory: what appear as different backgrounds in higher dimensions correspond to the same physics in lower dimensions, capturing T-duality, S-duality, and, more generally, their unification in U-duality.

These dualities are most familiar in toroidal compactifications, originating from the interplay between momentum and winding modes of strings (and, in M-theory, membranes). They form a central ingredient of string theory, relating all five perturbative string theories to M-theory. However, the structure outlined above suggests a much richer web of dualities extending beyond tori. Some of these were first identified in the context of classical non-linear $\sigma$-models describing closed string worldsheets, known as non-abelian T-duality \cite{delaOssa:1992vci} and Poisson--Lie T-duality \cite{Klimcik:1995ux}. Generalized Scherk--Schwarz reductions provide a natural framework to study all of them, and even to uncover new generalized U-dualities not yet derived from membrane worldvolume analyses. Formally, this intricate network of dualities can be understood by systematically enumerating the admissible subgroups $H$ for a given Lie group $G$.

\paragraph{(Exceptional) generalized cosets:}
The generalized Scherk--Schwarz reductions on $G/H$ discussed so far are restricted to spaces that are generalized parallelizable. They possess a trivial generalized structure group and this property guarantees the preservation of maximal supersymmetry. To go beyond this setting and capture scenarios with reduced or broken supersymmetry, it is natural to consider spaces with a non-trivial generalized structure group $\GS$. The guiding idea comes from ordinary geometry: Scherk--Schwarz reductions can be generalized from group manifolds $G$ to coset spaces (or homogeneous spaces) $\GS \backslash G$, where $\GS$ is a subgroup of $G$ that plays the role of the structure group.

On a group manifold, right-invariant vector fields $e_a$ satisfy the algebra \eqref{eq:LieLeftInv}. For a coset space $\GS \backslash G$, one must work with coset representatives $m \in \GS \backslash G$. Splitting the generators of $\mathrm{Lie}(G)$ into $t_\beta$ (generating $\GS$) and $t_a$ (coset directions), the Maurer--Cartan form decomposes as
\begin{align}
 \rmd m\, m^{-1} = e^a\, t_a + \omega^{\exA}\, t_{\exA} \,,
\end{align}
where $\omega^{\exA}$ is the connection 1-form of the principal $\GS$-bundle. The dual vector fields then satisfy
\begin{align}\label{eq:LieOnCoset}
 \Lie_{e_a} e_b = - f_{ab}{}^c\,e_c - 2\, \omega^{\exA}_{[a|} \, f_{\exA |b]}{}^c\,e_c\,,
\end{align}
generalizing the group-manifold formula \eqref{eq:LieLeftInv}. The connection $\omega^{\exA}$ encodes the curvature of the bundle structure introduced by the quotient.

The natural question now arises: can this construction be extended to generalized geometry? Specifically, given a generalized Scherk--Schwarz reduction on a coset space $G/H$ (with trivial generalized structure group), can we introduce a non-trivial structure group $\GS$ analogous to the ordinary structure group above? \cite{Demulder:2019vvh} initiated this line of investigation by constructing generalized frames that satisfy a relation similar to \eqref{eq:LieOnCoset}. It soon became clear that the answer is affirmative and leads to the notion of generalized cosets, which take the form of double cosets \cite{Butter:2022iza,Hassler:2023axp}
\begin{align}
 M = \GS \backslash G / H\,.
\end{align}
The appearance of the double coset reflects two independent identifications: the quotient by $\GS$ implements the generalized structure group (analogous to the frame bundle in conventional geometry), while the quotient by $H$ encodes the generalized duality symmetries. This reflects the extra coset structure that arises when the original Scherk--Schwarz reduction, defined on the group manifold $G$, is generalized to the quotient space $\GS \backslash G$. This shift in perspective is illustrated schematically in the diagram below, showing the transition from standard to generalized geometry and from group manifolds to (generalized) cosets:
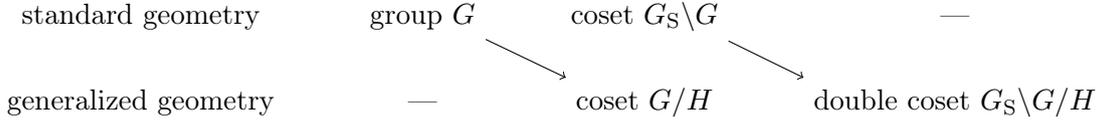
\begin{align*}
  \begin{tikzpicture}
    \matrix (m) [matrix of nodes, column sep=1cm, row sep=0.5cm,ampersand replacement=\&] {
      standard geometry \& group $G$ \& coset $\GS\backslash G$ \& --- \\
      generalized geometry \& --- \& coset $G/H$ \&  double coset $\GS\backslash G / H$\\
    };
    \draw[->] (m-1-2.south east) -- (m-2-3.north west);
    \draw[->] (m-1-3.south east) -- (m-2-4.north west);
  \end{tikzpicture}
\end{align*}
Our discussion of generalized dualities above suggests that the additional group $H$ in the second row encodes generalized dualities. To establish this framework requires a systematic treatment of covariant curvatures in extended field theories. First developments appeared in generalized geometry, where the tangent space is extended by the cotangent space, unifying diffeomorphisms with $B$-field gauge transformations in string theory. The corresponding duality group is $\OO(d,d)$, the setting of double field theory (DFT) \cite{Siegel:1993xq,Siegel:1993th,Siegel:1993bj,Hull:2009mi,Hohm:2010pp,Jeon:2011cn,Hohm:2011si,Geissbuhler:2013uka}. For higher-dimensional and more intricate scenarios, one must work with exceptional generalized geometry and exceptional field theory (ExFT) \cite{West:2000ga,West:2001as,Hillmann:2009ci,Berman:2010is,Berman:2011jh,Berman:2012vc,West:2012qz,Hohm:2013pua,Hohm:2013vpa,Hohm:2013uia,Hohm:2014fxa,Hohm:2015xna,Abzalov:2015ega,Musaev:2015ces,Berman:2015rcc}, which realize the full U-duality groups unifying all bosonic degrees of freedom of maximal supergravities. Just as the ordinary coset construction relies on the curvature of the principal bundle connection, exceptional generalized cosets require a systematic notion of covariant curvatures. The curvature hierarchy developed in \cite{Hassler:2023axp} provides precisely such a framework, enabling the construction and analysis of exceptional generalized cosets.

\paragraph{Consistent truncations:}
There remains the question of how exceptional generalized cosets connect to consistent truncations. A key result \cite{Cassani:2019vcl}, exemplifying approach~\ref{item:recast}, states that if the intrinsic torsion of a generalized $\GS$-structure is constant and transforms as a singlet under the structure group $\GS$, then one can expand all fields in terms of $\GS$-invariant tensors to obtain a consistent truncation. However, constructing a complete set of $\GS$-invariant tensors is highly non-trivial, since one generally needs to solve differential equations. Ensuring that the intrinsic torsion (computed via the generalized Lie derivatives of these tensor fields) satisfies the required constancy and $\GS$-singlet property is equally challenging.

Remarkably, for exceptional generalized cosets $\GS\backslash G / H$, both of these intertwined challenges are resolved algebraically. The intrinsic torsion is automatically constant and a $\GS$-singlet due to the algebraic structure of the double coset, and the $\GS$-invariant tensors can be constructed systematically from the generalized frame fields without solving any differential equations. This algebraic resolution simultaneously guarantees consistency of the truncation and enables a purely algebraic understanding of generalized dualities: replacing the subgroup $H$ with another admissible choice, while keeping the same $G$ and $\GS$, produces different higher-dimensional backgrounds that truncate consistently to the same lower-dimensional theory -- a concrete realization of generalized U-duality.

In summary, exceptional generalized cosets provide a unified algebraic framework that addresses all the aforementioned challenges simultaneously. They enable the construction of consistent truncations with arbitrary amounts of supersymmetry, the systematic generation of $\GS$-invariant tensors, and a complete understanding of generalized dualities, all without the need to solve differential equations.

We aim to present a comprehensive overview of generalized cosets and their dualities, with a particular focus on explicit examples, including
\begin{enumerate}[label=\arabic*)]
  \item\label{item:S2} the two-sphere and its various dual background spaces (section~\ref{sec:DFTex1}),
  \item the $\text{SU}(2)/\text{U}(1)$ coset Wess--Zumino--Witten model, with two distinct but dual target-space geometries (section~\ref{sec:SU2-WZW}),
  \item the two-dimensional Euclidean black hole, also known as the cigar, the trumpet, and flat space with a linear dilaton (section~\ref{sec:SL2-WZW}),
  \item the M-theory lift of \ref{item:S2} and its reduction to type IIB supergravity (section~\ref{sec:ex1-ExFT}),
  \item the coset space $\SO(3)\backslash\SO(1,3)$ and one of its duals (section~\ref{sec:ex2-ExFT}),
  \item a reduction on $\text{U}(1)\backslash \SO(5)\times \text{U}_G(1)/\SO(4)$, which results in a half-maximal lower-dimensional theory with four scalar fields, four vector fields for the gauge group $\SO(3)\times\text{U}(1)$, and a 2-form field (section~\ref{sec:ex3-ExFT})
  \item the Sasaki--Einstein manifold $T^{1,1}$, its non-abelian T-dual with respect to a $\text{SU}(2)$-isometry as presented in \cite{Itsios:2013wd}, and an additional background arising from a generalized duality \newline (section~\ref{sec:ex4-ExFT}),
  \item two quarter-maximal reductions to five dimensions with either
    \begin{itemize}
      \item 8 vector multiplets for the gauge groups $\text{SU}(3)\times\text{U}(1)$ or $\text{SU}(2,1)\times\text{U}(1)$ \newline (section~\ref{sec:ex5-ExFT}),
      \item 4 vector multiplets, 1 hyper multiplet, and the gauge group $\SO(3)\times\mathbb{R}^+\times \text{U}(1)^2$ (section~\ref{sec:ex6-ExFT}).
    \end{itemize}
\end{enumerate}
This wealth of examples illustrates the versatility of generalized cosets. One of the main technical challenges is understanding the representation theory of the exceptional Lie groups that underlie ExFT and its associated generalized geometry. To level the playing field for readers unfamiliar with these structures, we provide a concise review of the relevant material in section~\ref{sec:ExFT}. This section introduces the fundamental objects, such as the generalized Lie derivative and generalized fluxes, in a manner that is as agnostic as possible with respect to the choice of duality group. This unified language ensures that known results for the duality group $\OO(d,d)$ in DFT \cite{Butter:2022iza} integrate naturally into our framework, while even standard differential geometry emerges by considering the duality group $\text{GL}(d)$. For the latter, we explain in section~\ref{sec:GL} how the fundamental quantities of differential geometry, namely torsion and Riemann curvature, can be reconstructed within the \textit{mega-space} approach. This approach, closely related to Cartan geometry, unifies the frame field and spin connection into a single object, giving rise to both torsion and curvature. In sections~\ref{sec:GL-curvatures} and \ref{sec:algebra-GL} we reveal a close relation between curvatures and the algebra underlying the Lie derivative. Extending these algebraic structures to more general duality groups in section~\ref{sec:algebras-decom}, we introduce the aforementioned hierarchy of curvatures in generalized geometry and extended field theories in section~\ref{sec:frames-curvatures}. The conceptual ideas behind these results were already outlined in our letter \cite{Hassler:2023axp}, but here we work out all details. Equipped with these tools, we proceed in section~\ref{sec:consistent-tr-dualities} to construct generalized cosets. The central results are presented in section~\ref{sec:consistent-tr}, where we show how they lead to consistent truncations, and in section~\ref{sec:gen-dualities}, where we explain how generalized dualities arise from choosing different subgroups $\GS\backslash G/H$ in the double coset. Unlike cosets, double cosets may possess singular points when the left and right group actions coincide at certain locations. Since generalized dualities change the right group $H$, they generally identify smooth spaces with singular ones. This phenomenon is analyzed in section~\ref{sec:singularity}. From there we turn to the examples listed above. The remainder of section~\ref{sec:consistent-tr-dualities} is devoted to those based on the duality group $\OO(d,d)$. In section~\ref{sec:examples-ExFT} we study full exceptional generalized cosets, which break different amounts of supersymmetry. Finally, section~\ref{sec:conclusion} summarizes our findings and outlines avenues for future research.

\paragraph{Short cut:}
For the busy reader, \pagediff{start}{end} pages plus appendices may be too much to absorb on a first reading. As a shortcut, we suggest taking our letter \cite{Hassler:2023axp} as an executive summary and then proceeding directly to section~\ref{sec:examples-ExFT}.

\section{Extended field theories}\label{sec:ExFT}
Here we establish our notation by reviewing some basic results of extended field theory. Our discussion mainly concerns DFT and $\Edd[d]$ ExFT with $d\le 8$, though similar methods apply to other extended field theories (such as \cite{Ciceri:2016hup,Bossard:2023ajq}). A key emphasis will be on the gauge transformations, called generalized diffeomorphisms, since they play a central role in the following sections. Their infinitesimal version is mediated by the generalized Lie derivative, reviewed in section~\ref{sec:gen-Lie}. We introduce a unified framework for different duality groups by expressing them in terms of the $Y$-tensor, defined in section~\ref{sec:Y-tensor}, which allows the generalized Lie derivative to be formulated in a uniform way across all duality groups. Section~\ref{sec:fluxes} then presents the generalized fluxes that follow from the generalized Lie derivative. These fluxes are natural covariant objects and will later be shown to contain the full hierarchy of geometric curvatures.

\subsection{\texorpdfstring{$Y$}{Y}-tensor and various tools}\label{sec:Y-tensor}
Let us consider the duality algebra $\mathrm{Lie}(\GD)$ generated by
\begin{align}
 [t_{\adji},\,t_{\adjj}] = f_{\adji\adjj}{}^{\adjk}\,t_{\adjk}\,,
\end{align}
where $\adji,\adjj=1,\dotsc,\dim \GD$ are indices for the adjoint representation, $\irrep{adj}$. We introduce the following representations:
\begin{table}[t]
\centering{\small
 \begin{tabular}{c|c|ccccccc}
      & $\Odd$ & $\text{SL(2)}\times\mathbb{R}^+$ & $\text{SL(3)}\times\text{SL(2)}$ & SL(5) & Spin(5,5) & $\Edd[6]$ & $\Edd[7]$& $\Edd[8]$\\ \hline
      $\irrep{adj}$ & $\irrep{\binom{2d+\mathfrak{n}}{2}}^{\vphantom{|}}$ & $\irrep{3}\oplus\irrep{1}$ & $(\irrep{8},\irrep{1})\oplus(\irrep{1},\irrep{3})$ & $\irrep{24}$ & $\irrep{45}$ & $\irrep{78}$ & $\irrep{133}$ & $\irrep{248}$ \\[1mm] \hline
      $R_1$ & $\irrep{2d+\mathfrak{n}}$ & $\irrep{2}\oplus\irrep{1}$ & $(\irrepb{3},\irrep{2})$ &$\irrepb{10}$ & $\irrep{16}$ & $\irrep{27}$ & $\irrep{56}$ & $\irrep{248}$ \\
      $R_2$ & $\irrep{1}$ & $\irrep{2}$ & $(\irrep{3},\irrep{1})$ & $\irrep{5}$ & $\irrep{10}$ & $\irrepb{27}$ & $\irrep{133}$ & $\irrep{3875}\oplus \irrep{1}$ \\
      $R_3$ & --- & $\irrep{1}$ & $(\irrep{1},\irrep{2})$ & $\irrepb{5}$ & $\irrepb{16}$ & $\irrep{78}$ & $\irrep{912}$ & $\cdots$ \\ \hline
      $C_1$ & $\irrep{1}$ & $\irrep{2}$ & $(\irrep{3},\irrep{1})$ & $\irrep{5}$ & $\irrep{10}$ & $\irrepb{27}$ & $\irrep{133}\oplus\irrep{1}$ & $\genfrac{}{}{0pt}{0}{\irrep{3875}\oplus^{\vphantom{|}}}{\irrep{248}\oplus\irrep{1}}$ \\
      $C_2$ & --- & $\irrep{1}$ & $(\irrep{1},\irrep{2})$ & $\irrepb{5}$ & $\irrepb{16}$ & $\irrep{78}\oplus\irrep{1}$ & $\irrep{912}\oplus\irrep{56}$ & $\cdots$ \\ \hline
      $\Calpha$ & $2$ & --- & --- & $3$ & $4$ & $6$ & $12$ & $60^{\vphantom{|}}$ \\
      $\Cbeta$ & $0$ & $1/7$ & $1/6$ & $1/5$ & $1/4$ & $1/3$ & $1/2$ & $1$ 
\end{tabular}}
\caption{Relevant representations and constants for various duality groups $\GD$. }
\label{tab:representations}
\end{table}
\begin{itemize}
\item for the U-duality group $\GD=\Edd[d]$, we introduce the $R_p$-representations ($p=1,2,\cdots$) which are the representations of $p$-form gauge fields in maximal gauged supergravity \cite{Riccioni:2007au}. Moreover, we introduce $C_p$-representations ($p=1,2,\cdots$) which describe the $p$-brane charge multiplet in \cite{West:2003fc,Cook:2008bi} (the $C_0$-representation coincides with the $R_1$-representation). 

\item For the T-duality group $\GD=\Odd$, the $R_1$- and $C_0$-representation are defined as the $(2d+n)$-dimensional vector representation and the $R_2$- and $C_1$-representation are the singlet. 
The $R_p$- and $C_{p-1}$-representation with $p\geq 3$ are void in this case.
\end{itemize}
All relevant representations are summarized in Table \ref{tab:representations} for each duality group. The curved (flat) indices $I,J,\cdots$ ($A,B,\cdots$) are introduced for the $R_1$-representation. In general, the $R_2$- and $R_3$-representation consist of several irreducible representations. We use the curved (flat) indices $\II,\JJ,\cdots$ ($\AA,\BB,\cdots$) or $\III,\JJJ,\cdots$ ($\AAA,\BBB,\cdots$) to denote the largest one in the $R_2$- or $R_3$-representation, say $\hat{R}_2$- or $\hat{R}_3$-representation. For example, in the $\Edd[8]$ case, a vector in the $R_2$-representation consists of a vector $V_{\II}$ in $\irrep{3875}$ and a vector $V$ in $\irrep{1}$. 

For a simple duality group $\GD$, we define the (rescaled) Cartan--Killing form $\kappa_{\adji\adjj}$ as
\begin{align}
 \kappa_{\adji\adjj} \equiv - \frac{1}{\Calpha}\,\tr_{R_1}(t_{\adji}\,t_{\adjj}) = - \frac{1}{\Calpha}\,(t_{\adji})_I{}^J\,(t_{\adjj})_J{}^I \,,
\end{align}
where the constant $\Calpha$ is given for each group in Table \ref{tab:representations}. This metric is used to lower/raise adjoint indices via contraction with $\kappa_{\adji\adjj}$ and its inverse $\kappa^{\adji\adjj}$. Our (non-standard) choice of an overall minus sign ensures that, in a suitable basis of $\mathrm{Lie}(\GD)$, the dual generators
\begin{align}
 t^{\adji} \equiv \kappa^{\adji\adjj}\,t_{\adjj}\,,
\end{align}
satisfy the same commutation relations as $t_{\adji}$. Consequently, the map
\begin{align}
 \theta(t_{\adji}) = t^{\adji}\,,\qquad 
 \theta(t^{\adji}) = t_{\adji}\,,
\end{align}
defines a Cartan involution on $\mathrm{Lie}(\GD)$. When $\GD$ is semi-simple (i.e., $\Edd[2]$ or $\Edd[3]$), the dual generators $t^{\adji}$ can be obtained by truncating the corresponding higher-rank generators. In this convention, the components of $(\kappa_{\adji\adjj},\, f_{\adji\adjj}{}^{\adjk})$ and $(\kappa^{\adji\adjj},\, f^{\adji\adjj}{}_{\adjk})$ coincide in the chosen basis.

With the above definitions, we introduce the $Y$-tensor $Y^{IJ}_{KL}$, which defines the section condition that constrains the dependence of all fields
\begin{align}
 Y^{IJ}_{KL}\,\partial_I \otimes \partial_J = 0\,,
\label{eq:SC}
\end{align}
as
\begin{align}
 Y^{IJ}_{KL} \equiv \delta^I_K\,\delta^J_L + (t^{\adji})_K{}^J\,(t_{\adji})_L{}^I + \Cbeta\,\delta^I_L\,\delta^J_K\,.
\end{align}
It can be expanded using the projectors onto the $C_1$-representation in Table~\ref{tab:representations} as
\begin{align}
 Y^{IJ}_{KL} = \begin{cases}
(2\,d+\mathfrak{n})\,(P_{\hat{R}_2})^{IJ}{}_{KL} & (\Odd)
\\
2\,d\,(P_{\hat{R}_2})^{IJ}{}_{KL} & (\mathrm{E}_{d(d\leq 6)}) 
\\
 2\,d\,(P_{\hat{R}_2})^{IJ}{}_{KL} -28\,(P_{\irrep{1}})^{IJ}{}_{KL} & (\Edd[7])
\\
 2\,d\,(P_{\hat{R}_2})^{IJ}{}_{KL} -30\,(P_{\irrep{248}})^{IJ}{}_{KL} + 62\,(P_{\irrep{1}})^{IJ}{}_{KL} & (\Edd[8])
\end{cases},
\end{align}
where $d$ is the rank of the duality group $\GD$. 
The projector $(P_{\hat{R}_2})^{IJ}{}_{KL}$ to the $\hat{R}_2$-representation can be expressed as
\begin{align}
\begin{cases}
(2\,d+\mathfrak{n})\,(P_{\hat{R}_2})^{IJ}{}_{KL}= \eta^{IJ}\,\eta_{KL}\,, \qquad& (\Odd)
\\
 2\,d\,(P_{\hat{R}_2})^{IJ}{}_{KL} = \eta^{IJ;\II}\,\eta_{KL;\II}\,,\qquad& (\Edd)
\end{cases}
\end{align}
by using the so-called $\eta$-symbol $\eta_{IJ;\KK}$ \cite{Linch:2016ipx,Sakatani:2017xcn} (which reduces to $\eta^{IJ}$ in the $\Odd$ case because the $R_2$-representation is a singlet). 
They are the Clebsch--Gordan coefficients connecting the symmetric product of two $R_1$-representations and the $\hat{R}_2$-representation. In our convention, $\eta_{IJ;\KK}$ and $\eta^{IJ;\KK}$ are defined to have the same numerical components, and are normalized such that the following relations hold for the respective duality groups:
\begin{alignat}{3}
 \Odd &:&\quad \eta^{IJ}\,\eta_{IJ} &= 2\,d+\mathfrak{n}\,,&
 \eta^{IK}\,\eta_{KJ} &= \delta^I_J\,,
\\
 \Edd[d] &:&\quad \eta^{IJ;\II}\,\eta_{IJ;\JJ} &= 2\,(d-1) \,\delta^{\II}_{\JJ}\,,\qquad& 
 \eta^{IK;\LL}\,\eta_{KJ;\LL} &= \tfrac{2\,(d-1)\dim \hat{R}_2}{\dim R_1}\,\delta^I_J\,.
\end{alignat}
In the case of $\Edd[7]$, there is a well-known symplectic form $\omega_{IJ}=\omega_{[IJ]}$\,, satisfying
\begin{align}
 \omega_{IK}\,\omega^{KJ} = - \delta_I^J\,,
\end{align}
where $\omega_{IJ}$ and $\omega^{IJ}$ have the same components in our convention. With it, we can construct the projector $(P_{\irrep{1}})^{IJ}{}_{KL}$ in the $\Edd[7]$ case as
\begin{align}
 (P_{\irrep{1}})^{IJ}{}_{KL} = \tfrac{1}{56}\,\omega^{IJ}\,\omega_{KL}\,.
\end{align}
For $\Edd[8]$, the $R_1$-representation is isomorphic to $\irrep{adj}$, and there is an intertwiner $\chi_{I\adji}$ or $\chi^{I\adji}$ satisfying
\begin{align}
 \chi^{I\adji}\,\chi_{I\adjj} = \delta^{\adji}_{\adjj}\,,\qquad
 \chi^{I\adji}\,\chi_{J\adji} = \delta^I_J\,.
\end{align}
Again, $\chi_{I\adji}$ and $\chi^{I\adji}$ have the same components in our convention. They allow to convert the adjoint indices into the $R_1$ indices, for example by
\begin{align}
 f^{IJ}{}_K = \chi^{I\adji}\,\chi^{J\adjj}\,\chi_{K\adjk}\,f_{\adji\adjj}{}^{\adjk} = \chi^{I\adji}\,(t_{\adji})_K{}^J\,,\qquad
 \kappa^{IJ} = \chi^{I\adji}\,\chi^{I\adjj}\,\kappa_{\adji\adjj}\,.
\end{align}
Then $f^{IJ}{}_K$ and $\kappa^{IJ}$ have the same components as $f_{IJ}{}^K$ and $\kappa_{IJ}$\,, respectively, and satisfy
\begin{align}
 \kappa_{IJ} = -\tfrac{1}{\Calpha}\,f_{IK}{}^L\,f_{JL}{}^K\,,\qquad \kappa_{IK}\,\kappa^{KJ} = \delta_I^J\,.
\end{align}
Having defined these objects, we are able to express the $Y$-tensor for $\Edd[8]$ and the projectors it contains as
\begin{align}
 Y^{IJ}_{KL} &= 2\,\delta^{(I}_K\,\delta^{J)}_L + f_K{}^{JP}\,f_{PL}{}^I\,,
\\
 (P_{\irrep{248}})^{IJ}{}_{KL} &= \tfrac{1}{60}\,f^{IJ}{}_M\,f^M{}_{KL}\,,\qquad 
 (P_{\irrep{1}})^{IJ}{}_{KL} = \tfrac{1}{248}\,\kappa^{IJ}\,\kappa_{KL}\,.
\end{align}
Hence, it is possible to write the $Y$-tensor for each duality group listed in Table \ref{tab:representations} in a quadratic form as
\begin{align}
 Y^{IJ}_{KL} = \begin{cases}
 \eta^{IJ}\,\eta_{KL} & (\Odd) 
\\
 \eta^{IJ;\II}\,\eta_{KL;\II} & (\mathrm{E}_{d(d\leq 6)}) 
\\
 \eta^{IJ;\II}\,\eta_{KL;\II} -\tfrac{1}{2}\,\omega^{IJ}\,\omega_{KL} & (\Edd[7]) 
\\
 \eta^{IJ;\II}\,\eta_{KL;\II} -\tfrac{1}{2}\, f^{IJ}{}_M\,f^M{}_{KL} + \tfrac{1}{4}\,\kappa^{IJ}\,\kappa_{KL} & (\Edd[8])
\end{cases} \,.
\label{eq:Y-tensor2}
\end{align}
From \eqref{eq:Y-tensor2}, it follows that the section condition \eqref{eq:SC} now decomposes into
\begin{align}
\begin{cases}
 \eta^{IJ}\,\partial_I\otimes \partial_J=0 & (\Odd) 
\\
 \eta^{IJ;\II}\,\partial_I\otimes \partial_J=0 & (\mathrm{E}_{d(d\leq 6)}) 
\\
 \eta^{IJ;\II}\,\partial_I\otimes \partial_J=0\,,\quad \omega^{IJ}\,\partial_I\otimes \partial_J = 0 & (\Edd[7]) 
\\
 \eta^{IJ;\II}\,\partial_I\otimes \partial_J=0\,,\quad f^{IJ}{}_M\,\partial_I\otimes \partial_J=0\,,\quad \kappa^{IJ}\,\partial_I\otimes \partial_J=0 & (\Edd[8])
\end{cases} \,.
\end{align}

\subsection{Generalized Lie derivative}\label{sec:gen-Lie}
We now introduce the generalized Lie derivative, which generates infinitesimal generalized diffeomorphisms. The closure of its algebra is a crucial consistency requirement for the gauge structure. While this closure holds for most duality groups, it fails for $\Edd[8]$, necessitating a modification of the derivative itself.

For $\GD=\Odd$ or $\Edd[d]$ with $d\le 7$, the generalized Lie derivative is defined as
\begin{align}
 \gLie_{V} W^I \equiv V^J\,\partial_J W^I - W^J\,\partial_J V^I + Y^{IJ}_{KL}\,\partial_J V^K\, W^L + (w-\Cbeta)\,(\partial_K V^K)\, W^I \,,
\label{eq:gen-Lie-upto-E7}
\end{align}
where $w$ is the weight of $W^I$, equal to $\Cbeta$ for gauge parameters (and $-\Cbeta$ for $\partial_I$). For later convenience, we also define a derivative
\begin{align}
 L_{V} W^I \equiv V^J\,\partial_J W^I - W^J\,\partial_J V^I + Y^{IJ}_{KL}\,\partial_J V^K\, W^L \,,
\end{align}
where the last term in the generalized Lie derivative is dropped. Using the notation
\begin{align}
 \Delta_{V} \equiv \delta_V - \gLie_V\,,
\end{align}
the consistency condition, or the general covariance (i.e., the requirement that $\gLie_{V_2} V_3^I$ behaves as a generalized vector field), can be expressed as
\begin{align}
 \Delta_{123}{}^I \equiv \Delta_{V_1} \bigl(\gLie_{V_2} V_3^I\bigr)
 = \gLie_{\gLie_{V_1}V_2} V_3^I - \bigl[\gLie_{V_1},\, \gLie_{V_2}\bigr] V_3^I =0 \,.
\label{eq:diffeo-consistency}
\end{align}
As studied in detail in \cite{Berman:2012vc}, $\Delta_{123}{}^I$ contains terms involving $\partial_J V_s\,\partial_K V_t$ ($s,t=1,2,3$) and terms involving $\partial_J\partial_K V_1$\,. 
The former vanishes when the section condition \eqref{eq:SC} and certain algebraic identities are applied, leading to
\begin{align}\label{eq:diffeo-consistency-upto-E7}
 \Delta_{123}{}^I = \bigl(Y^{I(K|}_{LJ}\,Y^{L|M)}_{NP} - Y^{I(K}_{NP}\,\delta^{M)}_J\bigr)\,\partial_K\partial_MV_1^N\,V_2^P\,V_3^J\,.
\end{align}
This trivially vanishes for $\GD=\Odd$. For $\GD=\Edd[d]$ with $d\leq 7$ a bit more work is required. However, using the identity
\begin{align}
 Y^{I(K}_{LJ}\,Y^{|L|M)}_{NP} - Y^{I(K}_{NP}\,\delta^{M)}_J
 &= 
 Y^{(KM)}_{L(N}\,Y^{IL}_{P)J}
 - Y^{(KM)}_{J(N}\,\delta^I_{P)}
 + \tfrac{1}{2}\,\omega^{IL}\,Y^{(KM)}_{LJ}\,\omega_{NP} \,,
\label{eq:YY-id-7}
\end{align}
where $\omega^{IJ}=0=\omega_{IJ}$ for $d\leq 6$, we eventually find that $\Delta_{123}{}^I=0$ under the section condition. 

\subsubsection*{Generalized Lie derivative in the $\Edd[8]$ case}

\paragraph{General discussion:}
In the case of $\Edd[8]$, the consistency check fails if one assumes \eqref{eq:gen-Lie-upto-E7}. A direct calculation \cite{Berman:2012vc} shows that, even after imposing the section condition, the commutator does not close:
\begin{align}
 \Delta_{123}{}^I = f^{IK}{}_J\,f^{M}{}_{NP}\,\partial_K\partial_MV_1^N\,V_2^P\,V_3^J \neq 0\,.
\end{align}
This indicates that the generalized Lie derivative must be modified. The inconsistency is resolved by promoting the gauge parameter from a vector $V^I$ to a pair $(V^I,\, \Sigma_I)$, where\begin{align}
 \Sigma^I{}_J \equiv \Sigma^{\adji}\,(t_{\adji})_J{}^I
\end{align}
is a new field in the adjoint representation of the duality group \cite{Hohm:2014fxa}. Accordingly, the generalized Lie derivative is extended to
\begin{align}\label{eq:gen-Lie-E8}
 \gLie_{(V,\Sigma)} W^I \equiv \gLie_{V} W^I + \Sigma^I{}_J\,W^J\,.
\end{align}
From this definition, it is clear that $\Sigma^I{}_J$ should have zero weight ($w=0$). However, $\Sigma^I{}_J$ is not covariant under the generalized Lie derivative. In addition, its first index is constrained such that
\begin{align}
 \Sigma^I{}_J\,\partial_I = 0
\end{align}
holds under the section condition. Now, the consistency condition \eqref{eq:diffeo-consistency-upto-E7} can be expressed as
\begin{align}
 \widehat{\Delta}_{123}{}^I &\equiv \Delta_{(V_1,\Sigma_1)} \bigl(\gLie_{(V_2,\Sigma_2)} V_3^I\bigr)
\nn\\
 &= \gLie_{(\delta_{(V_1,\Sigma_1)}V_2,\delta_{(V_1,\Sigma_1)}\Sigma_2)} V_3^I - \bigl[\gLie_{(V_1,\Sigma_1)},\, \gLie_{(V_2,\Sigma_2)}\bigr] V_3^I =0 \,.
\label{eq:diffeo-consistencyE8}
\end{align}
Here, the first part of the extended parameter $V_2$ is still covariant, implying $\delta_{(V_1,\Sigma_1)}V_2=\gLie_{(V_1,\Sigma_1)}V_2$. However $\Sigma$ is not a covariant object and $\delta_{(V_1,\Sigma_1)}\Sigma_2$ is determined such that the consistency condition \eqref{eq:diffeo-consistencyE8} is satisfied under the section condition. The result is that
\begin{align}
 \bigl(\delta_{(V_1,\Sigma_1)}\Sigma_2\bigr)^I{}_J &= 2\,V_{[1}^K\,\partial_K (\Sigma_{2]})^I{}_J 
 + 2\,Z^{KL}_{MJ}\,(\Sigma_{[1})^I{}_K\,\partial_L V_{2]}^M
 - 2\,Z^{IK}_{LM}\,(\Sigma_{[1})^M{}_J\,\partial_K V_{2]}^L
 - \Delta_{12}{}^I{}_J
\nn\\
 &\quad + [\Sigma_1,\,\Sigma_2]^I{}_J
 - Z^{IK}_{LJ}\,\partial_K \bigl[(\Sigma_1)^L{}_M\,V_2^M\bigr]
\nn\\
 &= L_{V_1}(\Sigma_2)^I{}_J - L_{V_2}(\Sigma_1)^I{}_J
 + [\Sigma_1,\,\Sigma_2]^I{}_J - \Delta_{12}{}^I{}_J
\nn\\
 &\quad - Z^{IK}_{LJ}\,\partial_K \bigl[(\Sigma_1)^L{}_M\,V_2^M\bigr]\,,
\label{eq:delta-Sigma}
\end{align}
where
\begin{align}
\begin{split}
 Z^{IJ}_{KL} &\equiv Y^{IJ}_{KL} - \delta^I_K\,\delta^J_L\,,\qquad
 \gLie_{V}=\gLie_{(V,0)}\,, \qquad\text{and}
\\
 \Delta_{12}{}^I{}_J&\equiv (Y^{I(K|}_{LJ}\,Y^{L|M)}_{NP} - Y^{I(K}_{NP}\,\delta^{M)}_J)\,\partial_K\partial_MV_1^N\,V_2^P\,.
\end{split}
\end{align}
Taking into account the property
\begin{align}
 \gLie_{(\Sigma\,W,0)} V^I = Z^{IK}_{LJ}\,\partial_K \bigl[\Sigma^L{}_M\,W^M\bigr]\, V^L = \gLie_{(0, Z^{\bullet K}_{L\bullet}\,\partial_K [(\Sigma\,W)^L])} V^I\,,
\end{align}
of the extended generalized Lie derivative in \eqref{eq:gen-Lie-E8}, and assuming that $[\Sigma_1,\,\Sigma_2]=0$, we can rewrite $\widehat{\Delta}_{123}{}^I$ as
\begin{align}
 \widehat{\Delta}_{123}{}^I = \gLie_{(V_{12},\Sigma_{12})} V_3^I - \bigl[\gLie_{(V_1,\Sigma_1)},\, \gLie_{(V_2,\Sigma_2)}\bigr] V_3^I =0 \,,
\end{align}
where
\begin{align}
 V_{12} &\equiv \gLie_{V_1}V_2\,,
\\
 (\Sigma_{12})^I{}_J &\equiv L_{V_1}(\Sigma_2)^I{}_J - L_{V_2}(\Sigma_1)^I{}_J - \Delta_{12}{}^I{}_J \,,
\end{align}
and $\Delta_{12}{}^I{}_J$ is defined through $\Delta_{123}{}^I=\Delta_{12}{}^I{}_J\,V_3{}^J$. 
In particular, the antisymmetric part $\widehat{\Delta}_{[12]3}{}^I=0$ corresponds to the closure of the gauge algebra
\begin{align}
 \bigl[\delta_{(V_1,\Sigma_1)},\, \delta_{(V_2,\Sigma_2)}\bigr] = \delta_{(V_{[12]},\Sigma_{[12]})} \,,
\end{align}
and the expression for $\Sigma_{[12]}$ coincides with the one given in \cite{Hohm:2014fxa}. 

\paragraph{An explicit construction of $\Sigma$:}
While the introduction of $\Sigma$ restores gauge consistency, its origin remains somewhat formal. Here, we revisit the proposal of \cite{Rosabal:2014rga,Cederwall:2015ica}, which demonstrates that $\Sigma$ is not an independent field but can be explicitly constructed from the primary gauge parameter $V^I$ together with a geometric object known as the Weitzenb\"ock connection (also known as the flat connection). This elucidates the geometric underpinning of the extended gauge structure.

Assuming that the extended space admits globally defined generalized frames $E_A{}^I$, we define the associated Weitzenb\"ock connection as
\begin{align}
 W_{IJ}{}^K \equiv \partial_I E_J{}^A\,E_A{}^K\,.
\end{align}
It is straightforward to check that the Weitzenb\"ock connection transforms as
\begin{align}
 \delta_1 W_{IJ}{}^K&=\partial_I (\gLie_{(V_1,\,\widetilde{\Sigma}_1)}E_J{}^A)\,E_A{}^K
 + \partial_I E_J{}^A\,\gLie_{(V_1,\,\widetilde{\Sigma}_1)}E_A{}^K
\nn\\
 &= L_{(V_1,\,\widetilde{\Sigma}_1)} W_{IJ}{}^K + \bigl(\partial_I\partial_J V_1^K - Y^{KM}_{NJ}\,\partial_I\partial_M V_1^N\bigr) - f^{KM}{}_J \,\partial_I \bigl( V_1^N\,W_{MN} \bigr)\,,
\end{align}
by noting that the first index of $W_{IJ}{}^K$ is the derivative index and using the section condition. Looking at a particular component in the expansion
\begin{align}
 W_{IJ}{}^K = W_{I}{}^{\adji}\,(t_{\adji})_J{}^K + W_{I}{}^0 \,(t_0)_J{}^K
 = W_{IL} \,f^{LK}{}_J + W_{I}{}^0 \,(t_0)_J{}^K\,,
\end{align}
where $W_{IJ}\equiv \chi_{J\adji}\,W_{I}{}^{\adji}$ and $(t_0)_{I}{}^{J}=-\delta_I^J$, we find that $W_{IJ}$ transforms as
\begin{align}
 \delta_1 W_{IJ} = L_{(V_1,\,\widetilde{\Sigma}_1)} W_{IJ} + f^L{}_{JK} \,\partial_I\,\partial_L V_1^K + \partial_I \bigl( V_1^K\,W_{JK} \bigr)\,.
\end{align}
We then define
\begin{align}
 \widetilde{\Sigma}_s{}^I{}_J \equiv \chi^{I\adji}\,\chi_{\adjj L}W_{K}{}^{\adjj}\,(t_{\adji})_J{}^K\,V_s^L
 = f^{IK}{}_J\,W_{KL}\,V_s^L\,.
\label{eq:tilde-Sigma}
\end{align}
Under the section condition, we observe that it satisfies
\begin{align}
 \widetilde{\Sigma}_s{}^I{}_J\,\partial_I = f^{IK}{}_J\,W_{KL}\,V_s^L\,\partial_I = 0\,,
\end{align}
and
\begin{align}
 [\widetilde{\Sigma}_1,\,\widetilde{\Sigma}_2]^I{}_J = - f^{KL}{}_P\, f^{PI}{}_J\,W_{KM}\,W_{LN}\,V_1^M\,V_2^N = 0\,.
\end{align}
Moreover, the transformation behavior of $\widetilde{\Sigma}^I{}_J$,
\begin{align}
 (\delta_1\widetilde{\Sigma}_2)^I{}_J
 = f^{IK}{}_J\, W_{KL}\,\delta_{(V_1,\widetilde{\Sigma}_1)} V_2^L + f^{IK}{}_J\,\delta_1W_{KL}\,V_2^L\,,
\end{align}
matches \eqref{eq:delta-Sigma} when rewritten as
\begin{align}
 (\delta_1\widetilde{\Sigma}_2)^I{}_J = L_{V_1}(\widetilde{\Sigma}_2)^I{}_J - L_{V_2}(\widetilde{\Sigma}_1)^I{}_J - \Delta_{12}{}^I{}_J
 - Z^{IK}_{LJ}\,\partial_K \bigl[(\widetilde{\Sigma}_1)^L{}_M\,V_2^M\bigr] \,.
\end{align}
Hence, $\widetilde{\Sigma}$ exhibits the expected transformation behavior and can be consistently used as the gauge parameter $\Sigma$ in the generalized Lie derivative $\gLie_{(V,\Sigma)}$. 

In the presence of a globally defined frame, such as in generalized parallelizable spaces, we use the gauge parameter $\widetilde{\Sigma}^I{}_J$ to define the generalized fluxes $X_{AB}{}^C$. In more general contexts, we work with a formal gauge parameter $\Sigma^I{}_J$ for the generalized Lie derivative.

\subsection{Generalized fluxes}\label{sec:fluxes}
With the consistent generalized Lie derivative at hand, we can now introduce one of the central objects of this work: the generalized fluxes $X_{AB}{}^C$ associated with the generalized frame field $E_A{}^I$. For $\GD=\Odd$ or $\Edd[d]$ with $d\le 7$, they are defined as
\begin{align}
 \gLie_{E_A} E_B{}^I = -X_{AB}{}^C\,E_C{}^I\,.
\end{align}
For the $\Edd[8]$ case, we take \eqref{eq:tilde-Sigma} to define
\begin{align}
 \gLie_{(E_A,\widetilde{\Sigma}_A)} E_B{}^I = \gLie_{E_A} E_B{}^I + \chi^{I\adji}\,\chi_{\adjj L}W_{K}{}^{\adjj}\,(t_{\adji})_J{}^K\,E_A{}^L\,E_B{}^J = -X_{AB}{}^C\,E_C{}^I
\end{align}
by following \cite{Rosabal:2014rga}. This modification is in accordance with the modification of the generalized Lie derivative explained in the previous subsection. From these definitions, the generalized fluxes for $d\leq 8$ can be expressed as
\begin{align}
 X_{AB}{}^C = \Theta_A{}^{\adja}\,(t_{\adja})_{B}{}^C - \bigl[ \tfrac{1}{1+\Cbeta}\,(t^{\adja})_A{}^D\,(t_{\adja})_{B}{}^C + \delta_A^D \,\delta_B^C\bigr]\, \vartheta_D\,,
\label{eq:X-decomp}
\end{align}
where
\begin{align}
 &\Theta_A\equiv \mathbb{P}_{A}{}^{\adja B}{}_{\adjb}\,W_B{}^{\adjb}\, t_{\adja} \,,\qquad
 \vartheta_A \equiv (1+\Cbeta)\,W_A{}^0 - \Cbeta\,W_{D}{}^{\adja}\,(t_{\adja})_{A}{}^D\,,
\\
 &\mathbb{P}_{A}{}^{\adja B}{}_{\adjb} \equiv \delta_A^B\,\delta^{\adja}_{\adjb} + (t_{\adjb}\,t^{\adja})_{A}{}^B - \tfrac{\Cbeta}{1+\Cbeta}\,(t^{\adja}\,t_{\adjb})_A{}^B + \chi^{B\adja}\,\chi_{\adjb A}\,,
\end{align}
and $\chi^{A\adja}=0=\chi_{\adja A}$ for $d\leq 7$. We note that $\mathbb{P}_{A}{}^{\adja B}{}_{\adjb}$ is a sum of projectors to irreducible representations associated with the embedding tensor (see for example (2.55) of \cite{Sakatani:2020wah} for details). The expression \eqref{eq:X-decomp} shows that the matrix
\begin{align}
 (X_A)_{B}{}^C \equiv X_{AB}{}^C\,,
\end{align}
is spanned by the generators $t_{\adja}$ of the duality algebra $\mathrm{Lie}(\GD)$ and an additional generator $t_0$ of a scale symmetry. In the DFT case, where $\Cbeta=0$ and $W_A{}^0=0$, the generator $t_0$ does not appear. 

In $\Edd[8]$ ExFT, $X_A$ receives an additional contribution from the term $W_{D}{}^{\adjb}\,\chi_{\adjb A}\, \chi^{D\adja}\,t_{\adja}$. By explicitly computing $\chi^{D\adja}$ in the M-theory/type IIB section, we find that $\chi^{D\adja}\,t_{\adja}$ contains the $\Edd[8]$ generators $R^{i_1\cdots i_8,i'}$/$R^{m_1\cdots m_7,m'}$ that have mixed symmetry (and are associated with the dual graviton). If we consider $d\geq 9$, additional extra contributions arise. The required modifications of generalized Lie derivatives beyond the $\Edd[8]$ case have been explored in \cite{Bossard:2018utw,Bossard:2021jix,Bossard:2021ebg}. However we are not aware of a general expression beyond $d\leq 8$. Hence, in this work, we focus on the unmodified part and only discuss possible modifications indirectly.

\section{Riemann tensor in GR}\label{sec:GL}
In a recent work by the authors \cite{Hassler:2023axp}, the approach of Polacek and Siegel \cite{Polacek:2013nla} (see also \cite{Butter:2022iza}) was extended to construct a hierarchy of curvatures in exceptional geometry. To highlight the core methodology of the present work, we now apply our formalism to the familiar setting of Riemannian geometry, where the duality group is simply $\GD=\GL(d)$. We show that, by introducing an auxiliary ``mega-space,'' the fundamental geometric tensors, namely torsion and Riemann curvature, appear together as components of a single algebraic object, the generalized fluxes. This seemingly circuitous route provides a powerful framework that, in contrast to traditional methods, generalizes naturally to the exceptional geometries explored in subsequent sections. 

\subsection{Our approach}\label{sec:GL-approach}
In section \ref{sec:ExFT}, we considered the cases $\GD=\Odd$ and $\GD=\Edd[d]$. However, most of the steps in our method for constructing covariant curvatures can already be understood in the simple case $\GD=\GL(d)$. Here, the generators of $\mathrm{Lie}(\GD)$ are
\begin{align}
 t_{\adja} = K^a{}_b \quad (a,b=1,\dotsc,d)\,,
\end{align}
which satisfy the commutation relation
\begin{align}\label{eq:GL-algebra}
 [K^a{}_b,\,K^c{}_d] = \delta^c_b\,K^a{}_d - \delta^a_d\,K^c{}_b\,. 
\end{align}
The $R_1$-representation is taken to be the $d$-dimensional vector representation with matrix representation
\begin{align}
 (K^c{}_d)_a{}^b = \delta_a^c\,\delta_d^b\,. 
\label{eq:K-GL}
\end{align}
Here we take $\Calpha=1$ and $\Cbeta=0$ and then find $t^{\adja} = K_a{}^b \equiv -K^b{}_a$\,. Following the narrative of the last section, we express the standard Lie derivative in terms of the $Y$-tensor as
\begin{align}
 Y^{ab}_{cd} = \delta^a_c\,\delta^b_d + (K_e{}^f)_d{}^a\, (K^e{}_f)_c{}^b = 0\,.
\end{align}
In this case, the section condition becomes trivial and the generalized Lie derivative reduces to the expected form
\begin{align}
 \gLie_{v} w^I = v^J\,\partial_J w^I - w^J\,\partial_J v^I \,.
\end{align}
Similarly, the generalized fluxes reduce to the ordinary geometric fluxes
\begin{align}
 \gLie_{e_a} e_b = -X_{ab}{}^c\,e_c \,,
\label{eq:geometric-fluxes}
\end{align}
which are antisymmetric in their first two indices, $X_{ab}{}^c = -X_{ba}{}^c$, due to the antisymmetry of the ordinary Lie derivative, $\gLie_{e_a} e_b = - \gLie_{e_b} e_a$\,. 

\subsubsection{Setup}\label{sec:setup}
To obtain Riemannian geometry, we need a non-degenerate metric $g_{ij}$. In the language of frames, it can be defined by an invertible, constant matrix $g^{ab}$, the flat metric, such that it gives rise to the inverse metric after contraction with the frame: $g^{ij} = g^{ab}\,e_a{}^i\,e_b{}^j$. As the metric should exclusively contain all information about the geometry, the frame with $d^2$ components is redundant. To resolve this, an additional local symmetry is introduced to remove the redundant degrees of freedom. This symmetry is captured by the stabilizer group $H=\OO(d)$ of the flat metric $g^{ab}$. The structure group $\GS$ can be either the full $\GS=H$ or, in the case of reduced holonomy, only a subgroup of it.

Our main object of study is the $d$-dimensional physical manifold $M$. To analyze its geometry in a covariant manner, we first embed it within a larger, auxiliary manifold $\hat{M}$ of dimension $p=d+n$, which we refer to as the mega-space. This construction, a $p$-dimensional smooth and paracompact manifold, serves to geometrize the structure group $\GS$ (of dimension $n$), treating its degrees of freedom on an equal footing with the physical coordinates. As we shall see, this unification allows for a manifestly covariant treatment of geometric quantities.

\paragraph{Assumption 1.}
We assume that the Lie group $\GS$ acts freely and properly from the left on $\hat{M}$. The left action is generated by the right-invariant vector fields $\hat{e}_{\exA}$ ($\exA=1,\dotsc,n$) satisfying
\begin{align}
 \gLie_{\hat{e}_{\exA}} \hat{e}_{\exB} = - \hat{X}_{\exA\exB}{}^{\exC}\,\hat{e}_{\exC} \,,
\end{align}
where $\hat{X}_{\exA\exB}{}^{\exC}=-\hat{X}_{\exB\exA}{}^{\exC}$ are the structure constants of $\mathrm{Lie}(\GS)$. This assumption establishes $\hat{M}$ as a principal $\GS$-bundle over the physical space $M\equiv \GS\backslash \hat{M}$. Introducing local coordinates $(y^{\hat{i}})=(y^\mu,\,y^i)$ ($\hat{i}=1,\dotsc,p$, $\mu=1,\dotsc,n$, $i=n+1,\dotsc,p$) on $\hat{M}$ such that $y^\mu$ and $y^i$ are coordinates on $\GS$ and $M$, respectively, we express the right-invariant vector fields as
\begin{align}
 \hat{e}_{\exA}=\hat{e}_{\exA}{}^\nu(y^\mu)\,\partial_\nu\,.
\label{eq:hat-e-exA}
\end{align}
Using the right-invariant vector fields $\hat{e}_{\exA}$, we construct the frame field $\hat{e}_{\hat{a}} \equiv (\hat{e}_{\exA},\,\hat{e}_{a})$ ($\hat{a}=1,\dotsc,p$, $a=n+1,\dotsc,p$) on the mega-space such that the matrix
\begin{align}
 \hat{e}_{\hat{a}}{}^{\hat{i}} \equiv \begin{pmatrix}
 \hat{e}_{\exA}{}^{\hat{i}} \\
 \hat{e}_{a}{}^{\hat{i}}
\end{pmatrix}= \begin{pmatrix}
 \hat{e}_{\exA}{}^{\mu} & 0 \\
 \hat{e}_{a}{}^{\mu} & \hat{e}_{a}{}^{i}
\end{pmatrix}
\label{eq:hat-e-gauge}
\end{align}
is everywhere non-degenerate. 

\paragraph{Assumption 2.}
We assume that under the infinitesimal left action of $\GS$, generated by $\gLie_{\hat{e}_{\exA}}$, the frame field $\hat{e}_{\hat{a}}$ transforms linearly with a constant matrix
\begin{align}
 \gLie_{\hat{e}_{\exA}} \hat{e}_{\hat{b}}{}^{\hat{i}} = (\mathfrak{t}_{\exA})_{\hat{b}}{}^{\hat{c}}\,\hat{e}_{\hat{c}}{}^{\hat{i}}\,.
\label{eq:second-prop}
\end{align}
This crucial assumption identifies an infinitesimal diffeomorphism along a fiber direction with a purely algebraic operation on the frame, namely a gauge transformation of the structure group $\GS$ represented by the constant matrix $\mathfrak{t}_{\exA}$. In other words, the action of $\gLie_{\hat{e}_{\exA}}$ is assumed to rotate the flat indices of the frame field by $(\mathfrak{t}_{\exA})_{\hat{b}}{}^{\hat{c}}$.

In terms of the geometric fluxes \eqref{eq:geometric-fluxes}, this assumption means that the $(\hat{X}_{\exA})_{\hat{b}}{}^{\hat{c}}=-(\mathfrak{t}_{\exA})_{\hat{b}}{}^{\hat{c}}$ part of the flux is constant. Since the left action $\delta_{\exA}=\gLie_{\hat{e}_{\exA}}$ satisfies the Lie algebra of $\GS$,
\begin{align}
 [\delta_{\exA},\,\delta_{\exB}] = - \hat{X}_{\exA\exB}{}^{\exC}\,\delta_{\exC}\,,
\end{align}
the constant matrices $(\mathfrak{t}_{\exA})_{\hat{b}}{}^{\hat{c}}$ are required to obey the same commutation relations
\begin{align}\label{eq:t-commutator}
 [\mathfrak{t}_{\exA},\,\mathfrak{t}_{\exB}] = \hat{X}_{\exA\exB}{}^{\exC}\,\mathfrak{t}_{\exC}\,.
\end{align}
Namely, $(\mathfrak{t}_{\exA})_{\hat{b}}{}^{\hat{c}} = - \hat{X}_{\exA \hat{b}}{}^{\hat{c}}$ plays the role of a representation matrix for the generators of $\mathrm{Lie}(\GS)$. In our setup this generally takes the form
\begin{align}
 (\mathfrak{t}_{\exA})_{\hat{b}}{}^{\hat{c}} 
 = - \begin{pmatrix} \hat{X}_{\exA \exB}{}^{\exC} & 0 \\
 \hat{X}_{\exA b}{}^{\exC} & \hat{X}_{\exA b}{}^{c} 
\end{pmatrix},
\label{eq:t_exA}
\end{align}
because the lower-triangular form of $\hat{e}_{\exA}$ in \eqref{eq:hat-e-gauge} should be preserved. The submatrices $(\mathfrak{t}_{\exA})_{b}{}^{c} = -\hat{X}_{\exA b}{}^{c}$ can also be regarded as representation matrices for generators of $\mathrm{Lie}(\GS)$. In this paper, we always assume that $(\mathfrak{t}_{\exA})_{b}{}^{c}$ provides a faithful representation.

\paragraph{Assumption 3.}
So far, we have retained the component $\hat{X}_{\exA b}{}^{\exC}$ for generality, but it is truncated later as we discuss below. This is our final assumption:
\begin{align}
 \hat{X}_{\exA b}{}^{\exC} = 0\,.
\label{eq:ssumption-3}
\end{align}
This technical assumption corresponds to the condition for a reductive coset space, which ensures that the Lie algebra of the total group $G$ splits cleanly into the subalgebra generating $\GS$ and its complementary part. This condition is not overly restrictive: as noted in \cite{Castellani:1999fz} (see also p.~251 of Ref.[11] therein), whenever $\GS$ is compact or semisimple, one can choose the horizontal basis $\hat{e}_{a}$ such that condition \eqref{eq:ssumption-3} is satisfied. As we will see in section~\ref{sec:restricted-diffeos}, this assumption allows us to consistently define the spin connection on the physical manifold $M$.

\subsubsection{Frames and spin connection on \texorpdfstring{$M$}{M}}
For later convenience, we introduce a matrix
\begin{align}
 \widetilde{m}_{\hat{a}}{}^{\hat{b}}(y^\mu)\equiv \exp\bigl(y^\mu\, \mathfrak{t}_{\mu}\bigr)_{\hat{a}}{}^{\hat{b}} = \begin{pmatrix}
 \widetilde{m}_{\exA}{}^{\exB} & 0 \\
 \widetilde{m}_{a}{}^{\exB} & \widetilde{m}_{a}{}^{b} \end{pmatrix}\in \GS\quad \bigl(\mathfrak{t}_{\mu}\equiv \delta_\mu^{\exA}\,\mathfrak{t}_{\exA}\bigr)\,,
\end{align}
satisfying
\begin{align}
 \widetilde{m}^{-1}\,\widetilde{v}_{\exA}{}^{\mu}\,\partial_\mu \widetilde{m} = \mathfrak{t}_{\exA} \,,\qquad 
 \widetilde{m}_{\exA}{}^{\exB}\, \widetilde{v}_{\exB}{}^{\mu} = \hat{e}_{\exA}{}^{\mu}\,,
\end{align}
where $\widetilde{v}_{\exA}{}^{\mu}=\widetilde{v}_{\exA}{}^{\mu}(y^\mu)$ are the left-invariant vector fields on $\GS$, and $\widetilde{m}_{a}{}^{\exB}=0$ when $\hat{X}_{\exA b}{}^{\exC}=0$. We then parameterize $\hat{e}_{\hat{a}}{}^{\hat{i}}$ as
\begin{align}
 \hat{e}_{\hat{a}}{}^{\hat{i}} = \widetilde{m}_{\hat{a}}{}^{\hat{b}}\, e_{\hat{b}}{}^{\hat{i}}\,,\qquad
 e_{\hat{a}}{}^{\hat{i}} \equiv n_{\hat{a}}{}^{\hat{b}}\,v_{\hat{b}}{}^{\hat{i}}\,,
\label{eq:hate}
\end{align}
where
\begin{align}
 n_{\hat{a}}{}^{\hat{b}} \equiv \begin{pmatrix}
 \delta_{\exA}^{\exB} & 0 \\
 -\Omega_{a}^{\exB} & \delta_a^b
\end{pmatrix},\qquad
 v_{\hat{a}}{}^{\hat{i}} \equiv \begin{pmatrix}
 \widetilde{v}_{\exA}{}^{\mu} & 0 \\
 0 & e_a{}^i
\end{pmatrix}.
\label{eq:n-v-param}
\end{align}
Remarkably, under Assumption 1 and our choice of local coordinates $y^{\hat{i}}$ that realize \eqref{eq:hat-e-gauge}, eq.~\eqref{eq:second-prop} of Assumption 2 implies
\begin{align}
 \partial_\mu e_a{}^i=0\,,\qquad
 \partial_\mu \Omega_{a}^{\exB} = 0\,.
\end{align}
This shows that $e_a{}^i(y^i)$ and $\Omega_{a}^{\exB}(y^i)$ are fields on the physical space $M$, and the mega-space $\hat{M}$ admits a clean factorization into the gauge orbits $\GS$ and the physical space $M$.

\subsection{Restricted diffeomorphisms}\label{sec:restricted-diffeos}
Now let us consider diffeomorphisms on the mega-space $\hat{M}$. If we perform a general diffeomorphism, our gauge \eqref{eq:hat-e-gauge} is broken, and the distinction between the $y^\mu$-directions and the physical space becomes opaque. We thus consider a restricted class of diffeomorphisms generated by the diffeomorphism parameters
\begin{align}
 \hat{\xi} = \hat{\xi}^{\hat{a}}\,\hat{e}_{\hat{a}} 
 = \xi^{\exA}(y^i) \,\widetilde{v}_{\exA}
 + \xi^{a}(y^i) \,e_a \qquad \bigl(\widetilde{v}_{\exA} \equiv \widetilde{v}_{\exA}{}^{\mu}\,\partial_{\mu}\,,\quad e_a\equiv e_{a}{}^i\,\partial_i\bigr) \,,
\end{align}
where
\begin{align}
 \hat{\xi}^{\hat{a}} \equiv \xi^{\hat{c}}(y^i)\,(n^{-1})_{\hat{c}}{}^{\hat{b}}\,(\widetilde{m}^{-1})_{\hat{b}}{}^{\hat{a}}\,.
\end{align}
We then find
\begin{align}
 \delta_{\hat{\xi}} \hat{e}_{\hat{a}}
 &= \widetilde{m}_{\hat{a}}{}^{\hat{b}}\,\delta_{\hat{\xi}} e_{\hat{b}}\,,
\\
 \delta_{\hat{\xi}} e_{\hat{a}}{}^{\hat{i}}
 &= 
 \begin{pmatrix}
 0 & 0 \\ 
 -\bigl[\xi^b D_b\Omega_a^{\exA} + D_{a}\xi^{\exA} + \xi_a{}^b\,\Omega_b^{\exA} 
 + \xi^{\exB}\,(\Omega_a^{\exC}\,\hat{X}_{\exB\exC}{}^{\exA} + \hat{X}_{\exB a}{}^{\exA})\bigr]\,v_{\exA}{}^\mu 
& \mathbb{L}_{\hat{\xi}} e_{a}{}^i 
\end{pmatrix},
\end{align}
where $D_a \equiv e_{a}{}^{i}\,\partial_i$, $\xi\equiv \xi^a\,e_a$, $\xi_{\hat{a}}{}^{\hat{b}} \equiv \xi^{\exA} \,(\mathfrak{t}_{\exA})_{\hat{a}}{}^{\hat{b}}$, and
\begin{align}
 \mathbb{L}_{\hat{\xi}} e_{a}{}^i \equiv \gLie_{\xi} e_{a}{}^i + \xi_a{}^b\,e_b{}^i\,.
\end{align}
In general, we define $\mathbb{L}_{\hat{\xi}}\equiv \gLie_{\xi} + \xi^{\exA} \rho(\mathfrak{t}_{\exA})$ as the covariant transformation, where $\rho$ represents a suitable representation matrix of $\GS$. Defining a $\mathrm{Lie}(\GS)$-valued 1-form field
\begin{align}
 \Omega_{a\hat{b}}{}^{\hat{c}} \equiv \Omega_a^{\exA}\,(\mathfrak{t}_{\exA})_{\hat{b}}{}^{\hat{c}}\,,
\end{align}
and recalling the parameterization \eqref{eq:n-v-param}, we find that this gauge transformation corresponds to
\begin{align}
 \delta_{\hat{\xi}} e_{a}{}^i &= \mathbb{L}_{\hat{\xi}} e_{a}{}^i \,,\qquad 
 \delta_{\hat{\xi}} \widetilde{m}_{\hat{a}}{}^{\hat{b}} = 0\,,\qquad
 \delta_{\hat{\xi}} \widetilde{v}_{\exA}{}^\mu = 0\,, 
\\
 \delta_{\hat{\xi}} \Omega_{ab}{}^c &= \xi^d D_d\Omega_{ab}{}^c + D_{a}\xi_{b}{}^c + [\xi,\,\Omega_a]_{b}{}^c + \xi_a{}^d\,\Omega_{db}{}^c + \xi^{\exC}\,\hat{X}_{\exC a}{}^{\exA}\,(\mathfrak{t}_{\exA})_b{}^c
\nn\\
 &= \mathbb{L}_{\hat{\xi}} \Omega_{ab}{}^c + D_{a}\xi_{b}{}^c + \xi^{\exC}\,\hat{X}_{\exC a}{}^{\exA}\,(\mathfrak{t}_{\exA})_b{}^c\,.
\end{align}
Using assumption \eqref{eq:ssumption-3} to set $\hat{X}_{\exC a}{}^{\exA}=0$, we obtain
\begin{align}
 \delta_{\hat{\xi}} \Omega_{ab}{}^c = \mathbb{L}_{\hat{\xi}} \Omega_{ab}{}^c + D_{a}\xi_{b}{}^c \,,
\end{align}
which reproduces the standard transformation rule of the spin connection $\Omega_{ab}{}^c$. Here, $\xi^a$ is interpreted as the diffeomorphism parameter on the physical space, while $\xi_a{}^b$ corresponds to the parameter of local $\GS$ frame transformations. In general, one may keep $\hat{X}_{\exC a}{}^{\exA}$ non-vanishing, as in \cite{Hassler:2024yis}. However, in this paper we impose \eqref{eq:ssumption-3} throughout, so that $\Omega_{ab}{}^c$ can be identified with the spin connection. Then \eqref{eq:second-prop} decomposes into
\begin{align}
 \gLie_{\hat{e}_{\exA}} \hat{e}_{\exB}{}^{\hat{i}} = -\hat{X}_{\exA\exB}{}^{\exC}\,\hat{e}_{\exC}{}^{\hat{i}}\,,\qquad \text{and} \qquad
 \gLie_{\hat{e}_{\exA}} \hat{e}_{b}{}^{\hat{i}} = (\mathfrak{t}_{\exA})_{b}{}^{c}\,\hat{e}_{c}{}^{\hat{i}}\,,
\end{align}
and the frames $\hat{e}_{\exB}$ and $\hat{e}_{b}$ do not mix under the left action of $\GS$.

We note that the derivative $\mathbb{L}_{\hat{\xi}}$ forms a closed algebra
\begin{align}
 [\mathbb{L}_{\hat{\xi}_1},\, \mathbb{L}_{\hat{\xi}_2}] e_a{}^i
 = \mathbb{L}_{\hat{\xi}_{12}} e_a{}^i\,,\qquad
 \hat{\xi}_{12}
 \equiv \xi_{12}^{\exA} \,\widetilde{v}_{\exA} + \xi^{a}_{12} \,v_a\,,
\end{align}
where
\begin{align}
 \xi_{12}^{\exA} \equiv \xi_1\cdot \xi_2^{\exA} - \xi_2\cdot \xi_1^{\exA} + \xi_1^{\exB}\,\xi_2^{\exC}\, \hat{X}_{\exB\exC}{}^{\exA} \,,\qquad 
 \xi^{a}_{12} \equiv \gLie_{\xi_1} \xi_2^a
\end{align}
with $\xi_1\cdot \xi_2^\alpha = \xi_1^{a}\,D_a \xi_2^\alpha$. In the following, when a quantity transforms as $\delta_{\hat{\xi}} = \mathbb{L}_{\hat{\xi}}$, it is called covariant under the restricted diffeomorphisms. 

\subsection{Curvatures}\label{sec:GL-curvatures}
Thus far, we have decomposed the mega-space $\hat{M}$ into the physical space $M$ and orbits of the structure group $\GS$. By imposing $\hat{X}_{\exA b}{}^{\exC}=0$, we identified a restricted class of diffeomorphisms on $\hat{M}$ that unify diffeomorphisms on $M$ with gauge transformations associated with $\GS$. 

But how does this help to construct the Riemann tensor? The key quantities are the fluxes $\hat{X}_{\hat{a}\hat{b}}{}^{\hat{c}}$, which transform as scalar fields on the mega-space $\hat{M}$ and can be expressed as
\begin{align}
 \hat{X}_{\hat{a}\hat{b}}{}^{\hat{c}} 
 = \widetilde{m}_{\hat{a}}{}^{\hat{d}}\,\widetilde{m}_{\hat{b}}{}^{\hat{e}}\,(\widetilde{m}^{-1})_{\hat{f}}{}^{\hat{c}}\,
 X_{\hat{d}\hat{e}}{}^{\hat{f}} \,,
\end{align}
where
\begin{align}
 X_{\hat{a}\hat{b}}{}^{\hat{c}} = 2\,e_{\hat{a}}{}^{\hat{i}}\,e_{\hat{b}}{}^{\hat{j}}\,\partial_{[\hat{i}}e_{\hat{j}]}{}^{\hat{c}}
 + n_{\hat{a}}{}^\exA\, X_{\exA\hat{b}}{}^{\hat{c}}
 - n_{\hat{b}}{}^\exA\, X_{\exA\hat{a}}{}^{\hat{c}}\,.
\end{align}
As demonstrated below, $X_{\hat{a}\hat{b}}{}^{\hat{c}}$ depends only on $y^i$ and transforms covariantly under restricted diffeomorphisms. This can be confirmed through direct computation: since $\hat{X}_{\hat{a}\hat{b}}{}^{\hat{c}}$ transforms as a scalar field on $\hat{M}$, we find
\begin{align}
 \delta_{\hat{\xi}} \hat{X}_{\hat{a}\hat{b}}{}^{\hat{c}} = \xi^{\exA} \,D_{\exA} \hat{X}_{\hat{a}\hat{b}}{}^{\hat{c}} + \xi^{i}\,\partial_i \hat{X}_{\hat{a}\hat{b}}{}^{\hat{c}}
 = \widetilde{m}_{\hat{a}}{}^{\hat{d}}\,\widetilde{m}_{\hat{b}}{}^{\hat{e}}\,(\widetilde{m}^{-1})_{\hat{f}}{}^{\hat{c}}\,\mathbb{L}_{\hat{\xi}}X_{\hat{d}\hat{e}}{}^{\hat{f}}\,,
\end{align}
where
\begin{align}
 \mathbb{L}_{\hat{\xi}} X_{\hat{a}\hat{b}}{}^{\hat{c}} = \xi^{i}\,\partial_i X_{\hat{a}\hat{b}}{}^{\hat{c}}
 + \xi^{\exA} \,(\mathfrak{t}_{\exA})_{\hat{a}}{}^{\hat{d}}\,X_{\hat{d}\hat{b}}{}{}^{\hat{c}}
 + \xi^{\exA} \,(\mathfrak{t}_{\exA})_{\hat{b}}{}^{\hat{d}}\,X_{\hat{a}\hat{d}}{}{}^{\hat{c}}
 - \xi^{\exA} \,(\mathfrak{t}_{\exA})_{\hat{d}}{}^{\hat{c}}\,X_{\hat{a}\hat{b}}{}{}^{\hat{d}}\,.
\end{align}
This confirms that $X_{\hat{a}\hat{b}}{}^{\hat{c}}$ is a covariant object. Since it is constructed from derivatives of the frame field (which contains components of both the physical frame and the spin connection), we naturally expect $X_{\hat{a}\hat{b}}{}^{\hat{c}}$ to encode curvature tensors. We now demonstrate this explicitly.

A straightforward calculation shows that
\begin{align}
 X_{\exA\hat{b}}{}^{\hat{c}} = \hat{X}_{\exA\hat{b}}{}^{\hat{c}}\,.
\end{align}
Thus, the only non-trivial components of $X_{\hat{a}\hat{b}}{}^{\hat{c}}=X_{[\hat{a}\hat{b}]}{}^{\hat{c}}$ are $X_{ab}{}^{\hat{c}}$. A short calculation yields
\begin{align}\label{eq:Riemann-torsion}
 X_{ab}{}^{c} &= 
 2\,e_{a}{}^{i}\,e_{b}{}^{j}\,\partial_{[i} e_{j]}{}^c 
 + 2\,\Omega_{[ab]}{}^c \,,
\\
 X_{ab}{}^{\exC} &= - \Omega_{a}^{\exA}\,\Omega_{b}^{\exB}\,X_{\exA\exB}{}^{\exC} 
 + 2\,e_{a}{}^{i}\,e_{b}{}^{j}\,\partial_{[i} e_{j]}{}^d\,\Omega_d^{\exC} 
 + 2\,D_{[a} \Omega_{b]}^{\exC} 
 + 2\,\Omega_{[ab]}{}^{\exC} \,.
\end{align}
The first equation expresses the torsion of the physical space in terms of the spin connection $\Omega_{ab}{}^c$. To make this explicit, recall the Weitzenb\"ock connection
\begin{align}
 W_{ab}{}^c \equiv e_{a}{}^{i}\,e_{b}{}^{j}\,\partial_{i} e_{j}{}^c \,,
\end{align}
associated with $e_{a}{}^{i}$. This allows us to relate the spin connection to the affine connection via
\begin{align}
 \Gamma_{ab}{}^c \equiv W_{ab}{}^c + \Omega_{ab}{}^c \,,
\end{align}
which follows directly from the vielbein postulate
\begin{align}
 \tnabla_i e_a{}^j \equiv \partial_i e_a{}^j - \Omega_{ia}{}^b\,e_b{}^j + \Gamma_{ik}{}^j\,e_a{}^k = 0\,.
\end{align}
This relation is particularly useful, as it permits us to freely switch between flat and curved indices using $e_a{}^i$ and its inverse $e_i{}^a$. Rewriting \eqref{eq:Riemann-torsion} in terms of the affine connection, we see that $X_{ab}{}^c$ is precisely the torsion tensor
\begin{align}
 T_{ab}{}^c \equiv 2\,\Gamma_{[ab]}{}^c = X_{ab}{}^c
\end{align}
for the connection $\tnabla$.

We express the second non-trivial component $X_{ab}{}^{\exC}$ as
\begin{align}
 R_{abc}{}^{d} \equiv X_{ab}{}^{\exC} \,(\mathfrak{t}_{\exC})_{c}{}^{d}\,. 
\end{align}
Taking into account the result for the torsion already derived, we find\footnote{If we keep $X_{\exA b}{}^{\exC}$, we have an additional term $2\,\Omega_{[ab]}{}^{\exC}\,(\mathfrak{t}_{\exC})_{c}{}^{d}$ whose interpretation is not clear.}
\begin{align}
 R_{abc}{}^{d} = 2\,D_{[a} \Omega_{b]c}{}^d
 - \Omega_{ac}{}^e\,\Omega_{be}{}^d 
 + \Omega_{bc}{}^e\,\Omega_{ae}{}^d 
 - 2\,\Omega_{[ab]}{}^e\,\Omega_{ec}{}^{d}
 + T_{ab}{}^e\,\Omega_{ec}{}^{d} \,,
\label{eq:Rabcd}
\end{align}
which is the ordinary Riemann tensor in flat indices. Converting to curved indices, we obtain\footnote{To recover the standard convention, identify our $\Gamma_{ij}{}^k$, $\Omega_{ab}{}^c$, $R_{ijk}{}^l$ with $\Gamma_{i}{}^k{}_j$, $\Omega_{a}{}^c{}_b$, $R^l{}_{kij}$\,.}
\begin{align}
 R_{ijk}{}^{l} = 2\,\partial_{[i} \Gamma_{j]k}{}^l
 + \Gamma_{ip}{}^l\,\Gamma_{jk}{}^p
 - \Gamma_{jp}{}^l\,\Gamma_{ik}{}^p \,.
\end{align}
The Ricci tensor is defined as usual by
\begin{align}
 R_{ab} \equiv - R_{acb}{}^{c} = D_{c} \Omega_{ab}{}^c - D_{a} \Omega_{cb}{}^c
 + \Omega_{ab}{}^c\,\Omega_{dc}{}^d 
 - \Omega_{da}{}^c\,\Omega_{cb}{}^{d}
 - T_{ad}{}^c\,\Omega_{cb}{}^{d} \,.
\end{align}

To define the Ricci scalar, we return to the $\GS$-invariant metric $g_{ab}$ on $M$. As mentioned earlier, it is constant and stabilized by the structure group $\GS$. In equations, this means
\begin{align}
 \partial_i g_{ab} = 0\,,\qquad
 (\mathfrak{t}_{\exA})_a{}^c\,g_{cb} + (\mathfrak{t}_{\exA})_b{}^c\,g_{ac} =0\,.
\end{align}
Therefore, the flat metric satisfies
\begin{align}
 \mathbb{L}_{\hat{\xi}} g_{ab} = 0\,,\qquad \tnabla_a g_{bc}=0\,,
\end{align}
and gives rise to the metric $g_{ij} = e_i{}^a\,e_j{}^b\,g_{ab}$, which is compatible with the connection ($\tnabla_k g_{ij} = 0$). With this metric, we define the Ricci scalar as
\begin{align}
 R \equiv g^{ab} R_{ab} = g^{ij} R_{ij} \,.
\end{align}
We conclude that the torsion $T_{ab}{}^c$ and curvature $R_{abc}{}^d$ are contained in the non-trivial components of the covariant object $X_{\hat{a}\hat{b}}{}^{\hat{c}}$\,. In the simplest case $\GD=\GL(d)$, only the torsion tensor and Riemann tensor appear in $X_{\hat{a}\hat{b}}{}^{\hat{c}}$; in general, however, additional curvature tensors emerge, as discussed in subsequent sections.

\subsection{Algebraic rephrasing}\label{sec:algebra-GL}
While the case $\GD=\GL(d)$ admits an intuitive geometric decomposition of the mega-space into the physical manifold and the $y^\mu$ directions, this approach quickly becomes unwieldy for the exceptional groups $\Edd[d]$. To address this, we adopt a systematic algebraic reformulation based on the level decomposition of the extended duality algebra. Following \cite{Hassler:2023axp}, this procedure provides a step-by-step construction of the relevant curvature tensors. To illustrate the method in its simplest setting, we first present the decomposition for $\GD=\GL(d)$.

We start with the extended duality group $\GDM=\GL(p)$ associated with the mega-space and decompose the generators $\widehat{t}_{\adjah}\equiv \hat{K}^{\hat{a}}{}_{\hat{b}}$ ($\hat{a},\hat{b}=1,\dots,p$) under a subgroup $\GL(d)\times \GL(n)\subset\GDM$\,,
\begin{align}\label{eq:branching-GL}
 \widehat{t}_{\adjah} = \hat{K}^{\hat{a}}{}_{\hat{b}} = \{R^{a}_{\exB},\, K^{\exA}{}_{\exB},\,K^{a}{}_{b},\,R^{\exA}_{b}\}\,,
\end{align}
where $R^{a}_{\exB}\equiv K^{a}{}_{\exB}$ and $R^{\exA}_{b}=-K^{\exB}{}_b$\,. A more sophisticated description of this branching is given by removing the $n$-th node in the Dynkin diagram:
\begin{align}\label{eq:Dynkin-GL}
 \xygraph{
    *[F*:gray]\cir<5pt>{} ([]!{-(0,.4)} {1}) - [r]
    \cdots ([]!{+(0,-.4)} {}) - [r]
    *[F*:gray]\cir<5pt>{} ([]!{-(0,.4)} {n-1}) - [r]
    *[F*:gray]\cir<5pt>{}*{\scalebox{2}{$\times$}}{~} ([]!{-(0,.4)} {n}) - [r]
    *\cir<5pt>{} - [r]
    \cdots ([]!{+(0,-.4)} {}) - [r]
    *\cir<5pt>{} 
(
        - [u] *\cir<5pt>{}
)}\ .
\end{align}
At this stage, the decomposition \eqref{eq:branching-GL} is already evident from the index structure. However, for more complicated duality groups, Dynkin diagrams similar to the one shown here become extremely valuable, as demonstrated in section~\ref{sec:extended-algebra}. 

As readily observed, $K^{\exA}{}_{\exB}$ and $t_{\adja}\equiv K^{a}{}_{b}$ are generators of $\GL(n)$ and $\GD=\GL(d)$, respectively. From the commutation relations 
\begin{align}
 [\hat{K}^{\hat{a}}{}_{\hat{b}},\,\hat{K}^{\hat{c}}{}_{\hat{d}}] = \delta^{\hat{c}}_{\hat{b}}\,\hat{K}^{\hat{a}}{}_{\hat{d}} - \delta^{\hat{a}}_{\hat{d}}\,\hat{K}^{\hat{c}}{}_{\hat{b}}\,,
\end{align}
we obtain
\begin{align}
 [R^\exA_a,\,R^\exB_b] &= 0 \,, 
\\
 [K^\exA{}_\exB,\,R^\exC_c] &= \delta^\exC_\exB\,R^\exA_c \,, 
\\
 [K^a{}_b,\,R^{\exC}_c] &= - \delta^a_c\,R^{\exC}{}_b \,, 
\\
 [K^\exA{}_\exB,\,K^\exC{}_\exD] &= \delta^\exC_\exB\,K^\exA{}_\exD - \delta^\exA_\exD\,K^\exC{}_\exB \,, 
\\
 [K^\exA{}_\exB,\,K^c{}_d] &= 0 \,, 
\\
 [K^a{}_b,\,K^c{}_d] &= \delta^c_b\,K^a{}_d - \delta^a_d\,K^c{}_b \,, 
\\
 [R^a_\exA,\,R^\exB_b] &= \delta^a_b\,K^\exB{}_\exA - \delta^\exB_\exA\,K^a{}_b \,, 
\\
 [K^\exA{}_\exB,\,R^c_\exC] &= - \delta^\exA_\exC\,R^c{}_\exB \,, 
\\
 [K^a{}_b,\,R^c_{\exC}] &= \delta^c_b\,R^a_{\exC} \,, 
\\
 [R^a_{\exA},\,R^b_{\exB}] &= 0\,.
\end{align}
The corresponding matrix representations, in the $R_1$-representation, are
\begin{align}
\begin{alignedat}{2}
 (K^{\exC}{}_{\exD})_{\hat{a}}{}^{\hat{b}} &= \begin{pmatrix}
 \delta_{\exA}^{\exC}\,\delta_{\exD}^{\exB} & 0 \\
 0 & 0
\end{pmatrix},&\qquad 
 (K^{c}{}_d)_{\hat{a}}{}^{\hat{b}} &= \begin{pmatrix}
 0 & 0 \\
 0 & \delta_{a}^{c}\,\delta_d^b
\end{pmatrix},
\\
 (R^{\exC}_c)_{\hat{a}}{}^{\hat{b}} &= \begin{pmatrix}
 0 & -\delta_{\exA}^{\exC}\,\delta_c^b \\
 0 & 0
\end{pmatrix},&\qquad
 (R^{c}_{\exC})_{\hat{a}}{}^{\hat{b}} &= \begin{pmatrix}
 0 & 0 \\
 \delta_a^c\,\delta_{\exC}^{\exB} & 0
\end{pmatrix}.
\end{alignedat}
\end{align}
Next, we define an operator that measures the level by
\begin{align}
 L \equiv K^{\exA}{}_{\exA}\,.
\end{align}
This commutes with $K^{\exA}{}_{\exB}$ and $t_{\adja} = K^{a}{}_{b}$, implying that their level is $\ell=0$. The generators $R^{\exA}_a$ with an upper index $\exA$ have level $\ell=1$, while the generators $R_{\exA}^a$ with a lower index $\exA$ have level $\ell=-1$. By the Jacobi identity, the level is additive under commutators. We now rephrase the assumptions from section~\ref{sec:setup} using these definitions. 

\paragraph{Assumption 1.}
The frame field $\hat{e}_{\hat{a}}$ contains the right-invariant vector fields $\hat{e}_{\exA}$ of $\GS$ such that $\hat{e}_{\hat{a}}{}^{\hat{i}}$ takes the form \eqref{eq:hat-e-gauge}. Hence, $\hat{e}_{\hat{a}}{}^{\hat{i}}$ is given by an exponential of the negative-level generators $R^a_{\exA}$ and the level-$0$ generators. Moreover, the part generated by $K^{\exA}{}_{\exB}$ (the upper-left block of $\hat{e}_{\hat{a}}{}^{\hat{i}}$) depends only on $y^\mu$.

\paragraph{Assumptions 2 and 3.}
Due to the constraint \eqref{eq:second-prop} and $\hat{X}_{\exA b}{}^{\exC}=0$, $X_{\exA}$ must take the form
\begin{align}
 X_{\exA} = X_{\exA\exB}{}^{\exC} \,K^{\exB}{}_{\exC} + X_{\exA}{}^{\adja} \,t_{\adja}\,,
\end{align}
where $X_{\exA\exB}{}^{\exC}$ and $X_{\exA}{}^{\adja}$ are constants.

\noindent Using the matrix representations of the generators of $\mathfrak{gl}(p)$, the parameterization of $\hat{e}_{\hat{a}}{}^{\hat{i}}$ takes the form
\begin{align}
 \hat{e}_{\hat{a}}{}^{\hat{i}} = m_{\hat{a}}{}^{\hat{b}}\, e_{\hat{b}}{}^{\hat{i}}\,,\qquad \text{and} \qquad
 e_{\hat{a}}{}^{\hat{i}} \equiv n_{\hat{a}}{}^{\hat{b}}\,v_{\hat{b}}{}^{\hat{i}}\,,
\end{align}
where
\begin{align}
 n = \exp\bigl(-\Omega_c^{\exC}\,R_{\exC}^c\bigr)_{\hat{a}}{}^{\hat{b}}\,,\qquad
 v = \exp\bigl[(\ln \widetilde{v})_{\exC}{}^{\mu}\,\delta_\mu^{\exD}\,K^{\exC}{}_{\exD}\bigr]\,\exp\bigl[(\ln e)_a{}^i\,\delta_i^b\,K^a{}_b \bigr] \,.
\end{align}
Here, $\Omega_c^{\exC}$ and $e_a{}^i$ depend only on $y^i$, whereas $\widetilde{v}$ depends only on $y^\mu$. The torsion and curvature tensors appear in the nontrivial components of the untwisted fluxes $X_{a}$, which decompose as
\begin{align}
 X_a = X_{a}{}^{\adja}\,t_{\adja} + X_{ab}{}^{\exB}\,R_{\exB}^b 
 = X_{ab}{}^{c}\,K^b{}_c + X_{ab}{}^{\exB}\,R_{\exB}^b\,.
\end{align}
From this, the torsion tensor $T_{ab}{}^{c}=X_{ab}{}^{c}$ is identified with the level-$0$ component (spanned by $K^b{}_c$), while the Riemann tensor $R_{abc}{}^d = X_{ab}{}^{\exA}\,(\mathfrak{t}_{\exA})_c{}^d$ arises as the level-$(-1)$ component (spanned by $R_{\exB}^b$). For $\GD=\GL(d)$, $R_{\exA}^a$ is the only negative-level generator, but more generally, lower levels may appear, giving rise to an extended hierarchy of curvature tensors.

In this algebraic reformulation, familiar geometric quantities are neatly organized within the graded structure of the extended Lie algebra. Torsion appears at level $0$, Riemann curvature at level $-1$, and further tensors at lower levels.

\subsection{Riemann tensor of the Levi--Civita connection}\label{sec:torsionless-GL}
We have so far considered a general connection $\tnabla$ with non-vanishing torsion
\begin{align}
 (\gLie^{\tnabla}_v -\gLie_v) w^i = 2\,\Gamma_{[jk]}{}^i \,v^j\,w^k = T_{jk}{}^i \,v^j\,w^k\,.
\end{align}
In (super)gravity, it is also useful to work with the Levi--Civita connection $\bnabla$, the unique torsion-free, metric-compatible connection. It can be constructed as $\bnabla \equiv \tnabla - \SSS$, where $\SSS$ is the contorsion tensor \cite{Castellani:1999fz},
\begin{align}
 \SSS_{ab}{}^c \equiv \tfrac{1}{2}\,\bigl[ X_{ab}{}^c + g^{cd}\,\bigl(X_{da}{}^e\,g_{eb} +X_{db}{}^e\,g_{ea}\bigr)\bigr] \,,
\label{eq:contorsion-0}
\end{align}
which is related to the torsion tensor as
\begin{align}
 2\,\SSS_{[ab]}{}^c = X_{[ab]}{}^c = T_{ab}{}^c \,.
\label{eq:contorsion-torsion-0}
\end{align}
Correspondingly, we define
\begin{align}
 \omega_{ab}{}^c \equiv \Omega_{ab}{}^c - \SSS_{ab}{}^c\,,\qquad
 \gamma_{ij}{}^k \equiv \Gamma_{ij}{}^k - \SSS_{ij}{}^k\,,
\end{align}
and note that the matrix $\SSS_i$ defined by $(\SSS_i)_j{}^k \equiv \SSS_{ij}{}^k$ is an element of the Lie algebra $\mathrm{Lie}(H)$ that preserves the metric $g_{ij}$\,, i.e., $(\SSS_k)_i{}^l\,g_{lj}+(\SSS_k)_j{}^l\,g_{il}=0$. If we decompose $\mathrm{Lie}(H)$ as 
\begin{align}
 \mathrm{Lie}(H)=\mathrm{Lie}(\GS)+\mathrm{Lie}(\GS)^{\perp}\,,
\end{align}
the part $\SSS_i^\perp$ of $\SSS_i$ expanded by the generators of $\mathrm{Lie}(\GS)^{\perp}$ gives the intrinsic torsion $\SSS^\perp_{[ij]}{}^k$. 
From \eqref{eq:Rabcd}, we find
\begin{align}
 R_{abc}{}^{d}
&= 2\,D_{[a} \omega_{b]c}{}^d
 - \omega_{ac}{}^e\,\omega_{be}{}^d 
 + \omega_{bc}{}^e\,\omega_{ae}{}^d 
 - 2\,\omega_{[ab]}{}^e\,\omega_{ec}{}^{d} 
\nn\\
 &\quad
 + 2\,\tnabla_{[a}\SSS_{b]c}{}^d 
 + X_{ab}{}^{e} \,\SSS_{ec}{}^d 
 + 2\,\SSS_{[a|c}{}^e\,\SSS_{|b]e}{}^d \,,
\end{align}
and see that the Riemann tensor of the Levi--Civita connection $\bnabla$ can be expressed as
\begin{align}
 \bar{R}_{abc}{}^d &\equiv 2\,D_{[a} \omega_{b]c}{}^d
 - \omega_{ac}{}^e\,\omega_{be}{}^d 
 + \omega_{bc}{}^e\,\omega_{ae}{}^d 
 - 2\,\omega_{[ab]}{}^e\,\omega_{ec}{}^{d}
\nn\\
 &= R_{abc}{}^{d}
 - X_{ab}{}^{e} \,\SSS_{ec}{}^d 
 - 2\,\SSS_{[a|c}{}^e\,\SSS_{|b]e}{}^d 
 - 2\,\tnabla_{[a}\SSS_{b]c}{}^d\,.
\label{eq:bar-Riemann}
\end{align}

Let us now focus on the particular case where $\tnabla_d X_{ab}{}^c =0$ and $\nabla_c X_{ab}{}^{\exC}=0$\,. The former condition implies $\tnabla_d \SSS_{ab}{}^c=0$, while the latter gives $\tnabla_e R_{abc}{}^d=0$ due to the faithfulness of $(\mathfrak{t}_{\exA})_{b}{}^{c}$. The Bianchi identity for the fluxes then gives
\begin{align}
 [\mathfrak{t}_{\exA},\,X_{b}] = X_{\exA b}{}^c\,X_c\,,\qquad 
 [X_a,\,X_b]_{\exC}{}^{\exD} = - X_{ab}{}^{\exE}\,X_{\exE\exC}{}^{\exD} \,.
\label{eq:BI-GL}
\end{align}
These equations, together with
\begin{align}
 \tnabla_d X_{ab}{}^c &= D_d X_{ab}{}^c - [\Omega_d,\,X_a]_b{}^c - \Omega_{da}{}^e\,X_{eb}{}^c \,,
\\
 \tnabla_d X_{ab}{}^{\exC} &= D_d X_{ab}{}^{\exC} - \Omega_{da}{}^e\,X_{eb}{}^{\exC} - \Omega_{db}{}^e\,X_{ae}{}^{\exC} -\Omega_d^{\exE}\,X_{\exE\exD}{}^{\exC}\,X_{ab}{}^{\exD}\,,
\end{align}
show that $X_{ab}{}^c$ and $X_{ab}{}^{\exC}$ are constant. It follows that $X_{\hat{a}\hat{b}}{}^{\hat{c}}$ is constant, whence the generalized fluxes coincide with the constants $\hat{X}_{\hat{a}\hat{b}}{}^{\hat{c}}=X_{\hat{a}\hat{b}}{}^{\hat{c}}$, as required by the Bianchi identity. The mega-space then admits a set of linearly independent vector fields $\{\hat{e}_{\hat{a}}\}$ satisfying the Lie bracket relations $[\hat{e}_{\hat{a}},\hat{e}_{\hat{b}}]=-\hat{X}_{\hat{a}\hat{b}}{}^{\hat{c}}\,\hat{e}_{\hat{c}}$ which realize the Lie algebra of $G$. These vector fields act transitively on the physical space $M$, while the $n$-dimensional isotropy subalgebra $\mathrm{Lie}(\GS)$, generated by $\{\hat{e}_{\exA}\}$, corresponds to the subgroup that leaves each point of $M$ invariant. Hence, at least locally, the physical space may be identified with the homogeneous space $M \cong \GS \backslash G$. The Riemann tensor \eqref{eq:bar-Riemann} then reduces to the known expression \cite{Castellani:1999fz} for the Riemann tensor of the Levi--Civita connection on the coset space.

We now establish a close connection to a classical theorem by Ambrose and Singer \cite{Ambrose1958OnHR}:
\begin{theorem}
  Let $(M,\,g)$ be a simply connected and complete Riemannian manifold. Then the following statements are equivalent:
  \begin{enumerate}
    \item The manifold $M$ is homogeneous.
    \item The manifold $M$ admits a linear connection $\tnabla$ satisfying
      \begin{equation}\label{eqn:constTorsionAndCurvature}
        \tnabla R = 0\,, \qquad \tnabla \SSS = 0\,, \qquad \tnabla g = 0\,,
      \end{equation}
      where $R$ is the Riemann curvature tensor of $\tnabla$ and $\SSS=\tnabla - \bnabla$ is the contorsion tensor, with $\bnabla$ denoting the Levi--Civita connection.
  \end{enumerate}
\label{theorem:AS}
\end{theorem}
\noindent Up to eq.~\eqref{eq:bar-Riemann}, we have constructed a general framework in which a physical space $M$ carries a structure group $\GS$ and is endowed with a metric-compatible spin connection $\tnabla$. The requirement that this connection satisfy \eqref{eqn:constTorsionAndCurvature} is equivalent to the condition that $\hat{X}_{\hat{a}\hat{b}}{}^{\hat{c}}$ be constant on $M$. The physical manifold may then be identified with a homogeneous space $M \cong \GS\backslash G$ as discussed above. Conversely, if $M$ is given as a homogeneous space $\GS\backslash G$, then $\hat{X}_{\hat{a}\hat{b}}{}^{\hat{c}}$ are the structure constants of $G$ and hence constant, which ensures that \eqref{eqn:constTorsionAndCurvature} is automatically satisfied. In this sense, Theorem~\ref{theorem:AS} emerges naturally within our framework, although our discussion remains heuristic rather than fully rigorous. In section~\ref{sec:gen-coset}, we discuss an extension of this theorem to define generalized coset spaces. Remarkably, spaces satisfying the generalized version necessarily assume the form of a double coset $M = \GS \backslash G / H$ rather than a simple homogeneous space $\GS \backslash G$.

\section{Algebraic identities for reductions of extended spaces}\label{sec:algebras-decom}
Extending our construction from $\GD=\GL(d)$ to $\GD=\Odd$ or $\GD=\Edd[d]$ requires algebraic identities analogous to those in section \ref{sec:algebra-GL}. This involves three steps:
\begin{enumerate}
\item Construct the level decomposition of the extended Lie algebra for $\GDM = \GD \times \GL(n)$.

\item Obtain the commutation relations for all generators of this Lie algebra.

\item Construct their matrix representations in the $R_1$-representation.
\end{enumerate}
We address each step sequentially in the following subsections. These constructions provide the algebraic foundation for our subsequent construction of generalized curvatures.

\subsection{Extended algebra}
\label{sec:extended-algebra}

We combine the duality group $\GD$ with the $\GL(n) \supset \GS$ to form a single extended duality group $\GDM$. For example, when $\GD=\mathrm{O}(d,d)$, we take $\GDM=\mathrm{O}(p,p)$ with $p\equiv d+n$.\footnote{We focus on this natural extension, though other extensions may be possible.} The structure of this extension is captured by appending $n$ nodes to the Dynkin diagram of the Lie algebra of the duality group $\mathrm{O}(d,d)$:
\begin{align}
 \xygraph{
    *[F*:gray]\cir<5pt>{} ([]!{-(0,.4)} {1}) - [r]
    \cdots ([]!{+(0,-.4)} {}) - [r]
    *[F*:gray]\cir<5pt>{} ([]!{-(0,.4)} {n-1}) - [r]
    *[F*:gray]\cir<5pt>{}*{\scalebox{2}{$\times$}}{~} ([]!{-(0,.4)} {n}) - [r]
    *\cir<5pt>{} - [r]
    \cdots ([]!{+(0,-.4)} {}) - [r]
    *\cir<5pt>{} 
(
        - [u] *\cir<5pt>{} ,
        - [r] *\cir<5pt>{} 
)}\ .
\end{align}
Deleting the $n$-th node of this extended diagram recovers the subgroup $\GL(n)\times \mathrm{O}(d,d)$. For $\GD=\Odd$, the same pattern gives the extended duality group $\GDM=\Odd[p]$. Comparing this diagram with \eqref{eq:Dynkin-GL}, we recognize a pattern that we follow to construct $\GDM$ for $\GD=\Edd[d]$. We append one additional node to the right to obtain
\begin{align}
 \xygraph{
    *[F*:gray]\cir<5pt>{} ([]!{-(0,.4)} {1}) - [r]
    \cdots ([]!{+(0,-.4)} {}) - [r]
    *[F*:gray]\cir<5pt>{} ([]!{-(0,.4)} {n-1}) - [r]
    *[F*:gray]\cir<5pt>{}*{\scalebox{2}{$\times$}}{~} ([]!{-(0,.4)} {n}) - [r]
    *\cir<5pt>{} - [r]
    \cdots ([]!{+(0,-.4)} {}) - [r]
    *\cir<5pt>{} 
(
        - [u] *\cir<5pt>{} ,
        - [r] *\cir<5pt>{} 
        - [r] *\cir<5pt>{} 
)}\ .
\end{align}
This establishes that the extended duality group is $\GDM=\Edd[p]$ ($p\equiv d+n$) and, as required, it contains the subgroup $\GL(n)\times\Edd[d]$. 

To control the complexity, we grade the algebra of $\GDM$ by a level $\ell$, defined via the operator
\begin{align}
 L = K^{\exA}{}_{\exA}\,.
\end{align}
This level decomposition organizes the generators of $\GDM$ into a clear hierarchy, which we now present. The generators of the extended duality group $\GDM$ are denoted as $\widehat{t}_{\widehat{\adja}}$ and decompose into irreducible representations of $\GL(n)\times\GD$. At level $\ell=0$, the generators split as the direct sum $\mathrm{Lie}(\GL(n)) \oplus \mathrm{Lie}(\GD)$, where $K^\exA{}_\exB$ generate $\GL(n)$ and $t_{\adja}$ generate the duality group $\GD$. They satisfy
\begin{align}
 [K^\exA{}_\exB,\,K^\exC{}_\exD] = \delta^\exC_\exB\,K^\exA{}_\exD - \delta^\exA_\exD\,K^\exC{}_\exB \,,\qquad 
 [t_{\adja},\,t_{\adjb}] = f_{\adja\adjb}{}^{\adjc}\,t_{\adjc}\,,\qquad
 [K^\exA{}_\exB,\,t_{\adjb}] = 0\,,
\end{align}
where $f_{\adja\adjb}{}^{\adjc}$ are the structure constants of $\GD$. For $\Odd$, the levels are restricted to $-2\leq \ell\leq 2$, so that the generators are
\begin{align}
 \begin{tabular}{c||c|c|c|c|c|c}
 $\ell$ & $-2$ & $-1$ & $0$ & $0$ & $1$ & $2$ \\\hline
 $\widehat{t}_{\widehat{\adja}}$ & $R_{\exA_1\exA_2}$ & $R_\exA^A$ & $K^\exA{}_\exB$ & $t_{\adja}$& $R^\exA_A$ & $R^{\exA_1\exA_2}$ 
\end{tabular}\ .
\end{align}
For $\Edd$, an infinite number of generators may appear in principle, but we focus on the finite set of low-level generators. For any $\Edd$, $\widehat{t}_{\widehat{\adja}}$ always includes the following generators with $p$ antisymmetric indices ($\exA_1\cdots\exA_p$) and the index for the $\hat{R}_p$-representation,
\begin{align}
 \begin{tabular}{c||c|c|c|c|c|c|c|c|c|c}
 $\ell$ &$\cdots$& $-3$ & $-2$ & $-1$ & $0$ & $0$ & $1$ & $2$ & $3$&$\cdots$ \\\hline
 $\widehat{t}_{\widehat{\adja}}$ &$\cdots$& $R_{\alpha_1\exA_2\exA_3\AAA}$ & $R_{\exA_1\exA_2;\AA}$ & $R_\exA^A$ & $K^\exA{}_\exB$ & $t_{\adja}$& $R^\exA_A$ & $R^{\exA_1\exA_2;\AA}$ & $R^{\exA_1\exA_2\exA_3;\AAA}$&$\cdots$
\end{tabular}\ .
\end{align}
Additional mixed-symmetry generators appear at levels with $|\ell|\geq 9-d$. Positive-level generators up to $\ell=7$ are
\begin{align}
 \begin{tabular}{c||c|c|c|c|c|c|c}
 $\ell$ & $1$ & $2$ & $3$ & $4$ & $5$ & $6$ & $7$ \\ \hline\hline
 $\Edd[2]$ & --- & --- & --- & --- & --- & --- & $R^{\exA_1\cdots\exA_6,\exA'}$ \\\hline
 $\Edd[3]$ & --- & --- & --- & --- & --- & $R^{\exA_1\cdots\exA_5,\exA'}$ & $R_A^{\exA_1\cdots\exA_6,\exA'}$ \\\hline
 $\Edd[4]$ & --- & --- & --- & --- & $R^{\exA_1\cdots\exA_4,\exA'}$ & $R_A^{\exA_1\cdots\exA_5,\exA'}$ & $\cdots$ \\\hline
 $\Edd[5]$ & --- & --- & --- & $R^{\exA_1\exA_2\exA_3,\exA'}$ & $R_A^{\exA_1\cdots\exA_4,\exA'}$ & $\cdots$ & $\cdots$ \\\hline
 $\Edd[6]$ & --- & --- & $R^{\exA_1\exA_2,\exA'}$ & $R_A^{\exA_1\exA_2\exA_3,\exA'}$ & $\cdots$ & $\cdots$ & $\cdots$ \\\hline
 $\Edd[7]$ & --- & $R^{\exA,\exA'}$ & $R_A^{\exA_1\exA_2,\exA'}$ & $\cdots$ & $\cdots$ & $\cdots$ & $\cdots$ \\ \hline
 $\Edd[8]$ & --- & \begin{tabular}{c}
 $R^{\exA,\exA'}_A$ \\
 $R^{\exA_1\exA_2}$ \end{tabular} & $\cdots$ & $\cdots$ & $\cdots$ & $\cdots$ & $\cdots$ \\
\end{tabular}\ .
\label{eq:generators-mixed-symmetry}
\end{align}
Each generator transforms in an irreducible representation of $\GL(n)$. For instance, $R^{\exA,\exA'}$ transforms in the representation $\scalebox{0.5}{\young(~~)}$ while $R_A^{\exA_1\exA_2,\exA'}$ transforms in $\scalebox{0.5}{\young(~~,~)}$. In principle, we can extend to $|\ell|\geq 11-d$, where many more generators appear. However, these generators are omitted in \eqref{eq:generators-mixed-symmetry}, indicated by ellipses, and are not considered in this paper. While we have explicitly presented the positive-level generators, each has a corresponding negative-level counterpart with indices upside down.

For $\GD=\Edd$, we summarize the generators in each dimension with $d\leq 7$ and $|\ell| \leq 3$ by
\begin{align}
 d\leq 5 :&\quad
 \begin{tabular}{|c||c|c|c|c|c|c|c|c|} \hline
 $\ell$ & $-3$ & $-2$ & $-1$ & $0$ & $0$ & $1$ & $2$ & $3$ \\\hline
 $\widehat{t}_{\widehat{\adja}}$ & $R_{\alpha_1\exA_2\exA_3;\AAA}$ & $R_{\exA_1\exA_2;\AA}$ & $R_\exA^A$ & $K^\exA{}_\exB$ & $t_{\adja}$& $R^\exA_A$ & $R^{\exA_1\exA_2;\AA}$ & $R^{\exA_1\exA_2\exA_3;\AAA}$ \\\hline
\end{tabular}\ ,
\\
 d=6 :&\quad
 \begin{tabular}{|c||c|c|c|c|c|c|c|c|} \hline
 $\ell$ & $-3$ & $-2$ & $-1$ & $0$ & $0$ & $1$ & $2$ & $3$ \\\hline
 $\widehat{t}_{\widehat{\adja}}$ & $\begin{matrix}
R_{\alpha_1\exA_2\exA_3;\AAA} \\[1mm] R_{\exA_1\exA_2,\exA'} \end{matrix}$ & $R_{\exA_1\exA_2;\AA}$ & $R_\exA^A$ & $K^\exA{}_\exB$ & $t_{\adja}$& $R^\exA_A$ & $R^{\exA_1\exA_2;\AA}$ & $\begin{matrix}
R^{\exA_1\exA_2\exA_3;\AAA} \\ R^{\exA_1\exA_2,\exA'}\end{matrix}$ \\\hline
\end{tabular}\ ,
\\
 d=7 :&\quad
 \begin{tabular}{|c||c|c|c|c|c|c|c|c|} \hline
 $\ell$ & $-3$ & $-2$ & $-1$ & $0$ & $0$ & $1$ & $2$ & $3$ \\\hline
 $\widehat{t}_{\widehat{\adja}}$ & $\begin{matrix}
R_{\alpha_1\exA_2\exA_3;\AAA} \\[2mm] R_{\exA_1\exA_2,\exA'}^A \end{matrix}$ & $\begin{matrix}
R_{\exA_1\exA_2;\AA}\\[1mm] R_{\exA,\exA'}\end{matrix}$ & $R_\exA^A$ & $K^\exA{}_\exB$ & $t_{\adja}$& $R^\exA_A$ & $\begin{matrix}
R^{\exA_1\exA_2;\AA} \\[1mm] R^{\exA,\exA'}\end{matrix}$ & $\begin{matrix} R^{\exA_1\exA_2\exA_3;\AAA} \\ R^{\exA_1\exA_2,\exA'}_A \end{matrix}$ \\\hline
\end{tabular}\ .
\end{align}
For $d=8$, we consider only generators with $|\ell| \leq 2$ for simplicity,
\begin{align}
 \begin{tabular}{|c||c|c|c|c|c|c|} \hline
 $\ell$ & $-2$ & $-1$ & $0$ & $0$ & $1$ & $2$ \\\hline
 $\widehat{t}_{\widehat{\adja}}$ & $\begin{matrix}
R_{\exA_1\exA_2;\AA} \\[2mm] R_{\exA,\exA'}^A\\[1mm] R_{\exA_1\exA_2} \end{matrix}$ & $R_\exA^A$ & $K^\exA{}_\exB$ & $t_{\adja}$& $R^\exA_A$ & $\begin{matrix}
R^{\exA_1\exA_2;\AA} \\ R^{\exA,\exA'}_A\\[1mm] R^{\exA_1\exA_2} \end{matrix}$ \\\hline
\end{tabular}\ ,
\end{align}
where the level-$2$ generators $R^{\exA_1\exA_2;\AA}$ and $R^{\exA_1\exA_2}$ correspond to the $R_2$-representation $\irrep{3875}\oplus \irrep{1}$ of Table~\ref{tab:representations}.

\subsection{Commutation relations}\label{sec:commutation}

We now determine the commutation relations among these generators.

\paragraph{Low-level generators.}
Each generator transforms in a representation of $\GL(n)\times\GD$. Hence, the commutation relations between the level-$0$ generators and the positive/negative-level generators are governed by their respective representations, such as
\begin{align}
 [K^\exA{}_\exB,\,R_{\exC_1\exC_2\exC_3;\CCC}] &= -3\,\delta_{[\exC_1}^{\exA}\,R_{|\exB|\exC_2\exC_3];\CCC}\,,
\\
 [K^\exA{}_\exB,\,R_{\exC_1\exC_2;\CC}] &= -2\,\delta_{[\exC_1}^{\exA}\,R_{|\exB|\exC_2];\CC}\,,
\\
 [K^\exA{}_\exB,\,R_{\exC}^C] &= - \delta_\exC^\exA\,R_{\exB}^C\,,
\\
 [K^\exA{}_\exB,\,R^{\exC}_C] &= \delta_\exB^\exC\,R^{\exA}_C\,,
\\
 [K^\exA{}_\exB,\,R^{\exC_1\exC_2;\CC}] &= 2\,\delta^{[\exC_1}_{\exB}\,R^{|\exA|\exC_2];\CC}\,,
\\
 [K^\exA{}_\exB,\,R^{\exC_1\exC_2\exC_3;\CCC}] &= 3\,\delta^{[\exC_1}_{\exB}\,R^{|\exA|\exC_2\exC_3];\CCC}\,,
\\
 [t_{\adja},\,R^{\exB_1\exB_2\exB_3;\BBB}] &= (t_{\adja})_{\CCC}{}^{\BBB}\,R^{\exB_1\exB_2\exB_3;\CCC}\,,
\\
 [t_{\adja},\,R^{\exB_1\exB_2;\BB}] &= (t_{\adja})_{\CC}{}^{\BB}\,R^{\exB_1\exB_2;\CC}\,,
\\
 [t_{\adja}\,R_\exA^B] &= (t_{\adja})_C{}^B\,R_\exB^C\,,
\\
 [t_{\adja},\,R^\exB_B] &= -(t_{\adja})_B{}^C\,R^\exB_C\,,
\\
 [t_{\adja},\,R_{\exB_1\exB_2;\BB}] &= - (t_{\adja})_{\BB}{}^{\CC}\,R_{\exB_1\exB_2;\CC}\,,
\\
 [t_{\adja},\,R_{\exB_1\exB_2\exB_3;\BBB}] &= - (t_{\adja})_{\BBB}{}^{\CCC}\,R_{\exB_1\exB_2\exB_3;\CCC}\,.
\end{align}
Here, we define the matrix representations of the generators $t_{\adja}$ in the $R_1$-, $\hat{R}_2$-, and $\hat{R}_3$-representations such that the invariant tensors satisfy
\begin{align}
\begin{split}
 \eta_{AB;\EE}\,(t_{\adja})_{\CC}{}^{\EE} + \eta_{EB;\CC}\,(t_{\adja})_A{}^E + \eta_{AE;\CC}\,(t_{\adja})_B{}^E &= 0\,, \qquad \text{and}
\\
 (t_{\adja})_{\DDD}{}^{\EEE}\,Z_{B}{}^{\CC}{}_{\EEE} 
 + (t_{\adja})_B{}^E\,Z_E{}^{\CC}{}_{\DDD} - (t_{\adja})_{\EE}{}^{\CC}\,Z_A{}^{\EE}{}_{\DDD} &= 0\,,
\end{split}
\label{eq:invariant-tensors}
\end{align}
where $Z_{A}{}^{\BB}{}_{\CCC}$ is an invariant tensor connecting the $R_1$-, $\hat{R}_2$-, and $\hat{R}_3$-representation. In the $\Odd$ case, the $R_2$-representation is the singlet ($R^{\exA_1\exA_2;\AA}=R^{\exA_1\exA_2}$\,, $R_{\exA_1\exA_2;\AA}=R_{\exA_1\exA_2}$\,, $\eta_{AB;\CC}=\eta_{AB}$) and the level-$(\pm 3)$ generators are absent. Consequently, $Z_{A}{}^{\BB}{}_{\CCC}$ vanishes.

Non-trivial commutation relations arise for combinations of positive- and negative-level generators. Keeping in mind that the level provides a natural grading for the algebra, the commutator of a level-$p$ generator and a level-$q$ generator has level $(p+q)$. We therefore define
\begin{align}
 [R^{\exA}_{A},\,R^{\exB}_{B}] = \begin{cases}
 \eta_{AB;\CC}\,R^{\exA\exB;\CC} & (\Odd \text{ or } \mathrm{E}_{d(d\leq 6)}) \\
 \eta_{AB;\CC}\,R^{\exA\exB;\CC} + \omega_{AB}\,R^{\exA,\exB} & (\Edd[7]) \\
 \eta_{AB;\CC}\,R^{\exA\exB;\CC} + f_{AB}{}^C\,R_C^{\exA,\exB} + \kappa_{AB}\,R^{\exA\exB} \quad & (\Edd[8])
\end{cases}\,.
\label{eq:eta-def}
\end{align}
In our convention, the Cartan involution on $\mathrm{Lie}(\GDM)$ simply flips the position of the indices, and negative-level generators like $R_{\exA}^{A}$ are mapped to their positive-level counterparts, for example
\begin{align}
 \theta(R_{\exA}^{A}) = R^{\exA}_{A}\,.
\end{align}
Hence, negative-level generators, such as $R_{\exA}^{A}$, satisfy the same algebra as the corresponding positive-level generators with the position of the indices flipped, namely
\begin{align}
 [R_{\exA}^{A},\,R_{\exB}^{B}] = \begin{cases}
 \eta^{AB;\CC}\,R_{\exA\exB;\CC} & (\Odd \text{ or } \mathrm{E}_{d(d\leq 6)}) \\
 \eta^{AB;\CC}\,R_{\exA\exB;\CC} + \omega^{AB}\,R_{\exA,\exB} & (\Edd[7]) \\
 \eta^{AB;\CC}\,R_{\exA\exB;\CC} + f^{AB}{}_C\,R^C_{\exA,\exB} + \kappa^{AB}\,R_{\exA\exB} \quad & (\Edd[8])
\end{cases}\,.
\end{align}
In the following, we omit commutators among negative-level generators because they follow directly from this symmetry. The commutators between positive and negative generators are determined as
\begin{align}
 [R_\exA^A,\,R^\exB_B] &= \delta_\exA^\exB\,(t^{\adja})_B{}^A\, t_{\adja} + \delta^A_B\,\bigl(K^\exB{}_\exA -\Cbeta\,\delta^\exB_\exA\,K^\exC{}_\exC\bigr) \,,
\\
 [R^{\exA_1\exA_2;\AA},\,R_\exB^B] &= 2\,\delta^{\exA_1\exA_2}_{\exB\exC}\,\eta^{BC;\AA}\,R^{\exC}_C \,, 
\\
 [R^{\exA_1\exA_2;\AA},\,R_{\exB_1\exB_2;\BB}] &= 2\,\delta^{\exA_1\exA_2}_{\exB_1\exB_2}\,(t^{\adja})_{\BB}{}^{\AA}\,t_{\adja} -4\,\delta_{\BB}^{\AA}\,\bigl(\delta^{[\exA_1}_{[\exB_1}\,K^{\exA_2]}{}_{\exB_2]} - \Cbeta\,\delta^{\exA_1\exA_2}_{\exB_1\exB_2}\,K^{\exC}{}_{\exC}\bigr) \,, 
\end{align}
by imposing the Jacobi identity. For $\Edd[7]$ and $\Edd[8]$, we find additional commutation relations,
\begin{alignat}{2}
 \Edd[7]&:\quad& [R^{\exA,\exB},\,R_{\exC}^C] &= \delta^{(\exA}_{\exC}\,\delta^{\exB)}_{\exD}\,\omega^{CD}\,R^{\exD}_D\,,
\qquad
 [R^{\exA,\exB},\,R_{\exC_1\exC_2;\CC}] = 0\,,
\\
 && [R^{\exA,\exB},\,R_{\exC,\exD}] &= -2\,\delta_{(\exC}^{(\exA}\,K^{\exB)}{}_{\exD)} +2\,\Cbeta\,\delta^{(\exA}_{\exC}\,\delta^{\exB)}_{\exD}\,K^{\exE}{}_{\exE}\,,
\\
 \Edd[8]&:\quad& [R^{\exA,\exB}_A,\,R_{\exC}^C] &= \delta^{(\exA}_{\exC}\,\delta^{\exB)}_{\exD}\,f_{A}{}^{CD}\,R^{\exD}_D\,,
\qquad 
 [R^{\exA,\exB}_A,\,R_{\exC_1\exC_2;\CC}] =0\,,
\\
 && [R^{\exA\exB},\,R_{\exC}^C] &= \tfrac{1}{2}\,\delta^{\exA\exB}_{\exC\exD}\,\kappa^{CD}\,R^{\exD}_D\,,\qquad
 [R^{\exA\exB},\,R_{\exC_1\exC_2;\CC}] =0\,,
\\
 && [R^{\exA,\exB}_A,\,R_{\exC,\exD}^B] &= - f_A{}^{BC}\,t_C - 2\,\delta_A^B\,\bigl(\delta_{(\exC}^{(\exA}\,K^{\exB)}{}_{\exD)} -\Cbeta\,\delta^{(\exA}_{\exC}\,\delta^{\exB)}_{\exD}\,K^{\exE}{}_{\exE}\bigr)\,,
\\
 && [R^{\exA\exB},\,R_{\exC\exD}] &= - \delta_{[\exC}^{[\exA}\,K^{\exB]}{}_{\exD]} + \Cbeta\,\delta^{\exA\exB}_{\exC\exD}\,K^{\exE}{}_{\exE}\,.
\end{alignat}
Up to this level ($-2\leq \ell\leq 2$), all commutators (except those between $\ell=2$ and $\ell=-2$) for $\Edd[6]$ and $\Edd[7]$ have been summarized in Appendix A of \cite{Tumanov:2014pfa} using different notations that depend on the rank of $\Edd[d]$. To our knowledge, the $\Edd[8]$ algebra determined here has not been presented elsewhere. For $\Odd$, the above results are already complete because there is no level higher than $\ell=2$. 

\paragraph{Level-3 generators.}
We now extend the construction to the level-$3$ generators. Here we focus on the case of $\Edd[d]$ with $d\leq 7$, where the level-$3$ generators are constructed as
\begin{align}
 [R^\exA_A,\,R^{\exB_1\exB_2;\BB}] = Z_{A}{}^{\BB}{}_{\CCC}\,R^{\exA\exB_1\exB_2;\CCC} + \eta_A{}^{\lC;\BB}\,R_{\lC}^{\exB_1\exB_2,\exA} \,,
\label{eq:def-Z}
\end{align}
with
\begin{align}
 \eta_A{}^{\lC;\BB} \equiv \begin{cases}
 0 & (d\leq 5)
\\
 \chi_A^{\BB} & (d=6)
\\
 \eta_A{}^{C;\BB} & (d=7)
\end{cases}\,,\qquad
 R_{\lC}^{\exB_1\exB_2,\exA} \equiv \begin{cases}
 \ \text{---} & (d\leq 5)
\\
 R^{\exB_1\exB_2,\exA} & (d=6)
\\
 R_C^{\exB_1\exB_2,\exA} & (d=7)
\end{cases}\,.
\label{eq:etaAlCBB}
\end{align}
Here $\chi_A^{\BB}$ or $\chi^A_{\BB}$ ($\chi^A_{\CC}\,\chi^{\CC}_B=\delta^A_B$\,, $\chi_C^{\AA}\,\chi^C_{\BB}=\delta^{\AA}_{\BB}$\,, and $\chi_A^{\BB}$ has the same components as $\chi^A_{\BB}$) is the intertwiner between the $R_1$- and $R_2$-representation of $\Edd[6]$. In the $\Edd[7]$ case, we have defined
\begin{align}
 \eta_A{}^{C;\BB}\equiv \eta^{DC;\BB}\,\omega_{DA}\,,\qquad \eta^A{}_{C;\BB}\equiv \eta_{DC;\BB}\,\omega^{DA}\,.
\end{align}
Requiring the Jacobi identities, we can determine the commutators as follows:
\begin{align}
 [R^{\exA_1\exA_2\exA_3;\AAA},\,R_{\exB}^B] &= 3\,\delta^{\exA_1\exA_2\exA_3}_{\exB\exC_1\exC_2}\,Z^B{}_{\CC}{}^{\AAA}\,R^{\exC_1\exC_2;\CC}\,,
\\
 [R^{\exA_1\exA_2\exA_3;\AAA},\,R_{\exB_1\exB_2;\BB}] &= - 3!\,\delta^{\exA_1\exA_2\exA_3}_{\exB_1\exB_2\exC}\,Z^{C}{}_{\BB}{}^{\AAA}\, R^{\exC}_C\,,
\\
 [R^{\exA_1\exA_2\exA_3;\AAA},\,R_{\exB_1\exB_2\exB_3;\BBB}] &= -3!\,\delta^{\exA_1\exA_2\exA_3}_{\exB_1\exB_2\exB_3}\,(t^{\adja})_{\BBB}{}^{\AAA}\,t_{\adja}
\nn\\
 &\quad - 18\,\bigl(\delta^{\exA_1\exA_2\exA_3}_{\exC_1\exC_2\exD}\,\delta^{\exC_1\exC_2\exE}_{\exB_1\exB_2\exB_3}\,K^{\exD}{}_{\exE} - \delta^{\exA_1\exA_2\exA_3}_{\exB_1\exB_2\exB_3}\,L\bigr)\,.
\end{align}
In the $\Edd[6]$ case, we additionally have
\begin{align}
 [R^{\exA_1\exA_2,\exA'},\,R_{\exB}^B] &= \bigl(\delta^{\exA_1\exA_2}_{\exC_1\exC_2}\,\delta^{\exA'}_{\exB}-\delta^{\exA_1\exA_2\exA'}_{\exC_1\exC_2\exB}\bigr)\,\chi^B_{\CC}\,R^{\exC_1\exC_2;\CC}\,,
\\
 [R^{\exA_1\exA_2,\exA'},\,R_{\exB_1\exB_2;\BB}] &= -2\,\bigl(\delta^{\exA_1\exA_2}_{\exB_1\exB_2}\,\delta^{\exA'}_{\exC}-\delta^{\exA_1\exA_2\exA'}_{\exB_1\exB_2\exC}\bigr)\,\chi^C_{\BB}\,R^{\exC}_C\,,
\\
 [R^{\exA_1\exA_2,\exA'},\,R_{\exB_1\exB_2\exB_3;\BBB}] &=0\,,
\\
 [R^{\exA_1\exA_2,\exA'},\,R_{\exB_1\exB_2,\exB'}] &=
 -2\,\bigl(\delta^{\exA_1\exA_2}_{\exB_1\exB_2}\,K^{\exA'}{}_{\exB'} +2\,\delta^{[\exA_1}_{[\exB_1}\,K^{\exA_2]}{}_{\exB_2]}\,\delta^{\exA'}_{\exB'} -3\,\delta^{[\exA_1\exA_2}_{[\exB_1\exB_2}\,K^{\exA']}{}_{\exB']}\bigr)
\nn\\
 &\quad +6\,\Cbeta\,\bigl(\delta^{\exA_1\exA_2}_{\exB_1\exB_2}\,\delta^{\exA'}_{\exB'} - \delta^{\exA_1\exA_2\exA'}_{\exB_1\exB_2\exB'} \bigr)\,L\,,
\end{align}
while for $\Edd[7]$, we also have to take into account 
\begin{align}
 [R^{\exC}_C,\,R^{\exA,\exB}] &= - R^{\exC(\exA,\exB)}_C\,,
\\
 [R^{\exA_1\exA_2,\exA'}_A,\,R_{\exB}^B] &= \bigl(\delta^{\exA_1\exA_2}_{\exC_1\exC_2}\,\delta^{\exA'}_{\exB}-\delta^{\exA_1\exA_2\exA'}_{\exC_1\exC_2\exB}\bigr)\,\eta^B{}_{A\CC}\,R^{\exC_1\exC_2;\CC}
 -2\,\delta^{\exA_1\exA_2}_{\exB(\exC_1}\,\delta^{\exA'}_{\exC_2)}\,\delta_A^B\,R^{\exC_1,\exC_2}\,,
\\
 [R^{\exA_1\exA_2,\exA'}_A,\,R_{\exB_1\exB_2;\BB}] &= -2\,\bigl(\delta^{\exA_1\exA_2}_{\exB_1\exB_2}\,\delta^{\exA'}_{\exC}-\delta^{\exA_1\exA_2\exA'}_{\exB_1\exB_2\exC}\bigr)\,\eta^C{}_{A\BB}\,R^{\exC}_C \,,
\\
[R^{\exA_1\exA_2,\exA'}_A,\,R_{\exB_1\exB_2\exB_3;\BBB}] &=0\,,
\\
 [R^{\exA_1\exA_2,\exA'}_A,\,R_{\exB_1\exB_2,\exB'}^B] &= -2\,\bigl(\delta^{\exA_1\exA_2}_{\exB_1\exB_2}\,\delta^{\exA'}_{\exB'}-\delta^{\exA_1\exA_2\exA'}_{\exB_1\exB_2\exB'}\bigr)\,\bigl[(t^{\adja})_A{}^B\,t_{\adja} - 3\,\Cbeta\,\delta_A^B\,L\bigr]
\nn\\
 &\quad -2\,\bigl(\delta^{\exA_1\exA_2}_{\exB_1\exB_2}\,K^{\exA'}{}_{\exB'} +2\,\delta^{[\exA_1}_{[\exB_1}\,K^{\exA_2]}{}_{\exB_2]}\,\delta^{\exA'}_{\exB'} -3\,\delta^{[\exA_1\exA_2}_{[\exB_1\exB_2}\,K^{\exA']}{}_{\exB']}\bigr)\,\delta_A^B\,,
\\
 [R^{\exA_1\exA_2,\exA'}_A,\,R_{\exB,\exB'}] &= 2\,\delta^{\exA_1\exA_2}_{\exC(\exB}\,\delta^{\exA'}_{\exB')}\,R_A^{\exC}\,.
\end{align}

\paragraph{Identities.}
The Jacobi identities impose several relations on the invariant tensors. If we define the combination
\begin{align}
 \eta_{ABC;\DDD} \equiv \eta_{AB;\EE}\,Z_{C}{}^{\EE}{}_{\DDD}\,,
\end{align}
it should have no totally symmetric contribution, namely
\begin{align}
 \eta_{(ABC);\DDD} = 0\,.
\end{align}
For $\Edd[d]$ with $d\leq 6$, we have explicitly constructed the invariant tensor $Z_{A}{}^{\BB}{}_{\CCC}$ (see Appendix \ref{app:formulas}), and this identity is indeed satisfied. 

For $\Edd[6]$ and $\Edd[7]$, we define the combination
\begin{alignat}{2}
 \Edd[6]&:\quad& c_{ABC} &\equiv \eta_{AB;\DD}\,\chi_C^{\DD}\,,\quad\text{and}
\\
 \Edd[7]&:\quad& c_{ABC}{}^D &\equiv \eta_{AB;\EE}\,\eta_C{}^{D;\EE} + \omega_{C(A}\,\delta_{B)}^D \,,
\end{alignat}
which is totally symmetric in the lower indices $ABC$. We have verified this property using its explicit form. In the $\Edd[7]$ case, this is related to the familiar totally symmetric tensor as
\begin{align}
 \Edd[7] :\quad c_{ABCD} \equiv c_{ABC}{}^E\,\omega_{ED} \,.
\end{align}

Additionally, for the Jacobi identity to hold, the following identities must be satisfied:
\begin{align}
 &2\,\eta^{AE;\CC}\,\eta_{EB;\DD} + 3\, Z^{A}{}_{\DD}{}^{\EEE}\,Z_B{}^{\CC}{}_{\EEE} 
 + (t^{\adja})_A{}^B\,(t_{\adja})_{\CC}{}^{\DD} = 2\,(1+\Cbeta)\,\delta^A_B\,\delta^{\CC}_{\DD}\,,
\\
 &\eta^{AE;\CC}\,\eta_{EB;\DD} - (t^{\adja})_A{}^B\,(t_{\adja})_{\CC}{}^{\DD} + (2\,\Cbeta - 1)\,\delta^A_B\,\delta^{\CC}_{\DD}
 = \eta^A{}_{\lE;\DD}\,\eta_B{}^{\lE;\CC},
\end{align}
where $\eta^A{}_{\lE;\DD}$ is defined in \eqref{eq:etaAlCBB}. We have verified the second identity for $d\leq 7$ and the first for $d\leq 6$ using the explicit form of $Z_A{}^{\BB}{}_{\CCC}$\,. Both are equivalent to
\begin{align}
 Z^{A}{}_{\DD}{}^{\EEE}\,Z_B{}^{\CC}{}_{\EEE} + (t^{\adja})_A{}^B\,(t_{\adja})_{\CC}{}^{\DD} - 2\,\Cbeta \,\delta^A_B\,\delta^{\CC}_{\DD}
 = -\tfrac{2}{3}\,\eta^A{}_{\lE\DD}\,\eta_B{}^{\lE\CC} \,,
\\
 \eta^{AE;\CC}\,\eta_{EB;\DD} + Z^{A}{}_{\DD}{}^{\EEE}\,Z_B{}^{\CC}{}_{\EEE} - \delta^A_B\,\delta^{\CC}_{\DD}
 = \tfrac{1}{3}\,\eta^A{}_{\lE\DD}\,\eta_B{}^{\lE\CC} \,.
\label{eq:Y3-ZZ}
\end{align}
By contracting the indices in the first identity, we find for $3\leq d\leq 7$,
\begin{align}\label{eq:Z-contraction-R2-R3}
 Z^{A}{}_{\CC}{}^{\DDD}\,Z_B{}^{\CC}{}_{\DDD} 
 &= \left\{\begin{array}{ll}
2\,\Cbeta \,\dim \hat{R}_2 & (3\leq d\leq 5) \\
2\,\Cbeta \,\dim \hat{R}_2 -\frac{2}{3} & (d=6) \\
2\,\Cbeta \,\dim \hat{R}_2-\frac{4\,(d-1)\dim \hat{R}_2}{3\dim \hat{R}_1} & (d=7) \end{array}\right\}\,\,\delta^A_B
 = d\,\tfrac{\dim \hat{R}_3}{\dim \hat{R}_1}\,\delta_A^B\,, 
\\
 Z^{C}{}_{\BB}{}^{\DDD}\,Z_C{}^{\AA}{}_{\DDD} 
 &= \left\{\begin{array}{ll}
2\,\Cbeta \,\dim R_1 & (3\leq d\leq 5) \\
2\,\Cbeta \,\dim R_1-\frac{2}{3} & (d=6) \\
2\,\Cbeta \,\dim R_1-\frac{4\,(d-1)}{3} & (d=7) \end{array}\right\}\,\delta_{\BB}^{\AA}
 = d\,\tfrac{\dim \hat{R}_3}{\dim \hat{R}_2}\,\delta_{\BB}^{\AA}\,,
\end{align}
which implies (verified for $d\leq 6$)
\begin{align}
 Z^{A}{}_{\BB}{}^{\DDD}\,Z_{A}{}^{\BB}{}_{\CCC} = d\,\delta_{\CCC}^{\DDD}\,.
\end{align}
An exception is $\Edd[2]$, where \eqref{eq:Z-contraction-R2-R3} is replaced by $Z^{A}{}_{\CC}{}^{\DDD}\,Z_B{}^{\CC}{}_{\DDD} =\frac{2}{3}\,(P_{\irrep{3}})_B{}^C{}_C{}^A$, while the other two identities remain valid. Converting the $\hat{R}_2$-indices of \eqref{eq:Y3-ZZ} to $R_1$-indices, and defining
\begin{align}
 Y^{ABC}_{DEF} \equiv \eta^{ABC;\GGG}\,\eta_{DEF;\GGG}\,,
\end{align}
we further obtain
\begin{align}
 Y^{ABC}_{DEF} = Y^{(AB)}_{(DE)}\,\delta^C_F - Y^{(AB)}_{GF}\,Y^{GC}_{(DE)} + \begin{cases}
0 & (d\leq 5) \\
\tfrac{1}{3}\,c^{ABC}\,c_{DEF} & (d=6) \\
-\tfrac{1}{3}\,\omega^{CH}\,Y^{(AB)}_{HG}\,Y_{(DE)}^{GI}\,\omega_{IF}& (d=7) \end{cases}\,.
\end{align}
By definition, we have $Y^{(ABC)}_{DEF}=0$ (verified also for $d=7$). Symmetrizing, this yields
\begin{align}
 0 = 3\,Y^{(AB}_{(DE)}\,\delta^{C)}_F - 3\,Y^{(AB|}_{GF}\,Y^{G|C)}_{(DE)} + \begin{cases}
0 & (d\leq 5) \\
 c^{ABC}\,c_{DEF} & (d=6) \\
-\omega^{(C|H}\,Y^{|AB)}_{HG}\,Y_{(DE)}^{GI}\,\omega_{IF}& (d=7) \end{cases}\,.
\end{align}

\paragraph{Higher-level generators.}

In principle, we could extend these steps to determine the algebra of higher-level generators using the Jacobi identities, but we restrict our analysis to this level. 

\subsection{Representation}\label{sec:representation}
\paragraph{Basis.}
Here we consider the $R_1$-representation of the extended duality group $\GDM$.
We introduce a basis $\ket{{}^{\Ah}}$ which transforms in the $R_1$-representation,
\begin{align}
 \widehat{t}_{\adjah}\,\ket{{}^{\Ah}} = \ket{{}^{\Bh}}\,(\widehat{t}_{\adjah})_{\Bh}{}^{\Ah}\,.
\end{align}
In the $\Odd$ case, under the branching $\Odd[p]\to \GL(n)\times \Odd$, the $R_1$-representation decomposes as
\begin{align}
 \ket{{}^{\Ah}} = \bigl\{\ket{{}^\exA},\, \ket{{}^A},\, \ket{{}_{\exA}} \bigr\}\,.
\end{align}
For exceptional groups $\Edd$, under the branching $\Edd[p]\to \GL(n)\times \Edd$, the $R_1$-representation contains more components, namely
\begin{align}
 \bigl\{\ket{{}^\exA},\, \ket{{}^A},\, \ket{{}_{\exA\AA}},\, \ket{{}_{\exA_1\exA_2\AAA}}, \cdots\bigr\}\,.
\end{align}
In $\Edd[6]$ and $\Edd[7]$, we additionally have
\begin{align}
\begin{split}
 \Edd[6]:&\quad \{\ket{{}_{\exA_1\exA_2}},\,\ket{{}_{\exA_1,\exA_2}}\}\,,
\\
 \Edd[7]:&\quad \{\ket{{}_{\exA}},\,\ket{{}^A_{\exA_1\exA_2}},\,\ket{{}^A_{\exA_1,\exA_2}}\}\,.
\end{split}
\label{eq:basis-mixed-symmetry}
\end{align}
We embed these basis elements as follows, depending on the value of $d$ in the duality group:
\begin{align}
\begin{split}
 d\leq 5:&\quad \bigl\{\ket{{}^{\Ah}}\bigr\} = \bigl\{\ket{{}^\exA},\, \ket{{}^A},\, \ket{{}_{\exA\AA}},\, \ket{{}_{\exA_1\exA_2\AAA}}, \cdots\bigr\}\,,
\\
 d=6:&\quad \bigl\{\ket{{}^{\Ah}}\bigr\} =\{\ket{{}^\exA},\, \ket{{}^A},\, \ket{{}_{\exA\AA}},\, \ket{{}_{\exA_1\exA_2\AAA}},\,\ket{{}_{\exA_1\exA_2}},\,\ket{{}_{\exA_1,\exA_2}}, \cdots\}\,,
\\
 d=7:&\quad \bigl\{\ket{{}^{\Ah}}\bigr\} =\{\ket{{}^\exA},\, \ket{{}^A},\, \ket{{}_{\exA\AA}},\, \ket{{}_{\exA}},\, \ket{{}_{\exA_1\exA_2\AAA}},\, \ket{{}^A_{\exA_1\exA_2}},\,\ket{{}^A_{\exA_1,\exA_2}}, \cdots \}\,.
\end{split}
\end{align}
For $\Edd[8]$, we consider only the components
\begin{align}
 d=8: \quad \bigl\{\ket{{}^{\Ah}}\bigr\} =\{\ket{{}^\exA},\, \ket{{}^A},\, \ket{{}_{\exA\AA}},\, \ket{{}_{\exA}^A},\, \ket{{}_{\exA}}, \cdots \}\,.
\end{align}
Note that states with $p$-form indices transform in the $C_p$-representation (recall $C_{0}=R_1$). For example, in the $\Edd[7]$ case, $\ket{{}_{\exA_1\exA_2\AAA}}$ and $\ket{{}^A_{\exA_1\exA_2}}$ transform in the $C_2$-representation $\irrep{912}\oplus\irrep{56}$ while in the $\Edd[8]$ case, $\ket{{}_{\exA\AA}}$, $\ket{{}_{\exA}^A}$, and $\ket{{}_{\exA}}$ transform in the $C_1$-representation $\irrep{3875}\oplus \irrep{248}\oplus\irrep{1}$. 

\paragraph{Construction of the representation.}
The vector $\ket{{}^{\Ah}}$ has the weight
\begin{align}
 \Cbeta_p\equiv \begin{cases}0 & (\Odd)\\ \frac{1}{9-p} &(\Edd)
\end{cases}\,,
\end{align}
and the $\mathfrak{gl}(n)$ generators act therefore as
\begin{align}
 K^{\exB}{}_{\exC}\,\ket{{}^{\exA}} &= \delta^{\exA}_{\exC}\,\ket{{}^{\exB}} + \Cbeta_p\,\delta^{\exB}_{\exC}\,\ket{{}^{\exA}}\,,
\\
 K^{\exB}{}_{\exC}\,\ket{{}^{A}} &= \Cbeta_p\,\delta^{\exB}_{\exC}\,\ket{{}^{A}}\,,
\\
 K^{\exB}{}_{\exC}\,\ket{{}_{\exA\AA}} &= - \delta_{\exA}^{\exB}\,\ket{{}_{\exC\AA}} + \Cbeta_p\,\delta^{\exB}_{\exC}\,\ket{{}_{\exA\AA}}\,,
\end{align}
which can be straightforwardly extended to the other basis elements. The $\edd[d]$ generators also act, for example, as
\begin{align}
 t_{\adja}\,\ket{{}^{\exA}} &= 0\,, 
\\
 t_{\adja}\,\ket{{}^{A}} &= (t_{\adja})_B{}^A\, \ket{{}^{B}}\,,
\\
 t_{\adja}\,\ket{{}_{\exA\AA}} &= -(t_{\adja})_{\AA}{}^{\BB}\, \ket{{}_{\exA\BB}}\,.
\end{align}
Since the state $\ket{{}^{\exA}}$ has the highest level ($\ell=1$ for $\Odd$ and $\ell=1+n\,\Cbeta_p$ for $\Edd$), it is annihilated by the positive-level generators
\begin{align}
 R^{\exB}_B\, \ket{{}^{\exA}} = 0\,,\quad
 R^{\exB_1\exB_2;\BB}\, \ket{{}^{\exA}} = 0\,,\quad \cdots\,.
\end{align}
We now introduce the generalized transpose as
\begin{align}
 A^{\tp} \equiv -\theta(A)\qquad \forall A\in \mathrm{Lie}(\GDM)\,,
\end{align}
and introduce the dual basis $\bra{{}_{\Ah}}\equiv \ket{{}^{\Ah}}^{\tp}$ such that the normalization is given by
\begin{align}
 \braket{{}_{\Ah}}{{}^{\Bh}} = \delta^{\Bh}_{\Ah}\,,
\end{align}
where $\delta^{\Bh}_{\Ah}$ is the identity matrix. For both $\Odd$ and $\Edd$ with $d\leq 5$, it takes the form
\begin{align}
 \delta_{\Ah}^{\Bh} = \begin{pmatrix}
 \delta_{\exA}^{\exB} & 0 & 0 & 0 \\
 0 & \delta_A^B & 0 & 0 & \cdots \\
 0 & 0 & \delta^{\exA}_{\exB}\,\delta^{\AA}_{\BB} & 0 \\
 0 & 0 & 0 & 2! \,\delta^{\exA_1\exA_2}_{\exB_1\exB_2}\,\delta^{\AAA}_{\BBB} \\
 & \vdots & & & \ddots
\end{pmatrix}
\end{align}
under the convention for index contractions
\begin{align}
 V^{\Ah}\,W_{\Ah} 
 = V^{\exA}\,W_{\exA} 
 + V^{A}\,W_{A} 
 + V_{\exA\AA}\,W^{\exA\AA}
 + \tfrac{1}{2!}\,V_{\exA_1\exA_2\AAA}\,W^{\exA_1\exA_2\AAA} + \cdots\,.
\end{align}
That is, contractions of $p$ antisymmetric indices are always normalized by $1/p!$. Recalling that the Cartan involution exchanges the positive-level generators with their negative-level counterparts, we find for example
\begin{align}
 (R^{\exB}_B)^{\tp} = - R_{\exB}^B\,,\quad
 (R^{\exB_1\exB_2;\BB})^{\tp} = - R_{\exB_1\exB_2;\BB}\,.
\label{eq:R-transpose}
\end{align}
Hence, the dual basis $\bra{{}_{\exA}}$ should be annihilated by the negative-level generators
\begin{align}
 \bra{{}_{\exA}}\,R_{\exB}^B = -\bigl(R^{\exB}_B\,\ket{{}^{\exA}}\bigr)^{\tp} = 0\,,\quad
 \bra{{}_{\exA}}\,R_{\exB_1\exB_2;\BB} = 0\,,\quad \cdots\,.
\end{align}

Lower-level states arise by acting with negative-level generators on the highest-level states $\ket{{}^{\exA}}$. Following this idea, $R_{\exA}^A\,\ket{{}^{\exB}}$ should be proportional to $\delta_{\exA}^{\exB}\ket{{}^{A}}$\,. 
Considering the normalization of $\ket{{}^{A}}$\,, we have
\begin{align}
 R_{\exA}^A\,\ket{{}^{\exB}} = \delta_{\exA}^{\exB}\ket{{}^{A}}
\end{align}
in our convention. Then, by applying the relation
\begin{align}
 (\widehat{t}_{\adjah})_{\Ah}{}^{\Bh} = \bra{{}_{\Ah}}\,\widehat{t}_{\adjah}\,\ket{{}^{\Bh}} = -\bra{{}_{\Bh}}\,\widehat{t}^{\adjah}\,\ket{{}^{\Ah}}
 = -(\widehat{t}^{\adjah})_{\Bh}{}^{\Ah} \,,
\label{eq:transpose}
\end{align}
where $\widehat{t}^{\adjah}\equiv \theta(\widehat{t}_{\adjah})$, we can determine the explicit matrix elements of the generators $\widehat{t}_{\adjah}$ in the $R_1$-representation. As a result, we find
\begin{align}
 \bra{{}_A}R_{\exC}^C\,\ket{{}^{\exB}} = \delta_{\exC}^{\exB}\,\delta_A^C \,,\qquad
 \bra{{}_{\exA}} R^{\exC}_C \,\ket{{}^{B}} = - \delta_{\exA}^{\exC}\,\delta_C^B\,.
\end{align}
We can continue this computation to determine the matrix representations of the generators to any level we need.

For the next level, we should have
\begin{align}
 R_{\exC_1\exC_2;\CC}\,\ket{{}^{\exB}} = - 2\,\delta_{\exC_1\exC_2}^{\exC\exB}\,|_{\exC\CC} \rangle \,,
\end{align}
where the minus sign is a matter of convention, while the numerical factor is chosen such that $\braket{{}^{\exA\AA}}{{}_{\exB\BB}}=\delta^{\exA}_{\exB}\,\delta^{\AA}_{\BB}$ holds. We then obtain
\begin{align}
 - 2\,\delta_{\exC_1\exC_2}^{\exA\exB}\,\,\delta^{\AA}_{\CC} = \bra{{}^{\exA\AA}}\,R_{\exC_1\exC_2;\CC}\,\ket{{}^{\exB}}\,,\qquad
 2\,\delta^{\exC_1\exC_2}_{\exA\exB}\,\,\delta_{\AA}^{\CC} = \bra{{}_{\exB}}\,R^{\exC_1\exC_2;\CC}\,\ket{{}_{\exA\AA}}\,.
\end{align}
Recalling that $[R_{\exA}^{A},\,R_{\exB}^{B}] = \eta^{AB;\CC}\,R_{\exA\exB;\CC}$ for $\Odd$ or $\Edd[d]$ with $d\leq 6$, we further find
\begin{align}
 \tfrac{1}{d-1}\,\eta_{AB;\CC}\,\bra{{}^{\exC\CC}}\,R_{[\exC_1}^A \, \ket{{}^B}\,\delta^{\exB}_{\exC_2]} = - 2\,\delta_{\exC_1\exC_2}^{\exC\exB}\,,
\end{align}
which implies
\begin{align}
 \bra{{}^{\exA\AA}}\,R_{\exC}^C \, \ket{{}^B} = - \eta^{BC;\AA}\,\delta^{\exA}_{\exC}\,,\qquad
 \bra{{}_B}\,R^{\exC}_C \, \ket{{}_{\exA\AA}} = \eta_{BC;\AA}\,\delta_{\exA}^{\exC}\,.
\end{align}
By iterating this procedure, we finally obtain the matrices
\begin{align}\label{eq:R1-repr-first}
 (R^C_\exC)_{\Ah}{}^{\Bh} &= \begin{pmatrix}
 0 & 0 & 0 & 0 \\
 \delta_\exC^\exB\,\delta_A^C & 0 & 0 & 0 & \cdots \\
 0 & -\eta^{CB;\AA}\,\delta^\exA_\exC & 0 & 0 \\
 0 & 0 & 2!\,\delta^{\exA_1\exA_2}_{\exB\exC}\,Z^C{}_{\BB}{}^{\AAA} & 0 \\
 & \vdots & & & \ddots
\end{pmatrix},
\\
 (R_C^\exC)_{\Ah}{}^{\Bh} &= \begin{pmatrix}
 0 & -\delta^\exC_\exA\,\delta^B_C & 0 & 0 \\
 0 & 0 & \eta_{CA;\BB}\,\delta_\exB^\exC & 0 & \cdots \\
 0 & 0 & 0 & -2!\,\delta_{\exB_1\exB_2}^{\exA\exC}\,Z_C{}^{\AA}{}_{\BBB} \\
 0 & 0 & 0 & 0 \\
 & \vdots & & & \ddots
\end{pmatrix},
\\
 (R_{\exC_1\exC_2;\CC})_{\Ah}{}^{\Bh} &= \begin{pmatrix}
 0 & 0 & 0 & 0 \\
 0 & 0 & 0 & 0 & \cdots \\
 -2!\,\delta^{\exA\exB}_{\exC_1\exC_2}\,\delta^{\AA}_{\CC} & 0 & 0 & 0 \\
 0 & -2!\,\delta^{\exA_1\exA_2}_{\exC_1\exC_2}\,Z^B{}_{\CC}{}^{\AAA} & 0 & 0 \\
 & \vdots & & & \ddots
\end{pmatrix},
\\
 (R^{\exC_1\exC_2;\CC})_{\Ah}{}^{\Bh} &= \begin{pmatrix}
 0 & 0 & -2!\,\delta_{\exA\exB}^{\exC_1\exC_2}\,\delta_{\BB}^{\CC} & 0 \\
 0 & 0 & 0 & 2!\,\delta_{\exB_1\exB_2}^{\exC_1\exC_2}\,Z_A{}^{\CC}{}_{\BBB} & \cdots \\
 0 & 0 & 0 & 0 \\
 0 & 0 & 0 & 0 \\
 & \vdots & & & \ddots
\end{pmatrix},
\\
 (R_{\exC_1\exC_2\exC_3;\CCC})_{\Ah}{}^{\Bh} &= \begin{pmatrix}
 0 & 0 & 0 & 0 \\
 0 & 0 & 0 & 0 & \cdots \\
 0 & 0 & 0 & 0 \\
 3!\,\delta^{\exA_1\exA_2\exB}_{\exC_1\exC_2\exC_3}\,\delta^{\AAA}_{\CCC} & 0 & 0 & 0 \\
 & \vdots & & & \ddots
\end{pmatrix},
\\\label{eq:R1-repr-last}
 (R^{\exC_1\exC_2\exC_3;\CCC})_{\Ah}{}^{\Bh} &= \begin{pmatrix}
 0 & 0 & 0 & -3!\,\delta_{\exA\exB_1\exB_2}^{\exC_1\exC_2\exC_3}\,\delta_{\BBB}^{\CCC} \\
 0 & 0 & 0 & 0 & \cdots \\
 0 & 0 & 0 & 0 \\
 0 & 0 & 0 & 0 \\
 & \vdots & & & \ddots
\end{pmatrix} ,
\end{align}
for $\Odd$ or $\Edd[d]$ with $d\leq 5$. For $\Odd$ one has to restrict the matrices to the first three rows and columns. The antisymmetrized Kronecker delta $\delta^{\exA_1\cdots \exA_p}_{\exB_1\cdots \exB_p}$ always appears with the factor $p!\,\delta^{\exA_1\cdots \exA_p}_{\exB_1\cdots \exB_p}$ while other non-trivial numerical coefficients do not appear. As we see from \eqref{eq:transpose}, the generalized transpose $^{\tp}$ exchanges rows and columns, $\Ah\leftrightarrow \Bh$, and flips the index position upside-down. Moreover, an additional minus sign is needed due to \eqref{eq:R-transpose}. Examining the matrices in \eqref{eq:R1-repr-first}--\eqref{eq:R1-repr-last}, one easily observes the resulting relation between positive and negative level generators.

\paragraph{Larger duality groups: $\Edd[6]$, $\Edd[7]$, and $\Edd[8]$.}
In $\Edd[6]$ and $\Edd[7]$, the matrix representations are slightly more complicated. Here we find that it is convenient to introduce additional highlighted factors for the contraction of indices,
\begin{align}
 \Edd[6]:\quad 
 V^{\Ah}\,W_{\Ah} 
 &= V^{\exA}\,W_{\exA} 
 + V^{A}\,W_{A} 
 + V_{\exA\AA}\,W^{\exA\AA}
 + \tfrac{1}{2!}\,V_{\exA_1\exA_2\AAA}\,W^{\exA_1\exA_2\AAA}
\nn\\
 &\quad + \textcolor{red}{\tfrac{1}{12}}\,V_{\exA_1\exA_2}\,W^{\exA_1\exA_2}
 + \textcolor{red}{\tfrac{1}{4}}\,V_{\exA,\exA'}\,W^{\exA,\exA'}
 + (\text{higher levels}) \,,
\\
 \Edd[7]:\quad 
 V^{\Ah}\,W_{\Ah} 
 &= V^{\exA}\,W_{\exA} 
 + V^{A}\,W_{A} 
 + V_{\exA\AA}\,W^{\exA\AA}
 + \textcolor{red}{\tfrac{1}{2}}\,V_{\exA}\,W^{\exA}
 + \tfrac{1}{2!}\,V_{\exA_1\exA_2\AAA}\,W^{\exA_1\exA_2\AAA}
\nn\\
 &\quad + \textcolor{red}{\tfrac{1}{12}}\,V^A_{\exA_1\exA_2}\,W_A^{\exA_1\exA_2}
 + \textcolor{red}{\tfrac{1}{4}}\,V^A_{\exA,\exA'}\,W_A^{\exA,\exA'}
 + (\text{higher levels}) \,.
\end{align}
With them, the identity matrix $\delta_{\Ah}^{\Bh}$ and the level-$0$ generators in our convention are
\begin{align}
 \delta_{\Ah}^{\Bh} &= \left(\begin{array}{ccc|c|cccc}
 \delta_{\exA}^{\exB} & 0 & 0 & 0 &0 &0 &0 \\
 0 & \delta_A^B & 0 & 0 &0 &0 &0 & \cdots \\
 0 & 0 & \delta^{\exA}_{\exB}\,\delta^{\AA}_{\BB} & 0 &0 &0 &0 \\ \hline
 0 & 0 & 0 & 2 \,\delta^{\exA}_{\exB} &0 &0 &0 \\ \hline
 0 & 0 & 0 & 0 &2! \,\delta^{\exA_1\exA_2}_{\exB_1\exB_2}\,\delta^{\AAA}_{\BBB} &0 &0 \\
 0 & 0 & 0 & 0 &0 &12\,\delta^{\exA_1\exA_2}_{\exB_1\exB_2}\,\delta_{\lA}^{\lB} &0 \\
 0 & 0 & 0 & 0 &0 &0 &4\,\delta^{(\exA}_{(\exB}\delta^{\exA')}_{\exB')}\,\delta_{\lA}^{\lB} \\
 & \vdots & & & &&&\ddots
\end{array}\right) ,
\\
 (K^{\exC}{}_{\exD})_{\Ah}{}^{\Bh} &= \left(\begin{array}{ccc|c|cccc}
 \delta_{\exA}^{\exC}\,\delta^{\exB}_{\exD} & 0 & 0 & 0 &0 &0 &0 \\
 0 & 0 & 0 & 0 &0 &0 &0 & \cdots \\
 0 & 0 & -\delta^{\exA}_{\exD}\,\delta_{\exB}^{\exC}\,\delta^{\AA}_{\BB} & 0 &0 &0 &0 \\ \hline
 0 & 0 & 0 & -2\,\delta^{\exA}_{\exD}\,\delta_{\exB}^{\exC} &0 &0 &0 \\ \hline
 0 & 0 & 0 & 0 &\mathsf{K}_2\,\delta^{\AAA}_{\BBB} &0 &0 \\
 0 & 0 & 0 & 0 &0 &12\,\mathsf{K}_2\,\delta_A^B &0 \\
 0 & 0 & 0 & 0 &0 &0 &4\,\mathsf{K}_{1,1}\,\delta_A^B \\
 & \vdots & & & &&&\ddots
\end{array}\right) + \Cbeta_p\,\delta_{\Ah}^{\Bh} 
\nn\\
 &\quad \bigl[\mathsf{K}_2 \equiv -4\,\delta_{\exD\exE}^{\exA_1\exA_2}\,\delta^{\exC\exE}_{\exB_1\exB_2}\,,\quad
 \mathsf{K}_{1,1} \equiv - \delta^{\exA}_{\exD}\,\delta_{\exB}^{\exC}\,\delta^{\exA'}_{\exB'}
 - \delta^{\exA}_{\exB}\,\delta^{\exA'}_{\exD}\,\delta_{\exB'}^{\exC} \bigr]\,,
\\
 (t_{\adja})_{\Ah}{}^{\Bh} &= \left(\begin{array}{ccc|c|cccc}
 0 & 0 & 0 & 0 &0 &0 &0 \\
 0 & (t_{\adja})_A{}^B & 0 & 0 &0 &0 &0 & \cdots \\
 0 & 0 & -\delta^{\exA}_{\exB}\,(t_{\adja})_{\BB}{}^{\AA} & 0 &0 &0 &0 \\ \hline
 0 & 0 & 0 & 0 &0 &0 &0 \\ \hline
 0 & 0 & 0 & 0 &-2!\,\delta^{\exA_1\exA_2}_{\exB_1\exB_2}\,(t_{\adja})_{\BBB}{}^{\AAA} &0 &0 \\
 0 & 0 & 0 & 0 &0 & 12\,\delta^{\exA_1\exA_2}_{\exB_1\exB_2}\,(t_{\adja})_{\lA}{}^{\lB} &0 \\
 0 & 0 & 0 & 0 &0 &0 & 4\,\delta^{(\exA}_{(\exB}\delta^{\exA')}_{\exB')}\,(t_{\adja})_{\lA}{}^{\lB} \\
 & \vdots & & & &&&\ddots
\end{array}\right) ,
\end{align}
where the highlighted fourth row/column should be truncated in the $\Edd[6]$ case, and
\begin{align}
 \delta_{\lA}^{\lB} \equiv \begin{cases} 1 & (\Edd[6]) \\
 \delta_A^B & (\Edd[7])
\end{cases}\,,\qquad
 (t_{\adja})_{\lA}{}^{\lB} \equiv \begin{cases} 0 & (\Edd[6]) \\
 (t_{\adja})_{A}{}^{B} & (\Edd[7])
\end{cases}\,.
\end{align}
For the negative-level generators we find
\begin{align}
 (R^C_\exC)_{\Ah}{}^{\Bh} &= \left(\begin{array}{ccc|c|cccc}
 0 & 0 & 0 & 0 &0 &0 &0 \\
 \delta_\exC^\exB\,\delta_A^C & 0 & 0 & 0 &0 &0 &0 & \cdots \\
 0 & -\eta^{CB;\AA}\,\delta^\exA_\exC & 0 & 0 &0 &0 &0 \\ \hline
 0 & \omega^{BC}\,\delta^{\exA}_{\exC} & 0 & 0 &0 &0 &0 \\ \hline
 0 & 0 & 2\,\delta^{\exA_1\exA_2}_{\exB\exC}\,Z^C{}_{\BB}{}^{\AAA} & 0 &0 &0 &0 \\
 0 & 0 & 2\,\delta^{\exA_1\exA_2}_{\exB\exC}\,\eta^{C}{}_{\lA\BB} & 6\,\delta^{\exA_1\exA_2}_{\exB\exC}\,\delta_A^C &0 &0 &0 \\
 0 & 0 & 2\,\delta^{(\exA}_{\exB}\,\delta^{\exA')}_{\exC}\,\eta^{C}{}_{\lA\BB} & -2\,\delta^{(\exA}_{\exB}\delta^{\exA')}_{\exC}\,\delta_A^C &0 &0 &0 \\
 & \vdots & & & &&&\ddots
\end{array}\right),
\\
 (R_{\exC_1\exC_2;\CC})_{\Ah}{}^{\Bh} &= \left(\begin{array}{ccc|c|cccc}
 0 & 0 & 0 & 0 &0 &0 &0 &\\
 0 & 0 & 0 & 0 &0 &0 &0 & \cdots \\
 -2\,\delta^{\exA\exB}_{\exC_1\exC_2}\,\delta^{\AA}_{\CC} & 0 & 0 &0 &0 &0 & 0 \\ \hline
 0 & 0 & 0 & 0 &0 &0 &0 & \\ \hline
 0 & -2\,\delta^{\exA_1\exA_2}_{\exC_1\exC_2}\,Z^B{}_{\CC}{}^{\AAA} & 0 &0 &0 &0 & 0 \\
 0 & 4\,\delta^{\exA_1\exA_2}_{\exC_1\exC_2}\,\eta^B{}_{\lA\CC} & 0 & 0 &0 &0 &0 & \\
 0 & 0 & 0 & 0 &0 &0 &0 & \\
 & \vdots & & & &&&\ddots
\end{array}\right),
\\
 (R_{\exC,\exC'})_{\Ah}{}^{\Bh} &= \left(\begin{array}{ccc|c|cccc}
 0 & 0 & 0 & 0 &0 &0 &0 \\
 0 & 0 & 0 & 0 &0 &0 &0 & \cdots \\
 0 & 0 & 0 & 0 &0 &0 &0 \\ \hline
 -2\,\delta^{\exA}_{(\exC} \delta^{\exB}_{\exC')} & 0 & 0 & 0 &0 &0 &0 \\ \hline
 0 & 0 & 0 & 0 &0 &0 &0 \\
 0 & 0 & 0 & 0 &0 &0 &0 \\
 0 & 2\,\delta^{\exA}_{(\exC}\delta^{\exA'}_{\exC')}\,\delta_A^B & 0 & 0 &0 &0 &0 \\
 & \vdots & & & &&& \ddots
\end{array}\right)\qquad (\Edd[7]\text{ only}),
\\
 (R_{\exC_1\exC_2\exC_3;\CCC})_{\Ah}{}^{\Bh} &= \left(\begin{array}{ccc|c|cccc}
 0 & 0 & 0 & 0 &0 &0 &0 \\
 0 & 0 & 0 & 0 &0 &0 &0 &\cdots \\
 0 & 0 & 0 & 0 &0 &0 &0 \\ \hline
 0 & 0 & 0 & 0 &0 &0 &0 \\ \hline
 3!\,\delta^{\exA_1\exA_2\exB}_{\exC_1\exC_2\exC_3}\,\delta^{\AAA}_{\CCC} & 0 & 0 & 0 &0 &0 &0 \\
 0 & 0 & 0 & 0 &0 &0 &0 \\
 0 & 0 & 0 & 0 &0 &0 &0 \\
 & \vdots & & & &&&\ddots
\end{array}\right),
\\
 (R^{\lC}_{\exC_1\exC_2,\exC'})_{\Ah}{}^{\Bh} &= \left(\begin{array}{ccc|c|cccc}
 0 & 0 & 0 & 0 &0 &0 &0 \\
 0 & 0 & 0 & 0 &0 &0 &0 &\cdots \\
 0 & 0 & 0 & 0 &0 &0 &0 \\ \hline
 0 & 0 & 0 & 0 &0 &0 &0 \\ \hline
 0 & 0 & 0 & 0 &0 &0 &0 \\
 -3!\,(\delta^{\exA_1\exA_2}_{\exC_1\exC_2}\,\delta^{\exB}_{\exC'} - \delta^{\exA_1\exA_2\exB}_{\exC_1\exC_2\exC'})\,\delta_{\lA}^{\lC} & 0 & 0 & 0 &0 &0 &0 \\
 4\,\delta^{\exB(\exA}_{\exC_1\exC_2}\,\delta^{\exA')}_{\exC'}\,\delta_{\lA}^{\lC} & 0 & 0 & 0 &0 &0 &0 \\
 & \vdots & & & &&&\ddots
\end{array}\right)\,,
\end{align}
while the positive-level generators arise after taking the generalized transpose. In the $\Edd[8]$ case, we use the convention
\begin{align}
 \Edd[8]:\quad 
 V^{\Ah}\,W_{\Ah} 
 &= V^{\exA}\,W_{\exA} 
 + V^{A}\,W_{A} 
 + V_{\exA\AA}\,W^{\exA\AA}
 + \textcolor{red}{\tfrac{1}{2}}\,V^A_{\exA}\,W_A^{\exA}
 + \textcolor{red}{\tfrac{1}{4}}\,V_{\exA}\,W^{\exA}
\nn\\
 &\quad + (\text{higher levels}) \,,
\end{align}
and the relevant matrix representations of the generators are determined as
\begin{align}
 (R^C_\exC)_{\Ah}{}^{\Bh} &= \begin{pmatrix}
 0 & 0 & 0 & 0 &0 \\
 \delta_\exC^\exB\,\delta_A^C & 0 & 0 & 0 &0 & \cdots \\
 0 & -\eta^{BC;\AA}\,\delta^\exA_\exC & 0 & 0 &0 \\
 0 & f_A{}^{BC}\,\delta^{\exA}_{\exC} & 0 & 0 &0 \\
 0 & -\kappa^{BC}\,\delta^{\exA}_{\exC} & 0 & 0 &0 \\
 & \vdots & & & & \ddots
\end{pmatrix} ,
\\
 (R_{\exC_1\exC_2;\CC})_{\Ah}{}^{\Bh} &= \begin{pmatrix}
 0 & 0 & 0 & 0 &0 & \\
 0 & 0 & 0 & 0 &0 & \cdots \\
 -2\,\delta^{\exA\exB}_{\exC_1\exC_2}\,\delta^{\AA}_{\CC} & 0 & 0 &0 &0 \\
 0 & 0 & 0 & 0 &0 & \\
 0 & 0 & 0 & 0 &0 & \\
 & \vdots & & & & \ddots\end{pmatrix},
\\
 (R^C_{\exC,\exC'})_{\Ah}{}^{\Bh} &= \begin{pmatrix}
 0 & 0 & 0 & 0 &0 \\
 0 & 0 & 0 & 0 &0 & \cdots \\
 0 & 0 & 0 & 0 &0 \\
 -2\,\delta^{\exA}_{(\exC} \delta^{\exB}_{\exC')}\,\delta_A^C & 0 & 0 & 0 &0 \\
 0 & 0 & 0 & 0 &0 \\
 & \vdots & & & & \ddots
\end{pmatrix} ,
\\
 (R_{\exC_1\exC_2})_{\Ah}{}^{\Bh} &= 
\begin{pmatrix}
 0 & 0 & 0 & 0 &0 \\
 0 & 0 & 0 & 0 &0 & \cdots \\
 0 & 0 & 0 & 0 &0 \\
 0 & 0 & 0 & 0 &0 \\
 -2\,\delta^{\exA\exB}_{\exC_1\exC_2} & 0 & 0 & 0 &0 \\
 & \vdots & & & & \ddots
\end{pmatrix} .
\end{align}

\paragraph{Summary.}

In this section, we have unified $\GL(n)$ and $\GD$ into the extended duality group $\GDM$ and decomposed its generators into irreducible representations of $\GL(n)\times \GD$. In principle, the number of generators can be infinite. Focusing on the lowest few levels, we have determined all commutation relations. Moreover, we constructed the matrix representations of these generators in the $R_1$-representation. These serve as the building blocks for the generalized frame field discussed in section \ref{sec:frames-curvatures}. The matrices in other representations, such as $\hat{R}_2$- or $\hat{R}_3$-representations, can be determined by using the relations in \eqref{eq:invariant-tensors}. 

As a side remark, we note that similar matrix representations have been used in \cite{Aldazabal:2013via} to reproduce the tensor hierarchy of gauged supergravity from ExFT. However, in that work, the generators appearing in \eqref{eq:generators-mixed-symmetry} and the corresponding basis elements from \eqref{eq:basis-mixed-symmetry} were not considered, leading to certain issues. We expect that these issues can be resolved by using the matrix representations constructed here.

\subsection{Useful formulas for \texorpdfstring{$\Edd$}{Ed(d)}}
Before closing this section, we find some algebraic identities that are useful later. For $\GD=\Edd[d]$\,, the generalized Lie derivative mediates an infinitesimal $\Edd$ transformation along with a scale transformation generated by $t_0$\,. The latter complicates the computation of generalized fluxes. To simplify matters, we introduce shifted generators. 

The generalized Lie derivative can be expressed as
\begin{align}
 \gLie_{E_A} E_B{}^I = E_A{}^J\,\partial_J E_B{}^I + (t^{\adji})_{K}{}^J\,(t_{\adji})_L{}^I \,\partial_J E_A{}^K\, E_B{}^L
 + \Cbeta\, \partial_J E_A{}^J\, E_B{}^I\,,
\end{align}
and the generalized fluxes take the form
\begin{align}
 X_{AB}{}^C = W_{AB}{}^C 
 + (t^{\adja})_{D}{}^E \,W_{EA}{}^D\,(t_{\adja})_B{}^C
 - \Cbeta\, W_{DA}{}^D\, (t_0)_B{}^C\,,
\end{align}
where we have defined the Weitzenb\"ock connection in flat indices associated with $E_A{}^I$ as
\begin{align}
 W_{AB}{}^{C} \equiv - E_{A}{}^J\,\partial_J E_{B}{}^{I}\,E_{I}{}^C\,.
\end{align}
We now define the shifted generators $s_{\adja}$ and $s^{\adja}$ to rewrite the generalized fluxes as
\begin{align}
 X_{AB}{}^C = W_{AB}{}^C + (s^{\adja})_{D}{}^E \,W_{EA}{}^D\,(s_{\adja})_B{}^C \,,
\label{eq:X-simple}
\end{align}
which resembles the case of $\GD=\Odd$ or $\GD=\GL(d)$ where $\Cbeta=0$. 

To perform this rewriting explicitly, we first impose the section condition. For $\GD=\Edd[d]$\,, there are two solutions called the M-theory section and the type IIB section, where fields depend on $d$ coordinates or $(d-1)$ coordinates, respectively. In the M-theory section, we split the generalized coordinates $y^I$ as
\begin{align}
 y^I=\{y^i,\,y_{ij},\,y_{i_1\cdots i_5},\,\cdots \}\,,
\end{align}
and solve the section condition by imposing that fields depend only on $y^i$ ($i=1,\dotsc,d$). The coordinates are grouped into irreducible representations of the $\GL(d)$ subgroup generated by $K^i{}_j$\,. In the type IIB section, we decompose $y^I$ as
\begin{align}
 y^I=\{y^m,\,y_m^{\bm{\alpha}},\,y_{m_1m_2m_3},\,y_{m_1\cdots m_5}^{\bm{\alpha}},\,\cdots \}\quad (\bm{\alpha}=1,2)\,,
\end{align}
and fields depend only on $y^m$ ($m=1,\dotsc,d-1$). In this alternative arrangement, groups of coordinates transform in irreducible representations of the $\GL(d-1)\times \SL(2)$ subgroup, which are generated by $K^m{}_n$ and $R^{\bm{\alpha}}{}_{\bm{\beta}}$\,. The generators $t_{\adji}$ decompose accordingly in each solution of the section condition as
\begin{align}
 t_{\adji} &= \{\cdots,\,R_{i_1i_2i_3},\,K^i{}_j,\,R^{i_1i_2i_3},\,\cdots\}
 & & \text{(M-theory)}\,,
\\
 t_{\adji} &= \{\cdots,\,R_{m_1m_2}^{\bm{\alpha}},\,K^m{}_n,\,R^{\bm{\alpha}}{}_{\bm{\beta}},\,R^{m_1m_2}_{\bm{\alpha}},\,\cdots\}
 & & \text{(type IIB)}\,,
\intertext{and the dual generators are similarly decomposed as}
 t^{\adji} &= \{\cdots,\,R^{i_1i_2i_3},\,K_i{}^j,\,R_{i_1i_2i_3},\,\cdots\}
 & & \text{(M-theory)}\,,
\\
 t^{\adji} &= \{\cdots,\,R^{m_1m_2}_{\bm{\alpha}},\,K_m{}^n,\,R_{\bm{\alpha}}{}^{\bm{\beta}},\,R_{m_1m_2}^{\bm{\alpha}},\,\cdots\}
 & & \text{(type IIB)}\,.
\end{align}
If we construct their matrix representations, we find that the $i$--$j$ or $m$--$n$ components of $(K^k{}_l)_I{}^J$ or $(K^p{}_q)_I{}^J$ are
\begin{align}
 (K^k{}_l)_i{}^j = \delta^k_i\,\delta_l^j + \Cbeta\,\delta^k_l\,\delta_i^j \qquad \text{or} \qquad
 (K^p{}_q)_m{}^n = \delta^p_m\,\delta_q^n + \Cbeta\,\delta^p_q\,\delta_m^n\,. 
\end{align}
Compared to the standard $\GL(d)$ generators from \eqref{eq:K-GL}, this includes a shift term proportional to the identity matrix. To avoid carrying around this shift term explicitly, we introduce the refined generators
\begin{align}
 \widetilde{K}^i{}_j \equiv K^i{}_j + \Cbeta\,\delta^i_j\,t_0\,,\qquad
 \widetilde{K}^m{}_n \equiv K^m{}_n + \Cbeta\,\delta^m_n\,t_0
\end{align}
to define the shifted generators
\begin{align}
 s_{\adji} &= \{\cdots,\,R_{i_1i_2i_3},\,\widetilde{K}^i{}_j,\,R^{i_1i_2i_3},\,\cdots\}\,,
\\
 s_{\adji} &= \{\cdots,\,R_{m_1m_2}^{\bm{\alpha}},\,\widetilde{K}^m{}_n,\,R^{\bm{\alpha}}{}_{\bm{\beta}},\,R^{m_1m_2}_{\bm{\alpha}},\,\cdots\}\,,
\end{align}
and their shifted dual generators
\begin{align}
 s^{\adji} &= \{\cdots,\,R^{i_1i_2i_3},\,\widetilde{K}_i{}^j,\,R_{i_1i_2i_3},\,\cdots\}\,,
\\
 s^{\adji} &= \{\cdots,\,R^{m_1m_2}_{\bm{\alpha}},\,\widetilde{K}_m{}^n,\,R_{\bm{\alpha}}{}^{\bm{\beta}},\,R_{m_1m_2}^{\bm{\alpha}},\,\cdots\}\,,
\end{align}
where $\widetilde{K}_i{}^j\equiv -\widetilde{K}^j{}_i$ and $\widetilde{K}_m{}^n\equiv -\widetilde{K}^n{}_m$\,. 
Then, we find
\begin{align}
 (t^{\adji})_{L}{}^M \, (t_{\adji})_J{}^K - \Cbeta\,\delta_L^M \, (t_0)_J{}^K
 &= (s^{\adji})_{L}{}^M \, (s_{\adji})_J{}^K
\nn\\
 &\quad + \begin{cases} \frac{1}{9}\,(\widetilde{K}^i{}_i+t_0)_{L}{}^M\,(\widetilde{K}^j{}_j+t_0)_J{}^K & \text{(M-theory)} \\ 
 \frac{1}{8}\,(\widetilde{K}^m{}_m+t_0)_{L}{}^M\,(\widetilde{K}^m{}_m+t_0)_J{}^K & \text{(type IIB)} 
\end{cases}\,.
\end{align}
Under the section condition, fields depend only on $y^i$ or $y^m$. Thus, the second line does not contribute to the generalized Lie derivative because $(\widetilde{K}^i{}_i+t_0)_{L}{}^K\,\partial_K = (\widetilde{K}^i{}_i+t_0)_{L}{}^k\,\partial_k = 0$ for M-theory. A similar expression holds for type IIB, and we conclude that the simple expression \eqref{eq:X-simple} for the generalized fluxes is indeed correct.

When we consider the extended duality group $\GDM=\Edd[p]$ and decompose the generators under $\GL(n)\times \Edd$, we find 
\begin{align}
 \widehat{s}^{\adjah} \, \widehat{s}_{\adjah} 
 = \cdots + R_{\exA}^A\,R^{\exA}_A + \widetilde{K}_{\exA}{}^{\exB} \, \widetilde{K}^{\exA}{}_{\exB} + s^{\adja}\,s_{\adja} + R^{\exA}_A\,R_{\exA}^A + \cdots\,,
\label{eq:hat-s-decomp}
\end{align}
for the product of shifted generators $\widehat{s}_{\adjah}$, where $\widetilde{K}^{\exA}{}_{\exB}\equiv K^{\exA}{}_{\exB}+\Cbeta_p\,t_0$\,, $\widetilde{K}_{\exA}{}^{\exB} \equiv -\widetilde{K}^{\exB}{}_{\exA}$, and $s^{\adja}$ and $s_{\adja}$ are the shifted generators of $\Edd$ we just discussed. On their own, the latter do not form a closed algebra,\footnote{$s_{\adja}$ together with $t_0$ form the Lie algebra of $\Edd[d]\times \mathbb{R}^+$.} but $\widetilde{K}^{\exA}{}_{\exB}$ satisfy the $\mathfrak{gl}(n)$ algebra. We also note that the matrix representations of $\widetilde{K}^{\exA}{}_{\exB}$ and $s_{\adja}$ are block diagonal, and we have
\begin{align}
 (\widetilde{K}^{\exC}{}_{\exD})_{\exA}{}^{\exB} = \delta_{\exA}^{\exC}\,\delta_{\exD}^{\exB}\,,\qquad
 (\widetilde{K}^{\exC}{}_{\exD})_A{}^B = 0\,,\qquad
 (s_{\adja})_{\exA}{}^{\exB} = 0\,,\qquad
 (s^{\adja})_{\exA}{}^{\exB} = 0\,.
\label{eq:K-s-properties}
\end{align}
These two relations will prove very useful in the next section.

\section{Generalized frames and curvatures}\label{sec:frames-curvatures}
Having settled the algebraic part of the construction in the previous section, we now extend the discussion of section \ref{sec:GL} to double geometry or exceptional geometry with $\GD=\Odd$ or $\GD=\Edd[d]$, respectively. The layout of this section closely follows that of sections~\ref{sec:GL-approach}--\ref{sec:GL-curvatures} as we essentially repeat the steps discussed there within the setting of generalized geometry.

\subsection{Our approach}
We consider an extended duality group, $\Odd[p]$ or $\Edd[p]$\,. We assume that the section condition holds, and that all fields are defined on a mega-space $\hat{M}$ with coordinates $y^{\hat{i}}$, whose dimension is $p$ for $\Odd[p]$ or the M-theory section of $\Edd[p]$, while it is $p-1$ for the type IIB section of exceptional duality groups.

\subsubsection{Setup}
Following the discussion in section~\ref{sec:setup}, we begin with three assumptions.
\paragraph{Assumption 1.}
We assume that $\GS$ acts freely and properly from the left on $\hat{M}$. An infinitesimal left action is generated by the right-invariant vector fields $\Eh_{\exA}{}^{\Ih}$ ($\exA=1,\dotsc,n\equiv \dim \GS$), which satisfy the Lie algebra of $\GS$
\begin{align}
 \GLie_{\Eh_{\exA}} \Eh_{\exB}{}^{\Ih} = - \hat{X}_{\exA\exB}{}^{\exC}\,\Eh_{\exC}{}^{\Ih}\,,
\label{eq:GS-genLie}
\end{align}
where $\GLie$ is the generalized Lie derivative for $\GDM$ and $\hat{X}_{\exA\exB}{}^{\exC}=\hat{X}_{[\exA\exB]}{}^{\exC}$ are the structure constants of $\GS$. We also assume that they are isotropic and thus satisfy
\begin{align}
 \hat{Y}_{\Kh\Lh}^{\Ih\Jh}\,\Eh_{\exA}{}^{\Kh}\,\Eh_{\exB}{}^{\Lh} = 0\,.
\label{eq:isotropy}
\end{align}
This isotropy condition is trivial for the $\GL(d)$ case because $\hat{Y}_{\Kh\Lh}^{\Ih\Jh}=0$.
However, in general, this condition ensures that, up to a $\GD$ transformation, the matrix $\Eh_{\exA}$ takes the form
\begin{align}
 \Eh_{\exA}{}^{\Ih} = \begin{pmatrix} \widetilde{e}_{\exA}{}^{\mu} & 0 & \cdots & 0 \end{pmatrix},
\end{align}
where we have decomposed the local coordinates as $(y^{\hat{i}})=(y^\mu,\,y^i)$ and the $R_1$-representation is decomposed as $(\Eh_{\exA}{}^{\Ih})=\begin{pmatrix} \Eh_{\exA}{}^{\mu} & \Eh_{\exA}{}^{I} & \Eh_{\exA \mu \II} & \cdots \end{pmatrix}$. 
One can regard $\Eh_{\exA}$ as the $\exA$ components of the generalized frame field
\begin{align}
 \Eh_{\Ah}{}^{\Ih} &= \bar{E}_{\Ah}{}^{\Jh}\, \cN_{\Jh}{}^{\Ih}\,,
\end{align}
where $\bar{E}_{\exA}{}^{\Jh}$ is a lower-triangular matrix
\begin{align}
 \bar{E}_{\Ah}{}^{\Jh} = \begin{pmatrix} \widetilde{e}_{\exA}{}^{\mu} & 0 & \cdots \\ \vdots & \ddots \end{pmatrix},
\end{align}
and $\cN$ is an upper-triangular matrix generated by the positive-level generators
\begin{align}
 \cN = \exp\bigl(c_{\exA}^A\,R^{\exA}_A\bigr) \exp\bigl(\tfrac{1}{2}\,c_{\exA\exB;\AA}\,R^{\exA\exB;\AA}\bigr)\cdots\,.
\end{align}
If the potentials ($c_{\exA}^A$, $c_{\exA\exB;\AA}$, $\cdots$) have non-vanishing field strengths ($f_{\exA\exB}^A$, $f_{\exA\exB\exC;\AA}$, $\cdots$), then $\GLie_{\Eh_{\exA}} \Eh_{\exB}{}^{\Ih}$ would contain terms proportional to $f_{\exA\exB}^C\,E_{C}$ or $f_{\exA\exB\exC;\BB}\,E^{\exC;\BB}$.
However, the constraint \eqref{eq:GS-genLie} requires that these terms vanish. Hence, in our setup, these field strengths are zero, and we can eliminate the potentials by a suitable gauge choice. In this gauge, $\cN$ becomes the identity matrix, and the generalized frame field takes the lower-triangular form
\begin{align}
 \Eh_{\Ah}{}^{\Ih} = \begin{pmatrix} \widetilde{e}_{\exA}{}^{\mu} & 0 & \cdots \\ \vdots & \ddots \end{pmatrix}.
\end{align}
In particular, the vector fields $\Eh_{\exA}$ then take the form
\begin{align}
 \Eh_{\exA} = \widetilde{e}_{\exA}{}^\nu(y^\mu) \,\partial_\nu\qquad (\mu=1,\cdots,n)\,,
\label{eq:RI}
\end{align}
which matches the form already obtained in \eqref{eq:hat-e-exA}.
As in section \ref{sec:GL}, we regard $y^\mu$ as the coordinates on $\GS$, and $y^i$ as the coordinates on the physical space $M\equiv \GS\backslash \hat{M}$. 

\paragraph{Assumption 2.}
We further assume that the generalized frame field $\Eh_{\Ah}{}^{\Ih}$ satisfies
\begin{align}
 \GLie_{\Eh_{\exA}} \Eh_{\Bh}{}^{\Ih} = (\mathfrak{t}_{\exA})_{\Bh}{}^{\Ch}\,\Eh_{\Ch}{}^{\Ih}\,,
\label{eq:Lie-E-assumption}
\end{align}
where $(\mathfrak{t}_{\exA})_{\Bh}{}^{\Ch} = - (\hat{X}_{\exA})_{\Bh}{}^{\Ch} = - \hat{X}_{\exA\Bh}{}^{\Ch}$ is a constant matrix spanned by the (shifted) level-$0$ generators $\widetilde{K}^{\exA}{}_{\exB}$ and $s_{\adja}$ as
\begin{align}
 \hat{X}_{\exA} = \hat{X}_{\exA\exB}{}^{\exC}\,\widetilde{K}^{\exB}{}_{\exC} + \hat{X}_{\exA}{}^{\adja}\,s_{\adja}\,,\qquad 
 \widetilde{K}^{\exB}{}_{\exC} \equiv K^{\exB}{}_{\exC} +\Cbeta_p\,\delta^{\exB}_{\exC}\,t_0\,,
\label{eq:hatX-exA}
\end{align}
where $(\hat{X}_{\exA})_{\exB}{}^{\exC} = \hat{X}_{\exA\exB}{}^{\exC}$ due to \eqref{eq:K-s-properties}. No positive-level generators are allowed to appear here. Otherwise, the core property \eqref{eq:GS-genLie} of the construction would be violated.

Since we assume $\GS\subset \GD$\,, the matrix $(\hat{X}_{\exA})_B{}^C$ should be traceless, and this means that we have
\begin{align}
 \hat{X}_{\exA} = \hat{X}_{\exA\exB}{}^{\exC}\,\widetilde{K}^{\exB}{}_{\exC} + \hat{X}_{\exA}{}^{\adja}\,s_{\adja}
 = \hat{X}_{\exA\exB}{}^{\exC}\,\widetilde{K}^{\exB}{}_{\exC} + \hat{X}_{\exA}{}^{\adja}\,t_{\adja}\,,
\label{eq:traceless}
\end{align}
which is automatically satisfied in the $\Odd$ case.

\paragraph{Assumption 3.}
Furthermore, we assume that no negative-level generators appear in the expansion of $\hat{X}_{\exA}$ above. Their absence extends the assumption $X_{\exA a}{}^{\exC}=0$ discussed in section \ref{sec:setup}. As mentioned there, this assumption can, in principle, be relaxed.

\subsubsection{Parameterizations}
Due to our coordinate choice that realizes \eqref{eq:RI}, the matrix $\Eh_{\exA}{}^{\Ih}$ can be generated from the shifted level-$0$ generators and the negative generators. More precisely, we parameterize the generalized frames as
\begin{align}
 \Eh_{\Ah}{}^{\Ih} = \widetilde{M}_{\Ah}{}^{\Bh}\,E_{\Bh}{}^{\Ih}\,,\qquad
 E_{\Ah}{}^{\Ih} \equiv N_{\Ah}{}^{\Bh}\,V_{\Bh}{}^{\Ih}\,.
\end{align}
Several new quantities appear here, which we explain in the following. Let us begin with the matrix $V_{\Ch}{}^{\Ih}$: It is constructed from $\widetilde{K}^{\exA}{}_{\exB}$ and $s_{\adja}$, but only the part constructed from the generator $\widetilde{K}^{\exA}{}_{\exB}$ depends on the coordinates $y^\mu$,
\begin{align}
 V_{\Ah}{}^{\Ih} &= \begin{pmatrix}
 \widetilde{v}_{\exA}{}^\mu(y^\mu) & 0 & \cdots \\
 0 & E_A{}^I \\
 \vdots & & \ddots
\end{pmatrix}.
\label{eq:V-decomp}
\end{align}
Next comes the matrix $N_{\Ah}{}^{\Bh}$. It is constructed from the negative-level generators. For $\GD=\Odd$ and $\GD=\Edd[d]$ with $d\leq 7$, we parameterize it as
\begin{align}
 \Odd:\quad &N = \Exp{-\frac{1}{2!}\,\rho^{\exA\exB}\,R_{\exA\exB}}\,\Exp{-\Omega_A^\exA\,R_\exA^A}\,,
\label{eq:N-param-1}
\\
 \Edd[d]:\quad &N = \cdots \Exp{-\frac{1}{3!}\,\rho^{\exA\exB,\exC}_{\lC}\,R_{\exA\exB,\exC}^{\lC}} \,\Exp{-\frac{1}{3!}\,\rho^{\exA\exB\exC;\AAA}\,R_{\exA\exB\exC;\AAA}}\,\underbrace{\Exp{-\frac{1}{2!}\,\rho^{\exA,\exB}\,R_{\exA,\exB}}}_{\text{only for }\Edd[7]}\,\Exp{-\frac{1}{2!}\,\rho^{\exA\exB;\AA}\,R_{\exA\exB;\AA}}\,\Exp{-\Omega_A^\exA\,R_\exA^A}\,,
 \intertext{while for $\Edd[8]$, we restrict the discussion to generators with level equal to or greater than $-2$, so that}
 \Edd[8]:\quad &N =\cdots \Exp{-\frac{1}{2!}\,\rho^{\exA\exB}\,R_{\exA\exB}} \,\Exp{-\frac{1}{2!}\,\rho^{\exA,\exB}_C\,R_{\exA,\exB}^C} \,\Exp{-\frac{1}{2!}\,\rho^{\exA\exB;\AA}\,R_{\exA\exB;\AA}}\,\Exp{-\Omega_A^\exA\,R_\exA^A}\,.
\label{eq:N-param-3}
\end{align}
Finally, we introduce the block-diagonal matrix
\begin{align}
 \widetilde{M}_{\Ah}{}^{\Bh} \equiv \exp\bigl(y^\mu\, \mathfrak{t}_{\mu}\bigr)_{\Ah}{}^{\Bh}\,,
\end{align}
which depends only on $y^\mu$. It is an element of $\GS$ and satisfies
\begin{align}
 \widetilde{M}^{-1}\,D_\exA \widetilde{M} = -X_{\exA} \,,
\end{align}
where $D_{\Ah}\equiv E_{\Ah}{}^{\Ih}\,\partial_{\Ih}$\,. Under this construction, the assumption \eqref{eq:Lie-E-assumption} can be rewritten as
\begin{align}
 D_{\exA} N = 0 \,,\qquad D_{\exA} E_A{}^I = 0\,,
\end{align}
implying that $N_{\Ah}{}^{\Bh}$ and $E_A{}^I$ depend only on the coordinates $y^i$ on the physical space $M=\GS\backslash \hat{M}$. 

We present the explicit form of the matrix $N$ here for later convenience. For $\Odd$ and $\Edd[d]$ with $d\leq 5$, it is given by
\begin{align}
 N&= \begin{pmatrix}
 \delta_\exA^\exB & 0 & 0 & 0 \\
 -\Omega_A^\exB & \delta_A^B & 0 & 0 & \cdots \\
 \rho^{\exA\exB;\AA}-\frac{1}{2}\,\eta^{CD;\AA}\,\Omega^\exA_C\,\Omega^{\exB}_D & \eta^{BC;\AA}\,\Omega_C^\exA & \delta^{\exA}_{\exB}\,\delta^{\AA}_{\BB} & 0 \\
 N1 & N2 & -2\,\delta_\exB^{[\exA_1}\,\Omega_C^{\exA_2]}\,Z^C{}_{\BB}{}^{\AAA} & 2!\,\delta^{\exA_1\exA_2}_{\exB_1\exB_2}\,\delta^{\AAA}_{\BBB} \\
 & \vdots & & & \ddots
\end{pmatrix},
\\
 N1 &\equiv - \rho^{\exA_1\exA_2\exB;\AAA} - \rho^{\exA_1\exA_2;\CC}\,\Omega_D^\exB\,Z^D{}_{\CC}{}^{\AAA}
 + \tfrac{1}{3}\,\eta^{ECD;\AAA}\,\Omega_C^{[\exA_1}\,\Omega_D^{\exA_2]}\,\Omega_E^\exB\,,
\\
 N2 &\equiv \rho^{\exA_1\exA_2;\CC}\,Z^B{}_{\CC}{}^{\AAA}-\eta^{BCD;\AAA}\,\Omega_C^{[\exA_1}\,\Omega^{\exA_2]}_D\,,
\end{align}
where only the first three rows/columns appear in the $\Odd$ case. For $\Edd[6]$ and $\Edd[7]$, this matrix is extended to
\begin{align}
 N &= \left(\begin{array}{ccc|c|cccc}
 \delta_{\exA}^{\exB} & 0 & 0 & 0 &0 &0 &0 \\
 A_1 & \delta_A^B & 0 & 0 &0 &0 &0 & \cdots \\
 B_1 & A_2 & \delta^{\exA}_{\exB}\delta^{\AA}_{\BB} & 0 &0 &0 &0 \\ \hline
 B_2 & A_3 & 0 & 2\delta^{\exA}_{\exB} &0 &0 &0 \\ \hline
 C_1 & B_3 & A_4 & 0 & 2!\delta^{\exA_1\exA_2}_{\exB_1\exB_2}\delta^{\AAA}_{\BBB} &0 &0 \\
 C_2 & B_4 & A_5 & A_6 &0 &12\delta^{\exA_1\exA_2}_{\exB_1\exB_2}\delta_{\lA}^{\lB} &0 \\
 C_3 & B_5 & A_7 & A_8 &0 &0 &4\delta^{(\exA}_{(\exB}\delta^{\exA')}_{\exB')}\delta_{\lA}^{\lB} \\
 & \vdots & & & &&&\ddots
\end{array}\right),
\end{align}
with
\begin{align}
 A_1 &\equiv -\Omega_A^{\exB}\,,
\\
 A_2 &\equiv \eta^{BC;\AA}\,\Omega_C^{\exA}\,,
\\
 A_3 &\equiv -\omega^{BC}\,\Omega_C^{\exA}\,,
\\
 A_4 &\equiv -2\,\delta_{\exB}^{[\exA_1}\,\Omega_C^{\exA_2]}\,Z^C{}_{\BB}{}^{\AAA}\,,
\\
 A_5 &\equiv -2\,\delta_{\exB}^{[\exA_1}\,\Omega_C^{\exA_2]}\,\eta^C{}_{\lA\BB} \,,
\\
 A_6 &\equiv -6\,\delta_{\exB}^{[\exA_1}\,\Omega_A^{\exA_2]} \qquad (\Edd[7]\text{ only})\,,
\\
 A_7 &\equiv -2\,\delta_{\exB}^{(\exA}\,\Omega_C^{\exA')}\,\eta^C{}_{\lA\BB} \,,
\\
 A_8 &\equiv 2\,\delta_{\exB}^{(\exA_1}\,\Omega_A^{\exA_2)} \qquad (\Edd[7]\text{ only})\,,
\\
 B_1 &\equiv \rho^{\exA\exB;\AA}-\tfrac{1}{2}\,\eta^{CD;\AA}\,\Omega_C^{\exA}\,\Omega_D^{\exB}\,,
\\
 B_2 &\equiv \rho^{\exA,\exB}-\tfrac{1}{2}\,\omega^{CD}\,\Omega_C^{\exA}\,\Omega_D^{\exB}\,,
\\
 B_3 &\equiv \rho^{\exA_1\exA_2;\CC}\,Z^B{}_{\CC}{}^{\AAA} -\eta^{BCD;\AAA}\,\Omega_C^{[\exA_1}\,\Omega_D^{\exA_2]}\,,
\\
 B_4 &\equiv -2\,\rho^{\exA_1\exA_2;\CC}\,\eta^B{}_{\lA\CC} + \bigl(2\,\omega^{B[C}\,\delta^{D]}_{\lA}-\tfrac{1}{2}\,\omega^{CD}\,\delta_{\lA}^B\bigr)\,\Omega_C^{[\exA_1}\,\Omega_D^{\exA_2]}\,,
\\
 B_5 &\equiv -\rho^{\exA,\exA'}\,\delta^B_{\lA} -\bigl(\eta^{B}{}_{\lA\EE}\,\eta^{CD;\EE}+2\,\omega^{B(C}\,\delta_{\lA}^{D)}\bigr)\,\Omega_C^{(\exA}\,\Omega_D^{\exA')}\,,
\\
 C_1 &\equiv -\rho^{\exA_1\exA_2\exB;\AAA} - \rho^{\exA_1\exA_2;\CC}\,\Omega_D^{\exB}\,Z^D{}_{\CC}{}^{\AAA}
 + \tfrac{1}{3}\,\eta^{ECD;\AAA}\,\Omega_C^{[\exA_1}\,\Omega_D^{\exA_2]}\,\Omega_E^{\exB}\,,
\\
 C_2 &\equiv \rho^{\exA_1\exA_2,\exB}_{\lA} + 2\,\rho^{\exA_1\exA_2;\CC}\,\Omega_D^{\exB}\,\eta^D{}_{\lA\CC}
 + \tfrac{1}{2}\,\omega^{[CD}\,\delta_{\lA}^{E]}\,\Omega_C^{[\exA_1}\,\Omega_D^{\exA_2]}\,\Omega_E^{\exB}\,,
\\
 C_3 &\equiv -\tfrac{2}{3}\,\rho^{\exB(\exA,\exA')}_{\lA} + \delta_{\lA}^C\,\rho^{\exA,\exA'}\,\Omega_C^{\exB}
 + \tfrac{1}{3}\,\bigl(c^{E(CD)}{}_{\lA} + \omega^{E(C}\,\delta_{\lA}^{D)}\bigr)\,\Omega_C^{(\exA}\,\Omega_D^{\exA')}\,\Omega_E^{\exB}\,,
\\
 \delta_{\lA}^B &\equiv \begin{cases}
 0 & (\Edd[6]) \\ \delta_A^B & (\Edd[7])
\end{cases}.
\end{align}
Finally for $\Edd[8]$, we find
\begin{align}
 N &= \begin{pmatrix}
 \delta_{\exA}^{\exB} & 0 & 0 & 0 &0 \\
 -\Omega_A^\exB & \delta_A^B & 0 & 0 &0 & \cdots \\
 \rho^{\exA\exB;\AA} -\frac{1}{2}\,\eta^{CD;\AA}\,\Omega_C^{\exA}\,\Omega_D^{\exB} & \eta^{BC;\AA}\,\Omega_C^{\exA} & \delta^{\exA}_{\exB}\delta^{\AA}_{\BB} & 0 &0 \\
 \rho^{\exA,\exB}_A -\frac{1}{2}\,f_A{}^{CD}\,\Omega_C^{\exA}\,\Omega_D^{\exB} & -f_A{}^{BC}\,\Omega_C^{\exA} & 0 & 2\delta^{\exA}_{\exB}\delta_A^B &0 \\
 \rho^{\exA\exB} -\frac{1}{2}\,\kappa^{CD}\,\Omega_C^{\exA}\,\Omega_D^{\exB} & \kappa^{BC}\,\Omega_C^{\exA} & 0 & 0 &4\delta^{\exA}_{\exB} \\
 & \vdots & & & & \ddots
\end{pmatrix} .
\end{align}

\subsection{Gauge transformations}\label{sec:gauge-transf}
As in section~\ref{sec:restricted-diffeos}, we now consider a restricted class of generalized diffeomorphisms, where the gauge parameters take the form
\begin{align}
 \hat{\xi} = \hat{\xi}^{\Ah}\,\Eh_{\Ah} = \xi^{\Ah}(y^i)\,V_{\Ah}
\end{align}
with
\begin{align}
 \hat{\xi}^{\Ah} \equiv \xi^{\Ch}(y^i)\,(N^{-1})_{\Ch}{}^{\Bh}\,(\widetilde{M}^{-1})_{\Bh}{}^{\Ah}\,.
\end{align}
Once again, we find that the variation of the generalized frames $\delta_{\hat{\xi}} \Eh = \GLie_{\hat{\xi}} \Eh$ does not affect the $y^\mu$-dependent quantities. Thus, we have $\delta_{\hat{\xi}}\widetilde{M}=0$ and $\delta_{\hat{\xi}}\widetilde{v}=0$, and the variation of the generalized frame simplifies to
\begin{align}
 \delta_{\hat{\xi}} \Eh=\widetilde{M} \,\delta_{\hat{\xi}} E\,,\qquad 
 \delta_{\hat{\xi}} E \equiv \delta_{\hat{\xi}} N \,V + N \,\delta_{\hat{\xi}} V \,,
\end{align}
where indices have been omitted. For convenience, we untwist the variation as $N^{-1}\,\widetilde{M}^{-1}\,\delta_{\hat{\xi}} \Eh\,V^{-1}$ and compute this quantity. 
In the $\Odd$ case or the $\Edd[d]$ case with $d\leq 7$, this leads to the expansion
\begin{align}
 &N^{-1}\,\widetilde{M}^{-1}\,\delta_{\hat{\xi}} \Eh\,V^{-1} =\delta_{\hat{\xi}} V\,V^{-1} + N^{-1}\,\delta_{\hat{\xi}} N 
\nn\\
 &= \delta_{\hat{\xi}} V\,V^{-1} - \delta_{\hat{\xi}} \Omega^\exA_A\,R^A_\exA
 - \tfrac{1}{2}\,\bigl(\delta_{\hat{\xi}} \rho^{\exA\exB;\CC} + \eta^{AB;\CC}\,\Omega_A^{[\exA}\,\delta_{\hat{\xi}}\Omega_B^{\exB]}\bigr)\,R_{\exA\exB;\CC} 
\nn\\
 &\quad - \tfrac{1}{2}\,\bigl(\delta_{\hat{\xi}} \rho^{\exA,\exB} + \omega^{AB}\,\Omega_A^{(\exA}\,\delta_{\hat{\xi}}\Omega_B^{\exB)} \bigr)\,R_{\exA,\exB} 
\nn\\
 &\quad - \tfrac{1}{3!}\,\bigl(\delta_{\hat{\xi}} \rho^{\exA\exB\exC;\DDD} + 3\,Z^A{}_{\EE}{}^{\DDD}\,\Omega_A^{[\exA}\,\delta_{\hat{\xi}} \rho^{\exB\exC];\EE} - \eta^{ABC;\DDD}\,\delta_{\hat{\xi}} \Omega_A^{[\exA}\,\Omega_B^{\exB}\,\Omega_C^{\exC]}\bigr)\,R_{\exA\exB\exC;\DDD}
\nn\\
 &\quad - \tfrac{1}{3!}\,\bigl(\delta_{\hat{\xi}} \rho^{\exA\exB,\exC}_{\lD} + 3\,\eta^A{}_{\lD\CC}\,\Omega_A^{\exC}\,\delta_{\hat{\xi}} \rho^{\exA\exB;\CC} 
 -3\,\delta_{\lD}^E\,\Omega_{E}^{\exA}\,\delta_{\hat{\xi}} \rho^{\exB,\exC} 
\nn\\
 &\qquad\qquad - \eta^{AB;\EE}\,\eta^C{}_{\lD\EE}\,\delta_{\hat{\xi}}\Omega_A^{\exA}\,\Omega_B^{\exB}\,\Omega_C^{\exC} 
 +\omega^{AB}\,\delta_{\lD}^C\,\delta_{\hat{\xi}}\Omega_A^{(\exB}\,\Omega_B^{\exC)}\,\Omega_C^{\exA} \bigr)\,R_{\exA\exB,\exC}^{\lD}
 + \cdots\,,
\label{eq:delta-E-expansion}
\end{align}
where only the first line appears in the $\Odd$ case. For $\Edd[8]$, we have
\begin{align}
 &N^{-1}\,\widetilde{M}^{-1}\,\delta_{\hat{\xi}} \Eh\,V^{-1} =\delta_{\hat{\xi}} V\,V^{-1} + N^{-1}\,\delta_{\hat{\xi}} N 
\nn\\
 &= \delta_{\hat{\xi}} V\,V^{-1} - \delta_{\hat{\xi}} \Omega^\exA_A\,R^A_\exA
 - \tfrac{1}{2}\,\bigl(\delta_{\hat{\xi}} \rho^{\exA\exB;\CC} + \eta^{AB;\CC}\,\Omega_A^{[\exA}\,\delta_{\hat{\xi}}\Omega_B^{\exB]}\bigr)\,R_{\exA\exB;\CC} 
\nn\\
 &\quad - \tfrac{1}{2}\,\bigl(\delta_{\hat{\xi}} \rho^{\exA,\exB}_C + f^{AB}{}_C\,\Omega_A^{(\exA}\,\delta_{\hat{\xi}}\Omega_B^{\exB)} \bigr)\,R_{\exA,\exB}^C 
\nn\\
 &\quad - \tfrac{1}{2}\,\bigl(\delta_{\hat{\xi}} \rho^{\exA\exB} + \kappa^{AB}\,\Omega_A^{[\exA}\,\delta_{\hat{\xi}}\Omega_B^{\exB]}\bigr)\,R_{\exA\exB} +\cdots\,.
\label{eq:delta-E-expansion-E8}
\end{align}
Next, we compute the RHS of the relation $\delta_{\hat{\xi}} \Eh = \GLie_{\hat{\xi}} \Eh$ from the definition of the generalized Lie derivative to find
\begin{align}
 \bigl(N^{-1}\,\widetilde{M}^{-1}\,\GLie_{\hat{\xi}} \Eh\,V^{-1}\bigr)_{\Ah}{}^{\Bh}
 &= -\xi^{\exC}\,(N^{-1}X_\exC\,N)_{\Ah}{}^{\Bh}
 +\xi^{C}\,(N^{-1}\,D_{C}N)_{\Ah}{}^{\Bh}
 - \xi^{\cC}\,\check{W}_{\cC \Ah}{}^{\Bh} 
\nn\\
 &\quad + (\widehat{s}^{\adjah})_{\cD}{}^{\cC}\,\bigl(D_{\cC}\xi^{\cD} - \xi^{\cE}\,\check{W}_{\cC\cE}{}^{\cD}\bigr)\,(\widehat{s}_{\adjah})_{\Ah}{}^{\Bh} \,,
\label{eq:GLie-expansion}
\end{align}
where $\cA \equiv \{\exA,\, A\}$ and we have defined the Weitzenb\"ock connection associated with $V_{\Ah}{}^{\Ih}$ as
\begin{align}
 (\check{W}_{\Ah})_{\Bh}{}^{\Ch} \equiv - D_{\Ah} V_{\Bh}{}^{\Ih}\,(V^{-1})_{\Ih}{}^{\Ch}\,.
\end{align}
Using the expansion \eqref{eq:hat-s-decomp}, we find
\begin{align}
 V\,\Eh^{-1}\,\GLie_{\hat{\xi}} \Eh\,V^{-1} &= N^{-1}\,\widetilde{M}^{-1}\,\GLie_{\hat{\xi}} \Eh\,V^{-1} 
\nn\\
 &= -\xi^{\exC}\, N^{-1}X_\exC\,N 
 +\xi^{A}\, N^{-1}\,D_{A}N 
\nn\\
 &\quad - \xi^{\exC}\,\check{W}_{\exC \exA}{}^{\exB}\, \widetilde{K}^{\exA}{}_{\exB} - \xi^{C}\,\check{W}_{C}{}^{\adja}\, s_{\adja} 
\nn\\
 &\quad + (\widetilde{K}_{\exA}{}^{\exB})_{\exD}{}^{\exC}\,\bigl( - \xi^{\exE}\,\check{W}_{\exC\exE}{}^{\exD}\bigr)\, \widetilde{K}^{\exA}{}_{\exB} 
\nn\\
 &\quad + (s^{\adja})_{D}{}^{C}\,\bigl(D_{C}\xi^{D} - \xi^{E}\,\check{W}_{CE}{}^{D}\bigr)\, s_{\adja} 
\nn\\
 &\quad + (R^{\exA}_A)_{\exD}{}^{C}\,\bigl(D_{C}\xi^{\exD} \bigr)\, R_{\exA}^A \,,
\label{eq:Lie-E-expansion}
\end{align}
where we have used, for example, $D_{\exC}\xi^{\cD}=0$, $\check{W}_{\exC E}{}^{D}=0$, and $\check{W}_{C\exE}{}^{\exD}=0$. Collecting the generators $\widetilde{K}^{\exA}{}_{\exB}$ gives rise to
\begin{align}
 \bigl[-\xi^{\exC}\,X_{\exC\exA}{}^{\exB} + \xi^{\exC}\,\bigl(\check{W}_{\exA\exC}{}^{\exB}-\check{W}_{\exC\exA}{}^{\exB}\bigr)\bigr]\,\widetilde{K}^{\exA}{}_{\exB} =0\,.
\end{align}
This shows that the $y^\mu$-dependent part of $V$ (which is generated by $\widetilde{K}^{\exA}{}_{\exB}$) remains unchanged, as expected. Moreover, because the RHS of \eqref{eq:Lie-E-expansion} does not depend on $y^\mu$, it is clear that $\widetilde{M}(y^\mu)$ remains unchanged.

Since $\delta_{\hat{\xi}}\widetilde{v}=0$, the part of $\delta_{\hat{\xi}} V\,V^{-1}$ that is spanned by $\widetilde{K}^{\exA}{}_{\exB}$ does not appear, and we obtain
\begin{align}
 \delta_{\hat{\xi}} V\,V^{-1} = (\delta_{\hat{\xi}} V\,V^{-1})^{\adja}\,s_{\adja}\,.
\end{align}
Consequently, the RHS of \eqref{eq:delta-E-expansion} or \eqref{eq:delta-E-expansion-E8} contains only the generators $s_{\adja}$ and the negative-level generators, making it a perfect match with \eqref{eq:Lie-E-expansion}. By comparing \eqref{eq:delta-E-expansion} or \eqref{eq:delta-E-expansion-E8} with \eqref{eq:Lie-E-expansion}, we can successively determine the explicit form of the variations $\delta_{\hat{\xi}} V$\,, $\delta_{\hat{\xi}} \Omega^\exA_A$\,, $\delta_{\hat{\xi}} \rho^{\exA\exB;\CC}$\,, $\dotsc$\,, starting from the level-$0$ part. 
In the following, we discuss these variations for levels $0$ to $-3$.

We note that the generalized Lie derivative $\GLie_{\hat{\xi}}$ in \eqref{eq:Lie-E-expansion} does not include the necessary modification when $p=d+n$ is greater than seven. In this sense, the following results are not fully complete. However, as we explained at the end of section \ref{sec:fluxes}, this modification only affects generators at higher levels $\lvert\ell\rvert$. Since our primary interest lies in the lowest-level components of the curvature hierarchy, the unmodified derivative is sufficient for our purposes.

\subsubsection{Level \texorpdfstring{$0$}{0}}
For this part, it suffices to examine the terms in \eqref{eq:Lie-E-expansion} that are spanned by $s_{\adja}$\,. This gives
\begin{align}
 \bigl[-\xi^{\exC}\,X_{\exC}{}^{\adja} - \xi^{C}\,\check{W}_{C}{}^{\adja} + (s^{\adja})_{D}{}^{C}\,\bigl(D_{C}\xi^{D} - \xi^{E}\,\check{W}_{CE}{}^{D}\bigr)\bigr]\, s_{\adja} \,.
\end{align}
This should coincide with $\delta_{\hat{\xi}} V\,V^{-1}$ of \eqref{eq:delta-E-expansion} or \eqref{eq:delta-E-expansion-E8}. For this part, we need only examine the $A$--$B$ components. Using
\begin{align}
 \bigl(\delta_{\hat{\xi}} V\,V^{-1}\bigr)_A{}^B = \delta_{\hat{\xi}} E_A{}^I\,E_I{}^B \,,
\end{align}
we find
\begin{align}
 \delta_{\hat{\xi}} E_A{}^I = \gLie_{\xi} E_A{}^I + \xi_{A}{}^{B}\,E_B{}^I \equiv \mathbb{L}_{\hat{\xi}} E_A{}^I\,,
\label{eq:gauge-level-0}
\end{align}
where $\xi \equiv \xi^A\,V_A$ and $\xi_A{}^B \equiv \xi^{\exC}\,(\mathfrak{t}_{\exC})_A{}^B = -\xi^{\exC}\,X_{\exC}{}^{\adja}\,(s_{\adja})_A{}^B \in \mathrm{Lie}(\GS)$. This result shows that the gauge parameter $\xi^{\exA}$ generates local $\GS$ transformations of the physical frame, while $\xi^A$ generates generalized diffeomorphisms on the physical space $M$.

Note that the generalized Lie derivative $\gLie_{\xi}$ in \eqref{eq:gauge-level-0} is the original one for $\GD=\Edd[d]$ with $d\leq 7$, since the generalized Lie derivative $\GLie$ in \eqref{eq:GLie-expansion} remains unmodified. For $\GD=\Edd[8]$, the expression $\gLie_{\xi} E_A{}^I$ in \eqref{eq:gauge-level-0} should take a modified form $\gLie_{(E_A,\widetilde{\Sigma}_A)}$. This modification would arise from the modified generalized Lie derivative $\GLie$ in $\Edd[p]$ ExFT with $p>8$, whose explicit form we do not yet know.

\subsubsection{Level \texorpdfstring{$-1$}{-1}}
Next, we focus on the level-$(-1)$ generator $R^A_\exA$ and find
\begin{align}
 \delta_{\hat{\xi}} \Omega^\exA_A\,R^A_\exA
 &= \bigl[\Omega_A^\exB\,\xi^\exC\,f_{\exC\exB}{}^\exA - \xi^{\exC}\,X_{\exC}{}^{\adja}\,(t_{\adja})_A{}^C\,\Omega_C^\exA 
 + \gLie_\xi\Omega^\exA_A
 + D_{A}\xi^{\exA} \bigr]\,R^A_\exA \,.
\end{align}
Due to \eqref{eq:traceless}, we have
\begin{align}
 \xi_A{}^A=0\,,
\end{align}
implying $\xi_A{}^B = -\xi^{\exC}\,X_{\exC}{}^{\adja}\,(s_{\adja})_A{}^B = - \xi^{\exC}\,X_{\exC}{}^{\adja}\,(t_{\adja})_A{}^B$. Hence the variation of
\begin{align}
 \Omega_{AB}{}^C \equiv \Omega_{A}^\exA\, (\mathfrak{t}_{\exA})_{B}{}^{C}
\end{align}
can be expressed as
\begin{align}
 \delta_{\hat{\xi}} \Omega_{AB}{}^C &= \mathbb{L}_{\hat{\xi}} \Omega_{AB}{}^C + D_{A}\xi_{B}{}^C\,, \qquad \text{with}
\label{eq:delta-Omega}
\\
 \mathbb{L}_{\hat{\xi}} \Omega_{AB}{}^C &= \gLie_\xi \Omega_{AB}{}^C
 + \Omega_{AD}{}^C \,\xi_{B}{}^D
 + \xi_A{}^D\,\Omega_{DB}{}^C 
 - \Omega_{AB}{}^D \,\xi_{D}{}^C\,.
\end{align}
This shows that $\Omega_{AB}{}^C$ transforms as a generalized scalar field under generalized diffeomorphisms, while simultaneously transforming like a spin connection under local $\GS$ transformations. Introducing the anomalous part $\Delta_{\hat{\xi}} \equiv \delta_\xi - \mathbb{L}_{\hat{\xi}}$ for $\GS$ transformations, we obtain
\begin{align}\label{eq:OmegaABC}
 \Delta_{\hat{\xi}} \Omega_{AB}{}^C = D_{A}\xi_{B}{}^C \,.
\end{align}

Again, we do not claim that this result is complete for large $p$, since we have not introduced the modification in \eqref{eq:Lie-E-expansion}. However, this is the natural transformation law of the spin connection, and we expect that any modification will not affect this result at this level.

\subsubsection{Level \texorpdfstring{$-2$}{-2}}
Contributions at level $\ell \leq -2$ are governed by the relation
\begin{align}
 N^{-1}\,\delta_\xi N &= \xi^{A}\, N^{-1}\,D_{A}N -\xi^{\exC}\, N^{-1}X_\exC\,N \,.
 \intertext{Evaluating this for the level-$(-2)$ generator $R_{\exA\exB;\CC}$, the LHS gives}
                      &\phantom{=} - \bigl(\delta_\xi \rho^{\exA\exB;\CC} + \Omega_A^\exA\,\delta_\xi\Omega_B^\exB\,\eta^{AB;\CC}\bigr)\,R_{\exA\exB;\CC}\,,
\end{align}
while the RHS describes a covariant transformation -- the first term captures the generalized Lie derivative while the second term mediates a covariant $\GS$ transformation parameterized by $\xi_A{}^B$. Through direct computation, one confirms that the anomalous transformation of $\rho^{\exA\exB;\CC}$ is given by
\begin{align}
 \Delta_{\hat{\xi}} \rho^{\exA\exB;\CC} = - \eta^{AB;\CC} \,\Omega_A^{[\exA}\,\Delta_{\hat{\xi}}\Omega_B^{\exB]}
 = - \eta^{AB;\CC} \,\Omega_A^{[\exA}\, D_B \xi^{\exB]} \,.
\end{align}
Similarly, for $\Edd[7]$, we find
\begin{align}
 \Delta_{\hat{\xi}} \rho^{\exA,\exB} = -\omega^{AB}\,\Omega_A^{(\exA}\,\Delta_{\hat{\xi}}\Omega_B^{\exB)} 
 = -\omega^{AB} \,\Omega_A^{(\exA}\, D_B \xi^{\exB)}
\end{align}
whereas for $\Edd[8]$, the anomalous transformations read
\begin{align}
 \Delta_{\hat{\xi}} \rho^{\exA,\exB}_C &= - f^{AB}{}_C\,\Omega_A^{(\exA}\,\Delta_{\hat{\xi}}\Omega_B^{\exB)} 
 = -f^{AB}{}_C \,\Omega_A^{(\exA}\, D_B \xi^{\exB)} \,,
\\
 \Delta_{\hat{\xi}} \rho^{\exA\exB} &= -\kappa^{AB} \,\Omega_A^{[\exA}\,\Delta_{\hat{\xi}}\Omega_B^{\exB]} 
 = -\kappa^{AB} \,\Omega_A^{[\exA}\, D_B \xi^{\exB]} \,,
\end{align}
as now two fields contribute at this level.

In analogy with \eqref{eq:OmegaABC}, it is convenient to write all fields using only $R_1$-indices of the duality group $\GD$. To this end, we define for $\Edd[d]$ with $d\leq 7$
\begin{align}
 \rho_C^D{}_A^E{}_{;BF}&\equiv \rho^{\exB\exC}{}_{;BF}\,(\mathfrak{t}_{\exB})_{C}{}^D\,(\mathfrak{t}_{\exC})_{A}{}^E\,,
\\
 \rho_C^D{}_A^E&\equiv \rho^{\exB,\exC}\,(\mathfrak{t}_{\exB})_{C}{}^D\,(\mathfrak{t}_{\exC})_{A}{}^E\,,
\end{align}
while for $\Edd[8]$, we use
\begin{align}
 \rho_C^D{}_A^E{}_{;BF}&\equiv \rho^{\exB\exC}{}_{;BF}\,(\mathfrak{t}_{\exB})_{C}{}^D\,(\mathfrak{t}_{\exC})_{A}{}^E\,,
\\
 \rho_C^D{}^E_{AF}&\equiv \rho_F^{\exB,\exC} \,(\mathfrak{t}_{\exB})_{C}{}^D\,(\mathfrak{t}_{\exC})_{A}{}^E\,,
\\
 \rho_C^D{}_A^E &\equiv \rho^{\exB\exC} \,(\mathfrak{t}_{\exB})_{C}{}^D\,(\mathfrak{t}_{\exC})_{A}{}^E\,.
\end{align}
They can be combined into a single field, $\bm{\rho}_A^E{}_C^D{}_{;BE}$, as
\begin{align}
 \bm{\rho}_A^E{}_C^D{}_{;BE} 
 \equiv
 \begin{cases}
 \rho_A^E{}_C^D{}_{;BE} & (d\leq 6)
\\
 \rho_{AC;BE}^{ED} 
 - \tfrac{1}{2}\,\omega_{BE}\,\rho_{AC}^{ED} & (d=7)
\\
 \rho_{AC;BE}^{ED} 
 - \tfrac{1}{2}\,f_{BE}{}^F\,\rho_{ACF}^{ED} 
 + \tfrac{1}{4}\,\rho_{AC}^{ED} \,\kappa_{BE} &(d=8)
\end{cases}\,,
\end{align}
which has the compact anomalous transformation
\begin{align}
 \Delta_{\hat{\xi}} \bm{\rho}_A^E{}_C^D{}_{;BE} 
 = \tfrac{1}{2}\, Y^{GH}_{BE}\,\bigl(D_{G}\xi_{A}{}^E\,\Omega_{HC}{}^D - \Omega_{GA}{}^E\, D_{H}\xi_{C}{}^D\bigr)\,.
\label{eq:delta-rho}
\end{align}
Knowing this transformation is mandatory to show the covariance of the Riemann tensor later. 

\subsubsection{Level \texorpdfstring{$-3$}{-3}}
Following the same logic, we consider gauge transformations of the level-$(-3)$ fields $\rho^{\exA_1\exA_2\exA_3;\DDD}$ and $\rho^{\exA_1\exA_2,\exA'}_{\lC}$ for $\Edd[d]$ with $d\leq 7$. They are rewritten as
\begin{align}
 \rho^{PRT}_{QSU;ABC} &\equiv \rho^{\exA_1\exA_2\exA_3;\DDD}\,(\mathfrak{t}_{\exA_1})_Q{}^P\,(\mathfrak{t}_{\exA_2})_S{}^R\,(\mathfrak{t}_{\exA_3})_U{}^T\,\eta_{ABC;\DDD}\,,
\\
 \rho^{PRT}_{QSU;\lC} &\equiv \rho^{\exA_1\exA_2,\exA'}_{\lC}\,(\mathfrak{t}_{\exA_1})_Q{}^P\,(\mathfrak{t}_{\exA_2})_S{}^R\,(\mathfrak{t}_{\exA'})_U{}^T\,.
\end{align}
By direct computation, we can obtain the anomalous transformations. However, it seems that for $d=7$, the level-$(-2)$ curvature (to be introduced in \eqref{eq:R2}) fails to remain invariant under these transformations, indicating that a modification of the generalized Lie derivative becomes necessary. Since treating this modification properly is challenging, we restrict our analysis at this level to $d\leq 6$ only. The result is then
\begin{align}
 \Delta_{\hat{\xi}} \rho^{PRT}_{QSU;ABC}
 &= -\tfrac{2}{3}\, Y^{F[DE]}_{ABC}\,
 \bigl(\Omega_{DQ}{}^P\,\Omega_{ES}{}^R\,D_F\xi_{U}{}^T
 + \Omega_{DS}{}^R\,\Omega_{EU}{}^T\,D_F\xi_{Q}{}^P
\nn\\
 &\qquad\qquad\qquad\ 
 + \Omega_{DU}{}^T\,\Omega_{EQ}{}^P\,D_F\xi_{S}{}^R\bigr)\,,
\\
 \Delta_{\hat{\xi}} \rho^{PRT}_{QSU}
 &= c^{ABC}\,\Omega_{AU}{}^T\,\bigl(\Omega_{BQ}{}^P\,D_C\xi_S{}^R - \Omega_{BS}{}^R\,D_C\xi_Q{}^P\bigr) \,.
\end{align}

\subsection{Curvatures}
Following the spirit of section~\ref{sec:GL-curvatures}, we now compute the generalized fluxes $\hat{X}_{\Ah}$ associated with $\Eh_{\Bh}{}^{\Ih}$ to eventually extract curvature and torsion tensors that transform covariantly under the gauge transformations derived in the last section. As the first step in this computation, we express the Weitzenb\"ock connection on the mega-space as
\begin{align}
 - \Eh_{\Ah}{}^{\hat{J}}\, \partial_{\hat{J}} \Eh_{\Bh}{}^{\Ih}\,\Eh_{\Ih}{}^{\Ch}
 = \widetilde{M}_{\Ah}{}^{\Dh}\,\widetilde{M}_{\Bh}{}^{\Eh}\,(\widetilde{M}^{-1})_{\Fh}{}^{\Ch}\,\cW_{\Dh\Eh}{}^{\Fh}\,,
\end{align}
in terms of
\begin{align}
 \cW_{\Ah\Bh}{}^{\Ch} &\equiv W_{\Ah\Bh}{}^{\Ch} - \bigl(\widetilde{M}^{-1}\,D_{\Ah}\widetilde{M}\bigr)_{\Bh}{}^{\Ch} 
 = W_{\Ah\Bh}{}^{\Ch} - N_{\Ah}{}^\exA\, (\mathfrak{t}_{\exA})_{\Bh}{}^{\Ch}\,,
\label{eq:def-cW}
\\
 W_{\Ah\Bh}{}^{\Ch} &\equiv - D_{\Ah} E_{\Bh}{}^{\Ih}\, E_{\Ih}{}^{\Ch}
 = N_{\Ah}{}^{\Dh} \,\bigl(N\,\check{W}_{\Dh}\,N^{-1} - D_{\Dh}N\,N^{-1}\bigr)_{\Bh}{}^{\Ch}\,.
\label{eq:def-W}
\end{align}
The generalized fluxes associated with $\Eh_{\Bh}{}^{\Ih}$ are then given by
\begin{align}
 \hat{X}_{\Ah\Bh}{}^{\Ch} 
 = \widetilde{M}_{\Ah}{}^{\Dh}\,\widetilde{M}_{\Bh}{}^{\Eh}\,(\widetilde{M}^{-1})_{\Fh}{}^{\Ch}\,
 X_{\Dh\Eh}{}^{\Fh} \,,
\label{eq:X:def}
\end{align}
where the untwisted generalized fluxes $X_{\Ah\Bh}{}^{\Ch}$, 
\begin{align}
 X_{\Ah\Bh}{}^{\Ch} = 2\,\cW_{[\Ah\Bh]}{}^{\Ch} + \hat{Y}^{\Eh\Ch}_{\Bh\Dh}\,\cW_{\Eh\Ah}{}^{\Dh}
 = \cW_{\Ah\Bh}{}^{\Ch} + (\widehat{s}^{\adjah})_{\Dh}{}^{\Eh}\,\cW_{\Eh\Ah}{}^{\Dh}\,(\widehat{s}_{\adjah})_{\Bh}{}^{\Ch}\,,
\label{eq:pleq7}
\end{align}
are written in terms of the Weitzenb\"ock connection introduced above. Here we have omitted the modification of the generalized Lie derivative that is necessary for $p\geq 8$. Since our analysis focuses on components with small $\lvert\ell\rvert$, this modification will not play a significant role at this stage and can be included when necessary.

From \eqref{eq:X:def}, we can easily see that
\begin{align}
 X_{\exA\Bh}{}^{\Ch} 
 = (\widetilde{M}^{-1})_{\exA}{}^{\exD}\,(\widetilde{M}^{-1})_{\Bh}{}^{\Eh}\, \widetilde{M}_{\Fh}{}^{\Ch}\,
 \hat{X}_{\exD\Eh}{}^{\Fh}
 = - (\widetilde{M}^{-1})_{\exA}{}^{\exD}\, \bigl(\widetilde{M}^{-1}\,\mathfrak{t}_{\exD}\,\widetilde{M}\bigr)_{\Bh}{}^{\Ch}
 = \hat{X}_{\exA\Bh}{}^{\Ch} \,.
\end{align}
This shows that $X_{\exA}$ is constant. Consequently, we focus our analysis on the non-trivial components $X_{A}$, which encode various curvature quantities as anticipated in section~\ref{sec:GL-curvatures}. To compute $X_{A}$, it is useful to note that $X_{A}$ can be expanded in terms of the shifted generators $s_{\adja}$ and generators with negative levels as
\begin{align}
 X_{A} = X_{A}{}^{\adja}\,s_{\adja} + X_{A}{}^{\exB}_{B}\, R_\exB^B + \tfrac{1}{2!}\,X_{A}{}^{\exB_1\exB_2\BB} \,R_{\exB_1\exB_2;\BB} + \cdots\,.
\label{eq:XA-expansion}
\end{align}
This expansion is possible because $\cW_{\Ah\Bh}{}^{\Ch}$ gets only contributions from $\widetilde{K}^{\exA}{}_{\exB}$, $s_{\adja}$, and the generators with negative levels as they generate the mega-space frame. Thus, the only non-vanishing components of $\cW_{\Eh A}{}^{\Dh}$ are $\cW_{\Eh A}{}^{\cD}=\{\cW_{\Eh A}{}^{\exD},\,\cW_{\Eh A}{}^{D}\}$, and from \eqref{eq:pleq7} we obtain
\begin{align}
 X_{A} = \cW_{A} + (\widehat{s}^{\adjah})_{\cD}{}^{\Eh}\,\cW_{\Eh A}{}^{\cD}\, \widehat{s}_{\adjah} \,. 
\end{align}
The only component that could contribute with positive-level generators is
\begin{align}
 X_{A}\vert_{\text{positive}} = (\widehat{t}_{\text{negative}}^{\adjah})_{\cD}{}^{\Eh}\,\cW_{\Eh A}{}^{\cD}\, \widehat{s}^{\text{positive}}_{\adjah} \,, 
\end{align}
but the only non-vanishing $(\widehat{t}_{\text{negative}}^{\adjah})_{\cD}{}^{\Eh}$ that arises in this relation is $(R_{\exA}^B)_{D}{}^{\exE}=\delta_{\exA}^{\exE}\,\delta_D^B$ and we find
\begin{align}
 X_{A}\vert_{\text{positive}} = \cW_{\exE A}{}^{B}\,(R^{\exA}_B)_{\Bh}{}^{\Ch} = 0\,.
\end{align}
Therefore, $X_{A}$ does not contain any positive-level generators. Furthermore, one readily finds
\begin{align}
 X_{A\exB}{}^{\exC} = \cW_{A\exB}{}^{\exC} - \cW_{\exB A}{}^{\exC} = 0\,,
\end{align}
which confirms that $\widetilde{K}^{\exA}{}_{\exB}$ makes no contribution and establishing the validity of expansion \eqref{eq:XA-expansion}. We proceed to demonstrate that the coefficients of negative-level generators correspond to a hierarchy of torsion and curvature tensors.

\subsubsection{Level \texorpdfstring{$0$}{0}}

The level-$0$ component $X_{A}{}^{\adja}$ is determined through the $B$--$C$ components via
\begin{align}
 X_{AB}{}^C = X_{A}{}^{\adja}\,(s_{\adja})_B{}^C\,.
\end{align}
Employing \eqref{eq:pleq7}, we obtain
\begin{align}
 X_{AB}{}^{C} = \cW_{AB}{}^{C} + (s^{\adja})_{\Dh}{}^{\Eh}\,\cW_{\Eh A}{}^{\Dh}\,(s_{\adja})_{B}{}^{C}
 = \cW_{AB}{}^{C} + (s^{\adja})_{D}{}^{E}\,\cW_{EA}{}^{D}\,(s_{\adja})_{B}{}^{C},
\end{align}
where, in accordance with \eqref{eq:def-cW} and \eqref{eq:def-W},
\begin{align}
 \cW_{AB}{}^{C} = \check{W}_{AB}{}^{C} + \Omega_{AB}{}^{C}\,.
\label{eq:cW}
\end{align}

Since $\Omega_{AB}{}^C$ is identified with the spin connection associated with $\GS$, we employ it to define the generalized covariant derivative satisfying the vielbein postulate:
\begin{align}\label{eq:gen-vielbein-postulate}
 \tnabla_I E_A{}^J \equiv \partial_I E_A{}^J - \Omega_{IA}{}^B\,E_B{}^J + \Gamma_{IK}{}^J\,E_A{}^K = 0\,.
\end{align}
This fundamental constraint relates the spin connection to the affine connection $\Gamma_{IK}{}^J$, yielding
\begin{align}
 \Gamma_{AB}{}^C = \check{W}_{AB}{}^C + \Omega_{AB}{}^C = \cW_{AB}{}^{C} \,.
\end{align}

In conjunction with the $Y$-tensor, this enables us to express the level-$0$ contribution as
\begin{align}
 X_{AB}{}^C = 2\,\Gamma_{[AB]}{}^C + Y^{EC}_{BD}\,\Gamma_{EA}{}^D \equiv \cT_{AB}{}^C\,,
\label{eq:XABC}
\end{align}
thereby establishing its identification with the generalized torsion tensor for the connection $\tnabla$.

For $\GD=\Odd$ or $\GD=\Edd[d]$ with $d\leq 7$, considering \eqref{eq:gauge-level-0} and \eqref{eq:delta-Omega} gives the transformation law
\begin{align}
 \delta_{\hat{\xi}} X_{AB}{}^C = \mathbb{L}_{\hat{\xi}} X_{AB}{}^C + E_A{}^I\,E_B{}^J\,E_K{}^C\,\bigl(Y^{K(L}_{NI}\,\delta_J^{M)} - Y^{K(L|}_{PJ}\,Y^{P|M)}_{NI}\bigr)\,\partial_L\partial_M \xi^N\,.
\end{align}
Recalling identity \eqref{eq:YY-id-7}, the second term vanishes under the section condition. Hence $X_{AB}{}^C$ transforms covariantly, confirming the validity of \eqref{eq:gauge-level-0}, \eqref{eq:delta-Omega}, and \eqref{eq:XABC} for $\GD=\Odd$ or $\GD=\Edd[d]$ with $d\leq 7$, notwithstanding any modifications to the generalized Lie derivative.

For $\GD=\Edd[d]$ with $d\geq 8$, modifications in the generalized Lie derivative cannot be neglected. For $\Edd[8]$, the generalized torsion should become
\begin{align}
 X_{AB}{}^{C} &= 2\,\cW_{[AB]}{}^{C} + Y^{EC}_{BD}\,\cW_{EA}{}^{D} - \chi^{D\adja}\,\chi_{\adjb A}\,\cW_{D}{}^{\adjb}\,(t_{\adja})_{B}{}^{C} 
\nn\\
 &= 2\,\Gamma_{[AB]}{}^{C} + Y^{EC}_{BD}\,\Gamma_{EA}{}^{D} - \Gamma_{DA} \,f_{B}{}^{DC} \equiv \cT_{AB}{}^C\,,
\label{eq:E8-torsion}
\end{align}
as proposed in \cite{Cederwall:2015ica}. This expression is invariant under gauge transformations \eqref{eq:gauge-level-0} and \eqref{eq:delta-Omega}, provided that the appropriate generalized Lie derivative in $\Edd[8]$ ExFT is used. In principle, this result could be derived by starting from the generalized Lie derivative in $\Edd[p]$ ExFT with $p>8$, though we do not attempt this here.

\subsubsection{Level \texorpdfstring{$-1$}{-1}}
For the level-$(-1)$ part, we obtain
\begin{align}
 X_{AB}^{\phantom{A}\exB} &= \cW_{AB}^{\phantom{A}\exB} + (R^\exB_B)_{\Dh}{}^{\Eh}\,\cW_{\Eh A}{}^{\Dh}
\nn\\
 &= \cW_{AB}^{\phantom{A}\exB} 
 + (R^\exB_B)_{\exD}{}^{E}\,\cW_{EA}{}^{\exD}
 + (R^\exB_B)_{D\exE\EE}\,\cW^{\exE\EE}{}_{A}{}^D
 \underbrace{+\tfrac{1}{2}\, (R^\exB_B)_{D\exE}\,\cW^{\exE}{}_{A}{}^D}_{\Edd[7]}
 \nn\\\label{eq:X-level-1-raw}
 &= 2\,\cW_{[AB]}^{\phantom{[A}\exB} 
 + \eta_{BD;\EE}\,\cW^{\exB \EE}{}_{A}{}^D
 \underbrace{+\tfrac{1}{2}\,\omega_{BD}\,\cW^{\exB}{}_{A}{}^D}_{\Edd[7]}\,,
\end{align}
where the underbraced term appears only for $\Edd[7]$ (while $\Edd[8]$ is considered later) and $\cW_{AB}^{\phantom{A}\exB}$ arises as a coefficient in the expansion 
\begin{align}
 \cW_A = \cW_{A\exB}{}^{\exC}\,\widetilde{K}^{\exB}{}_{\exC} + \cW_{A}{}^{\adjb}\,s_{\adjb} + \cW_{AB}^{\phantom{A}\exB}\,R^B_{\exB} + \cdots\,.
\end{align}
We have employed the matrix elements $(R^\exB_B)_{\exD}{}^{E}= -\delta^\exB_{\exD}\,\delta^E_B$\,, $(R^\exB_B)_{D\exE \EE}=\eta_{BD;\EE}\,\delta^{\exB}_{\exE}$\,, and $(R^\exB_B)_{D\exE}=-\omega_{DB}\,\delta_{\exE}^{\exB}$\,, computed via the techniques of section~\ref{sec:representation}. The relevant components of the Weitzenb\"ock connection are given by
\begin{align}
 \cW_{AB}^{\phantom{A}\exB}
 &= D_A\Omega^\exB_B
 + \check{W}_{AB}{}^C\,\Omega_C^\exB
 + \Omega_A^\exA\,\Omega_B^\exC\,\check{W}_{\exA\exC}{}^\exB \,,
\\
 \cW^{\exB \EE}{}_{A}{}^D &= \eta^{BC;\EE}\,\Omega_B^\exB\,\check{W}_{CA}{}^D + \bigl(\rho^{\exB\exC;\EE}-\tfrac{1}{2}\,\Omega_C^\exB\,\eta^{CF;\EE}\,\Omega_F^\exC\bigr)\,X_{\exC A}{}^D\,,
\\
 \cW^{\exB}{}_{A}{}^D &= \omega^{BC}\,\Omega_B^\exB\,\check{W}_{CA}{}^D + \bigl(\rho^{\exB,\exC}-\tfrac{1}{2}\,\Omega_C^\exB\,\omega^{CE}\,\Omega_E^\exC\bigr)\,X_{\exC A}{}^D\,.
\end{align}
Substituting these into \eqref{eq:X-level-1-raw}, we find
\begin{align}
 X_{AB}^{\phantom{A}\exB} &= -2\,D_{[A}\Omega^\exB_{B]}
 - 2\,\check{W}_{[AB]}{}^C\,\Omega_C^\exB
 - 2\,\Omega_{[A|}^\exA\,\Omega_{|B]}^\exC\,\check{W}_{\exA\exC}{}^\exB
\nn\\\label{eq:X-level-1-refined}
 &\quad
 - Y_{DB}^{EC}\,\Omega_E^\exB\,\check{W}_{CA}{}^D
 - \bigl(\rho^{\exB\exC}{}_{;BC}+\tfrac{1}{2}\,\rho^{\exB,\exC}\,\omega_{BC}\bigr)\,X_{\exC A}{}^C 
 - \tfrac{1}{2}\,Y_{EB}^{CD}\,\Omega_C^\exB\,\Omega_D^\exC\,X_{\exC A}{}^E \,,
\end{align}
with $\rho^{\exB\exC}{}_{;BC}\equiv \rho^{\exB\exC;\EE}\,\eta_{BC;\EE}$\,. To further simplify this expression, we use
\begin{align}
 \Omega_A^\exA\,\check{W}_{\exA\exC}{}^\exB
 = \Omega_A^\exA\,v_\exA^\mu \,v_\exC^\nu\,\partial_\mu v_\nu^\exB\,,
\end{align}
to rewrite the last term on the first line of \eqref{eq:X-level-1-refined} as
\begin{align}
 2\,\Omega_{[A}^\exA\,\Omega_{B]}^\exC\,\check{W}_{\exA\exC}{}^\exB
 = 2\,\Omega_{[A}^\exA\,\Omega_{B]}^\exC\,v_\exA^\mu \,v_\exC^\nu\,\partial_{[\mu} v_{\nu]}^\exB
 = - \Omega_A^\exA\,\Omega_B^\exC\,f_{\exA\exC}{}^\exB \,.
\end{align}
Moreover, we define
\begin{align}
 R_{ABC}{}^D \equiv X_{AB}^{\phantom{A}\exC}\,(\mathfrak{t}_{\exC})_{C}{}^D\,,
\end{align}
to express the result purely in terms of $R_1$-indices of the duality group. In its final form, it reads
\begin{align}
 R_{ABC}{}^D
 &= 2\,D_{[A}\Omega_{B]C}{}^D
 + 2\,\check{W}_{[AB]}{}^E\,\Omega_{EC}{}^D
 - [\Omega_A,\,\Omega_B]_C{}^D 
\nn\\
 &\quad
 + Y_{GB}^{EF}\,\Omega_{EC}{}^D\,\check{W}_{FA}{}^G
 - \bigl(\rho^{\exB\exC}{}_{;EB}-\tfrac{1}{2}\,\rho^{\exB,\exC}\,\omega_{EB}\bigr)\,X_{\exC A}{}^E \,X_{\exB C}{}^D 
 + \tfrac{1}{2}\,Y_{EB}^{FG}\,\Omega_{FC}{}^D \,\Omega_{GA}{}^E 
\nn\\
 &= 2\,D_{[A}\Omega_{B]C}{}^D
 - 2\,\Omega_{[AB]}{}^E\,\Omega_{EC}{}^D
 - [\Omega_A,\,\Omega_B]_C{}^D 
\nn\\
 &\quad
 + \cT_{AB}{}^E\,\Omega_{EC}{}^D
 + \bm{\rho}_{AC;BE}^{ED} 
 - \tfrac{1}{2}\,Y_{EB}^{FG}\,\Omega_{FC}{}^D \,\Omega_{GA}{}^E \,.
\label{eq:gen-Riemann}
\end{align}
This is nothing else than the generalized Riemann tensor in flat indices. For $\GD=\Odd$ or $\GD=\Edd[d]$ with $d\leq 7$, it transforms covariantly after taking into account the transformation rules \eqref{eq:gauge-level-0}, \eqref{eq:delta-Omega}, and \eqref{eq:delta-rho} for its constituents. Hence, as expected, we are left with
\begin{align}
 \Delta_{\hat{\xi}} R_{ABC}{}^D = 0\,.
\end{align}
This suggests that the result \eqref{eq:gen-Riemann} is complete for $\GD=\Odd$ or $\GD=\Edd[d]$ with $d\leq 7$, although we have not introduced any modification in $X_A$. 

It follows directly from \eqref{eq:gen-Riemann} that the generalized Ricci tensor
\begin{align}
 R_{AB} \equiv - R_{(A|D|B)}{}^D
\end{align}
reads
\begin{align}
 R_{AB} &= D_{D}\Omega_{(AB)}{}^D - D_{(A|}\Omega_{D|B)}{}^D
 + \Omega_{(AB)}{}^E\,\Omega_{DE}{}^D
 - \Omega_{D(A|}{}^E\,\Omega_{E|B)}{}^D
\nn\\
 &\quad
 - \cT_{(A|C}{}^D\,\Omega_{D|B)}{}^C
 + \tfrac{1}{2}\,Y_{FE}^{CD}\,\Omega_{CA}{}^E \,\Omega_{DB}{}^F \,.
\end{align}
Remarkably, $\bm{\rho}_{AC;BE}^{ED}$ drops out under this projection. If we define the generalized fluxes associated with $E_A{}^I$ as
\begin{align}
 \check{X}_{AB}{}^C = 2\,\check{W}_{[AB]}{}^C + Y^{EC}_{BD}\,\check{W}_{EA}{}^D\,,
\label{eq:check-X}
\end{align}
we can rewrite the generalized Ricci tensor as
\begin{align}
 R_{AB} &= D_{D}\Omega_{(AB)}{}^D - D_{(A|}\Omega_{D|B)}{}^D
 + \Omega_{(AB)}{}^E\,\Omega_{DE}{}^D 
\nn\\
 &\quad
 - \bigl(\Omega_{(A|D}{}^E + \check{X}_{(A|C}{}^D\bigr)\,\Omega_{D|B)}{}^C
 - \tfrac{1}{2}\,Y_{FE}^{CD}\,\Omega_{CA}{}^E \,\Omega_{DB}{}^F \,.
\end{align}
As one might expect, this is the generalized Ricci tensor constructed in (5.20) of \cite{Aldazabal:2013mya} (for the $\Edd[7]$ case). Finally, there is the generalized Ricci scalar
\begin{align}
 R = \cM^{AB}\,R_{AB} \,,
\end{align}
which arises after contraction with the inverse of a (constant) $\GS$-invariant metric $\cM_{AB}\in \GD$ satisfying
\begin{align}
 (\mathfrak{t}_{\exA})_A{}^C\,\cM_{CB} + (\mathfrak{t}_{\exA})_B{}^C\,\cM_{AC} = 0\,.
\end{align}

As for the torsion at level $0$, it is instructive to switch to curved indices. After properly substituting the spin connection with the affine connection, the generalized Riemann/Ricci tensor become
\begin{align}
 R_{IJK}{}^L &= \partial_I\Gamma_{JK}{}^L - \partial_J\Gamma_{IK}{}^L +\Gamma_{IP}{}^L\,\Gamma_{JK}{}^P -\Gamma_{JP}{}^L\,\Gamma_{IK}{}^P
\nn\\
 &\quad +\tfrac{1}{2}\,Y^{RU}_{JS}\,\bigl(\Gamma_{RI}{}^S+\check{W}_{RI}{}^S\bigr)\,\bigl(\Gamma_{UK}{}^L-\check{W}_{UK}{}^L\bigr) 
 + \bm{\rho}_{IK;JM}^{ML} \,,
\\
 R_{IJ} &= \partial_{L}\Gamma_{(IJ)}{}^L
 - \partial_{(I|}\Gamma_{L|J)}{}^L
 + \Gamma_{(IJ)}{}^P\,\Gamma_{LP}{}^L
 - \Gamma_{(I|K}{}^L\,\Gamma_{L|J)}{}^K
\nn\\
 &\quad - \tfrac{1}{2}\,Y_{QP}^{KL}\,\bigl(\Gamma_{KI}{}^P \,\Gamma_{LJ}{}^Q - \check{W}_{KI}{}^P \,\check{W}_{LJ}{}^Q\bigr) \,.
\end{align}
For $\Odd$, raising or lowering indices with $\eta_{IJ}$ gives
\begin{align}
 R_{IJKL} &= \partial_I\Gamma_{JKL} - \partial_J\Gamma_{IKL} +\Gamma_{IPL} \,\Gamma_{JK}{}^P -\Gamma_{JPL}\,\Gamma_{IK}{}^P
\nn\\
 &\quad +\tfrac{1}{2}\, \bigl(\Gamma_{RIJ} +\check{W}_{RIJ} \bigr)\,\bigl(\Gamma^{R}{}_{KL} -\check{W}^R{}_{KL} \bigr) 
 + \rho_{IJ;KL}
\end{align}
with $\rho_{IJ;KL} = - \rho_{KL;IJ}$. This matches the generalized Riemann tensor constructed in \cite{Polacek:2013nla}, but we can eliminate $\rho$ by considering a symmetrized Riemann tensor $\cR_{IJKL}\equiv R_{IJKL}+R_{KLIJ}$\,,
\begin{align}
 \cR_{IJKL} = \mathring{R}_{IJKL} + \mathring{R}_{KLIJ} + \Gamma_{RIJ}\, \Gamma^{R}{}_{KL} - \check{W}_{RIJ}\,\check{W}^{R}{}_{KL} \,,
\end{align}
where $\mathring{R}_{IJKL}\equiv \partial_I\Gamma_{JKL} - \partial_J\Gamma_{IKL} +\Gamma_{IPL} \,\Gamma_{JK}{}^P -\Gamma_{JPL}\,\Gamma_{IK}{}^P$ is the ordinary Riemann tensor. This has been constructed in various earlier studies on DFT \cite{Siegel:1993xq,Siegel:1993th,Siegel:1993bj,Jeon:2011cn,Hohm:2011si}. The last term vanishes under the section condition, but was kept in (3.26) of \cite{Geissbuhler:2013uka} to also consider geometries where the section condition is replaced by the less restrictive closure constraint. For $\Edd[d]$, the generalized Riemann tensor has not been known until the present authors found it in \cite{Hassler:2023axp} for $\Edd[d]$ with $d\leq 6$. The expression for the generalized Ricci tensor matches (5.16) of \cite{Aldazabal:2013mya} (for $\Edd[7]$). Assuming the section condition, the last term vanishes and then (5.4) of \cite{Cederwall:2013naa} (for $\Edd[d]$ with $d\leq 7$) is reproduced.

Note that the generalized Riemann tensor $R_{IJK}{}^L$ is not totally determined by the supergravity fields. By imposing various conditions, such as metric compatibility and torsionlessness, we can determine most components of the connection $\Gamma_{IJ}{}^K$ in terms of the supergravity fields, but some components remain undetermined \cite{Jeon:2011cn,Hohm:2011si}. The supergravity equations of motion are expressed using components of the generalized Ricci tensor that do not contain the undetermined components. For example, in the $\Odd$ case, we can define two projectors
\begin{align}
 P_A{}^B \equiv \tfrac{1}{2}\,\bigl(\delta_A^B + \cM_A{}^B \bigr)\,,\qquad
 \bar{P}_A{}^B \equiv \tfrac{1}{2}\,\bigl(\delta_A^B - \cM_A{}^B \bigr)\,, 
\label{eq:P-Pbar}
\end{align}
and then $P_{A}{}^C\,\bar{P}_B{}^D\,R_{CD}$ are these components. Throughout this paper, we use the notation $\bar{A}$ and $\ubar{A}$ to denote indices on which the projectors $\bar{P}$ and $P$ have been applied, respectively; for example, $R_{\ubar{A}\bar{B}} \equiv P_A{}^C\,\bar{P}_B{}^D\,R_{CD}$\,. 

\paragraph{$\Edd[8]$ case:}
Up to this level, it is straightforward to repeat the same computation for $\Edd[8]$. Here we have
\begin{align}
 X_{AB}^{\phantom{A}\exB} &= \cW_{AB}^{\phantom{A}\exB} + (R^\exB_B)_{\Dh}{}^{\Eh}\,\cW_{\Eh A}{}^{\Dh}
\nn\\
 &= \cW_{AB}^{\phantom{A}\exB} 
 + (R^\exB_B)_{\exD}{}^{E}\,\cW_{EA}{}^{\exD}
 + (R^\exB_B)_{D\exE\EE}\,\cW^{\exE\EE}{}_{A}{}^D
 +\tfrac{1}{2}\, (R^\exB_B)_{D}{}^E_{\exE}\,\cW_E^{\exE}{}_{A}{}^D 
 +\tfrac{1}{4}\, (R^\exB_B)_{D\exE}\,\cW^{\exE}{}_{A}{}^D 
\nn\\
 &= 2\,\cW_{[AB]}^{\phantom{[A}\exB} 
 + \eta_{BD;\EE}\,\cW^{\exB \EE}{}_{A}{}^D
 +\tfrac{1}{2}\,f^E{}_{BD}\,\cW_E^{\exB}{}_{A}{}^D 
 + \tfrac{1}{4}\,\kappa_{BD}\, \cW^{\exB}{}_{A}{}^D 
\end{align}
with the relevant components of the Weitzenb\"ock connection:
\begin{align}
 \cW_{AB}^{\phantom{A}\exB}
 &= D_A\Omega^\exB_B
 + \check{W}_{AB}{}^C\,\Omega_C^\exB
 + \Omega_A^\exA \,\Omega_B^\exC\,\check{W}_{\exA\exC}{}^\exB \,,
\\
 \cW^{\exB \EE}{}_{A}{}^D &= \eta^{BC;\EE}\,\Omega_B^\exB\,\check{W}_{CA}{}^D + \bigl(\rho^{\exB\exC;\EE}-\tfrac{1}{2}\,\Omega_C^\exB\,\eta^{CF;\EE}\,\Omega_F^\exC\bigr)\,X_{\exC A}{}^D\,,
\\
 \cW_E^{\exB}{}_{A}{}^D &= f_E{}^{BC}\,\Omega_B^\exB\,\check{W}_{CA}{}^D + \bigl(\rho_E^{\exB,\exC}-\tfrac{1}{2}\,\Omega_C^\exB\,f_E{}^{CF}\,\Omega_F^\exC\bigr)\,X_{\exC A}{}^D\,,
\\
 \cW^{\exB}{}_{A}{}^D &= \kappa^{BC}\,\Omega_B^\exB\,\check{W}_{CA}{}^D + \bigl(\rho^{\exB\exC}-\tfrac{1}{2}\,\Omega_C^\exB\,\kappa^{CF}\,\Omega_F^\exC\bigr)\,X_{\exC A}{}^D\,.
\end{align}
Combining them, we obtain
\begin{align}
 X_{AB}^{\phantom{A}\exB} &= 2\,D_{[A}\Omega^\exB_{B]}
 + 2\,\check{W}_{[AB]}{}^C\,\Omega_C^\exB
 + 2\,\Omega_{[A|}{}^\exA\,\Omega_{|B]}^\exC\,\check{W}_{\exA\exC}{}^\exB
 + Y_{DB}^{EC}\,\Omega_E^\exB\,\check{W}_{CA}{}^D
\nn\\
 &\quad
 + \bigl(\rho^{\exB\exC}{}_{;BC}+\tfrac{1}{2}\,\rho_D^{\exB,\exC}\,f^D{}_{BC}+\tfrac{1}{4}\,\rho^{\exB\exC}\,\kappa_{BC}\bigr)\,X_{\exC A}{}^C 
 - \tfrac{1}{2}\,Y_{EB}^{CD}\,\Omega_C^\exB\,\Omega_D^\exC\,X_{\exC A}{}^E \,,
\end{align}
and finally
\begin{align}
 R_{ABC}{}^D
 &= 2\,D_{[A}\Omega_{B]C}{}^D
 - 2\,\Omega_{[AB]}{}^E\,\Omega_{EC}{}^D
 - [\Omega_A,\,\Omega_B]_C{}^D 
 + \cT_{AB}{}^E\,\Omega_{EC}{}^D
\nn\\
 &\quad
 + \bm{\rho}_{AC:BE}^{ED} 
 - \tfrac{1}{2}\,Y_{EB}^{FG}\,\Omega_{FC}{}^D \,\Omega_{GA}{}^E \,,
\label{eq:RABCD-E8}
\end{align}
where the generalized torsion $\cT_{AB}{}^C$ is given in \eqref{eq:XABC}. As we already mentioned there, this is not the covariant expression for $\Edd[8]$. However, if the expression \eqref{eq:XABC} is replaced by \eqref{eq:E8-torsion}, we find that the Riemann tensor transforms covariantly under the gauge transformations \eqref{eq:gauge-level-0}, \eqref{eq:delta-Omega}, and \eqref{eq:delta-rho}. We therefore conclude that \eqref{eq:gauge-level-0}, \eqref{eq:delta-Omega}, \eqref{eq:delta-rho}, and \eqref{eq:RABCD-E8}, together with the covariant generalized torsion $\cT_{AB}{}^C$ given in \eqref{eq:E8-torsion}, constitute the complete results. The somewhat cautious phrasing here is, again, due to the fact that we have not fully addressed the required modifications of the generalized Lie derivative on the mega-space.

\subsubsection{Level \texorpdfstring{$-2$}{-2}}
Let us finally consider the level-$(-2)$ part. For $\GD=\Odd$ or $\GD=\Edd[d]$ with $d\leq 7$, the relevant components of $X_A$ are
\begin{align}
 X_{A}^{\phantom{A}\exA\exB;\BB} &= \cW_{A}^{\phantom{A}\exA\exB;\BB} + (R^{\exA\exB;\BB})_{\Dh}{}^{\Eh}\,\cW_{\Eh A}{}^{\Dh}
\nn\\
&= \cW_{A}^{\phantom{A}\exA\exB;\BB}
 + (R^{\exA\exB;\BB})_{\exD\exE;\EE} \,\cW^{\exE;\EE}{}_{A}{}^{\exD}
 + \tfrac{1}{2!}\,(R^{\exA\exB;\BB})_{D\exE_1\exE_2;\EEE}\,\cW^{\exE_1\exE_2;\EEE}{}_{A}{}^{D} 
 + \tfrac{1}{12}\,(R^{\exA\exB;\BB})_{D}{}_{\exE_1\exE_2}^{\lE}\,\cW^{\exE_1\exE_2}_{\lE}{}_{A}{}^{D} 
\nn\\
 &= \cW_{A}^{\phantom{A}\exA\exB \BB}
 +2\,\delta^{\exA\exB}_{\exC\exD} \,\cW^{\exC;\BB}{}_{A}{}^{\exD}
 + Z_D{}^{\BB}{}_{\EEE}\,\cW^{\exA\exB;\EEE}{}_{A}{}^{D} 
 - \tfrac{1}{3}\,\eta_{D}{}^{\lE\BB}\,\cW^{\exA\exB}_{\lE}{}_{A}{}^{D} \,,
\end{align}
and for $\Edd[7]$, we also need
\begin{align}
 X_{A}^{\phantom{A}\exA,\exA'} &= \cW_{A}^{\phantom{A}\exA,\exA'} + (R^{\exA,\exA'})_{\Dh}{}^{\Eh}\,\cW_{\Eh A}{}^{\Dh}
\nn\\
 &= \cW_{A}^{\phantom{A}\exA,\exA'} + \tfrac{1}{2}\, (R^{\exA,\exA'})_{\exD\exE} \,\cW^{\exE}{}_{A}{}^{\exD} + \tfrac{1}{4}\,(R^{\exA,\exA'})_{D}{}_{\exE,\exE'}^E\,\cW^{\exE,\exE'}_{E}{}_{A}{}^{D}
\nn\\
 &= \cW_{A}^{\phantom{A}\exA,\exA'} + \cW^{(\exA}{}_{A}{}^{\exA')} - \tfrac{1}{2}\, \cW^{\exA,\exA'}_{D}{}_{A}{}^{D}\,.
\end{align}
Using $(W_A)_\exA{}^\exB = -\Omega_A^\exC\,\check{W}_{\exC\exA}{}^\exB$, we find
\begin{align}
 \cW_{A}^{\phantom{A}\exA\exB;\BB} 
 &= D_A\rho^{\exA\exB;\BB} -\eta^{CD;\BB}\,\Omega_C^{[\exA}\,D_A\Omega_D^{\exB]}
 - \rho^{\exA\exB;\CC}\,\check{W}_{A\CC}{}^{\BB} + 2\,(W_A)_{\exC}{}^{[\exA}\,\rho^{\exB]\exC;\BB}
\nn\\
 &\quad - \eta^{CF;\BB}\,\Omega_C^{[\exA}\,\Omega_D^{\exB]}\,\check{W}_{AF}{}^D 
 + \eta^{CD;\BB}\,\Omega_C^{[\exA|}\,\Omega_D^\delta\,(W_A)_{\delta}{}^{|\exB]} \,,
\\
 \cW^{\exC;\BB}{}_{A}{}^{\delta} 
 &= \eta^{CD;\BB}\,\Omega_{C}^\exC\,\bigl(D_D\Omega^\delta_A + \check{W}_{DA}{}^E\,\Omega_E^\delta \bigr) 
 -\bigl(\rho^{\exC\epsilon;\BB}-\tfrac{1}{2}\,\eta^{CD;\BB}\,\Omega^\exC_C\,\Omega_D^\epsilon\bigr)\,\Omega_A^\exB\,\check{W}_{\epsilon\exB}{}^\delta\,,
\\
 \cW^{\exA\exB;\EEE}{}_{A}{}^{D}
 &= \bigl(\rho^{\exA\exB;\FF}\,Z^B{}_{\FF}{}^{\EEE} - \Omega^{[\exA}_F\,\Omega^{\exB]}_G\,\eta^{BFG;\EEE} \bigr)\,\check{W}_{BA}{}^{D} 
\nn\\
 &\quad - \bigl(\rho^{\exA\exB\exC;\EEE} + \rho^{\exA\exB;\FF}\,\Omega_G^\exC\,Z^G{}_{\FF}{}^{\EEE}
 - \tfrac{1}{3}\,\eta^{FGH;\EEE}\,\Omega_G^{[\exA}\,\Omega_H^{\exB]}\,\Omega_F^\exC\bigr)\,X_{\exC A}{}^D\,,
\\
 \cW^{\exA\exB}_{\lE}{}_{A}{}^{D}
 &= -\bigl[2\,\rho^{\exA\exB;\CC}\,\eta^B{}_{\lE\CC} + \bigl(\tfrac{1}{2}\,\omega^{CD}\,\delta_{\lE}^B - 2\,\omega^{B[C}\,\delta^{D]}_{\lE}\bigr)\,\Omega_C^{[\exA}\,\Omega_D^{\exB]}\bigr]\,\check{W}_{BA}{}^{D} 
\nn\\
 &\quad + \bigl(\rho^{\exA\exB,\exC}_{\lE} + 2\,\rho^{\exA\exB;\CC}\,\Omega_F^{\exC}\,\eta^F{}_{\lE\CC}
 + \tfrac{1}{2}\,\omega^{CD}\,\delta_{\lE}^{F}\,\Omega_C^{[\exA}\,\Omega_D^{\exB}\,\Omega_F^{\exC]}\bigr)\,X_{\exC A}{}^D\,,
\\
 \cW_{A}^{\phantom{A}\exA,\exA'} &=
 D_A\rho^{\exA,\exA'}
 - \omega^{BC}\,\Omega_B^{(\exA}\,D_A\Omega_C^{\exA')} 
 + \omega^{BC}\,\Omega_B^{(\exA|}\,\Omega_C^{\exB}\,(W_A)_{\exB}{}^{|\exA')}
\nn\\
 &\quad -2\,\rho^{(\exA|,\exB}\,(W_A)_{\exB}{}^{|\exA')}
 -\omega^{CD}\,\Omega_C^{(\exA}\,\Omega_B^{\exA')}\,W_{AD}{}^B\,,
\\
 \cW^{\exA}{}_{A}{}^{\exB}&= \omega^{CD}\,\Omega_C^{\exA}\,\bigl(D_D\Omega_A^{\exB} + \check{W}_{DA}{}^{E}\,\Omega_E^{\exB}\bigr)
 -\Omega_A^{\exD}\, \bigl(\rho^{\exA,\exC}-\tfrac{1}{2}\,\omega^{CD}\,\Omega_C^{\exA}\,\Omega_D^{\exC}\bigr) \,\check{W}_{\exC\exD}{}^{\exB}\,,
\\
 \cW^{\exA,\exA'}_{C}{}_{A}{}^{D}&= -\bigl[\rho^{\exA,\exA'}\,\delta^B_{C} + \bigl(\eta^{B}{}_{C\EE}\,\eta^{FG;\EE}+2\,\omega^{B(F}\,\delta_{C}^{G)}\bigr)\,\Omega_F^{(\exA}\,\Omega_G^{\exA')}\bigr]\,\check{W}_{BA}{}^{D}
\nn\\
 &\quad +\bigl[ \rho^{\exA,\exA'}\,\Omega_C^{\exB} -\tfrac{2}{3}\,\rho^{\exB(\exA,\exA')}_{C}
 + \tfrac{1}{3}\,\bigl(c^{E(FG)}{}_{C} + \omega^{E(F}\,\delta_{C}^{G)}\bigr)\,\Omega_F^{(\exA}\,\Omega_G^{\exA')}\,\Omega_E^{\exB}\bigr]\,X_{\exB A}{}^D\,,
\end{align}
where $\check{W}_{A\BB}{}^{\CC}$ in the $R_2$-representation is defined to satisfy
\begin{align}
 \check{W}_{A\BB}{}^{\CC}\,\eta_{DE;\CC} + \check{W}_{AD}{}^{C}\,\eta_{CE;\BB} + \check{W}_{AE}{}^{C}\,\eta_{DC;\BB} = 0\,.
\end{align}

For $\GD=\Odd$ or $\GD=\Edd[d]$ with $d\leq 6$, combining the above results, we find that the level-$(-2)$ curvature
\begin{align}
 R_A{}^{PR}_{QS;FG} &\equiv X_{A}^{\phantom{A}\exA\exB \BB}\,(\mathfrak{t}_{\exA})_{Q}{}^P\,\,(\mathfrak{t}_{\exB})_{S}{}^R\,\eta_{FG;\BB}
\end{align}
takes the form
\begin{align}
 R_A{}^{PR}_{QS;FG}
 &= \tnabla_A\rho^{PR}_{QS;FG}
 + 2\,\cT_{A(F|}{}^H\,\rho^{PR}_{QS;|G)H}
 + \bm{\rho}^{PRD}_{QSA;FGD}
\nn\\
 &\quad + \tfrac{1}{2}\,Y^{CD}_{FG}\,\bigl[\bigl(R_{ACQ}{}^P - \bm{\rho}^{EP}_{AQ;CE} - D_C\Omega_{AQ}{}^P\bigr)\,\Omega_{DS}{}^R - ({}^P_Q\leftrightarrow {}^R_S)\bigr]
\nn\\
 &\quad + \tfrac{1}{6}\,Y^{I[X|}_{FG}\,Y^{Y]H}_{ID}\,\Omega_{XQ}{}^P\,\Omega_{YS}{}^R\,\Omega_{HA}{}^D
  - \tfrac{1}{3}\,\,Y^{H[X|}_{FG}\,\Omega_{XQ}{}^P\,\Omega_{YS}{}^R\,\Omega_{HA}{}^{|Y]}\,,
\label{eq:R2}
\end{align}
where
\begin{align}
 \tnabla_A\rho^{PR}_{QS;FG} &\equiv D_A\rho^{PR}_{QS;FG} + \Omega_{AT}{}^P\rho^{TR} + \Omega_{AT}{}^R\rho^{PT}
\nn\\
 &\quad - \Omega_{AQ}{}^T\rho^{PR}_{TS;FG}
 - \Omega_{AS}{}^T\rho^{PR}_{QT;FG}
 - \Omega_{AF}{}^T\rho^{PR}_{QS;TG}
 - \Omega_{AG}{}^T\rho^{PR}_{QS;FT}\,,
\\
 \bm{\rho}^{PRD}_{QSA;FGD} &\equiv
 \rho^{PRD}_{QSA;FGD} + \tfrac{1}{3}\,\eta_D{}^{\lE\BB}\,\eta_{FG;\BB}\,\rho^{PRD}_{QSA;\lE}\,.
\end{align}
A lengthy computation shows that this transforms covariantly under gauge transformations. For the $\Edd[7]$ case, however, we have not been able to obtain a consistent result, which suggests that the modification of the generalized Lie derivative needs to be properly incorporated.

In the $\Odd$ case, this marks the endpoint of the curvature hierarchy. For $\Edd[d]$, however, we can in principle consider lower levels where additional curvature tensors arise. Nonetheless, we do not pursue this direction here.

\subsection{Generalized Riemann tensor of the generalized Levi--Civita connection}
Following section~\ref{sec:torsionless-GL}, we also consider a torsionless connection. For the cases $\GD=\Odd$ or $\GD=\Edd[d]$ with $d\leq 7$, we begin by examining the generalized Riemann tensor of the generalized connection $\tnabla$, which takes the form
\begin{align}
 R_{ABC}{}^D
 &= 2\,D_{[A}\Omega_{B]C}{}^D
 - 2\,\Omega_{[AB]}{}^E\,\Omega_{EC}{}^D
 - [\Omega_A,\,\Omega_B]_C{}^D 
\nn\\
 &\quad
 + \cT_{AB}{}^E\,\Omega_{EC}{}^D
 + \bm{\rho}_{AC;BE}^{ED} 
 - \tfrac{1}{2}\,Y_{EB}^{FG}\,\Omega_{FC}{}^D \,\Omega_{GA}{}^E \,,
\end{align}
but this generalized connection generally has non-zero torsion $\cT_{AB}{}^C = X_{AB}{}^C$. Hence, we introduce the generalized Levi--Civita connection $\bnabla \equiv \nabla - \SSS$ and define the corresponding connections
\begin{align}
 \omega_{AB}{}^C &\equiv \Omega_{AB}{}^C - \SSS_{AB}{}^C \,, \qquad\text{and}
\\
 \bm{r}_{AC;BE}^{ED} &\equiv \bm{\rho}_{AC;BE}^{ED} + \tfrac{1}{2}\,Y^{FG}_{BE}\,\bigl( \SSS_{FA}{}^E\,\omega_{GC}{}^D - \SSS_{FC}{}^D\,\omega_{GA}{}^E \bigr)\,,
\end{align}
where the generalized contorsion tensor $(\SSS_A)_B{}^C = \SSS_{AB}{}^C$ is an element of $\GD$. For $\bnabla$ to be torsion-free, $\SSS_{AB}{}^C$ should satisfy
\begin{align}
 2\,\SSS_{[AB]}{}^E + Y^{EG}_{BF}\,\SSS_{GA}{}^F = \cT_{AB}{}^E\,,
\label{eq:cF-torsionless}
\end{align}
which generalizes the standard contorsion-torsion relation \eqref{eq:contorsion-torsion-0} from $\GL(d)$ to arbitrary $\GD$. With these new connections, we compute the generalized Riemann tensor of $\bnabla$ as
\begin{align}
 \bar{R}_{ABC}{}^D &\equiv
 2\,D_{[A}\omega_{B]C}{}^D
 - 2\,\omega_{[AB]}{}^E\,\omega_{EC}{}^D
 - [\omega_A,\,\omega_B]_C{}^D 
\nn\\
 &\quad
 + \bm{r}_{AC;BE}^{ED} 
 - \tfrac{1}{2}\,Y_{EB}^{FG}\,\omega_{FC}{}^D \,\omega_{GA}{}^E \,,
\end{align}
or equivalently,
\begin{align}
 \bar{R}_{ABC}{}^D &= R_{ABC}{}^D
 - \cT_{AB}{}^E\,\SSS_{EC}{}^D - 2\,\SSS_{[A|C}{}^E\,\SSS_{|B]E}{}^D
\nn\\
 &\quad + \tfrac{1}{2}\,Y^{FG}_{BE}\,\SSS_{FC}{}^D\,\SSS_{GA}{}^E - 2\,\tnabla_{[A}\SSS_{B]C}{}^D\,.
\label{eq:barR-SSS}
\end{align}
This reproduces \eqref{eq:bar-Riemann} for $\GD=\GL(d)$. Through appropriate contractions, this directly gives the generalized Ricci tensor
\begin{align}
 \bar{R}_{AB} &= \SSS_{(AB)}{}^D\,\SSS_{CD}{}^C - \SSS_{C(A}{}^D\,\SSS_{B)D}{}^C + \cT_{(A|C}{}^D\,\SSS_{D|B)}{}^C - \tfrac{1}{2}\,Y^{EF}_{CD}\,\SSS_{E(A|}{}^C\,\SSS_{F|B)}{}^D
\nn\\
 &\quad - \tnabla_{C}\SSS_{(AB)}{}^C + \tnabla_{(A|}\SSS_{C|B)}{}^C + R_{AB}\,.
\label{eq:R-omega}
\end{align}

In addition to \eqref{eq:cF-torsionless}, the generalized Levi--Civita connection should be metric compatible, requiring that $\SSS_{AB}{}^C$ satisfies
\begin{align}
 \SSS_{DA}{}^C\,\cM_{CB} + \SSS_{DB}{}^C\,\cM_{AC} = 0\,. 
\end{align}
For $\Odd$, a generalized contorsion tensor $\SSS_{AB}{}^C$ can be explicitly constructed as \cite{Geissbuhler:2013uka}
\begin{align}
 \SSS_{ABC} &= \tfrac{1}{3}\,\bigl(\cT_{ABC} + \cM_B{}^D\,\cM_C{}^E\,\cT_{ADE} - \tfrac{1}{2}\,\cM_A{}^D\,\cM_B{}^E\,\cT_{DEC} - \tfrac{1}{2}\,\cM_A{}^D\,\cM_C{}^E\,\cT_{DBE}\bigr)
\nn\\
 &\quad - \tfrac{2}{2\,d+\mathfrak{n}-2}\,\bigl(\eta_{A[B}\,\cT_{C]} + \cM_{A[B}\,\cM_{C]}{}^D\,\cT_D\bigr)\,. 
\label{eq:Sigma-semi-covariant}
\end{align}
Here, we use $\cT_A \equiv \SSS_{BA}{}^B$ which is related to the dilaton flux $F_A \equiv 2\,D_A d - \partial_I E_A{}^{I}$ through
\begin{align}
 \cT_A = \SSS_{BA}{}^B = \Omega_{BA}{}^B - \omega_{BA}{}^B = F_A + \Omega_{BA}{}^B\,,
\end{align}
where, in the last equality, we have used $F_A = -\omega_{BA}{}^B$, i.e., the covariant constancy of the dilaton $\bnabla_I d =0$. 

Note that the solution \eqref{eq:Sigma-semi-covariant} for the generalized contorsion tensor is not unique. This non-uniqueness reflects a fundamental feature of generalized connections discussed around \eqref{eq:P-Pbar}: they cannot be fully determined solely by covariant constraints on physical fields and inevitably contain undetermined components \cite{Jeon:2011cn,Hohm:2011si}. Crucially, however, physically relevant quantities are independent of these undetermined components. In particular, the generalized Ricci tensor and scalar are free of these undetermined components and are given by
\begin{align}
 \bar{R}_{\ubar{A}\bar{B}} &= - \bigl(\cT_{\ubar{C}} - \tnabla_{\ubar{C}}\bigr) \cT_{\ubar{A}\bar{B}}{}^{\ubar{C}}
 + \tnabla_{\bar{B}}\cT_{\ubar{A}}
 + \cT_{\ubar{A}\ubar{C}\bar{D}}\,\cT_{\bar{B}}{}^{\ubar{C}\bar{D}}
 + R_{\ubar{A}\bar{B}}\,,
\label{eq:DFT-Ricci-t}
\\
 \bar{R} &= -\tfrac{1}{12}\,\cM^{AD}\,\bigl(3\,\eta^{BE}\,\eta^{CF}-\cM^{BE}\,\cM^{CF}\bigr)\,\cT_{ABC}\,\cT_{DEF}
\nn\\
 &\quad + \cM^{AB}\,\bigl(2\,\tnabla_A \cT_A - \cT_A\,\cT_B \bigr) + R \,,
\label{eq:DFT-Ricci-s}
\end{align}
where $\bar{R} \equiv \cM^{AB}\,\bar{R}_{\ubar{A}\bar{B}}$\,. Of course, they can also be expressed in the familiar form as
\begin{align}
 \bar{R}_{\ubar{A}\bar{B}} &= - \bigl(F_{\ubar{C}} - D_{\ubar{C}}\bigr) \check{X}_{\ubar{A}\bar{B}}{}^{\ubar{C}}
 + D_{\bar{B}}F_{\ubar{A}}
 + \check{X}_{\ubar{A}\ubar{C}\bar{D}}\,\check{X}_{\bar{B}}{}^{\ubar{C}\bar{D}} \,,
\\
 \bar{R} &= -\tfrac{1}{12}\,\cM^{AD}\,\bigl(3\,\eta^{BE}\,\eta^{CF}-\cM^{BE}\,\cM^{CF}\bigr)\,\check{X}_{ABC}\,\check{X}_{DEF}
\nn\\
 &\quad + \cM^{AB}\,\bigl(2\,D_A F_A - F_A\,F_B \bigr) \,,
\end{align}
but the former expressions \eqref{eq:DFT-Ricci-t} and \eqref{eq:DFT-Ricci-s} are particularly advantageous when analyzing generalized dualities for generalized cosets in the next section. There, $\cT_{ABC}$, $\cT_A$, $R_{AB}$, and $R$ are constant and we can clearly see that $\bar{R}_{\ubar{A}\bar{B}}$ and $\bar{R}$, which appear in the equations of motion, are also constant. 

Finally, we note that in the $\Edd[d]$ case, similar expressions for $\SSS_{AB}{}^C$ have been found in \cite{Coimbra:2011ky,Coimbra:2012af,Park:2013gaj}, and it is again known that the undetermined parts of $\SSS_{AB}{}^C$ do not contribute to the generalized Ricci tensor or scalar.

\section{Consistent truncations and generalized duality}\label{sec:consistent-tr-dualities}
Finding solutions to supergravity's field equations is an important step in addressing many questions arising from string theory and holography. A powerful tool for making this complicated task more feasible is the use of consistent truncations. These simplify the problem considerably by removing decoupled degrees of freedom, typically in dimensional reduction. However, due to the non-linear nature of the field equations, constructing such truncations is challenging. Despite this difficulty, consistent truncations provide a systematic embedding of various gauged supergravities into ten- or eleven-dimensional theories, which elucidates their origins within string theory.

A well-established class of consistent truncations is based on generalized Scherk--Schwarz reductions, as explored in works such as \cite{Berman:2012uy,Lee:2014mla,Hohm:2014qga,Baguet:2015sma,duBosque:2015ehy,duBosque:2017dfc,Inverso:2017lrz,Bugden:2021wxg,Bugden:2021nwl,Hulik:2022oyc,Hassler:2022egz,Inverso:2024xok}. While initially challenging due to the non-linearity of the field equations, subsequent developments have reformulated this approach in purely algebraic terms, thereby circumventing these difficulties. However, it preserves the full supersymmetry of the original theory, so that truncations starting from maximal supergravity in ten or eleven dimensions necessarily lead to lower-dimensional gauged maximal supergravity without supersymmetry breaking.

Investigating consistent truncations that allow for supersymmetry breaking requires a more sophisticated framework. For instance, consistent truncations of maximal supergravities to half-maximal supergravities have been studied in \cite{Malek:2016vsh, Malek:2017njj}, and more recently a refined approach to constructing consistent truncations has been developed in \cite{Cassani:2019vcl}. This approach is governed by a structure group $\GS$ on the physical space $M$, which defines a set of $\GS$-invariant tensors denoted $\{Q_i\}$. Incorporating fermionic degrees of freedom requires that $\GS \subset K(\GD)$, where $K(F)$ denotes the maximal compact subgroup of a group $F$. If the intrinsic torsion is both $\GS$-invariant and constant, then a consistent truncation to lower dimensions can be achieved by keeping only the invariant tensors $\{Q_i\}$ in the reduction ansatz for the supergravity fields. The amount of supersymmetry preserved in the truncated theory is dictated by the number of supercharges that remain invariant under the action of the structure group $\GS$.

Although \cite{Cassani:2019vcl} provides a precise list of the necessary ingredients for constructing consistent truncations, it does not provide a systematic method for obtaining them. We address this problem here by introducing compactifications based on generalized cosets, a generalization of generalized group manifolds which underlie generalized Scherk--Schwarz reductions. They are characterized by the structure constants $\hat{X}_{\Ah\Bh}{}^{\Ch}$ of a Leibniz algebra and the embedding of a compatible structure group $\GS$. While they do not lead to the most general class of consistent truncations, they have the key advantage of providing a systematic construction for all invariant tensors $\{Q_i\}$ and the constant intrinsic torsion through an algebraic procedure which is exclusively based on the structure constants $\hat{X}_{\Ah\Bh}{}^{\Ch}$. Furthermore, they are directly related to generalized dualities and give rise to a highly versatile family of consistent truncations as demonstrated through various examples in sections~\ref{sec:examples-DFT} and \ref{sec:examples-ExFT}.

\subsection{Standard cosets}\label{sec:standard-coset}
Before addressing generalized cosets, let us first review standard cosets in general relativity as a preliminary example. We assume that the mega-space is a group manifold $G$ and construct the coset space $\GS\backslash G$. Moreover, the mega-space frame field $\hat{e}_{\hat{a}}{}^{\hat{i}}$ is built from the right-invariant vector fields satisfying
\begin{align}
 \Lie_{\hat{e}_{\hat{a}}} \hat{e}_{\hat{b}} = - \hat{X}_{\hat{a}\hat{b}}{}^{\hat{c}}\,\hat{e}_{\hat{c}}\,,
\end{align}
where $\hat{X}_{\hat{a}\hat{b}}{}^{\hat{c}}$ are the structure constants of $G$. To proceed, we parameterize group elements $\hat{m}_{\hat{a}}{}^{\hat{b}} \in G$ as
\begin{align}
 \hat{m}_{\hat{a}}{}^{\hat{b}}(y^{\hat{i}}) = \widetilde{m}_{\hat{a}}{}^{\hat{c}}(y^{\mu})\,m_{\hat{c}}{}^{\hat{b}}(y^{i})\,,
\label{eq:hatm-dec}
\end{align}
where $\widetilde{m}_{\hat{a}}{}^{\hat{b}}\in \GS$ and $m_{\hat{a}}{}^{\hat{b}}\in \GS\backslash G$. The left-invariant Maurer--Cartan 1-form $\hat{v}_{\hat{i}}{}^{\hat{a}}$ is defined by
\begin{align}
 \hat{m}^{-1}\,\partial_{\hat{i}} \hat{m} = \hat{v}_{\hat{i}}{}^{\hat{a}}\,\mathfrak{t}_{\hat{a}}\qquad \bigl(\mathfrak{t}_{\hat{a}}=-\hat{X}_{\hat{a}}\bigr)\,.
\end{align}
Using its inverse matrix $\hat{v}_{\hat{b}}{}^{\hat{i}}$, we can alternatively write $\hat{e}_{\hat{a}}{}^{\hat{i}}$ as
\begin{align}
 \hat{e}_{\hat{a}}{}^{\hat{i}} = \hat{m}_{\hat{a}}{}^{\hat{b}} \, \hat{v}_{\hat{b}}{}^{\hat{i}}\,.
\end{align}
Based on the parameterization of the group element in \eqref{eq:hatm-dec}, we obtain
\begin{align}
 \hat{m}^{-1}\,\partial_{\hat{i}} \hat{m} 
 = m^{-1}\,\bigl(\widetilde{m}^{-1}\,\partial_{\hat{i}}\widetilde{m} + \partial_{\hat{i}}m\,m^{-1}\bigr)\,m 
 = m^{-1} \begin{pmatrix}
 \widetilde{v}_{\mu}{}^{\exA}\,\mathfrak{t}_{\exA} \\
 \Omega_{i}^{\exA} \,\mathfrak{t}_{\exA} + e_{i}{}^{a} \,\mathfrak{t}_{a} \end{pmatrix} m\,,
\end{align}
after defining
\begin{align}
 \widetilde{m}^{-1}\,\partial_{\mu} \widetilde{m} = \widetilde{v}_{\mu}{}^{\exA}\,\mathfrak{t}_{\exA}\,,\qquad \text{and} \qquad
 \partial_i m\,m^{-1} = \Omega_{i}^{\exA}\,\mathfrak{t}_{\exA} + e_{i}{}^{a}\,\mathfrak{t}_{a}\,.
\end{align}
From here, we derive the expressions for $\hat{v}_{\hat{a}}{}^{\hat{i}}$ and eventually for $\hat{e}_{\hat{a}}{}^{\hat{i}}$, 
\begin{align}
 \hat{v}_{\hat{a}}{}^{\hat{i}} = (m^{-1})_{\hat{a}}{}^{\hat{b}} \begin{pmatrix} \widetilde{v}_{\exB}{}^{\mu} & 0 \\
 - \Omega_{a}^{\exB}\,\widetilde{v}_{\exB}{}^{\mu} & e_{b}{}^{i}
\end{pmatrix} ,\qquad
 e_{\hat{a}}{}^{\hat{i}} = \begin{pmatrix} \widetilde{v}_{\exA}{}^{\mu} & 0 \\
 - \Omega_{a}^{\exB}\,\widetilde{v}_{\exB}{}^{\mu} & e_{b}{}^{i}
\end{pmatrix} ,
\end{align}
where $\Omega_{a}^{\exB}\equiv e_{a}{}^{i}\,\Omega_{i}^{\exB}$. Remarkably, the resulting expression for $e_{\hat{a}}{}^{\hat{i}}$ is of the same form as given in \eqref{eq:hat-e-gauge}. Furthermore, one can confirm that all other assumptions made in section~\ref{sec:setup} are also satisfied. Hence, we have found a class of examples that fit within the general structures discussed in section~\ref{sec:GL}. A particularly convenient property of this setup is that we can show
\begin{align}
 X_{\hat{a}\hat{b}}{}^{\hat{c}} = \hat{X}_{\hat{a}\hat{b}}{}^{\hat{c}}\,.
\end{align}
Thus, the torsion and the Riemann tensor constructed from $X_{\hat{a}\hat{b}}{}^{\hat{c}}$ are constant. 

\subsection{Generalized cosets}\label{sec:gen-coset}
We now turn to the construction of generalized cosets \cite{Butter:2022iza,Hassler:2023axp}, assuming that the mega-space $\hat{M}$ is a generalized parallelizable space which admits a generalized frame field $\Eh_{\Ah}{}^{\Ih}$ satisfying
\begin{align}
 \gLie_{\Eh_{\Ah}} \Eh_{\Bh}{}^{\Ih} = - \hat{X}_{\Ah\Bh}{}^{\Ch}\,\Eh_{\Ch}{}^{\Ih} \,.
\label{eq:mega-parallel}
\end{align}
This is equivalent to assuming that $X_{\Ah\Bh}{}^{\Ch}$ is covariantly constant,
\begin{align}
 \nabla_D X_{\Ah\Bh}{}^{\Ch} = 0\,,
\end{align}
upon defining the covariant derivative $\nabla$ of a generalized tensor $T$ as
\begin{align}
 D_{\Ah} \widehat{T} &= D_{\Ah} (\widetilde{M}\cdot T) = \widetilde{M}\cdot \bigl(D_{\Ah} T + \widetilde{M}^{-1}\,D_{\Ah}\widetilde{M} \cdot T \bigr) 
\nn\\
 &= \widetilde{M} \cdot \bigl(D_{\Ah} T + N_{\Ah}{}^{\exA}\, \mathfrak{t}_{\exA} \cdot T \bigr) \equiv \widetilde{M} \cdot \nabla_{\Ah} T \,.
\end{align}
One motivation for this definition is that its restriction to the physical space coincides with the covariant derivative associated with the spin connection $\Omega_{AB}{}^C$,
\begin{align}
 \nabla_{D} X_{AB}{}^C = D_D X_{AB}{}^C - \Omega_{DA}{}^E\,X_{EB}{}^C - \Omega_{DB}{}^E\,X_{AE}{}^C + \Omega_{DE}{}^C\,X_{AB}{}^E \,.
\end{align}
Because $\hat{X}_{\Ah}=(\hat{X}_{\Ah\Bh}{}^{\Ch})$ is constant, the Bianchi identity for the generalized fluxes reduces to the Leibniz identity,
\begin{align}
 [\hat{X}_{\Ah},\,\hat{X}_{\Bh}] = - \hat{X}_{\Ah\Bh}{}^{\Ch}\,\hat{X}_{\Ch}\,,
\end{align}
which allows us to define the corresponding Leibniz algebra
\begin{align}
 \hat{T}_{\Ah}\circ \hat{T}_{\Bh} = \hat{X}_{\Ah\Bh}{}^{\Ch}\,\hat{T}_{\Ch}\,.
\label{eq:Leibniz-Alg}
\end{align}
We additionally define the matrices $\mathfrak{t}_{\Ah}=-\hat{X}_{\Ah}$\,, which satisfy the same algebra
\begin{align}
 [\mathfrak{t}_{\Ah},\, \mathfrak{t}_{\Bh}] = \hat{X}_{\Ah\Bh}{}^{\Ch}\,\mathfrak{t}_{\Ch}\,.
\end{align}
In general, the structure constants $\hat{X}_{\Ah\Bh}{}^{\Ch}$ contain a symmetric part. However, for this Leibniz algebra, it is possible to determine the associated Lie group $G$ (see, for example, \cite{Hassler:2022egz}). To construct generalized cosets, this $G$ must contain the structure group $\GS$ as an isotropic subgroup. 

By choosing a Lie group $H$ as an arbitrary minimally coisotropic subalgebra of $G$, we can construct the mega-space as the coset space $\hat{M} = G/H$. Following the procedure presented in \cite{Inverso:2017lrz,Bugden:2021wxg,Bugden:2021nwl,Hulik:2022oyc,Hassler:2022egz,Inverso:2024xok}, we can systematically construct the generalized frame field $\Eh_{\Ah}{}^{\Ih}$ satisfying \eqref{eq:mega-parallel} as
\begin{align}
 \Eh_{\Ah}{}^{\Ih} = \Mh_{\Ah}{}^{\Bh} \,\Vh_{\Bh}{}^{\Jh} \,\Nh_{\Jh}{}^{\Ih} \,,
\end{align}
where $\Mh_{\Ah}{}^{\Bh}$ is an element of $G/H$ and $\Vh_{\Ah}{}^{\Ih}$ is constructed as a block-diagonal matrix
\begin{align}
 \Vh_{\Ah}{}^{\Ih} = \diag \begin{pmatrix} \hat{v}_{\hat{a}}{}^{\hat{i}} & \cdots \end{pmatrix} ,
\label{eq:Vh}
\end{align}
with the right-invariant Maurer--Cartan form $\hat{v}_{\hat{i}}{}^{\Ah}$ as the only input. The latter satisfies
\begin{align}
 \Mh^{-1}\,\partial_{\hat{i}} \Mh = \hat{v}_{\hat{i}}{}^{\Ah}\,\mathfrak{t}_{\Ah} \,,
\label{eq:hat-v}
\end{align}
such that $\hat{v}_{\hat{a}}{}^{\hat{i}}$ is the inverse matrix of $\hat{v}_{\hat{i}}{}^{\hat{a}}$. Finally, the unipotent matrix $\Nh_{\Ih}{}^{\Jh}$ is constructed from $\hat{v}_{\hat{i}}{}^{\Ah}$ and $\hat{X}_{\Ah\Bh}{}^{\Ch}$. All necessary details are given in \cite{Hassler:2022egz}. For the moment, it suffices to note that it exists.

In order to identify the physical space $M=\GS\backslash \hat{M}$ and its curvatures, discussed in section \ref{sec:frames-curvatures}, we decompose $\Mh_{\Ah}{}^{\Bh}$ according to
\begin{align}
 \Mh_{\Ah}{}^{\Bh}(y^{\hat{i}}) = \widetilde{M}_{\Ah}{}^{\Ch}(y^{\mu})\,M_{\Ch}{}^{\Bh}(y^{i})\,.
\label{eq:Mh-dec}
\end{align}
This decomposition is also suggested by the insights gained from the standard coset in \eqref{eq:hatm-dec}. Next, we define
\begin{align}
 \widetilde{M}^{-1}\,\partial_{\mu} \widetilde{M} = - \widetilde{v}_{\mu}{}^{\exA}\,\hat{X}_{\exA}\,, 
\end{align}
and
\begin{align}\label{eq:hat-v-2}
 \Mh^{-1}\,\partial_{\hat{i}} \Mh = M^{-1}\,\bigl(\widetilde{M}^{-1}\,\partial_{\hat{i}} \widetilde{M} + \partial_{\hat{i}} M\,M^{-1}\bigr)\,M = - \hat{v}_{\hat{i}}{}^{\Ah} \hat{X}_{\Ah} \,,
\end{align}
which allows us to express the $\hat{v}_{\hat{i}}{}^{\hat{a}}$ component of $\hat{v}_{\hat{i}}{}^{\Ah}$ as
\begin{align}
 \hat{v}_{\hat{i}}{}^{\hat{a}} = \begin{pmatrix} \widetilde{v}_{\mu}{}^{\exB}(y^\mu) &0 \\
 \Omega_{i}^{\exB}(y^i) & e_{i}{}^{b}(y^i)\end{pmatrix} M_{\hat{b}}{}^{\hat{a}}(y^i) .
\label{eq:hat-v1-dec}
\end{align}
Its inverse, 
\begin{align}
 \hat{v}_{\hat{a}}{}^{\hat{i}} = (\bm{M}^{-1})_{\hat{a}}{}^{\hat{b}}\begin{pmatrix} \widetilde{v}_{\exB}{}^{\mu}(y^\mu) &0 \\
 -\Omega_{b}^{\exC}(y^i)\,\widetilde{v}_{\exC}{}^{\mu}(y^\mu) & e_{b}{}^{i}(y^i) \end{pmatrix} ,
\label{eq:hat-v-dec}
\end{align}
where $\Omega_{a}^{\exB}\equiv e_{b}{}^{i}\,\Omega_{i}^{\exB}$ and $(\bm{M}^{-1})_{\hat{a}}{}^{\hat{b}}$ is the inverse of $M_{\hat{a}}{}^{\hat{b}}$, is needed to eventually compute $\Vh_{\Ah}{}^{\Ih}$ in \eqref{eq:Vh}. With this result, we find that the $E_{\hat{a}}{}^{\hat{i}}$ component of $E_{\Ah}{}^{\Ih}$ reads
\begin{align}
 E_{\hat{a}}{}^{\hat{i}} = \begin{pmatrix} \widetilde{v}_{\exA}{}^{\mu}(y^\mu) &0 \\
 -\Omega_{a}^{\exB}(y^i)\,\widetilde{v}_{\exB}{}^{\mu}(y^\mu) & e_{a}{}^{i}(y^i) \end{pmatrix} .
\end{align}
The generalized frame field $\Eh_{\Ah}{}^{\Ih}$ is obtained by further acting with $\widetilde{M}_{\Ah}{}^{\Bh}$ to $\Eh_{\exB}{}^{\Ih}$. 
From this result, we can compute the top components $\Eh_{\exA}{}^{\hat{i}}$ of $\Eh_{\exA}{}^{\Ih}$ as
\begin{align}
 \Eh_{\exA}{}^{\hat{i}} =\begin{pmatrix} \widetilde{e}_{\exA}{}^{\mu} &0\end{pmatrix},\qquad
 \widetilde{e}_{\exA}{}^{\mu} \equiv \widetilde{M}_{\exA}{}^{\exB}(y^\mu)\,\widetilde{v}_{\exB}{}^{\mu}(y^\mu)
\end{align}
In general, other components of $\Eh_{\exA}{}^{\Ih}$ may be non-zero, but since $\Eh_{\exA}$ must satisfy the (isotropic) algebra
\begin{align}
 \gLie_{\Eh_{\exA}}\Eh_{\exB} = - \hat{X}_{\exA\exB}{}^{\exC}\,\Eh_{\exC}\,,
\end{align}
it is possible, by repeating the discussion above \eqref{eq:RI}, to choose a matrix $\Nh_{\Ih}{}^{\Jh}$ such that $\Eh_{\exA} = \widetilde{e}_{\exA}{}^{\mu} \,\partial_\mu$ holds. Then the entire setup discussed in section \ref{sec:frames-curvatures} is realized. The matrix $e_{a}{}^{i}$ and $\Omega_{a}^{\exB}$ introduced in \eqref{eq:hat-v1-dec} appear as the top components of $E_A{}^I$ in \eqref{eq:V-decomp} and the top components of $\Omega_A^{\exA}$ in \eqref{eq:N-param-1}--\eqref{eq:N-param-3}.

Using the Leibniz identity, one can show that the untwisted fluxes $X_{\Ah\Bh}{}^{\Ch}$ coincide with the structure constants:
\begin{align}
 X_{\Ah\Bh}{}^{\Ch} 
 = (\widetilde{M}^{-1})_{\Ah}{}^{\Dh}\,(\widetilde{M}^{-1})_{\Bh}{}^{\Eh}\, \widetilde{M}_{\Fh}{}^{\Ch}\,
 \hat{X}_{\Dh\Eh}{}^{\Fh} =\hat{X}_{\Ah\Bh}{}^{\Ch} \,.
\end{align}
Hence, the hierarchy of curvatures on the physical space $M=\GS\backslash \hat{M}$, such as the generalized torsion $\cT_{AB}{}^{C} =\hat{X}_{AB}{}^{C}$ and the generalized Riemann tensor $R_{ABC}{}^D = \hat{X}_{AB}{}^{\exC}\,(\mathfrak{t}_{\exC})_{C}{}^D$, constructed from $X_A=\hat{X}_A$, are (covariantly) constant. This type of physical space is similar to the coset space $M = \GS \backslash \hat{M}$ with $\hat{M} = G$ discussed in section \ref{sec:standard-coset}. However, a key difference is that a generalized parallelizable space is not necessarily a group manifold but rather a coset space $\hat{M} = G / H$ in its own right. Accordingly, the physical space takes the form of the double coset
\begin{align}
 M = \GS \backslash G / H \,.
\end{align}
We refer to these spaces as generalized cosets \cite{Demulder:2019vvh,Butter:2022iza,Hassler:2023axp}. In particular, when $\GD = \Edd$, they are called exceptional generalized cosets \cite{Hassler:2023axp}.

If we start with a generalized coset, the mega-space is a generalized parallelizable space characterized by a constant $\hat{X}_{\Ah\Bh}{}^{\Ch}$, and we can readily show $\nabla_{D} X_{\Ah\Bh}{}^{\Ch} =0$. This can be decomposed as $\nabla_D \cT_{AB}{}^{C}=0$, $\nabla_D R_{ABC}{}^D =0$, etc., and we can also show $\nabla_D \SSS_{AB}{}^C=0$ and $\nabla_C \cM_{AB}=0$. We thus expect an equivalence: (i) a manifold $M$ is a generalized coset $\GS \backslash G / H$ if and only if (ii) $M$ admits a linear connection $\tnabla$ satisfying $\nabla_{D} X_{\Ah\Bh}{}^{\Ch} =0$ and $\nabla_C \cM_{AB}=0$, which constitutes a natural extension of Theorem~\ref{theorem:AS}.

Because $\GS$ corresponds to the structure group of the internal space, a natural choice in practical applications is a subgroup $\GS \subset K(\GD)$ of the maximal compact subgroup of the duality group $\GD$. Moreover, the Lie group $G\subset \GD\times \mathbb{R}^+$ can be freely chosen, provided it contains $\GS$ as an isotropic subgroup. There is only one restriction: $G$ must originate from a Leibniz algebra whose structure constants $\hat{X}_{\Ah\Bh}{}^{\Ch}$ admit a geometric realization \cite{Hassler:2022egz}. That is, there must exist a subgroup $H\subset G$ with certain properties such that the generalized frame field $\Eh_{\Ah}{}^{\Ih}$ can be systematically constructed by following the procedure outlined in \cite{Inverso:2017lrz,Bugden:2021wxg,Bugden:2021nwl,Hassler:2022egz,Inverso:2024xok}.

In the following, we describe the construction of the invariant generalized tensors $\{Q_i\}$ on generalized cosets. Key elements in their construction are matrices for the generators of $\GS$ in various representations, such as $(\mathfrak{t}_{\exA})_A{}^B$ and $(\mathfrak{t}_{\exA})_{\BB}{}^{\AA}$, along with the generalized frame field $E_A{}^I$ on the generalized coset $M$ and the (constant) generalized torsion $X_{AB}{}^C$. All of these quantities are encoded in the structure constants $\hat{X}_{\Ah\Bh}{}^{\Ch}$. Furthermore, we note that the generalized frame field $E_A{}^I$ satisfies
\begin{align}
 \gLie_{E_A} E_B{}^I = - \check{X}_{AB}{}^C\,E_C{}^I\,.
\end{align}
Recalling \eqref{eq:cW} and \eqref{eq:check-X}, we compute the generalized fluxes $\check{X}_{AB}{}^C$ by
\begin{align}
 \check{X}_{AB}{}^C=X_{AB}{}^C -2\,\Omega_{[AB]}{}^C - Y^{EC}_{BD}\,\Omega_{EA}{}^D\,.
\end{align}
They are not constant, in contrast to those of generalized parallelizable spaces. Nevertheless, we demonstrate below that a constant embedding tensor can still be constructed from $\check{X}_{AB}{}^C$.

\subsection{Consistent truncation}\label{sec:consistent-tr}
In the extended field theory description of supergravity, the bosonic fields are usually organized into three categories:
\begin{itemize}
\item external metric $\mathfrak{g}_{\EXm\EXn}=\mathfrak{e}_{\EXm}{}^{\EXa}\,\mathfrak{e}_{\EXn}{}^{\EXb}\,\eta_{\EXa\EXb}$, which is a generalized scalar with weight $w=2\Cbeta$,
\item generalized metric $\cM_{IJ}\in \GD$, which is a generalized tensor with weight $w=0$, 
\item various $p$-form gauge fields $\cA_{\EXm}{}^I$, $\cB_{\EXm\EXn \II}$, $\cdots$, which are generalized tensors with weight $w=p\,\Cbeta$ transforming in the $R_{p}$-representation.
\end{itemize}
Here, we note that $\mathfrak{g}_{\EXm\EXn}$ is related to the standard\footnote{In DFT, the M-theory section of ExFT, and the type IIB section of ExFT, ``standard'' refers to the string-frame metric, the 11D metric, and the 10D Einstein-frame metric, respectively.} external metric $g_{\EXm\EXn}$ by
\begin{align}
 \mathfrak{g}_{\EXm\EXn} = \lvert \det g_{ij}\rvert^{\Cbeta}\,g_{\EXm\EXn}\,,
\label{eq:external-metric}
\end{align}
using the internal components $g_{ij}$ of the standard metric. To describe a consistent truncation, we must specify an ansatz for all of these fields. As we are dealing with dimensional reductions, this ansatz reconstructs the fields of the original theory in terms of the fields of the lower-dimensional, reduced theory. Take as an example the following reduction ansatz
\begin{align}
 \cA_{\EXm}{}^I = \cA_{\EXm}{}^{\cA}(x)\,K_{\cA}{}^{I}(y) \,, \qquad
 \cB_{\EXm\EXn \II} = \cB_{\EXm\EXn \overline{\cA}}(x)\,K_{\II}{}^{\overline{\cA}} (y) \,,\quad \dotsc
\end{align}
for the $p$-form gauge fields, where we have omitted $p(\geq 3)$-form fields and $(y)$ represents $(y^i)$. To write this ansatz, we introduced $\GS$-invariant generalized tensors, collectively denoted as $\{Q_i\}=\{K_{\cA}{}^I,\,K_{\II}{}^{\overline{\cA}},\,\cdots\}$\,, which are composed of
\begin{itemize}
\item $K_{\cA}{}^{I}$ ($\cA=1,\dotsc,N_1$)\,, a set of generalized vector fields with weight $w=\Cbeta$,
\item $K_{\II}{}^{\overline{\cA}}$ ($\overline{\cA}=1,\dotsc,N_2$), a set of generalized tensor fields in the $R_2$-representation with weight $w=2,\Cbeta$\,.
\end{itemize}
They satisfy the algebraic relations
\begin{align}
 \gLie_{K_{\cA}} K_{\cB}{}^I = - (X_{\cA})_{\cB}{}^{\cC}\,K_{\cC}{}^I\,,\qquad
 \gLie_{K_{\cA}} K_{\II}{}^{\overline{\cA}} = - (X_{\cA})_{\overline{\cB}}{}^{\overline{\cA}}\,K_{\II}{}^{\overline{\cB}}\,,
\label{eq:K-alg}
\end{align}
where $(X_{\cA})_{\cB}{}^{\cC}$ and $(X_{\cA})_{\overline{\cB}}{}^{\overline{\cA}}$ are constants that must be invariant under $\GS$. According to \cite{Cassani:2019vcl}, a consistent truncation arises by further expanding the external vielbein $e_{\EXm}{}^{\EXa}$ and the generalized metric $\cM_{IJ}$ using $\GS$-invariant generalized tensors.

In the following, we expand the discussion in \cite{Butter:2022iza}, which focuses on the case $\GD=\mathrm{O}(d,d)$. In particular, we systematically construct the $\GS$-invariant tensors that satisfy \eqref{eq:K-alg} with constant $(X_{\cA})_{\cB}{}^{\cC}$ and $(X_{\cA})_{\overline{\cB}}{}^{\overline{\cA}}$.

\subsubsection{Invariant generalized tensors}
Consider a generalized coset that admits a generalized frame field $\Eh_{\Ah}$ on the mega-space. We aim to decompose it into physical fields such as $E_A(y)$ and connection components $\Omega_A(y)$. Since $E_A{}^I(y)$ has weight $w = \Cbeta$, it is convenient to decompose it as
\begin{align}
 E_A{}^I(y) = \Exp{-2\Cbeta\Delta(y)} \cE_A{}^{I}(y)\,,\qquad 
 \Exp{-2\Delta} \equiv v \equiv \lvert \det e_{i}{}^a \rvert \,,
\end{align}
where $\cE_A{}^{I} \in \GD$, and $v$ originates from the determinant of $e_i{}^a$ in \eqref{eq:hat-v-dec}. The scale factor $\Exp{-2\Delta} = v$ plays the role of the volume density on the physical space when $X_{\Bh\Ah}{}^{\Bh} = 0$.\footnote{Following the discussion in Appendix C of \cite{Hassler:2023nht}, we can show that $\lvert \det \hat{v}_{\hat{i}}{}^{\hat{a}} \rvert\,\rmd^{p} x$, where $\hat{v}_{\hat{i}}{}^{\hat{a}}$ is defined in \eqref{eq:hat-v}, gives the left-invariant measure on the mega-space $\hat{M}$. In our construction, $\hat{v}_{\hat{a}}{}^{\hat{i}}$ takes the form \eqref{eq:hat-v-dec}, allowing us to factorize the invariant measure as $\lvert \det \hat{v}_{\hat{i}}{}^{\hat{a}} \rvert\,\rmd^p y = [\lvert \det \widetilde{v}_{\mu}{}^{\exA}(y^\mu) \rvert \,\rmd^n y^\mu]\, [\det M_{\hat{a}}{}^{\hat{b}}(y)\,v(y)\,\rmd^d y^i]$. When $X_{\Bh\Ah}{}^{\Bh} = 0$, we have $\det M_{\hat{a}}{}^{\hat{b}}(y) = 1$, and thus $v(y)\,\rmd^d y$ provides a natural measure on $M = \GS\backslash \hat{M}$.}

Next, let us construct a complete set of constant vectors $K_{\cA}{}^A$ that transform in the $R_1$ representation and are null eigenvectors of the generators of $\GS$,
\begin{align}
 K_{\cA}{}^C\,(\mathfrak{t}_{\exA})_C{}^B = 0\,.
\end{align}
This equation is purely algebraic, and for a given matrix representation $(\mathfrak{t}_{\exA})_B{}^A$ of the $\GS$ generators in the $R_1$ representation, we can readily find a complete set of solutions. The number of such null eigenvectors is denoted by $N_1$.

We also introduce a set of $\GS$-invariant dual vectors satisfying
\begin{align}
 (\mathfrak{t}_{\exA})_A{}^C\, K_C{}^{\cB} = 0\,.
\end{align}
Because $\GS$ is a subgroup of $K(\GD)$, the quantity $(\mathfrak{t}_{\exA})_{AB}=(\mathfrak{t}_{\exA})_A{}^C\,\delta_{CB}$ is antisymmetric, and the number of dual vectors $K^{\cB}$ is again $N_1$\,. For a given $K_{\cA}{}^B$, the corresponding $K_A{}^{\cB}$ is uniquely determined by imposing the condition\footnote{In this paper, we choose $K_{\cA}{}^B$ to be orthonormal, $K_{\cA}{}^C\,K_{\cB}{}^D\,\delta_{CD}=\delta_{\cA\cB}$, in which case $K_A{}^{\cB}$ has the same components as $K_{\cB}{}^A$.}
\begin{align}
 K_{\cA}{}^C\,K_C{}^{\cB} = \delta_{\cA}^{\cB}\,.
\end{align}
These vectors satisfy the compatibility conditions with the covariant derivative,
\begin{align}\label{eq:K-R1-cov-const}
 \tnabla_I K_{\cA}{}^B = \Omega_{IC}{}^B\,K_{\cA}{}^C = 0\,,\qquad
 \tnabla_I K_B{}^{\cA} = -\Omega_{IB}{}^C\,K_C{}^{\cA} = 0\,,
\end{align}
and are therefore covariantly constant. By contracting the Leibniz identity
\begin{align}
 (\mathfrak{t}_{\exA})_B{}^D\,X_{AD}{}^C
 - X_{AB}{}^D\,(\mathfrak{t}_{\exA})_{D}{}^C = -(\mathfrak{t}_{\exA})_{A}{}^D\,X_{DB}{}^C 
\label{eq:Bianchi}
\end{align}
with $K_{\cA}{}^A$, we find that the matrix $T_{\cA}\equiv -K_{\cA}{}^B\,X_B$ commutes with the generators $\mathfrak{t}_{\exA}$ of $\mathrm{Lie}(\GS)$. 
Furthermore, contracting \eqref{eq:Bianchi} with $K_{\cA}{}^A\,K_{\cB}{}^B$, we eventually obtain
\begin{align}
 K_{\cA}{}^A\,K_{\cB}{}^B\, X_{AB}{}^D\,(\mathfrak{t}_{\exA})_{D}{}^C = 0 \,.
\end{align}
This result implies that $K_{\cA}{}^A\,K_{\cB}{}^B\, X_{AB}{}^C$ can be expanded in terms of $K_{\cC}{}^C$, leading to
\begin{align}
 K_{\cA}{}^A\,K_{\cB}{}^B\, X_{AB}{}^C = X_{\cA\cB}{}^{\cC}\,K_{\cC}{}^C\,,
\end{align}
where $X_{\cA\cB}{}^{\cC}$ is a ($\GS$-invariant) constant satisfying
\begin{align}
 \nabla_I X_{\cA\cB}{}^{\cC} = \partial_I X_{\cA\cB}{}^{\cC} = 0\,.
\end{align}

This discussion applies to any other representation of the duality group $\GD$ as well. For example, we introduce a complete set of $\GS$-invariant constant vectors $K_{\CC}{}^{\overline{\cC}}$ in the $R_2$ representation, along with their duals $K_{\overline{\cC}}{}^{\CC}$, satisfying
\begin{align}
 (\mathfrak{t}_{\exA})_{\CC}{}^{\DD}\, K_{\DD}{}^{\overline{\cC}} = 0\,,\qquad 
 K_{\overline{\cC}}{}^{\DD}\, (\mathfrak{t}_{\exA})_{\DD}{}^{\CC} = 0\,,\qquad 
 K_{\overline{\cA}}{}^{\CC}\,K_{\CC}{}^{\overline{\cB}} = \delta_{\cA}^{\cB}\,.
\end{align}
These vectors also satisfy
\begin{align}
 \nabla_I K_{\CC}{}^{\overline{\cC}} = 0 \,,\qquad
 \nabla_I K_{\overline{\cC}}{}^{\CC} = 0 \,,
\end{align}
in analogy with \eqref{eq:K-R1-cov-const}. In the same vein, one can construct covariantly constant tensors in the remaining $R_n$-representations. From the $\GD\supset\GS$-invariance of $\eta_{AB;\CC}$, we find that
\begin{align}
 K_{\cA}{}^A\,K_{\cB}{}^B\,\eta_{AB;\DD}\,(\mathfrak{t}_{\exA})_{\CC}{}^{\DD} = 0
\end{align}
holds. Therefore, $K_{\cA}{}^A\,K_{\cB}{}^B\,\eta_{AB;\CC}$, and similarly $K_A{}^{\cA}\,K_B{}^{\cB}\,\eta^{AB;\CC}$, admits the expansion
\begin{align}
 K_{\cA}{}^A\,K_{\cB}{}^B\,\eta_{AB;\CC} = \eta_{\cA\cB;\overline{\cC}} \,K_{\CC}{}^{\overline{\cC}}\,,
\qquad
 K_A{}^{\cA}\,K_B{}^{\cB}\,\eta^{AB;\CC} = \eta^{\cA\cB;\overline{\cC}} \,K_{\overline{\cC}}{}^{\CC}\,,
\end{align}
where the coefficients are given by
\begin{align}
 \eta_{\cA\cB;\overline{\cC}} = K_{\cA}{}^A\,K_{\cB}{}^B\,\eta_{AB;\CC} \,K_{\overline{\cC}}{}^{\CC}\,,
\qquad
 \eta^{\cA\cB;\overline{\cC}} = K_A{}^{\cA}\,K_B{}^{\cB}\,\eta^{AB;\CC} \,K_{\CC}{}^{\overline{\cC}}\,.
\end{align}

Ultimately, we must promote the constant tensors discussed above to tensor fields defined over the full physical space. To this end, we employ the generalized frame field constructed on the generalized coset. In particular, we use it to define the generalized tensor fields,
\begin{align}
 K_{\cA}{}^I(y) \equiv K_{\cA}{}^B\,E_B{}^I(y)\,,\qquad
 K_{\II}{}^{\overline{\cA}}(y)\equiv K_{\BB}{}^{\overline{\cA}}\,E_{\II}{}^{\BB}(y)\,,
\end{align}
in the respective $R_n$-representation. For example, the generalized frame field $E_{\II}{}^{\BB}$ in the $R_2$-representation is defined via
\begin{align}
 E_A{}^I\,E_B{}^J\,\eta_{IJ;\KK} = \eta_{AB;\CC}\,E_{\KK}{}^{\CC}\,.
\end{align}
This is manifestly invariant under the action of $\GS$, as seen from the transformation rules
\begin{align}
 \delta_{\xi} E_A{}^I = \xi_{A}{}^{B}\,E_B{}^I\,,\qquad
 \delta_{\xi} E_{\II}{}^{\AA} = \xi_{\BB}{}^{\AA}\,E_{\II}{}^{\BB}\,,
\end{align}
and the fact that $K_{\cA}{}^B$ and $K_{\BB}{}^{\overline{\cA}}$ are null eigenvectors under the action of $\xi_{A}{}^{B}$ and $\xi_{\BB}{}^{\AA}$, respectively. Moreover, recalling the generalized vielbein postulate \eqref{eq:gen-vielbein-postulate}, it is also clear that these tensor fields are covariantly constant by construction,
\begin{align}
 \nabla_A K_{\cA}{}^I(y) = 0 \,,\qquad
 \nabla_A K_{\II}{}^{\overline{\cA}}(y) = 0\,.
\end{align}

For an arbitrary covariantly constant generalized tensor $Q_i$, generalized diffeomorphisms on the physical space act as
\begin{align}
 \gLie_{K_{\cA}} Q_i = \gLie_{K_{\cA}} Q_i - \gLie^\tnabla_{K_{\cA}} Q_i = - (\cT_{\cA})\cdot Q_i
\end{align}
due to the definition of the generalized torsion. These transformations close as a consequence of \eqref{eq:K-alg} and the identification $\cT_{\cA}=X_{\cA}$\,. Note that although the generalized fluxes $\check{X}_{AB}{}^C$ associated with $E_A{}^I$ are not always constant, the projected components $X_{\cA\cB}{}^{\cC}$ are constant. This important property follows from
\begin{align}
 X_{\cA\cB}{}^{\cC} &= K_{\cA}{}^A\,K_{\cB}{}^B\, X_{AB}{}^C\,K_C{}^{\cC}
\nn\\
 &= K_{\cA}{}^A\,K_{\cB}{}^B\, \bigl[\cW_{AB}{}^{C} + (s^{\adja})_{D}{}^{E}\,\cW_{EA}{}^{D}\,(s_{\adja})_{B}{}^{C} \bigr]\,K_C{}^{\cC}
\nn\\
 &= K_{\cA}{}^A\,K_{\cB}{}^B\, \bigl[ \check{W}_{AB}{}^{C} 
 + (s^{\adja})_{D}{}^{E}\, \check{W}_{EA}{}^{D} \,(s_{\adja})_{B}{}^{C} \bigr]\,K_C{}^{\cC}
\nn\\
 &= K_{\cA}{}^A\,K_{\cB}{}^B\, \check{X}_{AB}{}^C\,K_C{}^{\cC}\,.
\end{align}
Furthermore, the constants $X_{\cA\cB}{}^{\cC}$ can be interpreted as certain projected components of the generalized torsion $\cT_{AB}{}^C = X_{AB}{}^C$. Since they are unaffected by shifts in the connection of the form $\Delta\Omega_{AB}{}^C=\Delta\Omega_A^{\exA}\,(\mathfrak{t}_{\exA})_B{}^C$, they encode the intrinsic torsion of the generalized coset. Consequently, we have established that the intrinsic torsion $X_{\cA\cB}{}^{\cC}$ is constant and invariant under the action of $\GS$ -- precisely the condition required in \cite{Cassani:2019vcl} for consistent truncations.

Moreover, $X_{\cA\cB}{}^{\cC}$ plays a fundamental role as the embedding tensor or, equivalently, as the structure constants of the gauge group $\cG$ in the gauged supergravity that arises from the truncation. The matrices $(T_{\cA})_{\Bh}{}^{\Ch}$, which commute with $\mathrm{Lie}(\GS)$ and satisfy
\begin{align}
 [T_{\cA},\,T_{\cB}] = X_{[\cA\cB]}{}^{\cC}\,T_{\cC}\,,
\end{align}
describe how this gauge group $\cG$ is embedded into the duality group $\GD$.

Before closing this subsection, let us summarize the construction of the $\GS$-invariant generalized tensors:
\begin{itemize}
\item We start from a Leibniz algebra characterized by the structure constants $\hat{X}_{\Ah\Bh}{}^{\Ch}$. This leads to a Lie group $G$, and by selecting a minimally coisotropic subgroup $H$, we systematically construct the corresponding generalized frame field $\Eh_{\Ah}{}^{\Ih}$. 

\item After selecting the structure group $\GS$ as an isotropic subgroup of $G$, we decompose $\Eh_{\Ah}{}^{\Ih}$ into connection contributions and the generalized frame $E_A{}^I(y)$ on the physical space.

\item Finally, we identify all null eigenvectors of $\mathfrak{t}_{\exA}$ in the $R_n$-representation, such as $K_{\cA}{}^A$. Contracting them with the generalized frame fields in the respective $R_n$-representation, we construct the invariant tensors, such as $K_{\cA}{}^I = K_{\cA}{}^A\,E_A{}^I$. They automatically satisfy the relation \eqref{eq:K-alg} with constant intrinsic torsion.
\end{itemize}
In conclusion, the generalized coset provides a systematic framework for constructing consistent truncations. In what follows, we explore further details of this construction.

\subsubsection{Reduction ansatz}
Having constructed the $\GS$-invariant generalized tensors, we now specify the truncation ansatz:
\begin{align}\label{eq:reduction-ansatz-frame}
 \mathfrak{e}_{\EXm}{}^{\EXa} &= \Exp{-2\Cbeta\Delta(y)} \eD_{\EXm}{}^{\EXa}(x)\,,
\\
 \cM_{IJ} &= \cE_I{}^{A}(y)\,\cE_J{}^{B}(y)\,\cM_{AB}(x)= \Exp{-4\Cbeta\Delta(y)} E_I{}^{A}(y)\,E_J{}^{B}(y)\,\cM_{AB}(x)\,,
\\
 \cA_{\EXm}{}^I &= \cA_{\EXm}{}^{\cA}(x)\,K_{\cA}{}^{I}(y) \,,
\\\label{eq:reduction-ansatz-2form}
 \cB_{\EXm\EXn \II} &= \cB_{\EXm\EXn \overline{\cA}}(x)\,K_{\II}{}^{\overline{\cA}} (y) \,.
\end{align}
As already mentioned, all fields must be invariant under the structure group $\GS$. For $\cM_{IJ}$, this requirement imposes additional constraints: the scalar fields $\cM_{AB}(x)$ must take values in
\begin{align}
 \cM_{AB}(x) \in \frac{\mathrm{C}_{\GD}(\GS)}{K(\mathrm{C}_{\GD}(\GS))}\,,
\end{align}
where $\mathrm{C}_{\GD}(\GS)$ denotes the commutant group of $\GS$ in $\GD$. With this refinement, the reduction ansatz is complete. One can then show that the equations of motion (or action) of ExFT reduce precisely to those of a $D$-dimensional gauged supergravity. The resulting theory admits a remarkable geometric formulation: it can be expressed entirely in terms of the constant generalized torsion tensor $\cT_{AB}{}^C$ and the generalized Ricci tensor $R_{AB}$ of the physical space $M$. While the explicit form depends on the dimension $d$, we demonstrate this structure through representative terms of the reduced action.

Let us focus on the case $d\leq 6$. The 2-form field strength $\cF_2^I = \frac{1}{2}\,\cF_{\EXm\EXn}{}^I\,\rmd x^{\EXm}\wedge\rmd x^{\EXn}$ reduces to
\begin{align}
 \cF_{\EXm\EXn}{}^I &= 2\,\partial_{[\EXm}\cA_{\EXn]}{}^I - [\cA_{\EXm},\,\cA_{\EXn}]^I + \eta^{IJ;\KK} \,\partial_J \cB_{\EXm\EXn \KK}
\nn\\
 &= \bigl( 2\,\partial_{[\EXm}\cA_{\EXn]}{}^{\cA} + X_{[\cB\cC]}{}^{\cA}\,\cA_{\EXm}{}^{\cB}\,\cA_{\EXn}{}^{\cC} - Z_{\cB\cC}{}^{\cA}\,\cB_{\EXm\EXn}{}^{\cB\cC} \bigr)\, K_{\cA}{}^{I} \equiv \cF_{\EXm\EXn}{}^{\cA}\, K_{\cA}{}^{I}\,,
\end{align}
where
\begin{align}
 Z_{\cA\cB}{}^{\cC} \equiv 2\,X_{(\cA\cB)}{}^{\cC}\,.
\end{align}
This expression is manifestly determined by the constant generalized torsion through $X_{\cA\cB}{}^{\cC}$. When $X_{\Bh\Ah}{}^{\Bh}=0$, the physical space $M$ admits a well-defined volume form, and the kinetic term for the 1-form potential becomes
\begin{align}
 S_{1\text{-form}}= -\frac{1}{2}\int \rmd^D x\,\rmd^dy\, \mathfrak{e}\,\mathfrak{g}^{\EXm\EXp}\,\mathfrak{g}^{\EXn\EXq}\,\cM_{IJ}\, \cF_{\EXm\EXn}^I\,\cF_{\EXp\EXq}^J
 = -\frac{V_{\text{int}}}{2} \int \cM_{\cA\cB}\,\cF_2^{\cA}\wedge *_{\gD} \cF_2^{\cB}\,,
\end{align}
where the Hodge star operator $*_{\gD}$ is associated with the $D$-dimensional metric $\gD_{\EXm\EXn}=\eD_{\EXm}{}^{\EXa}\,\eD_{\EXn}{}^{\EXb}\,\eta_{\EXa\EXb}$, the internal volume is $V_{\text{int}}\equiv \int_M\rmd^dy\,\Exp{-2\Delta}$, and
\begin{align}
 \cM_{\cA\cB}(x^{\EXm})\equiv K_{\cA}{}^C\,K_{\cB}{}^D\,\cM_{CD}(x^{\EXm})\,.
\end{align}
In particular, for $\Edd$ with $d\leq 5$ or the $\Odd$ case, the 3-form field strength is given by
\begin{align}
 \cF_{\EXm\EXn\EXp\II} &= 3\,D_{[\EXm}B_{\EXn\EXp]\II} - 3\,\eta_{JK;\II}\,A_{[\EXm}{}^J\,\partial_{\EXn} A_{\EXp]}{}^K + \eta_{JK;\II}\,A_{[\EXm}{}^J\,[A_{\EXn},\,A_{\EXp]}]^K 
 + Z^L{}_{\II}{}^{\JJJ}\, \partial_L C_{\EXm\EXn\EXp \JJJ} 
\nn\\
 &= \cF_{\EXm\EXn\EXp\overline{\cA}}\,K_{\II}{}^{\overline{\cA}}\,,
\end{align}
where
\begin{align}
 \cF_{3\overline{\cA}} &= \rmd B_{2\overline{\cA}} + 2\,X_{\cB\cG}{}^{\cC} \,\eta_{\cC\cH;\overline{\cA}} \,A_1^{\cB}\wedge B_{2}{}^{\cG\cH} 
 - \tfrac{1}{2}\,\eta_{\cB\cC;\overline{\cA}}\,A_1^{\cB}\wedge \bigl(\rmd A_1^{\cC} + \tfrac{1}{3}\,X_{\cD\cE}{}^{\cC}\,A_1^{\cD}\wedge A_1^{\cE}\bigr) 
\nn\\
 &\quad - \tfrac{1}{d(d-1)}\,\bigl(X_{\cB\cC}{}^{\cF}\,\eta_{\cE\cF;\overline{\cA}} - X_{\cC\cE}{}^{\cF}\,\eta_{\cB\cF;\overline{\cA}}\bigr)\,\eta^{\cC\cE\cB;\overline{\overline{\cD}}}\,C_{3\overline{\overline{\cD}}}\,,
\end{align}
with the 3-form potential $C_{\EXm\EXn\EXp \II}$ (or $C_{3\overline{\overline{\cD}}}$) absent in the $\Odd$ case. Once again, this is expressed purely in terms of the constant generalized torsion. The kinetic term reads
\begin{align}
 S_{2\text{-form}}&= -\frac{1}{2}\int \rmd^D x\,\rmd^dy\, \mathfrak{e}\,\mathfrak{g}^{\EXm_1\EXn_1}\,\mathfrak{g}^{\EXm_2\EXn_2}\,\mathfrak{g}^{\EXm_3\EXn_3}\, \cM^{\overline{IJ}}\,\cF_{\EXm_1\EXm_2\EXm_3\overline{I}}\,\cF_{\EXn_1\EXn_2\EXn_3\overline{J}}
\nn\\
 &= -\frac{V_{\text{int}}}{2} \int \cM_{\overline{\cA\cB}}\,\cF_3^{\overline{\cA}}\wedge *_{\gD} \cF_3^{\overline{\cB}}\,.
\end{align}

For $4\leq d\leq 8$, the kinetic term for scalar fields takes the form
\begin{align}
 S_{\text{kin}} = \frac{1}{4\,\Calpha}\int \rmd^D x\,\rmd^dy\, \mathfrak{e}\,\mathfrak{g}^{\EXm\EXn}\, \cD_{\EXm} \cM^{IJ}\, \cD_{\EXn} \cM_{IJ}  
 = \frac{V_{\text{int}}}{4\,\Calpha}\int D \cM^{AB}\wedge *_{\gD} D \cM_{AB} \,.
\end{align}
For $d=3$, the duality group is $\GD=\text{SL}(3) \times \text{SL}(2)$ and the index $A$ decomposes as $A = (p,\alpha)$, where $p,q \in \mathbf{3}$ of $\text{SL}(3)$ and $\alpha,\beta \in \mathbf{2}$ of $\text{SL}(2)$. The generalized metric accordingly splits into two independent blocks $\cM_{pq}$ and $\cM_{\alpha\beta}$, and the kinetic term takes the form
\begin{align}
 S_{\text{kin}} = \frac{V_{\text{int}}}{4}\int \bigl(\cD \cM^{pq}\wedge *_{\gD} \cD \cM_{pq} + \cD \cM^{\alpha\beta}\wedge *_{\gD} \cD \cM_{\alpha\beta}\bigr)\,.
\end{align}
The covariant derivative $\cD = \rmd x^{\EXm}\wedge \cD_{\EXm}$ with $\cD_{\EXm}\equiv \partial_{\EXm} - \gLie_{A_{\EXm}}$ reduces to $\cD = \rmd - \cA_1^{\cA}\,X_{\cA}\,\cdot\,$, which is again determined by the constant generalized torsion.

The scalar potential requires more careful analysis. For DFT or $\Edd$ EFT with $d\leq 7$, the relevant terms in the higher-dimensional action are
\begin{align}
 S_{\text{pot}} 
 &= \int \rmd^Dx\,\rmd^dy\,\mathfrak{e}\,\Bigl[\tfrac{1}{4\Calpha}\,\cM^{IJ}\,\partial_I\cM^{KL}\,\partial_J \cM_{KL} 
 -\tfrac{1}{2}\,\cM^{IJ}\,\partial_I \cM^{KL}\,\partial_L \cM_{JK} 
\nn\\
 &\qquad\qquad\qquad\quad + \partial_I \ln \mathfrak{e}\,\partial_J\cM^{IJ} 
   + \cM^{IJ}\,
 \bigl(\partial_I \ln \mathfrak{e} \, \partial_J \ln \mathfrak{e} + \tfrac{1}{2}\,\partial_I \mathfrak{g}^{\EXm\EXn}\,\partial_J \mathfrak{g}_{\EXm\EXn}\bigr)\Bigr]\,.
\end{align}
This expression has a natural geometric interpretation as the generalized Ricci scalar $\bar{R}$ associated with the generalized Levi--Civita connection. According to the identity \eqref{eq:barR-SSS}, $\bar{R}$ decomposes into the generalized Ricci scalar $R$ and terms involving the generalized contorsion tensor $\SSS_{AB}{}^C$. Since $\SSS_{AB}{}^C$ can be expressed in terms of $\cT_{AB}{}^C$ as in \eqref{eq:Sigma-semi-covariant}, the scalar potential must ultimately be expressible purely in terms of the constant generalized torsion $\cT_{AB}{}^C$ and the generalized Ricci tensor $R_{AB}$. We verify this explicitly for the DFT case in \eqref{eq:reduced-Lagrangian-ODD}, where the complete reduction is worked out in detail.

\subsection{Generalized dualities}\label{sec:gen-dualities}
We have learned in the previous subsection that a consistent truncation on a generalized coset space $M=\GS \backslash G/H$ results in a gauged supergravity characterized by $\cT_{AB}{}^C$ and $R_{AB}$\,. Crucially, both quantities depend only on the Lie groups $G$ and $\GS$, and are independent of the choice of subgroup $H$. The role of $H$ becomes important only in constructing the generalized frame field $\Eh_{\Ah}{}^{\Ih}$ (which contains the generalized frame field $E_A{}^I$ on $M$) from a coset representative of $G/H$. In other words, the role of $H$ is crucial in constructing the reduction ansatz that connects the lower-dimensional theory to its higher-dimensional origin.

A key feature of generalized dualities is that the Lie group $G$ may admit multiple choices for the subgroup $H$, enabling the construction of generalized frame field $\Eh_{\Ah}{}^{\Ih}$ from coset representatives of different cosets, such as $G/H$, $G/H'$, and so on. Because we perform a consistent truncation, any solution found in the lower-dimensional theory can be uplifted to a higher-dimensional solution using the reduction ansatz defined by $E_A{}^I$, or any of its counterparts on a different coset. Given that generally multiple uplifts exist, a single lower-dimensional solution can lead to a family of distinct higher-dimensional solutions. If $G$ is an abelian group, these solutions are related by T- or U-duality. For non-abelian $G$, the mapping between different uplifted solutions is referred to as generalized (T- or U-)duality.

The generalized dualities discussed here encompass many well-known dualities as special cases. These include
\begin{enumerate}
  \item If $\GS$ is trivial (and therefore consists only of the identity element), one obtains
  \begin{enumerate}
    \item\label{item:worldsheet} For $\GD=\OO(d,d)$, both non-abelian T-duality \cite{delaOssa:1992vci} and Poisson--Lie T-duality \cite{Klimcik:1995ux} emerge as special cases. Recent developments \cite{Hassler:2017yza,Severa:2018pag,Demulder:2018lmj,Sakatani:2019jgu,Catal-Ozer:2019hxw,Borsato:2021vfy} have further clarified and extended these dualities by formulating them in a duality-covariant framework.
    \item\label{item:target1} For $\GD=\mathbb{R}^+\times \Odd$ or $\GD=\SL(2) \times \Odd[6]$, similar generalized dualities have been introduced in \cite{Fernandez-Melgarejo:2021zyj,Sakatani:2021eqo,Hassler:2023nht}. 
    \item\label{item:target2} For $\GD=\Edd$, generalized U-dualities studied in \cite{Sakatani:2019zrs,Malek:2019xrf,Malek:2020hpo,Sakatani:2020wah,Musaev:2020nrt,Blair:2022gsx,Blair:2022ahh,Blair:2022ndb,Hassler:2022egz} arise.
  \end{enumerate}
\item\label{item:dressing} If $\GS$ is non-trivial, generalized duality has been studied primarily for $\GD=\OO(d,d)$. Known cases include non-abelian T-duality where the isometry group acts with isotropy \cite{delaOssa:1992vci,Giveon:1993ai,Lozano:2011kb} and Poisson--Lie T-duality for dressing cosets \cite{Klimcik:1996np,Sfetsos:1999zm,Severa:2018pag,Sakatani:2021skx}. These are special cases of the Poisson--Lie T-duality of generalized cosets \cite{Butter:2022iza}, which corresponds to the generalized duality discussed here for $\GD=\OO(d,d)$.
\end{enumerate}
At this point one might appreciate that the dualities in \ref{item:worldsheet}) were originally derived from a worldsheet perspective. Only later were they recovered in the language of consistent truncations, we advertise here. By contrast, for \ref{item:target1}) and \ref{item:target2}), this new perspective helped to define them, while their analysis from the worldsheet/worldvolume perspective has only recently begun. Even greater is the potential for discovering new dualities in \ref{item:dressing}, as there is nearly nothing known about the duality groups $\Odd$ (except for \cite{Hassler:2024yis}) or $\Edd$ in this context. An important distinction for $\GD=\Edd$, as opposed to $\GD=\OO(d,d)$, is that the dimension of the generalized coset space $M=\GS \backslash G/H$ can be different depending on the choice of $H$. In general, the dimension of $M$ can be either $d$ or $d-1$. The former corresponds to an uplift to eleven-dimensional supergravity, while the latter leads to an uplift to ten-dimensional type IIB supergravity. When two such subgroups $H$ with different dimensions exist, the same lower-dimensional theory can be uplifted to both 11D and type IIB supergravity. This is a fascinating feature of generalized U-duality.

\subsubsection{Generalized dualities for different \texorpdfstring{$\GS$}{Gs}}\label{sec:differentGS}
From the target space perspective we adopt in this article, generalized dualities are based on the invariance of the truncated, lower-dimensional theory under different choices of the subgroup $H$. In other words, the same lower-dimensional theory emerges after the consistent truncation regardless of the choice of $H$. However, it is also possible to consider different realizations of the structure group $\GS$ which still produce the same lower-dimensional theory. This type of generalized duality has not been systematically studied in the literature because generalized dualities have been mainly discussed for generalized parallelizable spaces $G/H$ where $\GS$ is trivial.

To see how this new class of generalized dualities arises, consider an automorphism of $\hat{X}_{\Ah\Bh}{}^{\Ch}$, given by
\begin{align}
 \hat{X}_{\Ah\Bh}{}^{\Ch} = C_{\Ah}{}^{\Dh}\,C_{\Bh}{}^{\Eh}\,(C^{-1})_{\Fh}{}^{\Ch}\,\hat{X}_{\Dh\Eh}{}^{\Fh}\,,
\label{eq:Aut}
\end{align}
where $C_{\Ah}{}^{\Dh}\in \GDM$ is a constant matrix. Upon introducing new generators $\mathfrak{t}'_{\Ah}$ as
\begin{align}
 \mathfrak{t}'_{\Ah} \equiv C_{\Ah}{}^{\Bh}\,\mathfrak{t}_{\Bh}\,,
\end{align}
we find that their algebra
\begin{align}\label{eq:automorphismC}
 [\mathfrak{t}'_{\Ah},\,\mathfrak{t}'_{\Bh}] = \hat{X}_{\Ah\Bh}{}^{\Ch}\,\mathfrak{t}'_{\Ch}\,,
\end{align}
still has the same structure constants because of \eqref{eq:Aut}. Additionally, we perform the similarity transformation
\begin{align}
 (\mathfrak{t}'_{\Ah})_{\Bh'}{}^{\Ch'} = C_{\Bh'}{}^{\Bh}\,(\mathfrak{t}'_{\Ah})_{\Bh}{}^{\Ch}\,(C^{-1})_{\Ch}{}^{\Ch'}\,,
\label{eq:t'-matrix}
\end{align}
which again does not change the structure constants. Note that we introduce here a new class of indices, namely $\Ah'$, $\Bh'$, etc., to distinguish between $(\mathfrak{t}'_{\Ah})_{\Bh}{}^{\Ch}$ before and $(\mathfrak{t}'_{\Ah})_{\Bh'}{}^{\Ch'}$ after the similarity transformation. While in this new basis the Lie algebra of the structure group $\GS$ remains unchanged, its action on the mega-space is modified. We encode this new action by transitioning from $\GS$ to $\GS'$. The same argument applies to $H$. As a result, the consistent truncations based on $\GS\backslash G/H$, $\GS'\backslash G/H'$, and $\GS'\backslash G/H$ lead to the same lower-dimensional theory because they share the same $\hat{X}_{\Ah\Bh}{}^{\Ch}$ on the mega-space. However, the action of $\GS$ on $G$ is relevant for constructing the truncation ansatz. Hence, we may obtain different uplifts, which give rise to a generalized duality.

In sections \ref{sec:examples-DFT} and \ref{sec:examples-ExFT}, we further explain how this works by constructing the generalized frame fields $\Eh_{\Ah}{}^{\Ih}$ for simple examples. To do this, we decompose the generators $\mathfrak{t}_{\Ah} =(\mathfrak{t}_{\hat{a}},\, \mathfrak{t}_{\tilde{\alpha}})$ and adopt a convention where $H$ is contained in the Leibniz subalgebra generated by $\mathfrak{t}_{\tilde{\alpha}}$.\footnote{In this convention, the generators of $\GS$ are not necessarily contained in $\mathfrak{t}_{\hat{a}}$\,. In general, they are expressed as $\mathfrak{t}_{\exA}=k_{\exA}{}^{\Bh}\,\mathfrak{t}_{\Bh}$\,. Correspondingly, we have $\mathfrak{t}'_{\exA}=k_{\exA}{}^{\Bh}\,\mathfrak{t}'_{\Bh}$\,.} Accordingly, a change of basis corresponds to changing the subgroup $H$. Defining $H'$ as the subgroup generated by $\mathfrak{t}'_{\tilde{\alpha}}$ from $\mathfrak{t}'_{\Ah} =(\mathfrak{t}'_{\hat{a}},\, \mathfrak{t}'_{\tilde{\alpha}})$, we find that the generalized frame fields constructed from $\GS\backslash G/H$ and $\GS'\backslash G/H'$ are identical. This is not very surprising because their structure constants on the mega-space are precisely the same. Eventually, we seek to find two different frames which give rise to an alternative truncation ansatz and thus a generalized duality. Hence, we follow the initial idea that suggests replacing $H'$ with another subgroup. For example, $\GS\backslash G/H$ and $\GS'\backslash G/H$ produce the same lower-dimensional theory, but their respective frame fields and coset representatives are different. This shows that while we cannot change the Lie algebra of $\GS$, we can modify its action on the mega-space in a systematic way to obtain new generalized dualities. Remarkably, one can achieve the same results by keeping $\GS$ completely unchanged and merely modifying $H$. To see how this works, recall that the generalized frame fields constructed for $\GS\backslash G/H$ and $\GS'\backslash G/H'$ are identical. In the same vein, those constructed from $\GS\backslash G/H_1$ and $\GS'\backslash G/H'_1$ are also identical, where $H'_1$ is defined by $H_1$ and $C_{\Ah}{}^{\Bh}$ in the same way that $H'$ is defined by $H$ and $C_{\Ah}{}^{\Bh}$. By setting $H'_1 = H$, we find that the generalized frame field for $\GS'\backslash G/H$ coincides with that for $\GS\backslash G/H_1$. Consequently, the duality between $\GS\backslash G/H$ and $\GS'\backslash G/H$ can be reformulated as a duality between $\GS\backslash G/H$ and $\GS\backslash G/H_1$. This is a standard form of generalized duality involving only a modification of $H$. In this sense, choosing whether to keep $H$ or $\GS$ fixed is a matter of preference. However, the former viewpoint is more useful, at least in the context of discussing the axial-vector duality \cite{Giveon:1991sy,Dijkgraaf:1991ba,Kiritsis:1991zt,Rocek:1991ps,Kiritsis:1992uz,Kiritsis:1993ju,Giveon:1993ph} as a generalized T-duality in section \ref{sec:SU2-WZW}.

\subsubsection{Singularities at fixed points of the \texorpdfstring{$\GS$}{Gs} action}\label{sec:singularity}
A key distinction between generalized cosets and their standard counterparts is the possibility of singularities emerging at fixed points of the $\GS$-action. At these points, the orbits of the $\GS$-action have dimensions smaller than $n=\dim \GS$, leading to the natural expectation that the coset space $M=\GS \backslash G/H$ may exhibit singularities. In what follows, we propose a condition for identifying such fixed points and demonstrate that the generalized frame field becomes singular at these locations. 

Consider the coset representative $M(y^i) \in \GS \backslash G/H$ introduced in \eqref{eq:Mh-dec}. In particular, we examine an infinitesimal left multiplication by $\GS$, which corresponds to a gauge transformation studied in section \ref{sec:gauge-transf}, generated by $\xi^{\exA}$.\footnote{This correspondence has been explicitly confirmed in \cite{Butter:2022iza} for the case of $\GD=\OO(d,d)$.} We express the condition for a point $y^i=y_0^i$ to be a fixed point as the existence of a non-trivial subgroup $G_0$ with $f \in G_0 \subset \GS$ and $h \in G_0 \subset H$ satisfying
\begin{align}
 f\,M(y_0^i) = M(y_0^i)\,h \,.
\end{align}
Because $H$ is the gauge group acting from the right, this implies that left-multiplication by any element of $G_0 \subset \GS$ leaves the coset representative, and with it $y_0^i$, unchanged. This implies that $y^i=y_0^i$ is a fixed point. In terms of generators $\mathfrak{t}_{\Ah}\equiv -\hat{X}_{\Ah}$, this condition can be reformulated as the existence of $a^{\exA}_m\in\mathrm{Lie}(G_0)$ with $m=1,\dots,\dim G_0$ satisfying
\begin{align}
 a^{\exA}_m\,M^{-1}(y_0^i)\,\mathfrak{t}_{\exA}\,M(y_0^i) \in \mathrm{Lie}(H)\,.
\label{eq:fixed-point}
\end{align}
Rewriting the left-hand side, we obtain
\begin{align}
 a^{\exA}_m\,M^{-1}(y_0^i)\,\mathfrak{t}_{\exA}\,M(y_0^i) = a^{\exA}_m\,M_{\exA}{}^{\Bh}(y_0^i)\,\mathfrak{t}_{\Bh}\,,
\end{align}
and because by convention $\mathrm{Lie}(H)$ is orthogonal to the subspace spanned by $\mathfrak{t}_{\hat{a}}$, this leads to
\begin{align}
 a^{\exA}_m\,M_{\exA}{}^{\hat{b}}(y_0^i) = 0\,.
\end{align}

Now, recall that the matrix $\hat{v}_{\hat{i}}{}^{\hat{a}}$ in \eqref{eq:hat-v1-dec} must be invertible to construct the generalized frame field $\Eh_{\Ah}{}^{\Ih}$. In our parameterization \eqref{eq:hat-v1-dec}, we have
\begin{align}
 \hat{v}_{\mu}{}^{\hat{a}}(y^\mu,y_0^i) = \widetilde{v}_{\mu}{}^{\exB}(y^\mu) \,M_{\exB}{}^{\hat{a}}(y^i) \,.
\end{align}
For arbitrary $y^\mu$, it follows that
\begin{align}
 b^\mu_m(y^\mu)\,\hat{v}_{\mu}{}^{\hat{a}} = a^{\exA}_m\,M_{\exA}{}^{\hat{b}}(y_0^i) = 0\,,
\end{align}
where we have defined $b^\mu_m(y^\mu)\equiv a^{\exA}_m\,\widetilde{v}_{\exA}{}^\mu(y^\mu)$. This shows that $\hat{v}_{\hat{i}}{}^{\hat{a}}(y^\mu,y_0^i)$ has $\dim G_0$ independent null eigenvector $b^{\hat{i}}_m=(b^\mu,0)$, implying that $\hat{v}_{\hat{i}}{}^{\hat{a}}(y^\mu,y_0^i)$ is not invertible. Consequently, at a fixed point, the generalized frame field $\Eh_{\Ah}{}^{\Ih}(y^\mu,y_0^i)$, and with it $E_{A}{}^{I}(y_0^i)$, become singular. The condition \eqref{eq:fixed-point} remains invariant under the transformation of the coset representative
\begin{align}
 M \to f\,M\,h \qquad (f\in \GS,\quad h\in H)\,,
\end{align}
which implies that this singularity is not merely a coordinate singularity. Indeed, as we will see in examples later, the ordinary curvature tensor diverges at fixed points. In particular, if $\GS$ and $H$ have non-trivial overlap, \eqref{eq:fixed-point} is trivially satisfied: the identity element with $M_{\Ah}{}^{\Bh}(y_0^i)=\delta_{\Ah}^{\Bh}$ is a fixed point. In our examples, $\Mh_{\Ah}{}^{\Bh}$ is constructed as $\Mh=\exp(x^{\hat{i}}\,\delta_{\hat{i}}^{\hat{a}}\,\mathfrak{t}_{\hat{a}})$ and $M_{\Ah}{}^{\Bh}=\delta_{\Ah}^{\Bh}$ is realized at the origin $x^{\hat{i}}=0$. This confirms that a singularity always appears if $\GS$ overlaps with $H$.

Now, let us consider the case of a standard coset $\GS \backslash G/H$ where $G/H$ forms a Lie group $F$ and $\GS$ is a subgroup of $F$. Here, $H$ is a normal subgroup of $G$, and the condition \eqref{eq:fixed-point} reduces to the condition that $\GS$ and $H$ overlap. However, since $\GS$ is defined as a subgroup of the quotient group $F=G/H$, they do not overlap. Thus, the singularity discussed here does not arise in the standard coset space. In our conclusions in section~\ref{sec:conclusion}, we discuss a possible resolution of this type of singularity. 

\subsection{Consistent truncations and generalized duality in DFT}\label{sec:examples-DFT}
To gain insight into how the previously outlined construction of consistent truncations works in practice, we begin with the simplest case where $\GD=\OO(d,d)$. In this setting, it is also possible to observe some well-known generalized dualities. Our starting point is a $2d$-dimensional Lie algebra (which corresponds to the algebra \eqref{eq:Leibniz-Alg})
\begin{align}
 [\hat{T}_{\Ah},\,\hat{T}_{\Bh}] = \hat{X}_{\Ah\Bh}{}^{\Ch}\,\hat{T}_{\Ch}
\end{align}
equipped with an adjoint-invariant metric of split signature,
\begin{align}
 \langle \hat{T}_{\Ah},\,\hat{T}_{\Bh}\rangle =\eta_{\Ah\Bh}\,,
\end{align}
where $\hat{T}_{\Ah}=(\hat{T}_{\hat{a}},\, \hat{T}^{\hat{a}})$ ($\hat{a}=1,\dotsc,p=d+n$). Due to the invariance of the metric, the structure constants $\hat{X}_{\Ah\Bh\Ch}\equiv \hat{X}_{\Ah\Bh}{}^{\Dh}\,\eta_{\Dh\Ch}$ are totally antisymmetric. In general, the structure constants $\hat{X}_{\Ah\Bh\Ch}$ can be decomposed as
\begin{align}
 \hat{X}_{\hat{a}\hat{b}\hat{c}} = f_{\hat{a}\hat{b}\hat{c}}\,,\qquad \hat{X}_{\hat{a}\hat{b}}{}^{\hat{c}} = f_{\hat{a}\hat{b}}{}^{\hat{c}}\,,\qquad \hat{X}_{\hat{a}}{}^{\hat{b}\hat{c}}=f_{\hat{a}}{}^{\hat{b}\hat{c}}\,,\qquad \hat{X}^{\hat{a}\hat{b}\hat{c}} = f^{\hat{a}\hat{b}\hat{c}}\,.
\end{align}
Choosing the generators of the maximally isotropic (or minimally coisotropic) subgroup $H$ as $\{\hat{T}^{\hat{a}}\}$, we set $f^{\hat{a}\hat{b}\hat{c}}=0$. Consequently, the remaining non-zero structure constants are $\{f_{\hat{a}\hat{b}\hat{c}}\,,\ f_{\hat{a}\hat{b}}{}^{\hat{c}} \,,\ f_{\hat{a}}{}^{\hat{b}\hat{c}}\}$. 

For $\GD=\OO(d,d)$, there is also the T-duality-invariant dilaton $d(x,y)$, along with its corresponding single-index flux $F_{\Ah}$\,. Similar to the generalized frame fields $\Eh_{\Ah}{}^{\Ih}(y^\mu,\,y^i)$ and $E_A{}^I(y)$, we introduce the hatted dilaton $\hat{\sfd}(y^\mu,\,y^i)$ on the mega-space and the generalized dilaton $\sfd(y)$ on the physical space $M$ (see \cite{Butter:2022iza} for details). They are related by
\begin{align}
 \hat{\sfd}(y^\mu,\,y^i) = \sfd(y) -\tfrac{1}{2}\,\ln\det (\widetilde{v}_{\mu}{}^{\exA}) + \tfrac{1}{2}\,\ln\det (\widetilde{M}_{\exA}{}^{\exB})\,,
\label{eq:dilaton-twist}
\end{align}
and their associated single-index fluxes
\begin{align}
 \Fh_{\Ah}\equiv 2\,\Eh_{\Ah}{}^{\Ih}\,\partial_{\Ih} \hat{\sfd} - \partial_{\Ih} \Eh_{\Ah}{}^{\Ih}\,,\qquad
 \check{F}_{\Ah}\equiv 2\,E_{\Ah}{}^{I}\,\partial_{I} \sfd - \partial_{I} E_{\Ah}{}^{I}
\end{align}
are related through
\begin{align}
 \Fh_{\Ah} = \widetilde{M}_{\Ah}{}^{\Bh}\, \bigl[\check{F}_{\Bh} -E_{\Ch}{}^{\exA}\, (\mathfrak{t}_{\exA})_{\Bh}{}^{\Ch} \bigr]\,.
\end{align}
On generalized cosets, we choose $\hat{\sfd}(y^\mu,\,y^i)$ as
\begin{align}
 \hat{\sfd}(y^\mu,\,y^i) = \lambda(y^\mu,\,y^i) -\tfrac{1}{2}\,\ln\det (\hat{v}_{\hat{i}}{}^{\hat{a}})\,,
\end{align}
where $\lambda(y^\mu,\,y^i)$ must satisfy
\begin{align}
 \partial_{\hat{i}} \lambda = \tfrac{1}{2}\,v_{\hat{i}}{}^{\Ah}\,\cT_{\Ah}
\end{align}
with the constants
\begin{align}
 \cT_{\Ah} \equiv \begin{pmatrix} \cT_{\hat{a}} & \cT^{\hat{a}} \end{pmatrix} ,\qquad
 \cT^{\hat{a}} = - f_{\hat{b}}{}^{\hat{b}\hat{a}}\,.
\end{align}
Since $\lambda$ is implicitly given by a differential equation, integrability requires an additional constraint, namely
\begin{align}
 \hat{X}_{\Ah\Bh}{}^{\Ch}\,\cT_{\Ch} = 0\,.
\label{eq:X-F}
\end{align}
It ensures that
\begin{align}
 \Fh_{\Ah} = \Mh_{\Ah}{}^{\Bh}\, \cT_{\Bh} = \cT_{\Ah}
\end{align}
holds under the assumption
\begin{align}
 \cT_{\exA} = - f_{\exA\exB}{}^{\exB}\,.
\end{align}
We then obtain the expression for $\sfd(y^i)$ as
\begin{align}
 \sfd(y^i) = \bigl[\lambda -\tfrac{1}{2}\,\ln\det (\hat{v}_{\hat{i}}{}^{\hat{a}})\bigr]_{y^\mu=0} +\text{const.} 
\end{align}
However, this only describes the generalized dilaton for the internal space of the dimensional reduction we wish to perform. It has to be complemented by the generalized dilaton $\Phi(x)$ for the external directions $x^{\EXm}$ that remain after the consistent truncation, resulting in
\begin{align}\label{eq:reduction-ansatz-dilaton}
 d(x,y) = \Phi(x) + \sfd(y)\,.
\end{align}

By combining this ansatz for the generalized dilaton with \eqref{eq:reduction-ansatz-frame}--\eqref{eq:reduction-ansatz-2form}, we obtain the Lagrangian 
\begin{align}
 \cL &= \Exp{-2\,\phiD}\bigl(*R_g+4\,\rmd \phiD \wedge *\rmd \phiD
\nn\\
 &\qquad\qquad -\tfrac{1}{2}\,\cM_{AB}\, F_2^{\cA}\wedge * F_2^{\cB} 
 -\tfrac{1}{2}\, H_3\wedge * H_3
 + \tfrac{1}{8}\, \cD \cM^{AB}\wedge * \cD \cM_{AB} - V\bigr)
\label{eq:reduced-Lagrangian-ODD}
\end{align}
of the reduced theory, where $\Exp{-2\,\Phi}=\sqrt{-g}\,\Exp{-2\,\phiD}$, $R_g$ and $*$ are the ordinary Ricci scalar and the Hodge star operator associated with $g_{\EXm\EXn}$, and
\begin{align}
 F_2^{\cA} &= \rmd A^{\cA} + \tfrac{1}{2}\, \cT_{\cB\cC}{}^{\cA} \, A^{\cB}\wedge A^{\cC} \,,
\\
 H_3 &= \rmd B_2 - \tfrac{1}{2}\, A^{\cA}\wedge \rmd A_{\cA} -\tfrac{1}{3!}\, \cT_{\cA\cB\cC} \,A^{\cA}\wedge A^{\cB}\wedge A^{\cC}\,,
\\
 \cD \cM_{AB} & = \rmd \cM_{AB} -2\,A^{\cC}\, \cT_{\cC(A}{}^D \,\cM_{B)D} 
\end{align}
with $\cD=\rmd x^{\EXm}\wedge \cD_{\EXm}$\,. Additionally, the scalar potential
\begin{align}
 V = -\cM^{AB}\,\bigl( \tfrac{1}{12}\,\cM^{DE}\, \cM_{CF}\, \cT_{AD}{}^C\,\cT_{BE}{}^F + \tfrac{1}{4} \, \cT_{AC}{}^{D}\,\cT_{BD}{}^C + \cT_A\,\cT_B + R_{AB}\bigr) 
\end{align}
arises from the torsion and curvature of the internal generalized coset. In the following, we present several explicit generalized cosets and the resulting truncations.

\subsubsection{Example 1: \texorpdfstring{$\SO(2)\backslash \SO(3)$}{SO(2)\textbackslash SO(3)}}\label{sec:DFTex1}
Consider the Drinfel'd double $G=\text{ISO}(3)=\SO(3)\ltimes \text{T}_3$. This describes the isometry group of three-dimensional Euclidean space, where $\text{T}_3$ represents the group of translations in three dimensions. In this case, both $\SO(3)$ and $\text{T}_3$ are maximally isotropic subgroups of $G$. This Drinfel'd double can be constructed by introducing the non-vanishing structure constants
\begin{align}
 f_{12}{}^3 = 1\,,\quad
 f_{23}{}^1 = 1\,,\quad
 f_{31}{}^2 = 1\,.
\end{align}
Here, $\hat{T}_{\hat{a}}$ generate the Lie algebra of $\SO(3)$ while $\hat{T}^{\hat{a}}$ generate the subgroup $H=\text{T}_3$.

Now, let us take $\GS=\SO(2)$ as the isotropic subgroup generated by $\hat{T}_{\exA}= (\hat{T}_1)$.
This leads to the generalized coset
\begin{align}
 \GS\backslash G/H = \SO(2)\backslash \SO(3) \cong \text{S}^2\,,
\end{align}
which corresponds to a standard coset space.

\paragraph{\underline{8D theory:}}
For this setup, we find
\begin{align}
 f_{\exA\exB}{}^{\exC} &= 0\,,\qquad 
 (\mathfrak{t}_{1})_A{}^B = - (\hat{X}_{1})_A{}^B = \begin{pmatrix}
 0 & -1& 0 & 0 \\
 1 & 0 & 0 & 0 \\
 0 & 0 & 0 & -1 \\
 0 & 0 & 1 & 0 \end{pmatrix} ,
\\
 \hat{X}_{AB}{}^C &= 0\,,\qquad
 \hat{X}_{AB}{}^{1} = \begin{pmatrix}
 0 & 1& 0 & 0 \\
 -1 & 0 & 0 & 0 \\
 0 & 0 & 0 & 0 \\
 0 & 0 & 0 & 0 \end{pmatrix}, 
\end{align}
and thereby extract
\begin{align}
 \cT_{AB}{}^C = 0\,,\qquad R_{AB} = \begin{pmatrix} \delta_{ab} & 0 \\ 0 & 0 \end{pmatrix}.
\end{align}
Furthermore, there are no null eigenvectors $K_{\cA}$ of $(\mathfrak{t}_{1})_A{}^B$. Hence, the reduced theory does not contain vector fields. The commutant group is given by $\mathrm{C}_{\GD}(\GS)=\SL(2)\times \text{U}(1)$ and its maximal compact subgroup is $K(\mathrm{C}_{\GD}(\GS))=\SO(2)\times \text{U}(1)$. 
Consequently, all scalar fields retained in the truncation are parameterized by
\begin{align}
 \cM_{AB} = \begin{pmatrix} \delta_a^c & b\,\epsilon_{ac} \\
 0 & \delta^a_c \end{pmatrix}
 \begin{pmatrix}
 \Exp{a}\,\delta_{cd} & 0 \\ 0 & \Exp{-a}\,\delta^{cd}
\end{pmatrix}
 \begin{pmatrix} \delta_b^d & 0 \\
 -b\,\epsilon_{db} & \delta^b_d \end{pmatrix} \in \frac{\SL(2)}{\SO(2)}\,.
\label{eq:M-ex1}
\end{align}
In this parameterization, the generalized Ricci scalar takes the simple form
\begin{align}
 R = \cM^{AB}\,R_{AB} = 2\,\Exp{-a}\,.
\end{align}
Moreover, the constraint \eqref{eq:X-F} requires that the single-index flux $\cT_{\Ah}$ vanishes. This is all we need to truncate the NS--NS part of the ten-dimensional type II supergravity Lagrangian to
\begin{align}
 \cL_8 &= \Exp{-2\,\phiD} \bigl(* R_g+4\,\rmd\phiD\wedge *\rmd \phiD -\tfrac{1}{2}\, H_3\wedge * H_3
 - \tfrac{1}{2}\, \rmd a\wedge * \rmd a
 - \tfrac{1}{2}\,\Exp{-2a}\, \rmd b\wedge *\rmd b + * 2\,\Exp{-a}\bigr)
\label{eq:action-DFT-ex1}
\end{align}
in eight dimensions.

If we consider type II supergravities, we can also include the Ramond--Ramond fields. These fields decompose into $p$-form fields in 8D, but only the odd-form fields remain invariant under the $\GS$ transformations. For example, in type IIA theory, the components $C_{\EXm_1\cdots \EXm_p}$ and $C_{\EXm_1\cdots \EXm_p ab}$ ($p:\text{odd}$) give rise to $p$-form fields that transform as spinors under $\mathrm{O}(2,2)$. The fermions transform as spinors under one of the double Lorentz groups, $\OO(2)\times \OO(2)$. However, since $\GS$ is the diagonal subgroup of the double Lorentz group, there are no fermions invariant under $\GS$. As a result, all fermions must be truncated, leading to a complete breaking of supersymmetry. While we do not compute the explicit reduction ansatz for the Ramond--Ramond fields here, it is obtained in section \ref{sec:ExFT-ex1} by considering the U-duality group. 

In the following, we present two embeddings of this eight-dimensional theory into ten-dimensional supergravity, both of which are related via non-abelian T-duality. These embeddings have been studied in the literature \cite{delaOssa:1992vci}, focusing on specific choices for the moduli $a$ and $b$. We then find two additional embeddings that are connected to the original one through generalized T-dualities.

\paragraph{\underline{Generalized frame field:}}
To construct the generalized frame field on the mega-space, we choose the parameterization
\begin{align}
 l = f\,m\,h\,,\qquad f=\Exp{x\,\hat{T}_1}\in \GS\,,\qquad
 m =\Exp{\theta\,\hat{T}_2} \Exp{-\psi\,\hat{T}_1} \in \GS\backslash G/H\,,\qquad
 h\in H
\end{align}
for the elements $l\in G$ to obtain
\begin{align}\label{eq:genFrameSO31}
 \Eh_{\Ah}{}^{\Ih} = \left(\begin{array}{c|cccc|c}
 1 & 0 & 0 & 0 & 0 & 0 \\ \hline
 -\frac{\sin x}{\tan \theta} & \cos x & -\frac{\sin x}{\sin \theta} & 0 & 0 & 0 \\
 \frac{\cos x}{\tan \theta} & \sin x & \frac{\cos x}{\sin \theta} & 0 & 0 & 0 \\
 0 & 0 & 0 & \cos x & -\sin x \sin \theta & 0 \\
 0 & 0 & 0 & \sin x & \cos x \sin \theta & 0 \\ \hline
 0 & 0 & 0 & 0 & -\cos \theta & 1
\end{array}\right),
\end{align}
in the coordinates $y^{\hat{i}}=(x,\, \theta,\, \psi)$. Keeping in mind that the twist $\widetilde{M}$ in this case has the form
\begin{align}
 \widetilde{M}_{\Ah}{}^{\Bh} &= \exp(x\,\mathfrak{t}_1)_{\Ah}{}^{\Bh}= \left(\begin{array}{c|cccc|c}
 1 & 0 & 0 & 0 & 0 & 0 \\ \hline
 0 & \cos x & -\sin x & 0 & 0 & 0 \\
 0 & \sin x & \cos x & 0 & 0 & 0 \\
 0 & 0 & 0 & \cos x & -\sin x & 0 \\
 0 & 0 & 0 & \sin x & \cos x & 0 \\ \hline
 0 & 0 & 0 & 0 & 0 & 1
\end{array}\right) ,
\end{align}
we obtain the untwisted generalized frame field $E_{\Ah}{}^{\Ih} = (\widetilde{M}^{-1})_{\Ah}{}^{\Bh}\,\Eh_{\Bh}{}^{\Ih}$. Focusing on the inner components of this matrix, we extract the physical generalized frame field
\begin{align}\label{eq:frame-ex1-1}
  E_{A}{}^{I} &= \begin{pmatrix} 
 1 & 0 & 0 & 0 \\
 0 & \frac{1}{\sin \theta} & 0 & 0 \\
 0 & 0 & 1 & 0 \\
 0 & 0 & 0 & \sin \theta \end{pmatrix},
\intertext{and the generalized spin connection}
 \Omega_A^{\exB} &= \begin{pmatrix}
 0 \\ -\frac{1}{\tan \theta} \\ 0 \\ 0 \end{pmatrix}.
\end{align}
With the parameterization \eqref{eq:M-ex1}, the relations \eqref{eq:reduction-ansatz-frame}--\eqref{eq:reduction-ansatz-2form} between the ten- and eight-dimensional fields take the form
\begin{align}\label{eq:spin-connection-ex1-1}
 g_{\EXm\EXn}(x,y) &= g_{\EXm\EXn}(x)\,,\qquad
 A_{\EXm}{}^I(x,y) = 0\,,\qquad 
 B_{\EXm\EXn}(x,y) = b_{\EXm\EXn}(x)\,,
 \intertext{while for the internal directions we have}
 g_{ij}(x,y) &= \Exp{a(x)}\begin{pmatrix} 1 & 0 \\ 0 & \sin^2 \theta \end{pmatrix},\qquad
 B_{ij}(x,y) = b(x)\begin{pmatrix} 0 & \sin \theta \\ -\sin \theta & 0 \end{pmatrix}.
\end{align}
For the duality group $\GD=\OO(d,d)$, we also need to take the dilaton $d$ into account. Since a non-trivial $\cT_{\Ah}$ satisfying \eqref{eq:X-F} does not exist, the reduction ansatz \eqref{eq:reduction-ansatz-dilaton} simplifies to
\begin{align}
 d = \Phi -\tfrac{1}{2} \ln\sin\theta\,,
\end{align}
and the ten-dimensional dilaton $\phiS$ is then given by
\begin{align}
 \phiS(x) = \phiD(x) + \tfrac{1}{2}\,a(x)\,.
\end{align}

\paragraph{\underline{Non-abelian T-duality:}}
Now, we consider another maximal isotropic subgroup, namely $H'=\SO(3)$. It is realized by swapping the generators
\begin{align}
 \hat{T}'_a = \hat{T}^a\,,\qquad \hat{T}'^a = \hat{T}_a\,,
\end{align}
resulting in a new set of non-vanishing structure constants
\begin{align}
 f'_1{}^{23} =1\,,\qquad
 f'_2{}^{31} =1\,,\qquad
 f'_3{}^{12} =1\,.
\end{align}
Again, we employ the parameterization
\begin{align}
 l = f\,m'\,h'\,,\qquad f=\Exp{x\,\hat{T}_1}\in \GS\,,\qquad
 m' =\Exp{y\,\hat{T}^1} \Exp{z\,\hat{T}^2} \in \GS\backslash G/H'\,,\qquad
 h'\in H'
\end{align}
for the elements $l\in G$, and rotate the resulting generalized frame field on the mega-space back to the original basis $\hat{T}_{\Ah}$ to obtain
\begin{align}
 \Eh'_{\Ah}{}^{\Ih} = \left(\begin{array}{c|cccc|c}
 1 & 0 & 0 & 1 & 0 & 0 \\ \hline
 -\frac{y \cos x}{z} & z \sin x & -y \sin x & 0 & \cos x & -z \sin x \\
 -\frac{y \sin x}{z} & -z \cos x & y \cos x & 0 & \sin x & z \cos x \\
 -\frac{\sin x}{z} & 0 & \cos x & 0 & 0 & 0 \\
 \frac{\cos x}{z} & 0 & \sin x & 0 & 0 & 0 \\ \hline
 0 & 1 & 0 & 0 & 0 & 0 \end{array} \right),
\end{align}
in the coordinates $y^{\hat{i}}=(x,\,y,\,z)$. By construction, it satisfies the same algebra as before, namely
\begin{align}
 \gLie_{\Eh'_{\Ah}} \Eh'_{\Bh}{}^{\Ih} = - \hat{X}_{\Ah\Bh}{}^{\Ch}\,\Eh'_{\Ch}\,,
\label{eq:ELie-ex1}
\end{align}
where $\hat{X}_{\Ah\Bh}{}^{\Ch}$ are the same structure constants found for the previous frame \eqref{eq:genFrameSO31}. However, the naively constructed frame field does not have a lower-triangular form. To bring it into lower-triangular gauge, we perform a constant $B$-field gauge transformation, and obtain
\begin{align}
 \Eh'_{\Ah}{}^{\Ih} = \left(\begin{array}{c|cccc|c}
 1 & 0 & 0 & 0 & 0 & 0 \\ \hline
 -\frac{y \cos x}{z} & z \sin x & -y \sin x & \frac{y \cos x}{z} & \cos x & 0 \\
 -\frac{y \sin x}{z} & -z \cos x & y \cos x & \frac{y \sin x}{z} & \sin x & 0 \\
 -\frac{\sin x}{z} & 0 & \cos x & \frac{\sin x}{z} & 0 & 0 \\
 \frac{\cos x}{z} & 0 & \sin x & -\frac{\cos x}{z} & 0 & 0 \\ \hline
 0 & 1 & 0 & 0 & 0 & 1 \end{array} \right).
\end{align}
We note that this transformed frame still satisfies \eqref{eq:ELie-ex1}. Thus, in this form, we can compute the untwisted generalized frame field $E'_{\Ah}{}^{\Ih}$ and identify the physical generalized frame field 
\begin{align}
  E'_{A}{}^{I} &= \begin{pmatrix} 
 0 & 0 & \frac{y}{z} & 1 \\
 -z & y & 0 & 0 \\
 0 & 1 & 0 & 0 \\
 0 & 0 & -\frac{1}{z} & 0 \end{pmatrix},
 \intertext{and the corresponding spin connection as}
 \Omega'_A{}^{\exB} &= \begin{pmatrix}
 -\frac{y}{z} \\ 0 \\ 0 \\ -\frac{1}{z} \end{pmatrix}.
\end{align}
Both are clearly different from \eqref{eq:frame-ex1-1} and \eqref{eq:spin-connection-ex1-1}. Note that this difference cannot be removed by a diffeomorphism or a gauge transformation. As before, we use $E'_{A}{}^{I}$ to explicitly present the truncation ansatz
\begin{align}
 g_{\EXm\EXn}(x,y) &= g_{\EXm\EXn}(x)\,,\qquad
 A_{\EXm}{}^I(x,y) = 0\,,\qquad 
 B_{\EXm\EXn}(x,y) = b_{\EXm\EXn}(x)\,,
 \intertext{for the external and}
 g'_{ij}(x,y) &= 
 \frac{1}{r^2 z^2}\begin{pmatrix}
 r^4 + (y+b)^2 \quad & z(y+b) \\
   z(y+b) & z^2 \end{pmatrix} ,\qquad
 B'_{ij}(x,y) = 0\,,\qquad
 \phiS' = \phiD - \ln z\,,
\label{eq:natd-p}
\end{align}
for the internal directions with $r(x)\equiv \Exp{a(x)/2}$\,. To better compare this result with the existing literature, we replace the coordinate $z$ by $x$ through
\begin{align}
 z = \sqrt{x^2-y^2}\,.
\end{align}
Now, the internal metric and the standard dilaton read
\begin{align}
 \rmd s'^2 = \frac{r^2 \,\rmd y^2 + (x\,\rmd x + b\,\rmd y)^2}{r\,(x^2-y^2)}\,,\qquad 
 \phiS' = \phiD - \ln \sqrt{x^2-y^2}\,.
\end{align}
They match with the expressions found in \cite{delaOssa:1992vci} by using non-abelian T-duality on the worldsheet. To be more precise: In that work, the discussion was focused on the special case with $b=0$ and $\phiD=-\frac{1}{2}\,a(x)=-\ln r$.

We note that the region $x=y$, which corresponds to our $z=0$, has a curvature singularity. This is exactly the singularity discussed in section \ref{sec:singularity} with \eqref{eq:fixed-point} as the condition for the fixed point, namely
\begin{align}
 M^{-1}\,\mathfrak{t}_{1}\,M = \left(
\begin{array}{c|cccc|c}
 0 & 0 & 0 & -z & 0 & 0 \\ \hline
 0 & 0 & -1 & 0 & 0 & z \\
 0 & 1 & 0 & 0 & 0 & 0 \\
 0 & 0 & 0 & 0 & -1 & 0 \\
 0 & 0 & 0 & 1 & 0 & 0 \\ \hline
 0 & 0 & 0 & 0 & 0 & 0  
\end{array}
\right) \in \mathrm{Lie}(H)\,,
\end{align}
which results in a fixed point for $z=0$. 

\paragraph{\underline{Generalized T-duality:}}
In the context of Poisson--Lie T-duality, where $f_{abc}=0$ is required to obtain a Drinfel'd double, the previously discussed non-abelian T-dual is the only dual background. However, by exchanging the third generators, 
\begin{align}
 \hat{T}''_3 = \hat{T}^3\,,\qquad \hat{T}''^3 = \hat{T}_3\,,
\end{align}
we can consider more general T-dualities. As a result, the structure constants become
\begin{align}
 f''_1{}^{23} =1\,,\qquad
 f''_2{}^{31} =1\,,\qquad
 f''_{123} =1\,, 
\end{align}
where the three-dimensional Lie group $H''$ is of Bianchi type VII$_0$\,. We again construct the mega frame field using the parameterization
\begin{align}
 l = f\,m''\,h''\,,\qquad f=\Exp{x\,\hat{T}_1}\in \GS\,,\qquad
 m'' =\Exp{z\,\hat{T}^3} \Exp{y\,\hat{T}_2} \in \GS\backslash G/H''\,,\qquad
 h''\in H''\,,
\end{align}
to obtain the untwisted generalized frame field
\begin{align}
 E''_{\Ah}{}^{\Ih} = \left(\begin{array}{c|cccc|c}
 1 & 0 & 0 & 0 & 0 & 0 \\ \hline
 0 & 1 & z \tan y & 0 & 0 & 0 \\
 -\tan y & 0 & 0 & -z \tan y & 1 & 0 \\
 0 & 0 & 0 & 1 & 0 & 0 \\
 0 & 0 & 1 & 0 & 0 & 0 \\ \hline
 0 & 0 & \tan y & 0 & 0 & 1 \end{array} \right),
\end{align}
in the coordinates $y^{\hat{i}}=(x,\,y,\,z)$. By repeating the steps outlined before, we eventually obtain the internal metric, $B$-field, and the dilaton,
\begin{align}
\begin{split}
 g''_{ij} &= 
 \frac1{r^2}\begin{pmatrix}
 r^4 + (b-z \tan y)^2 \quad & b - z \tan y\\
 b-z \tan y & 1 
\end{pmatrix} ,\qquad
 B''_{ij} = 0\,,\qquad
 \phiS'' = \phiD - \ln \cos y\,.
\end{split}
\label{eq:gpp-DFT-ex1}
\end{align}
This result appears to be similar to \eqref{eq:natd-p}, but we expect that they are not related by a coordinate transformation. At least if the scalar fields $r$ (or $a$) and $b$ are constant, the ordinary Ricci scalars for $g'_{ij}$ and $g''_{ij}$ are given by
\begin{align}
 R' = - 2\,r^{-2}\,\bigl\{1-2\,z^{-2}\,[r^4+(y+b)^2] \bigr\}\,,\qquad 
 R'' = 2\,r^{-2}\, \bigl(1- 2 \cos^{-2} y \bigr)\,.
\end{align}
These expressions cannot be matched by a two-dimensional coordinate transformation that is independent of the scalar fields $r$ and $b$. Thus, we conclude that we have found two distinct geometries.

For completeness, we also determine the fixed points by solving the condition
\begin{align}
 M''^{-1}\,\mathfrak{t}_{1}\,M'' = \left(
\begin{array}{c|cccc|c}
 0 & -\sin y & 0 & 0 & -z & 0 \\ \hline
 \sin y & 0 & -\cos y & 0 & 0 & 0 \\
 0 & \cos y & 0 & 0 & 0 & z \\
 0 & 0 & 0 & 0 & -\cos y & \sin y \\
 0 & 0 & 0 & \cos y & 0 & 0 \\ \hline
 0 & 0 & 0 & -\sin y & 0 & 0 
\end{array}
\right) \in \mathrm{Lie}(H) \,.
\end{align}
This condition implies $\cos y=0$, and indeed singularities appear at these points. In particular, if $a$ and $b$ are constant, we find that the ordinary Ricci scalar $R''$ and the dilaton $\phiD''$ diverge at the fixed points $\cos y=0$.

\paragraph{\underline{Further generalized T-duality:}}
The previously chosen subgroup $H''$ does not overlap with $\GS$, but alternatively, we can choose another subgroup $H'''$, which contains $\GS$ and is isomorphic to $H''$. This can be realized by swapping the generators
\begin{align}
 \hat{T}'''_1 = \hat{T}^1\,,\qquad \hat{T}'''^1 = \hat{T}_1\,,
\end{align}
after which the structure constants
\begin{align}
 f'''_2{}^{31} =1\,,\qquad
 f'''_3{}^{12} =1\,,\qquad
 f'''_{123} =1\,,
\end{align}
arise. We then construct the mega frame field using the parameterization
\begin{align}
 l = f\,m''\,h''\,,\qquad f=\Exp{x\,\hat{T}_1}\in \GS\,,\quad
 m''' =\Exp{\theta\,\hat{T}_2} \Exp{-\psi\,\hat{T}^1} \in \GS\backslash G/H''\,,\quad
 h'''\in H'''
\end{align}
to obtain
\begin{align}
 E'''_{\Ah}{}^{\Ih} = \left(\begin{array}{c|cccc|c}
 1 & 0 & 0 & 0 & 0 & 0 \\ \hline
 0 & 1 & 0 & 0 & 0 & 0 \\
 \frac{1}{\tan \theta} & 0 & 0 & 0 & \frac{1}{\sin \theta} & 0 \\
 0 & 0 & 0 & 1 & 0 & 0 \\
 0 & 0 & \sin \theta & 0 & 0 & 0 \\ \hline
 0 & 0 & -\cos \theta & 0 & 0 & 1 \end{array} \right),
\end{align}
in the coordinates $y^{\hat{i}}=(x,\,\theta,\,\psi)$. Following the steps explained above, we eventually obtain
\begin{align}
 g'''_{ij} &= 
\begin{pmatrix}
 \Exp{-a}\,b^2+\Exp{a} & \frac{\Exp{-a}\,b}{\sin \theta} \\
 \frac{\Exp{-a}\,b}{\sin \theta} & \frac{\Exp{-a}}{\sin^2\theta} \end{pmatrix} ,\qquad
 B'''_{ij} = 0\,,\qquad
 \phiS''' = \phiD - \ln \sin \theta\,.
\end{align}
By performing a factorized T-duality along the $\psi$-direction, we find that this background (for arbitrary $a$ and $b$) is mapped back to the original one, which demonstrates the duality equivalence explicitly.

Once again, we confirm that fixed points appear at $\sin\theta=0$.  At these points, the ordinary Ricci scalar 
\begin{align}
 R'''= 2\,\Exp{-a}\,\bigl(1-2\sin^{-2}\theta\bigr)
\end{align}
and the dilaton diverge. In all of the examples presented here, we find that the curvature singularities appear only at the fixed points of $\GS$.

\subsubsection{Example 2: \texorpdfstring{$\text{U}(1)\backslash \text{SU}(2)\times \text{SU}(2)/\text{SU}(2)_{\text{diag}}$}{U(1)\textbackslash SU(2)xSU(2)/SU(2)diag}}\label{sec:SU2-WZW}
Our next example is based on the six-dimensional Lie group $G = \text{SU}(2) \times \text{SU}(2)$, which admits a maximally isotropic subgroup $H = \text{SU}(2)_{\text{diag}}$. Choosing this subgroup, the mega-space corresponds to the $\text{SU}(2)$ Wess--Zumino--Witten (WZW) model. The algebra $\mathfrak{su}(2)\oplus\mathfrak{su}(2)\cong \mathfrak{so}(4)$ is captured by the structure constants
\begin{align}
 f_3{}^{12} = 1\,,\quad
 f_1{}^{23} = 1\,,\quad
 f_2{}^{31} = 1\,,\quad
 f_{123} = 1\,.
\end{align}
In our chosen basis, the two $\mathfrak{su}(2)$ factors are generated by
\begin{align}
 \{t_1,\,t_2,\,t_3\}=\Bigl\{\frac{\hat{T}^1-\hat{T}_1}{2},\,\frac{\hat{T}^2-\hat{T}_2}{2},\,\frac{\hat{T}^3-\hat{T}_3}{2}\Bigr\}
\end{align}
and
\begin{align}
 \{\tilde{t}_1,\,\tilde{t}_2,\,\tilde{t}_3\}=\Bigl\{\frac{\hat{T}^1+\hat{T}_1}{2},\,\frac{\hat{T}^2+\hat{T}_2}{2},\,\frac{\hat{T}^3+\hat{T}_3}{2}\Bigr\}\,,
\end{align}
respectively, while the diagonal subalgebra $\mathfrak{su}(2)_{\text{diag}}$ is generated by their sum
\begin{align}\label{eq:diagonal-su2}
  \{t_1 + \tilde{t}_1,\,t_2 + \tilde{t}_2,\,t_3 + \tilde{t}_3\} =  \{\hat{T}^1,\,\hat{T}^2,\,\hat{T}^3\}\,.
\end{align}
By choosing a non-trivial structure group $\GS$, one obtains the target space of a gauged WZW model. In particular, we consider two different embeddings of $\text{U}(1)$ structure groups, namely
\begin{enumerate}
  \item $\mathfrak{u}(1)_{\text{V}}$, generated by $\hat{T}^1=\tilde{t}_1+t_1$, and
  \item $\mathfrak{u}(1)_{\text{A}}$, generated by $\hat{T}_1=\tilde{t}_1-t_1$. 
\end{enumerate}
Taking the coset representative $g \in G/H$ as an element of the right $\text{SU}(2)$ factor, generated by $\{\tilde{t}_1,\,\tilde{t}_2,\,\tilde{t}_3\}$, these choices for $\GS$ correspond to
\begin{enumerate}
\item gauging the vectorial transformation $g\to f\,g\,f^{-1}$, $f\in \text{U}(1)$, and
\item gauging the axial transformation $g\to f\,g\,f$, $f\in \text{U}(1)$.
\end{enumerate}
An automorphism \eqref{eq:automorphismC} of the Lie algebra $\gLie(G)$ is given by
\begin{align}
 \hat{T}'_1 = \hat{T}^1\,,\quad \hat{T}'_2 = \hat{T}^2\,,\quad \hat{T}'_3 = \hat{T}_3\,,\quad
 \hat{T}'^1 = \hat{T}_1\,,\quad \hat{T}'^2 = \hat{T}_2\,,\quad \hat{T}'^3 = \hat{T}^3\,.
\end{align}
It exchanges $\mathfrak{u}(1)_{\text{V}}$ and $\mathfrak{u}(1)_{\text{A}}$ and therefore, following the argumentation in section~\ref{sec:differentGS}, suggests that the two generalized cosets $\text{U}(1)_{\text{V}}\backslash \text{SU}(2)\times \text{SU}(2)/\text{SU}(2)_{\text{diag}}$ and $\text{U}(1)_{\text{A}}\backslash \text{SU}(2)\times \text{SU}(2)/\text{SU}(2)_{\text{diag}}$ result in the same eight-dimensional theory.

This type of duality is known as the axial-vector duality \cite{Giveon:1991sy,Dijkgraaf:1991ba,Kiritsis:1991zt,Rocek:1991ps,Kiritsis:1992uz,Kiritsis:1993ju,Giveon:1993ph}, and refers to the equivalence between $\text{U}(1)_{\text{V}}\backslash G\times G/G_{\text{diag}}$ and $\text{U}(1)_{\text{A}}\backslash G\times G/G_{\text{diag}}$ for a Lie group $G$. Here we take $G = \text{SU}(2)$, and the case $G = \SL(2)$ is studied in section \ref{sec:SL2-WZW}. We can also consider $G = \text{SU}(3)$ with the structure constants
\begin{align}
 f_{123}=1\,,\quad
 f_{147}=f_{246}=f_{257}=f_{345}= \tfrac{1}{2}\,,\quad
 f_{156}=f_{367}=-\tfrac{1}{2} \,,\quad
 f_{458}=f_{678}= \tfrac{\sqrt{3}}{2}\,,
\end{align}
where $f_{abc}=f_{ab}{}^d\,\delta_{dc}=f_{[abc]}$. Then, we find that the transformation
\begin{align}
 \hat{T}'_i = \hat{T}^i\,,\quad 
 \hat{T}'^i = \hat{T}_i\quad (i=4,5,6,7)\,,\quad 
 \hat{T}'_s = \hat{T}_s\,,\quad 
 \hat{T}'^s = \hat{T}^s\quad (s=1,2,3,8)\,,
\end{align}
is an automorphism. By taking $\mathfrak{u}(1)_{\text{V}}$ and $\mathfrak{u}(1)_{\text{A}}$ as the subalgebras generated, for example, by $\hat{T}^4 = \tilde{t}_4 + t_4$ and $\hat{T}_4 = \tilde{t}_4 - t_4$, respectively, we can again realize the axial-vector duality as a generalized T-duality. While we have not proven that axial-vector duality can always be realized as generalized T-duality, it at least arises in many concrete examples.

\paragraph{\underline{8D theory:}}
To identify the eight-dimensional theory, we concretely take $\GS = \text{U}(1)_{\text{V}}$ and obtain
\begin{align}
  f_{\exA\exB}{}^{\exC} &= 0 = \hat{X}_{AB}{}^C\,,\qquad &
  (\mathfrak{t}_{1})_A{}^B &= \begin{pmatrix}
 0 & -1 & 0 & 0 \\
 1 & 0 & 0 & 0 \\
 0 & 0 & 0 & -1 \\
 0 & 0 & 1 & 0 \end{pmatrix} = - \hat{X}_{AB}{}^{1}\,.
\end{align}
Since there is no non-vanishing $\cT_{A}$ satisfying \eqref{eq:X-F}, we find the eight-dimensional theory with
\begin{align}
 \cT_{AB}{}^C = 0 \,,\qquad \cT_{A} =0\,,\qquad R_{AB} = \delta_{AB}\,.
\end{align}
Since $(\mathfrak{t}_{1})_A{}^B$ takes the same form as in the previous example, the commutant group $\mathrm{C}_{\GD}(\GS)=\SL(2)\times \text{U}(1)$ is exactly the same, and the scalar fields can again be parameterized by \eqref{eq:M-ex1}. The truncation of fermions and the Ramond--Ramond fields is also the same, and supersymmetry is completely broken. From the general equation for the reduced Lagrangian \eqref{eq:reduced-Lagrangian-ODD}, we find 
\begin{align}
 \cL_8 &= \Exp{-2\,\phiD}\bigl[*R_g+4\,\rmd \phiD\wedge *\rmd \phiD
 -\tfrac{1}{2}\, H_3\wedge * H_3
\nonumber\\
 & \qquad\qquad - \tfrac{1}{2}\,(\rmd a\wedge *\rmd a
 +\Exp{-2a}\, \rmd b\wedge *\rmd b) + * 4 \cosh a + * 2\,\Exp{-a}\,b^2 \bigr]
\end{align}
in eight dimensions for the fields in the NS--NS sector.

In the following, we construct the generalized frame field for $H=\text{SU}(2)_{\text{diag}}$, and $\GS=\text{U}(1)_{\text{V}}$ or $\GS=\text{U}(1)_{\text{A}}$\,. They provide different ten-dimensional configurations corresponding to the same eight-dimensional theory presented above.

\paragraph{\underline{$\GS=\text{U}(1)_{\text{V}}$ and $H=\text{SU}(2)_{\text{diag}}$:}}
To consider the vectorial gauging, we use the parameterization
\begin{align}\label{eq:param-vector-sl2}
 l = f\,m \,h \,,\qquad f=\Exp{x\,\hat{T}^1}\in \text{U}(1)_{\text{V}}\,,\quad
 m =\Exp{\theta\,(\hat{T}_1+\hat{T}^1)}\,\Exp{r\,(\hat{T}_3+\hat{T}^3)} \in \text{U}(1)_{\text{V}}\backslash G/H \,,
\end{align}
where $h \in H$. Then the untwisted generalized frame field is found as
\begin{align}
 E_{\Ah}{}^{\Ih} = \left(\begin{array}{c|cccc|c}
 1 & 0 & 0 & 0 & 0 & 0 \\ \hline
 -\frac{\cos (2 r)+\cos (2 \theta)}{\sin (2 r)} & -\sin \theta \cos \theta & -\tan r \sin^2 \theta & -\sin \theta \cos \theta & \frac{\cos r \cos^2 \theta}{\sin r} & 0 \\
 -\frac{\sin \theta \cos \theta}{\sin r \cos r} & -\sin^2 \theta & \tan r \sin \theta \cos \theta & \cos^2 \theta & \frac{\sin \theta \cos \theta}{\tan r} & 0 \\
 \frac{\cos (2 r)- \cos (2 \theta)}{\sin (2 r)} & -\sin \theta \cos \theta & \tan r \cos^2 \theta & -\sin \theta \cos \theta & -\frac{\cos r \sin^2 \theta}{\sin r} & 0 \\
 -\frac{\sin \theta \cos \theta}{\sin r \cos r} & \cos^2 \theta & \tan r \sin \theta \cos \theta & -\sin^2 \theta & \frac{\sin \theta \cos \theta}{\tan r} & 0 \\ \hline
 -1 & 0 & 1 & 0 & 1 & 1 
\end{array}\right),
\end{align}
in the coordinates $y^{\hat{i}}=(x,\,r,\,\theta)$. Using the parameterization \eqref{eq:M-ex1}, we eventually obtain
\begin{align}
\begin{split}
 g_{ij} &= \frac{\Exp{a}}{\Delta_{\text{c}}}
\begin{pmatrix}
 1 & 0 \\
 0 & \tan^{-2} r 
\end{pmatrix},\qquad \phiS = \phiD + \tfrac{1}{2}\,a - \ln \lvert\sin r\rvert + \tfrac{1}{2}\, \ln \Delta_{\text{c}} \,,
\\
 B_{ij}&=\frac{b \cos (2 \theta)- (\Exp{2a}+b^2-1) \sin \theta \cos \theta}{\Delta_{\text{c}}}
\begin{pmatrix}
 0 & \tan^{-1} r \\
 -\tan^{-1} r & 0
\end{pmatrix},
\end{split}
\label{eq:geometry-tan-2}
\end{align}
where $\Delta_{\text{c}} = 1+b \sin (2 \theta) + (\Exp{2a}+b^2-1) \cos^2 \theta$\,. In particular, if we set $a=b=0$ and $\phiD= \text{const.}$, this corresponds to the familiar geometry
\begin{align}
 \rmd s^2 = \rmd r^2 + \tan^{-2}r\,\rmd \theta^2\,,\qquad 
 \phiS = \phiD - \ln \lvert\sin r\rvert\,.
\end{align}
For general $a,b$, this geometry has a singularity at
\begin{align}
 \sin r = 0\,,
\end{align}
which is the region where the $\GS$-action has its fixed point. 

\paragraph{\underline{$\GS=\text{U}(1)_{\text{A}}$ and $H=\text{SU}(2)_{\text{diag}}$:}}
To consider the axial gauging, we adapt the parameterization \eqref{eq:param-vector-sl2} according to
\begin{align}
 l = f'\,m \,h \,,\qquad f'=\Exp{x\,\hat{T}_1}\in \text{U}(1)_{\text{A}}\,,\quad
 m =\Exp{\theta\,(\hat{T}_1+\hat{T}^1)}\,\Exp{r\,(\hat{T}_3+\hat{T}^3)} \in \text{U}(1)_{\text{V}}\backslash G/H \,,
\end{align}
where $h\in H$. Since the subgroup $H$ is the same as in the previous case, the generalized frame field $\Eh_{\Ah}{}^{\Ih}$ satisfies the same equation, $\gLie_{\Eh'_{\Ah}} \Eh'_{\Bh}{}^{\Ih} = - \hat{X}_{\Ah\Bh}{}^{\Ch}\,\Eh'_{\Ch}$\,, but its decomposition is different. The untwisted generalized frame field now becomes
\begin{align}
 E'_{\Ah}{}^{\Ih} = \left(\begin{array}{c|cccc|c}
 1 & 0 & 0 & 0 & 0 & 0 \\ \hline
 \frac{\cos (2 r)+\cos (2 \theta)}{\sin (2 r)} & -\sin \theta \cos \theta & -\frac{\cos^2 \theta}{\tan r} & -\sin \theta \cos \theta & \tan r \sin^2 \theta & 0 \\
 \frac{\sin \theta \cos \theta}{\sin r \cos r} & \cos^2 \theta & -\frac{\sin \theta \cos \theta}{\tan r} & -\sin^2 \theta & -\tan r \sin \theta \cos \theta & 0 \\
 \frac{\cos (2 \theta)- \cos (2 r)}{\sin (2 r)} & -\sin \theta \cos \theta & \frac{\sin^2 \theta}{\tan r} & -\sin \theta \cos \theta & -\tan r \cos^2 \theta & 0 \\
 \frac{\sin \theta \cos \theta}{\sin r \cos r} & -\sin^2 \theta & -\frac{\sin \theta \cos \theta}{\tan r} & \cos^2 \theta & -\tan r \sin \theta \cos \theta & 0 \\ \hline
 -1 & 0 & 1 & 0 & 1 & 1 
\end{array}\right),
\end{align}
in the coordinates $y^{\hat{i}}=(x,\,r,\,\theta)$. Note that the generator of $\GS$ takes the form
\begin{align}
 (\mathfrak{t}_1)_A{}^B = \begin{pmatrix}
 0 & 0 & 0 & -1 \\
 0 & 0 & 1 & 0 \\
 0 & -1 & 0 & 0 \\
 1 & 0 & 0 & 0\end{pmatrix} ,
\end{align}
and the basis of the scalar field \eqref{eq:M-ex1} should be changed by the similarity transformation 
\begin{align}
 \cM_{AB} \to \cM'_{AB} = C_A{}^C\,C_B{}^D\,\cM_{CD}\,,\qquad C_A{}^B = \begin{pmatrix}
 1 & 0 & 0 & 0 \\
 0 & 0 & 0 & 1 \\
 0 & 0 & 1 & 0 \\
 0 & 1 & 0 & 0\end{pmatrix},
\end{align}
such that $\cM'_{AB}$ becomes a $\GS$ singlet. The dual geometry is then captured by
\begin{align}
  g'_{ij} &=\scalebox{0.9}{$
\begin{pmatrix}
 \Exp{a} \sin^2 \theta + \Exp{-a} (\cos \theta-b \sin \theta)^2 & \Exp{-a} [(\Exp{2 a}+b^2-1) \sin \theta \cos \theta-b \cos (2 \theta)] \tan r \\
 \Exp{-a} [(\Exp{2 a}+b^2-1) \sin \theta \cos \theta-b \cos (2 \theta)] \tan r & \Exp{-a} [\Exp{2 a} \cos^2 \theta +(b \cos \theta+\sin \theta)^2] \tan^2 r
\end{pmatrix}$},
\nn\\
 B'_{ij}&=0\,,\qquad \phiS' = \phiD - \ln \lvert\cos r\rvert\,.
\end{align}
It is related to the geometry in \eqref{eq:geometry-tan-2} through a factorized T-duality along the $\theta$-direction, followed by a sign flip $\theta \to -\theta$. Therefore, the duality equivalence is clearly understood for general $a$ and $b$. Again, a fixed point, and with it a singularity, appears when $\cos r=0$.

We note that this generalized coset can alternatively be interpreted as $\text{U}(1)_{\text{V}}\backslash G/H'$ where the Lie algebra of $H'$ is spanned by $\{\hat{T}_1,\,\hat{T}^2,\,\hat{T}_3\}$. From this viewpoint, the equivalence between the two generalized cosets $\text{U}(1)_{\text{V}}\backslash G/H$ and $\text{U}(1)_{\text{V}}\backslash G/H'$ follows from standard generalized T-duality. In this example, we do not find any further inequivalent subgroups $H$, and therefore only two distinct embeddings of the eight-dimensional theory into the ten-dimensional theory are possible.

\subsubsection{Example 3: \texorpdfstring{$\text{U}(1)\backslash \SL(2)\times \SL(2)/\SL(2)_{\text{diag}}$}{U(1)\textbackslash SL(2)xSL(2)/SL(2)diag}}\label{sec:SL2-WZW}
Here we consider a similar setup to the previous one, but with the non-compact groups $G=\SL(2)\times \SL(2)$ and $H=\SL(2)_{\text{diag}}$. The algebra $\mathfrak{sl}(2)\oplus\mathfrak{sl}(2)\cong \mathfrak{so}(2,2)$ for each of the two factors is captured by the structure constants
\begin{align}
 f_3{}^{12} = 1\,,\quad
 f_1{}^{23} = -1\,,\quad
 f_2{}^{13} = -1\,,\quad
 f_{123} = -1\,.
\end{align}
In particular, the left $\mathfrak{sl}(2)$ algebra is spanned by the generators
\begin{align}
 \{t_1,\,t_2,\,t_3\}=\Bigl\{\frac{\hat{T}_1+\hat{T}^1}{2},\,\frac{\hat{T}_2-\hat{T}^2}{2},\,\frac{\hat{T}_3-\hat{T}^3}{2}\Bigr\}\,,
\end{align}
while the right $\mathfrak{sl}(2)$ algebra is generated by
\begin{align}
 \{\tilde{t}_1,\,\tilde{t}_2,\,\tilde{t}_3\}=\Bigl\{\frac{-\hat{T}_1+\hat{T}^1}{2},\,\frac{-\hat{T}_2-\hat{T}^2}{2},\,\frac{-\hat{T}_3-\hat{T}^3}{2}\Bigr\}\,.
\end{align}
Their diagonal subalgebra $\mathfrak{sl}(2)_{\text{diag}}$ arises again from the sum
\begin{align}
   \{t_1 + \tilde{t}_1,\,t_2 + \tilde{t}_2,\,t_3 + \tilde{t}_3\} = \{\hat{T}^1,\,-\hat{T}^2,\,-\hat{T}^3\}\,,
\end{align}
as in \eqref{eq:diagonal-su2}. Now the axial and vectorial $\mathfrak{u}(1)$ algebras arise from
\begin{enumerate}
  \item $\mathfrak{u}(1)_{\text{V}}$ is generated by $\hat{T}^1=t_1+\tilde{t}_1$, and
  \item $\mathfrak{u}(1)_{\text{A}}$ is generated by $\hat{T}_1=t_1-\tilde{t}_1$. 
\end{enumerate}
They are exchanged by the automorphism
\begin{align}
 \hat{T}'_1 = \hat{T}^1\,,\quad \hat{T}'_2 = -\hat{T}^2\,,\quad \hat{T}'_3 = \hat{T}_3\,,\quad
 \hat{T}'^1 = \hat{T}_1\,,\quad \hat{T}'^2 = -\hat{T}_2\,,\quad \hat{T}'^3 = \hat{T}^3\,,
\end{align}
and thus both lead to the same eight-dimensional theory.

\paragraph{\underline{8D theory:}}
For $\mathfrak{u}(1)_{\text{V}}$, the $\GS$-action on the physical space is fixed by
\begin{align}
  f_{\exA\exB}{}^{\exC} &= 0 = \hat{X}_{AB}{}^C\,,\qquad &
  (\mathfrak{t}_{1})_A{}^B &= \begin{pmatrix}
 0 & -1 & 0 & 0 \\
 1 & 0 & 0 & 0 \\
 0 & 0 & 0 & -1 \\
 0 & 0 & 1 & 0 \end{pmatrix} = \hat{X}_{AB}{}^{1} .
\end{align}
Since the matrix $(\mathfrak{t}_{1})_A{}^B$ has the same form as in previous cases, the matter content of the eight-dimensional theory is the same as before. The only difference from the previous discussion lies in the sign of the generalized Ricci tensor
\begin{align}
 \cT_{AB}{}^C = 0 \,,\qquad \cT_{A} =0\,,\qquad R_{AB} = -\delta_{AB}\,.
\end{align}
Consequently, the sign of the relevant part in the scalar potential is flipped, leading to
\begin{align}
 \cL_8 &= \Exp{-2\,\phiD}\bigl[*R_g+4\,\rmd \phiD\wedge *\rmd \phiD
 -\tfrac{1}{2}\, H_3\wedge * H_3 
\nonumber\\
 & \qquad\qquad - \tfrac{1}{2}\, (\rmd a\wedge *\rmd a
 +\Exp{-2a}\, \rmd b\wedge * \rmd b) - * 4 \cosh a - * 2\,\Exp{-a}\,b^2 \bigr]\,.
\end{align}

In the following, we construct the frame field for $H=\text{SL}(2)_{\text{diag}}$, and $\GS=\text{U}(1)_{\text{V}}$ or $\GS=\text{U}(1)_{\text{A}}$. The key difference from the previous $\text{SU}(2)$ case is that there are now more inequivalent maximally isotropic subgroups $H$. We will explore them below.

\paragraph{\underline{$\GS=\text{U}(1)_{\text{V}}$ and $H=\SL(2)_{\text{diag}}$:}}
We adopt the parameterization
\begin{align}
 l = f\,m \,h \,,\qquad f=\Exp{x\,\hat{T}^1}\in \text{U}(1)_{\text{V}}\,,\quad
 m =\Exp{r\,(\hat{T}_2+\hat{T}^2)} \,\Exp{\theta\,(\hat{T}_1-\hat{T}^1)} \in \text{U}(1)_{\text{V}}\backslash G/H \,,
\end{align}
where $h\in H$. The untwisted generalized frame field is then given by
\begin{align}
 E_{\Ah}{}^{\Ih} =\left(
\begin{array}{c|cccc|c}
 1 & 0 & 0 & 0 & 0 & 0 \\
 -\frac{\sin\theta \cos\theta}{\sinh r \cosh r} & \cos^2 \theta & -\sin\theta \cos\theta \tanh r & \sin^2 \theta & \frac{\sin\theta \cos\theta}{\tanh r} & 0 \\ \hline
 \frac{\cos (2 \theta )-\cosh (2 r)}{\sinh (2 r)} & \sin\theta \cos\theta & \cos^2 \theta \tanh r & -\sin\theta \cos\theta & \frac{\sin^2 \theta}{\tanh r} & 0 \\
 \frac{\sin\theta \cos\theta}{\sinh r \cosh r} & \sin^2 \theta & \sin\theta \cos\theta \tanh r & \cos^2 \theta & -\frac{\sin\theta \cos\theta}{\tanh r} & 0 \\
 -\frac{\cos (2 \theta )+\cosh (2 r)}{2 \sinh r \cosh r} & -\sin\theta \cos\theta & \sin^2 \theta \tanh r & \sin\theta \cos\theta & \frac{\cos^2\theta}{\tanh r} & 0 \\ \hline
 -1 & 0 & 1 & 0 & 1 & 1 
\end{array}
\right)\,,
\end{align}
in the coordinates $y^{\hat{i}}=(x,\,r,\,\theta)$. This frame field results in the internal geometry
\begin{align}
\begin{split}
 g_{ij} &= \frac{\Exp{a}}{\Delta_{\text{s}}}
\begin{pmatrix}
 1 & 0 \\
 0 & \tanh^{-2} r 
\end{pmatrix},\qquad \phiS = \phiD + \tfrac{1}{2}\,a - \ln \lvert\sinh r\rvert + \tfrac{1}{2}\, \ln \Delta_{\text{s}} \,,
\\
 B_{ij}&=\frac{b \cos (2 \theta) + (\Exp{2a}+b^2-1) \sin \theta \cos \theta}{\Delta_{\text{s}}}
\begin{pmatrix}
 0 & \tanh^{-1} r \\
 -\tanh^{-1} r & 0
\end{pmatrix},
\end{split}
\label{eq:geometry-tanh-2}
\end{align}
where $\Delta_{\text{s}} = 1+b \sin (2\theta) + (\Exp{2a}+b^2-1) \sin^2\theta$. It exhibits a singularity/fixed point at $r=0$. In particular, by setting $a=0$ and $b=0$, we recover the trumpet geometry
\begin{align}
 \rmd s^2 = \rmd r^2 + \tanh^{-2} r\,\rmd \theta^2\,,\qquad
 \phiS = \phiD - \ln \lvert\sinh r\rvert\,.
\end{align}

\paragraph{\underline{$\GS=\text{U}(1)_{\text{A}}$ and $H=\SL(2)_{\text{diag}}$:} }
Next, we consider the parameterization
\begin{align}
 l = f\,m \,h \,,\qquad f=\Exp{x\,\hat{T}_1}\in \text{U}(1)_{\text{A}}\,,\quad
 m =\Exp{r\,(\hat{T}_2+\hat{T}^2)} \,\Exp{\theta\,(\hat{T}_1-\hat{T}^1)} \in \text{U}(1)_{\text{V}}\backslash G/H \,,
\end{align}
to obtain the untwisted generalized frame field
\begin{align}
 E'_{\Ah}{}^{\Ih} =\left(
\begin{array}{c|cccc|c}
 1 & 0 & 0 & 0 & 0 & 0 \\
 -\frac{\sin\theta \cos\theta}{\sinh r \cosh r} & \cos^2 \theta & -\sin\theta \cos\theta \tanh r & \sin^2 \theta & \frac{\sin\theta \cos\theta}{\tanh r} & 0 \\ \hline
 \frac{\cos (2 \theta )-\cosh (2 r)}{\sinh (2 r)} & \sin\theta \cos\theta & \cos^2 \theta \tanh r & -\sin\theta \cos\theta & \frac{\sin^2 \theta}{\tanh r} & 0 \\
 \frac{\sin\theta \cos\theta}{\sinh r \cosh r} & \sin^2 \theta & \sin\theta \cos\theta \tanh r & \cos^2 \theta & -\frac{\sin\theta \cos\theta}{\tanh r} & 0 \\
 -\frac{\cos (2 \theta )+\cosh (2 r)}{2 \sinh r \cosh r} & -\sin\theta \cos\theta & \sin^2 \theta \tanh r & \sin\theta \cos\theta & \frac{\cos^2\theta}{\tanh r} & 0 \\ \hline
 -1 & 0 & 1 & 0 & 1 & 1 
\end{array}
\right)\,,
\end{align}
in the coordinates $y^{\hat{i}}=(x,\,r,\,\theta)$. After again applying the transformation
\begin{align}
 \cM_{AB} \to \cM'_{AB} = C_A{}^C\,C_B{}^D\,\cM_{CD}\,,\qquad C_A{}^B = \begin{pmatrix}
 1 & 0 & 0 & 0 \\
 0 & 0 & 0 & -1 \\
 0 & 0 & 1 & 0 \\
 0 & -1 & 0 & 0\end{pmatrix},
\end{align}
the dual geometry
\begin{align}
  g'_{ij} &=\scalebox{0.8}{$
\begin{pmatrix}
 \Exp{-a} (\sin \theta -b \cos \theta)^2+ \Exp{a} \cos^2 \theta & -\frac{1}{2}\,\Exp{-a}\,[ (\Exp{2 a}+b^2-1) \sin (2 \theta)+2 b \cos (2 \theta )]\tanh r \\
 -\frac{1}{2}\,\Exp{-a}\,[ (\Exp{2 a}+b^2-1) \sin (2 \theta)+2 b \cos (2 \theta )]\tanh r & \Exp{-a} \,[\Exp{2 a} \sin^2 \theta +(b \sin \theta +\cos \theta)^2]\tanh^2 r 
\end{pmatrix}$},
\nn\\
 B'_{ij}&=0\,,\qquad \phiS' = \phiD - \ln \lvert\cosh r\rvert
\end{align}
arises. In this case, there are no fixed points, and consequently, no curvature singularities appear. For general values of $a$ and $b$, this geometry is related to the previous one by a factorized T-duality along the $\theta$-direction, followed by the sign flip $\theta\to -\theta$. Thus, the duality equivalence is evident. In particular, when $a=b=0$ and $\phiD= \text{const.}$, we recover the cigar geometry \begin{align}
 \rmd s'^2 = \rmd r^2 + \tanh^{2} r\,\rmd \theta^2\,,\qquad 
 \phiS' = \phiD - \ln \lvert\cosh r\rvert
\end{align}
described in \cite{Witten:1991yr}.

\paragraph{\underline{$\GS=\text{U}(1)_{\text{A}}$ and $H'$:}}
In contrast to $G=\text{SU}(2)\times \text{SU}(2)$, we now have more flexibility in selecting maximally isotropic subgroups $H$. As an example, consider the transformation
\begin{align}
 T_{\Ah} \to T'_{\Ah} = C_{\Ah}{}^{\Bh}\,T_{\Bh} 
\end{align} 
with
\begin{align}
 C_{\Ah}{}^{\Bh} \equiv \begin{pmatrix}
 -1 & 0 & -1 & 0 & 0 & 0 \\
 0 & 0 & 0 & 1 & 0 & -1 \\
 0 & 0 & 0 & 0 & 1 & 0 \\
 0 & 0 & 0 & -\frac{1}{2} & 0 & -\frac{1}{2} \\
 \frac{1}{2} & 0 & -\frac{1}{2} & 0 & 0 & 0 \\
 0 & 1 & 0 & 0 & 0 & 0\end{pmatrix}. 
\end{align}
Afterwards, we discover two maximally isotropic subalgebras generated by $T'_{\hat{a}}$ and $T'^{\hat{a}}$. Their structure constants are
\begin{align}
 f_{13}{}^1 = -1\,,\qquad
 f_{23}{}^2 = -1\,,\qquad
 f_1{}^{23} = 1\,,\qquad
 f_2{}^{13} = 1\,.
\end{align}
The former is of Bianchi type V, while the latter is of Bianchi type VI$_0$\,. They give rise to a six-dimensional Drinfel'd algebra corresponding to the second Drinfel'd double, $(8\vert 5.i\vert b=1)\cong (6_0\vert 5.iii\vert b=1)$, as classified in Theorem 1 of \cite{Snobl:2002kq}. In principle, we can consider more isotropic subgroups, but here we focus only on the $H'$ generated by $T'^{\hat{a}}$ and construct the generalized coset $\text{U}(1)_{\text{A}}\backslash G/H'$.

With the parameterization
\begin{align}
 l = f\,m \,h \,,\qquad f=\Exp{x\,\hat{T}_1}\in \text{U}(1)_{\text{V}}\,,\quad
 m =\Exp{r\,(\hat{T}_2+\hat{T}^2)} \,\Exp{\theta\,(\hat{T}_1-\hat{T}^1)} \in \text{U}(1)_{\text{V}}\backslash G/H \,,
\end{align}
for $h\in H'$ and $\hat{T}_1 =\hat{T}'^2-\frac{1}{2}\,\hat{T}'_1$\,, we construct the generalized frame field
\begin{align}
 E''_{\Ah}{}^{\Ih} =\left(
\begin{array}{c|cccc|c}
 1 & 0 & 0 & 0 & 0 & 0 \\\hline
 0 & \sin r \cos r & 0 & 0 & 1 & 0 \\
 1 & \sin^2r & \tan r & -\cos^{-2} r & \tan r & 0 \\
 0 & \sin r \cos r & 1 & 0 & 0 & 0 \\
 -1 & -\cos^2 r & 0 & 0 & 0 & 0 \\\hline
 1 & 1 & \tan r & -\cos^{-2}r & \tan r & 1 
\end{array}
\right),
\end{align}
in the coordinates $y^{\hat{i}}=(x,\,r,\,\theta)$. Taking the parameterization of the scalar fields $\cM'_{AB}$ from before, we obtain the target space geometry
\begin{align}
 g''_{ij} &= \begin{pmatrix}
 \frac{\Exp{-a} \,(\Exp{2 a}+b^2)}{\cos^2 r} & -\Exp{-a}\, [ (\Exp{2 a}+b^2) \tan r +b] \\
 -\Exp{-a}\, [ (\Exp{2 a}+b^2) \tan r +b] & \frac{1}{2}\, \Exp{-a}\,[ 1 + \Exp{2 a} - (\Exp{2 a}+b^2-1) \cos (2 r) +b^2+2 b \sin (2 r)] \end{pmatrix} ,
\\
 B''_{ij} &= \begin{pmatrix} 0 & -\tan r \\ \tan r & 0\end{pmatrix} ,\qquad
 \phiS'' = \phiD + \theta + \ln \lvert\cos r\rvert
\end{align}
in the internal space of the truncation. Although there is no fixed point, we observe a singularity at $\cos r = 0$. In fact, this is merely a coordinate singularity. To simplify the analysis, let us focus on the configuration $a = b = 0$ and $\phiD = \text{const.}$, where the geometry corresponds to flat space with a non-trivial dilaton,
\begin{align}
 \rmd s''^2 = \rmd r^2 + (\rmd \theta -\tan r\,\rmd r)^2\,,\quad B''_2 = -\tan r\,\rmd r \wedge \rmd \theta \,,\quad \phiS'' = \phiD + \theta + \ln \lvert\cos r\rvert\,.
\end{align}
For the new coordinate $\Exp{\rho} = \cos r$, we have
\begin{align}
 \rmd s''^2 = \tfrac{\rmd \rho^2}{\Exp{-2\rho}-1} + \rmd \tilde{\theta}^2\,,\quad B''_2 = \rmd \rho \wedge \rmd \tilde{\theta} \,,\quad \phiS'' = \phiD + \tilde{\theta}\,.
\end{align}
where $\tilde{\theta}=\theta+\rho$, and the singularity disappears. This example demonstrates that the cigar and trumpet geometries are both generalized T-dual to flat space with a linear dilaton. While further dual geometries can be explored by choosing different subgroups $H$ or $\GS$, such investigations may not offer additional insight. We shall therefore conclude here and move on to the exceptional case.

\section{Consistent truncations and generalized duality in ExFT}\label{sec:examples-ExFT}
In the following, we present various examples of exceptional generalized cosets. As before, the starting point for our discussion is generalized parallelizable spaces on the mega-space. For exceptional duality groups, these spaces have additional features compared to the $\Odd$ setting of the previous section. Over the last decade, extensive work has been done on their systematic construction \cite{duBosque:2017dfc,Inverso:2017lrz,Sakatani:2019zrs,Malek:2019xrf,Malek:2020hpo,Sakatani:2020wah,Bugden:2021wxg,Bugden:2021nwl,Hassler:2022egz}. Here, we adapt the approach of \cite{Hassler:2022egz}, which is based on Exceptional Geometric Algebras.

\subsection{A reminder on Exceptional Geometric Algebras}\label{sec:EGA}
An important difference compared to $\GD = \Odd$ is that the structure constants for exceptional duality groups do not satisfy the antisymmetry property $\hat{X}_{\Ah\Bh}{}^{\Ch} = - \hat{X}_{\Bh\Ah}{}^{\Ch}$. As a result, they describe a Leibniz algebra rather than a Lie algebra,
\begin{align}
 \hat{X}_{\Ah\Ch}{}^{\Eh}\,\hat{X}_{\Bh\Eh}{}^{\Dh} - \hat{X}_{\Bh\Ch}{}^{\Eh}\,\hat{X}_{\Ah\Eh}{}^{\Dh} + \hat{X}_{\Ah\Bh}{}^{\Eh}\,\hat{X}_{\Eh\Ch}{}^{\Dh} = 0\,.
\end{align}
The structure constants define the $\circ$-product,
\begin{align}
  \hat{T}_{\Ah} \circ \hat{T}_{\Bh} = \hat{X}_{\Ah\Bh}{}^{\Ch},
\end{align}
between the Leibniz algebra's generators $\hat{T}_{\Ah}$. A central question is whether a given Leibniz algebra can be realized by a generalized parallelizable space. It is answered in the affirmative if and only if there exists a similarity transformation in the duality groups which brings the Leibniz algebra into a canonical form called Exceptional Geometric Algebra (EGA) \cite{Hassler:2022egz}. Its explicit form depends on the imposed solution of the section condition. Hence, there are EGAs for either the M-theory or type IIB section. Up to $n=7$ dimensions, they are presented in Appendix~\ref{app:algebra} for the reader's convenience.

For any Leibniz algebra, we can uniquely identify the Lie subalgebra $\mathrm{Lie}(G)$ through the following procedure: Let $E$ denote the vector space spanned by $\hat{T}_{\Ah}$, and consider the subspace $\cI \subset E$ spanned by the generators of the form $\hat{X}_{(\Ah\Bh)}{}^{\Ch}\,\hat{T}_{\Ch}$. Owing to the Leibniz identity, we find that
\begin{align}
  \forall t\in \cI: \qquad t\circ \hat{T}_{\Ah}=0\,,\qquad \hat{T}_{\Ah}\circ t \in \cI
\end{align}
holds. This shows that $\cI$ is an ideal in $E$, and we can therefore define the quotient space $E / \cI$. For $a=a^{\Ah}\,T_{\Ah}\in E/\cI$ and $b=b^{\Ah}\,T_{\Ah}\in E/\cI$, we then have
\begin{align}
 a\circ b + b\circ a = 2\,a^{\Ah}\,b^{\Bh}\,X_{(\Ah\Bh)}{}^{\Ch}\,T_{\Ch} \sim 0\,,
\end{align}
which implies that the generators of $E/\cI$ form a Lie subalgebra of the full Leibniz algebra, identified as $\mathrm{Lie}(G)$. 

In order to define a generalized parallelizable space, we also need a second Lie group $H$ to form the coset $G/H$. Finding $H \subset G$ depends on the section of the EGA. For instance, in the M-theory section of $\GDM=\Edd[p]$, we decompose the generators as $\hat{T}_{\Ah} = (\hat{T}_{\hat{a}},\, \hat{T}_{\tilde{\alpha}})$, where the $\hat{T}_{\tilde{\alpha}}$ span a $(\dim E - p)$-dimensional subspace $V$. Looking at the algebra in Appendix~\ref{app:algebra}, one finds that $\hat{X}_{\tilde{\alpha}\tilde{\gamma}}{}^{\hat{a}}=0$ holds, indicating that the generators $\hat{T}_{\tilde{\alpha}}$ form a Leibniz subalgebra. Using $\hat{X}_{\tilde{\alpha}\tilde{\gamma}}{}^{\hat{a}}=0$, it is furthermore straightforward to check the Leibniz identity
\begin{align}
 \hat{X}_{\tilde{\alpha}\tilde{\gamma}}{}^{\tilde{\epsilon}}\,\hat{X}_{\tilde{\beta}\tilde{\epsilon}}{}^{\tilde{\delta}} - \hat{X}_{\tilde{\beta}\tilde{\gamma}}{}^{\tilde{\epsilon}}\,\hat{X}_{\tilde{\alpha}\tilde{\epsilon}}{}^{\tilde{\delta}} + \hat{X}_{\tilde{\alpha}\tilde{\beta}}{}^{\tilde{\epsilon}}\,\hat{X}_{\tilde{\epsilon}\tilde{\gamma}}{}^{\tilde{\delta}} = 0\,.
\end{align}
Finally, we also note that $X_{(\Ah\Bh)}{}^{\hat{c}}=0$ implies that the ideal $\cI$ is contained in the subspace $V$. Therefore, the quotient space $V/\cI$ spans a Lie algebra which we call $\mathrm{Lie}(H)$. By construction, it follows that $\dim (G/H) = \dim G - \dim H = p$. A similar argument holds for type IIB sections, but with the coset dimension reduced by one.

\subsection{Non-supersymmetric theories}\label{sec:ex1-ExFT}
Let us begin with two examples in which supersymmetry is completely broken under the consistent truncation.

\subsubsection{Example 1: \texorpdfstring{$\SO(2)\backslash \SO(3)\times \text{U}(1)$}{SO(2)\textbackslash SO(3)xU(1)}}\label{sec:ExFT-ex1}
First, we consider a simple example to highlight the difference between $\GD = \OO(d,d)$ and $\GD=\Edd$. Its mega-space is governed by an EGA in the M-theory section for the duality group $\GDM = \Edd[4] = \SL(5)$ with the only non-vanishing structure coefficients
\begin{align}
 f_{12}{}^3 = 1\,,\qquad
 f_{23}{}^1 = 1\,,\qquad
 f_{31}{}^2 = 1\,.
\label{eq:ExFT-ex1-fabc}
\end{align}
All resulting Leibniz products are given in Appendix~\ref{app:algebra}. From a physical point of view, this setup corresponds to the M-theory uplift of the example discussed in section \ref{sec:DFTex1}. However, in contrast to the $\OO(3,3)$ case, the resulting EGA is a Leibniz algebra rather than a Lie algebra. To write the latter out explicitly, we denote its ten generators by
\begin{align}
 \hat{T}_{\Ah} = \begin{pmatrix} \hat{T}_1&\cdots &\hat{T}_{10}\end{pmatrix} = \begin{pmatrix}
\hat{T}_1&\hat{T}_2&\hat{T}_3&\hat{T}_4&\hat{T}^{12}&\hat{T}^{13}&\hat{T}^{14}&\hat{T}^{23}&\hat{T}^{24}&\hat{T}^{34}\end{pmatrix} .
\end{align}
In this notation, the subspace $\cI$ is spanned by
\begin{align}
 \begin{pmatrix} \hat{T}_5&\hat{T}_6&\hat{T}_8\end{pmatrix}
\end{align}
and the quotient space $E/\cI$ is seven-dimensional. It corresponds to the Lie algebra of a seven-dimensional Lie group $G$. In particular, $\hat{T}_4$ commutes with all other generators, while the remaining six generators span the Lie algebra $\mathfrak{iso}(3)$,
\begin{align}
 [J_i,\,J_j] = \epsilon_{ij}{}^k\,J_k\,,\qquad
 [J_i,\,P_j] = \epsilon_{ij}{}^k\,P_k\,,\qquad
 [P_i,\,P_j] = 0\,,
\end{align}
where
\begin{align}
 J_i = \hat{T}_i\,,\qquad P_i = \begin{pmatrix} \hat{T}_{7} & \hat{T}_{9} & \hat{T}_{10} \end{pmatrix}\quad (i=1,2,3)\,.
\end{align}
Consequently, the group $G$ is identified as $\text{ISO}(3) \times \text{U}(1)$. At this point, the embedding of the Drinfel'd double $\text{ISO}(3)$ from section~\ref{sec:DFTex1} becomes obvious. The subspace $V$, spanned by $\begin{pmatrix} \hat{T}_5&\dotsc&\hat{T}_{10}\end{pmatrix}$, is reduced to the three-dimensional space $V/\cI$ after modding out the ideal $\cI$. The resulting Lie algebra $\mathrm{Lie}(H)$ corresponds to a three-dimensional Lie group $H$. It is straightforward to show that $H$ is the group of translations $\text{T}_3$ generated by $P_i$\,, and hence the coset $G/H$ is given by $G/H=\SO(3)\times \text{U}(1)$.

Similar to the example discussed in section \ref{sec:DFTex1}, we take the structure group to be $\GS = \SO(2)$, generated by $\hat{T}_1$, and decompose the generators as
\begin{align}
 \hat{T}_{\Ah} = \begin{pmatrix} \hat{T}_{\exA}&\hat{T}_{A}&\hat{T}^{\exA \AA}\end{pmatrix} \quad (\alpha=1)\,,
\end{align}
where $A$ and $\AA$ label the representations $R_1 = (\irrepb{3},\irrep{2})$ and $R_2 = (\irrep{3},\irrep{1})$ of $\GD = \SL(3)\times \SL(2)$, respectively. More explicitly, we adopt the basis
\begin{align}
 \hat{T}_{A} = \begin{pmatrix} \hat{T}_2&\hat{T}_3&\hat{T}_4&\hat{T}_8&\hat{T}_9&T_{10}\end{pmatrix} ,\qquad
 \hat{T}^{\exA \AA} = \begin{pmatrix} \hat{T}_5 &\hat{T}_6 &\hat{T}_7\end{pmatrix}.
\end{align}
and compute the matrix representations of the generator $\hat{T}_1$ in the $R_1$- and $R_2$-representations,
\begin{align}
 (\mathfrak{t}_1)_A{}^B = \begin{pmatrix}
 0 & -1 & 0 & 0 & 0 & 0 \\
 1 & 0 & 0 & 0 & 0 & 0 \\
 0 & 0 & 0 & 0 & 0 & 0 \\
 0 & 0 & 0 & 0 & 0 & 0 \\
 0 & 0 & 0 & 0 & 0 & -1 \\
 0 & 0 & 0 & 0 & 1 & 0\end{pmatrix},\qquad 
 (\mathfrak{t}_1)_{\AA}{}^{\BB} = \begin{pmatrix}
 0 & -1 & 0 \\
 1 & 0 & 0 \\
 0 & 0 & 0 \end{pmatrix},
\end{align}
given by $(\mathfrak{t}_1)_A{}^B = - \hat{X}_{1A}{}^B$ and $(\mathfrak{t}_1)_{\AA}{}^{\BB}$, respectively. With them, we obtain the generalized torsion/Ricci tensor, which read
\begin{align}
 \cT_{AB}{}^C = 0\,,\qquad 
 \cR_{AB} = \begin{pmatrix}
 1 & 0 & 0 & 0 & 0 & 0 \\
 0 & 1 & 0 & 0 & 0 & 0 \\
 0 & 0 & 0 & 0 & 0 & 0 \\
 0 & 0 & 0 & 0 & 0 & 0 \\
 0 & 0 & 0 & 0 & 0 & 0 \\
 0 & 0 & 0 & 0 & 0 & 0\end{pmatrix}.
\end{align}
In contrast to the analysis of section \ref{sec:DFTex1}, we now find two null
eigenvectors $K_{\cA}$ ($\cA = 1,2$) and $K_{\overline{\cA}}$ ($\overline{\cA} = \bar{1}$) which are given by
\begin{align}
 K_{1}{}^A = (0, 0, 1, 0, 0, 0)\,,\qquad
 K_{2}{}^A = (0, 0, 0, 1, 0, 0)\,,\qquad
 K_{\bar{1}}{}^{\AA} = (0,0,1)\,.
\end{align}
They correspond to Ramond--Ramond fields and therefore do not appear for $\Odd$. Their dual vectors $K_A{}^{\cB}$ and $K_{\AA}{}^{\overline{\cB}}$ take the same form, while the associated components of the $\eta$-symbol vanish,
\begin{align}
 \eta_{\cA\cB;\overline{\cC}} = 0\,,\qquad 
 \eta^{\cA\cB;\overline{\cC}} = 0\,.
\end{align}

After the consistent truncation, the eight-dimensional theory contains two 1-form fields corresponding to $K_{\cA}$, and the gauge group $\cG$ is abelian due to $\cT_{AB}{}^C = 0$. The reduction ansatz for the 1-form and 2-form fields is given by
\begin{align}
 A_1^I = A_1^{\cA}(x)\,K_{\cA}{}^{I}(y) \,, \qquad
 B_{2\II} = B_{2}(x)\,K_{\II}{}^{\bar{1}} (y) \,,
\end{align}
where the explicit forms of $K_{\cA}{}^{I}(y)$ and $K_{\II}{}^{\bar{1}}(y)$ are presented later, after constructing the generalized frame field. For higher $p$-form fields, the reduction ansatz can be constructed similarly. Because the gauge group $\cG$ is abelian, its field strengths simply become
\begin{align}
 F_{2}^{\cA} = \rmd A_{1}^{\cA}\,,\qquad 
 F_{3} = \rmd B_{2} \,.
\end{align}

Here $\GS$ is embedded into the $\SL(3)$ subgroup of $\GD = \SL(3)\times \SL(2)$, and the commutant group is $\mathrm{C}_{\Edd[3]}(\GS) = \mathbb{R}^+\times \SL(2)$. Accordingly, we parameterize the scalar fields $\cM_{AB}$ as
\begin{align}
  \cM_{AB} = \Exp{-\frac{3\,a+2\,\phiD}{3}}\scalebox{0.9}{$\begin{pmatrix}
 \Exp{2 a}+b^2 & 0 & 0 & 0 & 0 & b \\
 0 & \Exp{2 a}+b^2 & 0 & 0 & -b & 0 \\
 0 & 0 & \Exp{2 \phiD} (\Exp{2 a}+b^2) & b\,\Exp{2 \phiD} & 0 & 0 \\
 0 & 0 & b\,\Exp{2 \phiD} & \Exp{2 \phiD} & 0 & 0 \\
 0 & -b & 0 & 0 & 1 & 0 \\
 b & 0 & 0 & 0 & 0 & 1
\end{pmatrix}$} \in \mathbb{R}^+\times \frac{\SL(2)}{\SO(2)}\,.
\end{align}
In a suitable basis, it can be written as a product of two matrices,
\begin{align}
 m_{ab} = \begin{pmatrix}
 \Exp{-\frac{2}{3} \phiD} & 0 & 0 \\
 0 & \Exp{-\frac{2}{3} \phiD} & 0 \\
 0 & 0 & \Exp{\frac{4}{3}\phiD}\end{pmatrix}\in \SL(3) ,\qquad
 m_{\alpha\beta} = \Exp{-a} \begin{pmatrix} \Exp{2 a}+b^2 & b \\ b & 1 \end{pmatrix} \in \SL(2) .
\end{align}
As kinetic term for the scalar fields we find
\begin{align}
 \cL_{\text{kin}} &= \tfrac{1}{4}\, \cD m^{ab}\wedge *_{\gD} \cD m_{ab} + \tfrac{1}{4}\, \cD m^{\alpha\beta}\wedge *_{\gD} \cD m_{\alpha\beta} 
\nn\\
 &= - \tfrac{1}{2}\, \rmd a\wedge *_{\gD}\rmd a
 - \tfrac{1}{2}\,\Exp{-2a}\, \rmd b\wedge *_{\gD}\rmd b 
 - \tfrac{2}{3}\, \rmd \phiD\wedge *_{\gD}\rmd\phiD
\end{align}
with $\cD=\rmd x^{\EXm}\wedge \cD_{\EXm}$\,, while the kinetic term for the 1-form field is given by
\begin{align}
 \cL_2 = -\tfrac{1}{2}\,m_{\cA\cB}\, F_2^{\cA}\wedge *_{\gD}F_2^{\cB}
 = -\tfrac{1}{2}\,m_{\cA\cB}\, F_2^{\cA}\wedge *_{\gD}F_2^{\cB}
 = -\tfrac{1}{2}\,\Exp{\frac{4\,\phiD}{3}}\,m_{\mathtt{i}\mathtt{j}}\,F_2^{\mathtt{i}}\wedge *_{\gD}F_2^{\mathtt{j}}\,,
\end{align}
where $\mathtt{i},\mathtt{j}=3,4$ and the matrix $(m_{\mathtt{i}\mathtt{j}})$ has the same form as $(m_{\alpha\beta})$. For the 2-form potential, we work in the basis $B_{2\AA}= (B_{2 a_1a_2})$. By converting the indices via $B_{2}^a =\frac{1}{2}\, \epsilon^{ab_1b_2}\,B_{2 b_1b_2}$, the vector $K_{\AA}{}^{\overline{\cA}}$ becomes $K^{a\overline{\cA}} = (1,\,0,\,0)$, and the 2-form field takes the form $B_2^a = (B_2,\,0,\,0)$. With this convention, the kinetic term
\begin{align}
 \cL_3 = -\tfrac{1}{2}\,m_{ab}\, F_3^a\wedge *_{\gD} F_3^b 
 = -\tfrac{1}{2}\,\Exp{-\frac{2}{3}\,\phiD}\,F_3 \wedge *_{\gD} F_3
\end{align}
arises. Finally, we compute the generalized Ricci scalar
\begin{align}
 R= \cM^{AB}\,R_{AB} = 2\,\Exp{-a}\, \Exp{\frac{2}{3}\,\phiD}\,,
\end{align}
which is related to the scalar potential by $V = -R$ due to the absence of generalized torsion.

As we see below, it turns out that the metric $\gD_{\EXm\EXn}$ is related to the string-frame metric $g^{\text{s}}_{\EXm\EXn}$ in type IIA/IIB supergravity (which corresponds to $g_{\EXm\EXn}$ in section \ref{sec:DFTex1}) as
\begin{align}
 \gD_{\EXm\EXn}(x) = \Exp{-\frac{2\phiD(x)}{3}}\,g^{\text{s}}_{\EXm\EXn}(x)  \,.
\label{eq:string-frame}
\end{align}
Using this relation, the Einstein--Hilbert action in the external part becomes
\begin{align}
 *_{\gD}\bm{R} = \Exp{-2 \phiD} *_{\text{s}} \big(R_{\text{s}} + \tfrac{14}{3}\,\gD^{\EXm\EXn}\,\partial_{\EXm}\phiD\,\partial_{\EXn}\phiD\bigr) + \text{total deriv.}, 
\end{align}
and the potential term becomes
\begin{align}
 *_{\gD}R = \Exp{- 2\phiD} *_{\text{s}}(2\,\Exp{-a}) \,.
\end{align}
By combining all these results, we eventually obtain
\begin{align}
 &*_{\gD} \bm{R} + \cL_{\text{kin}} + *_{\gD} R + \cL_2 + \cL_3
\nn\\
 &= \Exp{-2 \phiD} \,\big(*_{\text{s}} R_{\text{s}}  + 4\, \rmd \phiD\wedge *_{\text{s}} \rmd \phiD - \tfrac{1}{2}\,F_3 \wedge *_{\text{s}}F_3
\nn\\
  &\qquad\qquad -\tfrac{1}{2}\,\rmd a\wedge *_{\text{s}} \rmd a - \tfrac{1}{2}\, \Exp{-2a}\,\rmd b\wedge *_{\text{s}} \rmd b  + *_{\text{s}} 2\,\Exp{-a}\bigr)
 -\tfrac{1}{2}\, \cM_{\mathtt{i}\mathtt{j}}\,F_2^{\mathtt{i}}\wedge *_{\text{s}} F_2^{\mathtt{j}}\,.
\end{align}
Note that the first term matches the Lagrangian in \eqref{eq:action-DFT-ex1} (which is expected because we are dealing with an uplift of this example to M-theory), while the second term originates from the Ramond--Ramond sector. We still need to explain the relation between $A_1^{\cA}$ and the Ramond--Ramond fields, which is done below.

\paragraph{\underline{M-theory section:}}
As already hinted above, we take the minimally coisotropic subgroup $H$ as the translation group $\text{T}_3$, and adopt a basis in which the structure constants are given by \eqref{eq:ExFT-ex1-fabc}. For any group element in $G$, we employ the parameterization
\begin{align}
 l = f\,m \,h \,,\qquad f=\Exp{s\,\hat{T}_1}\in \GS \,,\quad
 m =\Exp{\theta\, \hat{T}_2}\,\Exp{-\psi\, \hat{T}_1}\,\Exp{\zeta\, \hat{T}_4} \in \GS\backslash G/H \,,
\end{align}
which allows us to obtain the untwisted generalized frame field
\begin{align}
 E_{\Ah}{}^{\Ih} &= \begin{pmatrix}
 E_{\exA}{}^{\mu} & 0 & 0 \\
 -\Omega_A^{\mu} & E_A{}^I & 0 \\
 0 & \eta^{BC;\AA}\,\Omega_C^{\exA}\,E_B{}^I & E_{\mu}{}^{\exA}\,E_{\II}{}^{\AA}
\end{pmatrix}
\nn\\
 &= \left(\begin{array}{c|cccccc|ccc}
 1 & 0 & 0 & 0 & 0 & 0 & 0 & 0 & 0 & 0 \\ \hline
 0 & 1 & 0 & 0 & 0 & 0 & 0 & 0 & 0 & 0 \\
 \frac{1}{\tan\theta} & 0 & \frac{1}{\sin\theta} & 0 & 0 & 0 & 0 & 0 & 0 & 0 \\
 0 & 0 & 0 & 1 & 0 & 0 & 0 & 0 & 0 & 0 \\
 0 & 0 & 0 & 0 & \sin\theta & 0 & 0 & 0 & 0 & 0 \\
 0 & 0 & 0 & 0 & 0 & 1 & 0 & 0 & 0 & 0 \\
 0 & 0 & 0 & 0 & 0 & 0 & \sin\theta & 0 & 0 & 0 \\ \hline
 0 & 0 & 0 & 0 & \cos\theta & 0 & 0 & 1 & 0 & 0 \\
 0 & 0 & 0 & 0 & 0 & 0 & 0 & 0 & \sin\theta & 0 \\
 0 & 0 & 0 & 0 & 0 & 0 & -\cos\theta & 0 & 0 & 1 
\end{array}
\right),
\end{align}
in the coordinates $y^{\hat{i}}=(s,\,\theta,\,\psi,\,\zeta)$. After rescaling it to construct a matrix $\cE_A{}^I(y)$ with unit determinant, we are left with
\begin{align}
 \cE_A{}^{I}(y) = \Exp{2\Cbeta\Delta}\,E_A{}^I(y) = (\sin\theta)^{-\frac{1}{6}}\,E_A{}^I(y)\,,\qquad 
 \Exp{-2\Delta} =\sin\theta\,.
\end{align}
The resulting matrix $\cE_A{}^I(y)$ belongs to the duality group $\GD = \SL(3) \times \SL(2)$ and allows us to obtain the generalized metric
\begin{align}
 \cM_{IJ} &= \cE_I{}^A\,\cE_J{}^B\,\cM_{AB} 
\nn\\
          &=\Exp{-\frac{3a+ 2\phiD}{3}} (\sin \theta)^{\frac{1}{3}} \scalebox{0.9}{$\begin{pmatrix}
 \Exp{2 a}+b^2 & 0 & 0 & 0 & 0 & \frac{b}{\sin\theta} \\
 0 & (\Exp{2 a}+b^2) \sin^2\theta & 0 & 0 & -b \sin\theta & 0 \\
 0 & 0 & \Exp{2 \phiD} (\Exp{2 a}+b^2) & \frac{b\,\Exp{2\phiD}}{\sin\theta} & 0 & 0 \\
 0 & 0 & \frac{b\,\Exp{2 \phiD }}{\sin\theta} & \frac{\Exp{2\phiD}}{\sin^2 \theta} & 0 & 0 \\
 0 & -b \sin\theta & 0 & 0 & 1 & 0 \\
 \frac{b}{\sin\theta} & 0 & 0 & 0 & 0 & \frac{1}{\sin^2 \theta} \end{pmatrix}$}.
\end{align}
From this expression, we can extract the internal components of the metric and the 3-form potential as
\begin{align}
\begin{split}
 \rmd s^2 &= \Exp{-\frac{2 \phiS}{3}}\,\bigl[\Exp{a}\, (\rmd \theta^2 + \sin^2 \theta\,\rmd\psi^2)\bigr] + \Exp{\frac{4}{3}\,\phiS}\,\rmd \zeta^2 \,,
\\
 C_3 &=b\sin\theta\,\rmd\theta\wedge \rmd\psi\wedge \rmd\zeta\,, \qquad
 \phiS = \phiD + \tfrac{1}{2}\,a\,.
\end{split}
\end{align}
Taking $\zeta$ to be the coordinate along the M-theory circle, we see that $\phiS$ plays the role of the dilaton field in type IIA supergravity. 

Using \eqref{eq:external-metric} and the ansatz \eqref{eq:reduction-ansatz-frame}, the external metric takes the form
\begin{align}
 g_{\EXm\EXn} = \frac{1}{\lvert \det g_{ij}\rvert^{\Cbeta}}\, \mathfrak{g}_{\EXm\EXn} = \frac{\Exp{-4\Cbeta\Delta}}{\lvert \det g_{ij}\rvert^{\Cbeta}}\, \gD_{\EXm\EXn} = \Exp{-\frac{a(x)}{3}} \gD_{\EXm\EXn}(x) \,,
\end{align}
which after being transformed to the string-frame metric in type IIA supergravity becomes
\begin{align}
 g^{\text{s}}_{\EXm\EXn}(x)=\Exp{\frac{2 \phiS}{3}}\,g_{\EXm\EXn} = \Exp{\frac{2\phiD(x)}{3}}\, \gD_{\EXm\EXn}(x)\,,
\end{align}
and thus reproduces the aforementioned relation \eqref{eq:string-frame}.

Both $\GS$-invariant generalized vector fields $K_{\cA}{}^I=K_{\cA}{}^B\,E_B{}^I$ are given by
\begin{align}
 K_{1}{}^I = (0, 0, 1, 0, 0, 0)\,,\qquad
 K_{2}{}^I = (0, 0, 0, \sin\theta, 0, 0)\,.
\end{align}
They hint that the Ramond--Ramond fields are related to the two vector fields $A_{\EXm}{}^{\cA}(x)$ as
\begin{align}
 C_{\EXm} = - A_{\EXm}{}^{1}(x)\,,\qquad 
 C_{\EXm \theta\psi} = \sin\theta\,A_{\EXm}{}^2(x)\,.
\end{align}

\paragraph{\underline{Type IIB section:}}
After performing the change of basis
\begin{align}
 \hat{T}_{\Ah} \to \hat{T}'_{\Ah} = C_{\Ah}{}^{\Bh}\,\hat{T}_{\Bh}
\end{align}
with
\begin{align}
 C_{\Ah}{}^{\Bh} =
\begin{pmatrix}
 1 & 0 & 0 & 0 & 0 & 0 & 0 & 0 & 0 & 0 \\
 0 & 1 & 0 & 0 & 0 & 0 & 0 & 0 & 0 & 0 \\
 0 & 0 & 0 & 0 & 0 & 0 & 0 & 0 & 0 & 1 \\
 0 & 0 & 0 & 0 & 0 & 0 & 1 & 0 & 0 & 0 \\
 0 & 0 & 0 & 0 & 0 & 0 & 0 & 0 & 1 & 0 \\
 0 & 0 & 1 & 0 & 0 & 0 & 0 & 0 & 0 & 0 \\
 0 & 0 & 0 & 0 & 0 & -1 & 0 & 0 & 0 & 0 \\
 0 & 0 & 0 & 0 & 0 & 0 & 0 & -1 & 0 & 0 \\
 0 & 0 & 0 & 1 & 0 & 0 & 0 & 0 & 0 & 0 \\
 0 & 0 & 0 & 0 & 1 & 0 & 0 & 0 & 0 & 0\end{pmatrix},
\end{align}
the M-theory EGA is mapped to a type IIB EGA with the following structure constants
\begin{align}
 f'_1{}^{23}_{\bm{1}} = 1\,,\qquad
 f'_2{}^{13}_{\bm{1}} = -1\,,\qquad
 f'^{\bm{1}}_{123} = 1\,.
\end{align}
Following a similar argument as before, we find that the Lie subalgebra $\mathrm{Lie}(H')$ is generated by $\{T'_4,\,T'_5,\,T'_6,\,T'_9\}$, where the first three generators span a Bianchi type VI$_0$ Lie algebra, and the last one commutes with all others. An important point is that the dimension of $H'$ is four, whereas the dimension of $H$ is three. Accordingly, the coset $G/H'$ is three-dimensional, and we parameterize the group element using the three coordinates $\{x,\,y,\,z\}$ as
\begin{align}
 l = f\,m' \,h' \,,\qquad f=\Exp{x\,\hat{T}_1}\in \GS \,,\quad
 m' =\Exp{z\, \hat{T}_{10}}\,\Exp{y\, \hat{T}_{2}} \in \GS\backslash G/H'\,,\quad h'\in H' \,.
\end{align}
Now, the generalized frame field is given by
\begin{align}
 E'_{\Ah}{}^{\Ih} &=  \left(\begin{array}{c|cccccc|ccc}
 1 & 0 & 0 & 0 & 0 & 0 & 0 & 0 & 0 & 0 \\ \hline
 0 & 1 & z \tan y & 0 & 0 & 0 & 0 & 0 & 0 & 0 \\
 -\tan y & 0 & 0 & -z \tan y & 1 & 0 & 0 & 0 & 0 & 0 \\
 0 & 0 & 0 & 0 & 0 & -z \sin y & \cos y & 0 & 0 & 0 \\
 0 & 0 & 0 & 0 & 0 & -\cos y & 0 & 0 & 0 & 0 \\
 0 & 0 & 0 & 1 & 0 & 0 & 0 & 0 & 0 & 0 \\
 0 & 0 & 1 & 0 & 0 & 0 & 0 & 0 & 0 & 0 \\ \hline
 0 & 0 & 0 & 0 & 0 & \sin y & 0 & 0 & 0 & \cos y \\
 0 & 0 & 0 & 0 & 0 & 0 & 0 & 0 & -\cos y & 0 \\
 0 & 0 & \tan y & 0 & 0 & 0 & 0 & 1 & 0 & 0 \end{array}\right) ,
\end{align}
in the coordinates $y^{\hat{i}}=(x,\,y,\,z)$. Next, one computes the generalized metric $\cM'_{IJ}$, from which the supergravity fields can be identified. These fields precisely match those obtained in \eqref{eq:gpp-DFT-ex1}. Once again, the string-frame metric is given by
\begin{align}
 g^{\text{s}}_{\EXm\EXn} = \Exp{\frac{\phi}{2}}\, \sfg_{\EXm\EXn} = \frac{\Exp{\frac{\phi}{2}}}{\lvert\det \sfg_{ij}\rvert^{\Cbeta}}\, \mathfrak{g}_{\EXm\EXn} 
 = \frac{\Exp{\frac{\phi}{2}} \Exp{-4\Cbeta\Delta}}{\lvert\det \sfg_{ij}\rvert^{\Cbeta}}\, \gD_{\EXm\EXn} 
 = \Exp{\frac{2}{3}\,\phiD}\,\gD_{\EXm\EXn} \,,
\end{align}
confirming that the relation \eqref{eq:string-frame} is indeed satisfied. Moreover, we compute the $\GS$-invariant generalized vector fields $K'_{\cA}{}^I=K_{\cA}{}^B\,E'_B{}^I$ as
\begin{align}
 K'_{1}{}^I = (0, 0, 0, 0, -z\sin y, \cos y)\,,\qquad
 K'_{2}{}^I = (0, 0, 0, 0, -\cos y, 0)\,.
\end{align}
Their expressions show that the Ramond--Ramond fields are given by
\begin{align}
 C_{\EXm y} = -z\sin y\,A_{\EXm}{}^1 -\cos y\,A_{\EXm}{}^2\,,\qquad
 C_{\EXm z} = \cos y\,A_{\EXm}{}^1\,.
\end{align}

We could consider other subgroups $H$ as well, similar to the $\OO(3,3)$ case studied in section~\ref{sec:DFTex1}, but we shall conclude here and rather proceed to the next example. Due to the large rank of the exceptional group, we do not perform a detailed identification of the supergravity fields or compute the full Lagrangian. Instead, we focus on the algebraic structure or some specific configurations.

\subsubsection{Example 2: \texorpdfstring{$\SO(3)\backslash \SO(1,3)$}{SO(3)\textbackslash SO(1,3)}}\label{sec:ex2-ExFT}
Now, we consider a type IIB $\Edd[7]$ EDA with only the geometric-flux components
\begin{align}
\begin{split}
 f_{12}{}^3 &= 1\,,\quad
 f_{13}{}^2 = -1\,,\quad
 f_{15}{}^6 = 1\,,\quad
 f_{16}{}^5 = -1\,,\quad
 f_{23}{}^1 = 1\,,\quad
 f_{24}{}^6 = -1\,,
\\
 f_{26}{}^4 &= 1\,,\quad
 f_{34}{}^5 = 1\,,\quad
 f_{35}{}^4 = -1\,,\quad
 f_{45}{}^3 = -1\,,\quad
 f_{46}{}^2 = 1\,,\quad
 f_{56}{}^1 = -1
\end{split}
\end{align}
being non-zero. The resulting Leibniz algebra (all details can be found in Appendix~\ref{app:algebra}) has a 27-dimensional ideal $\cI$. Eventually, it gives rise to the Lie group $G$ with dimension 29, which decomposes as $G = \SO(1,3) \ltimes H$, where $H$ is a 23-dimensional nilpotent Lie group. Consequently, the mega-space $\hat{M} = G/H$ is locally diffeomorphic to the group manifold $\SO(1,3)$. To form an exceptional generalized coset, we moreover consider the $\mathfrak{so}(3)$ subalgebra generated by $T_{\exA}$ ($\exA=1,2,3$) and take the structure group to be the corresponding Lie group $\GS = \SO(3)$. Hence, the internal space for the compactification is the standard coset $M = \SO(3) \backslash \SO(1,3)$, or equivalently the hyperbolic space $H^3$. 

We decompose the $R_1$ representation of the mega-space as
\begin{align}
 \hat{T}_{\Ah} = \begin{pmatrix} \hat{T}_{\exA} & \hat{T}_{A} & \hat{T}^{\exA \AA} & \cdots \end{pmatrix} ,
\end{align}
and find that the matrices $(\mathfrak{t}_{\exA})_B{}^C$, representing the action of the structure group on the $R_1$ representation of $\GD=\text{SL}(5)$, possess a single null eigenvector $K_{\cA}{}^A$ ($\cA=1$) given by
\begin{align}
 K_{1}{}^A = (0,0,0,0,0,0,0,0,0,1)\,.
\end{align}
Similarly, the null eigenvectors $K_{\overline{\cA}}{}^{\AA}$ ($\overline{\cA}=1,2$) of the matrices $(\mathfrak{t}_{\exA})_{\BB}{}^{\CC}$ are found to be
\begin{align}
 K_{\bar{1}}{}^{\AA} = (1,0,0,0,0)\,,\qquad
 K_{\bar{2}}{}^{\AA} = (0,1,0,0,0)\,.
\end{align}
Thus, the truncated theory contains one vector field and two 2-form fields. 

To determine the number of preserved supersymmetries, it is useful to recall that the supersymmetry parameters in the original maximal supergravity transform in the $\irrep{4}$ of the maximal compact subgroup $K(\SL(5))=\text{USp}(4)$. This $\text{USp}(4)$ contains a subgroup $\text{SU}(2)\times \text{SU}(2)$, under which the $\irrep{4}$ decomposes as $(\irrep{2},\irrep{1})\oplus (\irrep{1},\irrep{2})$. Since our $\SO(3)$ corresponds to the diagonal subgroup of $\text{SU}(2)\times \text{SU}(2)$, there is no singlet under $\GS =\SO(3)$. Consequently, all supersymmetry parameters (as well as the fermions) are truncated, resulting in a non-supersymmetric theory.

Taking the commutant group $\mathrm{C}_{\Edd[4]}(\GS) = \mathbb{R}^+ \times \SL(2)$ into account, the generalized metric $\cM_{AB} \in \mathbb{R}^+ \times \frac{\SL(2)}{\SO(3)}$ can be parameterized in terms of three scalar fields $a(x)$, $b(x)$, and $c(x)$ as
\begin{align}
 \cM_{AB} = \Exp{-\frac{a}{5}} \begin{pmatrix} \Exp{a}\bm{1}_3 & 0 & 0 \\
 0 & \bm{m}\otimes \bm{1}_3 & 0\\
 0 & 0 & \Exp{-a}
\end{pmatrix} ,\qquad 
 \bm{m} = \Exp{c} \begin{pmatrix} \Exp{-2c} + b^2 & -b \\
 -b & 1 
\end{pmatrix}.
\end{align}
In this example, the generalized Ricci tensor is
\begin{align}
 R_{AB} = -2 \, \diag \begin{pmatrix} 1&1&1&0&0&0&0&0&0&0 \end{pmatrix},
\end{align}
and the generalized torsion vanishes. Consequently, the scalar potential is given by $V = -R$, which evaluates to
\begin{align}
 R = - 6\,\Exp{-\frac{4}{5}\,a}\,,
\end{align}
using the above parameterization.

\paragraph{\underline{Generalized frame fields:}}
To construct the generalized frame field, we introduce the parameterization
\begin{align}
 l = f\,m\,,\qquad f=\Exp{x\,\hat{T}_1}\Exp{z\,\hat{T}_2}\Exp{y\,\hat{T}_1}\in \GS\,,\qquad
 m =\Exp{u\,\hat{T}_4} \Exp{v\,\hat{T}_5} \Exp{w\,\hat{T}_6} \in \GS\backslash G\,,
\end{align}
with the coordinates $\mathsf{y}^{\sfi}=\{x,\,y,\,z,\,u,\,v,\,w\}$. After repeating the steps detailed in section~\ref{sec:ExFT-ex1}, the components of the untwisted generalized frame field arise as
\begin{align}
 \Omega_A^{\exB} &= \begin{pmatrix} \Omega_a^{\exB} \\ 0 \\ \vdots \\ 0 \end{pmatrix},\qquad
 \Omega_a^{\exB} = \begin{pmatrix}
 0 & 0 & 0 \\
 0 & 0 & -\tanh u \\
 - \frac{\tanh v}{\cosh u} & \tanh u & 0 \\\end{pmatrix} ,
\\
 E_A{}^I &= \diag\begin{pmatrix} e_a{}^i & e^i{}_a & e^i{}_a & 3!\, e^{[i_1}{}_{[a_1}\,e^{i_2}{}_{a_2}\,e^{i_3]}{}_{a_3]}\end{pmatrix} ,\qquad
 e_a{}^i = \begin{pmatrix}
 1 & 0 & 0 \\
 0 & \frac{1}{\cosh u} & 0 \\
 0 & 0 & \frac{1}{\cosh u\,\cosh v} \end{pmatrix} .
\end{align}
Moreover, we need the frame in the $R_2$ representation. It follows direct from $E_A{}^I$ as 
\begin{align}
 E_{\II}{}^{\AA} = \begin{pmatrix}
 1 & 0 & 0 & 0 & 0 \\
 0 & 1 & 0 & 0 & 0 \\
 0 & 0 & \cosh u & 0 & 0 \\
 0 & 0 & 0 & \cosh u \cosh v & 0 \\
 0 & 0 & 0 & 0 & \cosh^2 u \cosh v
\end{pmatrix}.
\end{align}
Finally we compute the generalized metric
\begin{align}
 \cM_{IJ} = \Exp{-4\Cbeta\,\Delta}\,E_I{}^A\,E_J{}^B\,\cM_{AB}\,,\qquad 
 \Exp{-2\,\Delta} = \cosh^2 u\cosh v \,,
\end{align}
and extract the internal part of the Einstein-frame metric, the dilaton, and the Ramond--Ramond 0-form potential as
\begin{align}
 g_{mn} = \Exp{\frac{a}{2}}\,g^{\text{H}^3}_{mn}\,,\qquad \phi = c\,,\qquad C_0 = b\,.
\end{align}
where
\begin{align}
 \rmd s^2_{\text{H}^3} \equiv \rmd u^2 + \cosh^2 u\,\bigl(\rmd v^2 + \cosh^2 v\,\rmd w^2\bigr) \,.
\end{align}

The basis for the 1-form and 2-form fields is given by
\begin{align}
 K_1{}^I &= \bigl(0,0,0,0,0,0,0,0,0,\text{vol}_{\text{H}^3}\bigr)\,,\qquad 
 \text{vol}_{\text{H}^3} \equiv \cosh^2 u \cosh v\,,
\\
 K_{\bar{1}}{}^{\II} &= (1,0,0,0,0)\,, \qquad
 K_{\bar{2}}{}^{\II} = (0,1,0,0,0)\,.
\end{align}
In terms of type IIB supergravity, this means that the 1-form field in the seven-dimensional theory corresponds to the Ramond--Ramond 4-form potential as 
\begin{align}
 C_{\EXm uvw} = A_{\EXm}(x)\,\text{vol}_{\text{H}^3} \,,
\end{align} 
while the two 2-form fields correspond to the $B$-field and the Ramond--Ramond 2-form field through
\begin{align}
 C_{\EXm\EXb}=A_{1\EXm\EXb}(x)\,,\qquad B_{\EXm\EXb}=A_{2\EXm\EXb}(x)\,. 
\end{align}

\paragraph{\underline{Generalized duality:}}
By performing a generalized T-duality along the 6\textsuperscript{th} direction, which, for example, swaps $\hat{T}_6$ and $\hat{T}_{\bm{1}}^6$, we obtain an M-theory EGA with
\begin{align}
\begin{split}
f_{12}{}^3 &= 1\,,\quad
f_{13}{}^2 = -1\,,\quad
f_{23}{}^1 = 1\,,\quad
f_{34}{}^5 = 1\,,\quad
f_{35}{}^4 = -1\,,\quad
f_{45}{}^3 = -1\,,\quad
\\
f_1{}^{567} &= 1\,,\quad
f_2{}^{467} = -1\,,\quad
f_4{}^{267} = -1\,,\quad
f_5{}^{167} = 1\,,\quad
h_{1567} = 1\,,\quad
h_{2467} = -1\,.
\end{split}
\end{align}
In this case, the generalized coset $\GS\backslash G/H'$ is not a standard coset. However, by choosing an appropriate parameterization, we can construct the generalized frame field straightforwardly. We then find that the same seven-dimensional theory can be embedded into 11D supergravity. Since the detailed analysis is not particularly illuminating, we do not pursue it further here.

\subsection{Half-maximal theories}
To find examples which preserve half of the original supersymmetry after the reduction, we first recall that the U-duality group $\Edd$ contains the T-duality group $\OO(d-1,d-1)$ as a subgroup. Its maximally compact subgroup is $\OO(d-1)\times \OO(d-1)$, and for $d \leq 6$, we take $\GS = \text{U}(1)$ as a subgroup of one of the two $\OO(d-1)$ factors. Then, the commutant $\mathrm{C}_{\GD}(\GS)$ becomes $\mathbb{R}^+ \times \OO(d-1,d-3)$, which is the duality group of half-maximal supergravity. In this way, we construct two explicit examples in which the resulting truncated theory corresponds to a half-maximal supergravity.

\subsubsection{Example 1: \texorpdfstring{$\text{U}(1)\backslash \SO(5)\times \text{U}_G(1)/\SO(4)$}{U(1)\textbackslash SO(5)xUG(1)/SO(4)}}\label{sec:ex3-ExFT}
Taking $\GDM = \Edd[5]$, we consider an M-theory EGA characterized by the structure constants
\begin{align}
 f_{25}{}^3 = 
 - f_{35}{}^2 = 1\,,\qquad 
 f_1{}^{234} = - f_2{}^{134} = f_3{}^{124} = - f_4{}^{123} = 1\,,\qquad 
 f_{1234} = -1\,.
\label{eq:so5-fabc}
\end{align}
The five generators
\begin{align}
 \{\hat{T}_9,\,\hat{T}_{12},\,\hat{T}_{14},\,\hat{T}_{15},\,\hat{T}_{16} \}\,,
\end{align}
span the ideal $\cI$, while the remaining 11 generators form the Lie algebra $\mathrm{Lie}(G) = \mathfrak{so}(5) \oplus \mathfrak{u}_G(1)$. A minimally coisotropic subalgebra $\mathrm{Lie}(H) = \mathfrak{so}(4)$ is generated by
\begin{align}
\{\hat{T}_6,\,\hat{T}_7,\,\hat{T}_8,\,\hat{T}_{10},\,\hat{T}_{11},\,\hat{T}_{13} \}\,,
\end{align}
while the abelian factor in $G$, $\text{U}_G(1)$, is generated by $\hat{T}_5+\hat{T}_8$\,. Next, we choose the structure group $\GS$ to be the $\text{U}(1)$ generated by
\begin{align}
 \mathfrak{t}_1 = \hat{T}_1+\hat{T}_5\,,
\end{align}
which is isotropic and, moreover, a linear combination of generators from $\mathfrak{so}(5)$ and $\mathfrak{u}_G(1)$. 

To reduce the mega-space to the physical space, we perform a change of basis,
\begin{align}
 \check{T}_{\Ah} = C_{\Ah}{}^{\Bh}\,\hat{T}_{\Bh}\,,\qquad
 C_{\Ah}{}^{\Bh} = \scalebox{0.8}{$\left(\begin{array}{cccccccccccccccc}
 1 & 0 & 0 & 0 & 1 & 0 & 0 & 0 & 0 & 0 & 0 & 0 & 0 & 0 & 0 & 0 \\
 0 & 1 & 0 & 0 & 0 & 0 & 0 & 0 & 0 & 0 & 0 & 0 & 0 & 0 & 0 & 0 \\
 0 & 0 & 1 & 0 & 0 & 0 & 0 & 0 & 0 & 0 & 0 & 0 & 0 & 0 & 0 & 0 \\
 0 & 0 & 0 & 1 & 0 & 0 & 0 & 0 & 0 & 0 & 0 & 0 & 0 & 0 & 0 & 0 \\
 0 & 0 & 0 & 0 & 1 & 0 & 0 & 0 & 0 & 0 & 0 & 0 & 0 & 0 & 0 & 0 \\
 0 & 0 & 0 & 0 & 0 & 1 & 0 & 0 & 0 & 0 & 0 & 0 & 0 & 0 & 0 & 0 \\
 0 & 0 & 0 & 0 & 0 & 0 & 1 & 0 & 0 & 0 & 0 & 0 & 0 & 0 & 0 & 0 \\
 0 & 0 & 0 & 0 & 0 & 0 & 0 & 1 & 0 & 0 & 0 & 0 & 0 & 0 & 0 & 0 \\
 0 & 0 & 0 & 0 & 0 & 0 & 0 & 0 & 1 & 0 & 0 & 0 & 0 & 0 & 0 & 0 \\
 0 & 0 & 0 & 0 & 0 & 0 & 0 & 0 & 0 & 1 & 0 & 0 & 0 & 0 & 0 & 0 \\
 0 & 0 & 0 & 0 & 0 & 0 & 0 & 0 & 0 & 0 & 1 & 0 & 0 & 0 & 0 & 0 \\
 0 & 0 & 0 & 0 & 0 & 1 & 0 & 0 & 0 & 0 & 0 & 1 & 0 & 0 & 0 & 0 \\
 0 & 0 & 0 & 0 & 0 & 0 & 0 & 0 & 0 & 0 & 0 & 0 & 1 & 0 & 0 & 0 \\
 0 & 0 & 0 & 0 & 0 & 0 & 1 & 0 & 0 & 0 & 0 & 0 & 0 & 1 & 0 & 0 \\
 0 & 0 & 0 & 0 & 0 & 0 & 0 & 1 & 0 & 0 & 0 & 0 & 0 & 0 & 1 & 0 \\
 0 & 0 & 0 & 0 & 0 & 0 & 0 & 0 & 0 & 0 & 0 & 0 & 0 & 0 & 0 & 1 
\end{array}\right)$} ,
\end{align}
which brings the $\mathfrak{u}(1)$ generator to $\check{T}_{1}$\,. In this new basis, the structure constants $\check{X}_{\Ah\Bh}{}^{\Ch}$ are given by
\begin{align}
\begin{split}
 \check{f}_{25}{}^3 &= 
 - \check{f}_{35}{}^2 = 
 -\check{f}_{12}{}^3 =
 \check{f}_{13}{}^2 = 1\,,\qquad 
 \check{f}_{1234} = -1\,,
\\
 \check{f}_1{}^{234} &= - \check{f}_2{}^{134} = \check{f}_2{}^{345} = \check{f}_3{}^{124} = - \check{f}_3{}^{245} = - \check{f}_4{}^{123} = \check{f}_4{}^{235} = 1\,,
\end{split}
\end{align}
and the generators decompose as
\begin{align}
 \check{T}_{\Ah} = \begin{pmatrix} \check{T}_{\exA}&\check{T}_{A}&\check{T}^{\exA \AA}\end{pmatrix} ,
\end{align}
where $\exA=1$ and $A$ or $\AA$ is $R_1 =\irrepb{10}$ or $R_2 = \irrep{5}$ of $\GD=\SL(5)$, respectively. Their components are denoted by
\begin{align}
 \check{T}_{A} = \begin{pmatrix} \check{T}_{a}&\check{T}^{a_1a_2}\end{pmatrix} ,\qquad 
 \check{T}^{\exA \AA} = \begin{pmatrix} \check{T}^{\exA a}&\check{T}^{\exA 2345}\end{pmatrix} \qquad (a=2,3,4,5)\,.
\end{align}
As in the previous examples, we need the matrix representation of the $\mathfrak{u}(1)$ generator, $(\mathfrak{t}_{1})_{\Ah}{}^{\Bh} = -\check{X}_{1\Ah}{}^{\Bh}$. It takes the form
\begin{align}
 (\mathfrak{t}_{1})_{\Ah}{}^{\Bh} &= 
 \begin{pmatrix}
 (\mathfrak{t}_{1})_{\exA}{}^{\exB} & 0 & 0 \\
 0 & (\mathfrak{t}_{1})_A{}^B & 0 \\
 0 & 0 & \delta^{\exA}_{\exB}\,(\mathfrak{t}_{1})_{\BB}{}^{\AA}
\end{pmatrix},
\end{align}
with $(\mathfrak{t}_{1})_{\exA}{}^{\exB}=0$,
\begin{align}
  (\mathfrak{t}_{1})_A{}^B = \scalebox{0.8}{$\begin{pmatrix}
 0 & 1 & 0 & 0 & 0 & 0 & 0 & -1 & 0 & 0 \\
 -1 & 0 & 0 & 0 & 0 & 1 & 0 & 0 & 0 & 0 \\
 0 & 0 & 0 & 0 & -1 & 0 & 0 & 0 & 0 & 0 \\
 0 & 0 & 0 & 0 & 0 & 0 & 0 & 0 & 0 & 0 \\
 0 & 0 & 1 & 0 & 0 & 0 & 0 & 0 & 0 & 0 \\
 0 & -1 & 0 & 0 & 0 & 0 & 0 & 1 & 0 & 0 \\
 0 & 0 & 0 & 0 & 0 & 0 & 0 & 0 & 1 & 0 \\
 1 & 0 & 0 & 0 & 0 & -1 & 0 & 0 & 0 & 0 \\
 0 & 0 & 0 & 0 & 0 & 0 & -1 & 0 & 0 & 0 \\
 0 & 0 & 0 & 0 & 0 & 0 & 0 & 0 & 0 & 0 \end{pmatrix}$},
\ \text{ and }\
(\mathfrak{t}_{1})_{\AA}{}^{\BB} = \scalebox{0.8}{$
\begin{pmatrix}
 0 & 1 & 0 & 0 & 0 \\
 -1 & 0 & 0 & 0 & 0 \\
 0 & 0 & 0 & 0 & 0 \\
 0 & 0 & 0 & 0 & 1 \\
 0 & 0 & 0 & -1 & 0\end{pmatrix}$} .
\end{align}
This is all that is needed to compute the generalized Ricci tensor
\begin{align}
 R_{AB} &= \begin{pmatrix}
 1 & 0 & 0 & 0 & 0 & -1 & 0 & 0 & 0 & 0 \\
 0 & 1 & 0 & 0 & 0 & 0 & 0 & -1 & 0 & 0 \\
 0 & 0 & 1 & 0 & 0 & 0 & 0 & 0 & 0 & 0 \\
 0 & 0 & 0 & 0 & 0 & 0 & 0 & 0 & 0 & 0 \\
 0 & 0 & 0 & 0 & 1 & 0 & 0 & 0 & 0 & 0 \\
 -1 & 0 & 0 & 0 & 0 & 1 & 0 & 0 & 0 & 0 \\
 0 & 0 & 0 & 0 & 0 & 0 & 0 & 0 & 0 & 0 \\
 0 & -1 & 0 & 0 & 0 & 0 & 0 & 1 & 0 & 0 \\
 0 & 0 & 0 & 0 & 0 & 0 & 0 & 0 & 0 & 0 \\
 0 & 0 & 0 & 0 & 0 & 0 & 0 & 0 & 0 & 0\end{pmatrix}
\end{align}
from the structure constants. Furthermore, there are non-vanishing components of the generalized torsion which we have omitted here.

In total, the four null eigenvectors $K_{\cA}{}^{A}$ ($\cA=1,2,3,4$),  
\begin{align}
 K_{\cA}{}^{A} = \frac{1}{\sqrt{2}} \begin{pmatrix}
 1 & 0 & 0 & 0 & 0 & 1 & 0 & 0 & 0 & 0 \\
 0 & 0 & 0 & 1 & 0 & 0 & 0 & 0 & 0 & -1 \\
 0 & 1 & 0 & 0 & 0 & 0 & 0 & 1 & 0 & 0 \\
 0 & 0 & 0 & 1 & 0 & 0 & 0 & 0 & 0 & 1\end{pmatrix},
\end{align}
of the matrix $(\mathfrak{t}_{1})_A{}^B$ can be found, while in the $R_2$ representation there is only one invariant $K_{\AA}{}^{\overline{\cA}}$ under $(\mathfrak{t}_{1})_{\AA}{}^{\BB}$. It is given by
\begin{align}
 K_{\AA}{}^{1} = \begin{pmatrix} 0& 0& 1& 0& 0\end{pmatrix} .
\end{align}
Applying them to the $\eta$-symbol, it becomes the $\OO(3,1)$-invariant metric
\begin{align}\label{eq:etaFromEtaSymbol}
 \eta_{\cA\cB} = \eta_{\cA\cB;\overline{\cC}=1}= \diag( +1,\, +1,\,+1,\,-1)\,,
\end{align}
and the components of the intrinsic torsion (or embedding tensor) $\cT_{\cA\cB}{}^{\cC}$ are found to be
\begin{align}
 \cT_{\cA\cB}{}^{\cC} = \cT_{[\cA\cB]}{}^{\cC}\,,\qquad
 \cT_{12}{}^{3} = \cT_{23}{}^{1} = \cT_{31}{}^{2} = \sqrt{2} \,.
\end{align}
They indicate that the gauge group of the resulting gauged supergravity is $\cG = \SO(3) \times \text{U}(1)$. After constructing the generalized frame field, we obtain the generalized vector fields $K_{\cA}{}^I = K_{\cA}{}^B\,E_B{}^I$, satisfying
\begin{align}
 \gLie_{K_{\cA}}K_{\cB} = - X_{\cA\cB}{}^{\cC}\,K_{\cC}\,.
\end{align}
We then check that the commutant group is $\mathrm{C}_{\SL(5)}(\text{U}(1)) = \mathbb{R}^+\times \OO(3,1)$, and accordingly, the generalized metric $\cM_{AB} \in \mathbb{R}^+\times \frac{\OO(3,1)}{\OO(3)}$ can be parameterized by
\begin{align}
 \cM_{AB} &= L_A{}^C\,L_B{}^D\,M_{CD}\,,
\\
 M_{AB} &= \diag \begin{pmatrix} \Exp{a}&\Exp{a}&\Exp{b}&\Exp{4 a+2 b}&\Exp{b}&\Exp{a}&\Exp{-3 a-b}&\Exp{a}&\Exp{-3 a-b}&\Exp{-2 (a+b)}\end{pmatrix} ,
\\
 L_A{}^B &= \scalebox{0.9}{$\begin{pmatrix}
 1 & 0 & 0 & d & 0 & 0 & 0 & 0 & 0 & 0 \\
 0 & 1 & 0 & c & 0 & 0 & 0 & 0 & 0 & 0 \\
 0 & 0 & 1 & 0 & 0 & 0 & 0 & 0 & 0 & 0 \\
 0 & 0 & 0 & 1 & 0 & 0 & 0 & 0 & 0 & 0 \\
 0 & 0 & 0 & 0 & 1 & 0 & 0 & 0 & 0 & 0 \\
 0 & 0 & 0 & d & 0 & 1 & 0 & 0 & 0 & 0 \\
 0 & 0 & -d & 0 & -c & 0 & 1 & 0 & 0 & 0 \\
 0 & 0 & 0 & c & 0 & 0 & 0 & 1 & 0 & 0 \\
 0 & 0 & -c & 0 & d & 0 & 0 & 0 & 1 & 0 \\
 d & c & 0 & c^2+d^2 & 0 & d & 0 & c & 0 & 1\end{pmatrix}$},
\end{align}
in terms of the four parameters $a,\,b,\,c,\,d$ that are scalar fields in seven dimensions. With this parameterization, the generalized Ricci scalar becomes
\begin{align}
 R = \cM^{AB}\,\cR_{AB} = 2\,\bigl[2\,\Exp{-a}+\Exp{-b} + \Exp{3 a+b}\,(c^2+d^2)\bigr]\,.
\end{align}

In summary, the bosonic field content of the resulting 7D half-maximal theory consists of
\begin{itemize}
\item the metric $\mathfrak{g}_{\EXm\EXn}$\,,

\item four scalar fields $\cM_{IJ}\in \mathbb{R}^+\times \frac{\OO(3,1)}{\OO(3)}$\,, 

\item four 1-form fields $A_{\EXm}{}^{\cA}$,

\item a 2-form field $B_{\EXm\EXn}$\,,
\end{itemize}
and the gauge group is $\cG =\SO(3)\times \text{U}(1)$. Higher $p$-form fields are not required for writing down the action. However, for instance, the 3-form and 4-form fields are electric-magnetic dual to the 2-form and 1-form fields, respectively.

To obtain the explicit form of the generalized frame field, we parameterize
\begin{align}
\begin{split}
 l &= f\,m\,h\,,\qquad f=\Exp{p\,(\hat{T}_1+\hat{T}_5)} \in \GS\,,\qquad h\in H\,,
\\
 m &=\Exp{x\,\hat{T}_4} \Exp{y\,\hat{T}_5} \Exp{z\,\hat{T}_2} \Exp{w\,\hat{T}_5} \in \GS\backslash G/H\,,
\end{split}
\end{align}
in the basis where the structure constants are given by \eqref{eq:so5-fabc}. With this choice of the coset representative, the generalized frame field is given by
\begin{align}
 E_A{}^I &= (\cos x \cos z)^{\frac{1}{3}} \begin{pmatrix}
 \delta_a^b & 0 \\ \Pi^{a_1a_2b} & 2\,\delta_{b_1b_2}^{a_1a_2} \end{pmatrix}
 \begin{pmatrix} e_b{}^i & 0 \\ 0 & 2\, e^{[b_1}{}_{[i_1}\,e^{b_2]}{}_{i_2]} \end{pmatrix},
\\
 \Exp{-2\Delta} &= \cos x\cos^2 z\sin z\,,
\end{align}
where we are using the coordinates $y^i=\begin{pmatrix} x, y, z, w \end{pmatrix}$ and
\begin{align}
 e_a{}^i &= (\cos x \cos z)^{-\frac{1}{3}}\begin{pmatrix}
 \sin x \cos y \tan z & -\frac{\cos x \sin y}{\tan z} & \cos x \cos y & \frac{\cos x \sin y}{\tan z} \\
 -\sin x \sin y \tan z & -\frac{\cos x \cos y}{\tan z} & -\cos x \sin y & \frac{\cos x \cos y}{\tan z} \\
 1 & 0 & 0 & 0 \\
 0 & 1 & 0 & 0 \end{pmatrix},
\\
 \Pi^{234} &= \tan x\,,\qquad
 \Pi^{245} = \tfrac{\sin y \tan z}{\cos x}\,,\qquad
 \Pi^{345} = \tfrac{\cos y \tan z}{\cos x}\,.
\end{align}
In the same vein, the generalized vector fields are obtained as
\begin{align}
 K_1{}^I+i\,K_3{}^I &= \tfrac{1}{\sqrt{2}}\,\Exp{-i\,y}\,\bigl(\sin x \tan z ,\ i \bigl(\tfrac{\tan z}{\cos x}-\tfrac{\cos x}{\tan z}\bigr) ,\ \cos x,\ \tfrac{i \cos x}{\tan z},
\nn\\
 &\qquad\qquad\quad\  0,\ -\cos z,\ -i \sin z,\ 0,\ 0,\ i \tan x \sin z \tan z\bigr)\,,
\\
 K_2{}^I+ K_4{}^I &= \bigl(0,\ \sqrt{2},\ 0,\ 0,\ 0,\ 0,\ 0,\ 0,\ 0,\ 0\bigr)\,,
\\
 K_4{}^I &=\tfrac{1}{\sqrt{2}}\,\bigl(0,\ 0,\ 0,\ 1,\ \cos x \cos z,\ 0,\ \cos x \cos z,\ \sin x \sin z,\ 0,\ -\sin x \sin z \bigr)\,.
\end{align}

In particular, for the specific configuration $\cM_{AB}=\delta_{AB}$ (i.e., $a=b=c=d=0$), the internal part of the supergravity fields is given by
\begin{align}
\begin{split}
 \rmd s^2 &= \cos^2 z \,\rmd x + \cos^2 x\,\cos^2 z \,  \rmd \tilde{y}^2 + \rmd z^2 + \sin^2 z\,\rmd w^2 \,,
\\
 C_{3} &= - \bigl(\cos x\cos z \sin^2z\,\rmd x\wedge \rmd y + \sin x\sin z\,\rmd y\wedge \rmd z\bigr)\wedge \rmd w\,,
\end{split}
\end{align}
where $\tilde{y}=y+w$. This metric corresponds to $S^4$, and the 4-form field strength becomes
\begin{align}
 F_4 = - 3\,\text{vol}_{S^4} \,,\qquad 
 \text{vol}_{S^4} = \cos x  \cos^2 z \sin z \,\rmd x\wedge \rmd y\wedge \rmd z\wedge w\,.
\end{align}
This demonstrates that the gauged supergravity admits the AdS$_7$ solution, where the external metric $\gD_{\EXm\EXn} = g_{\EXm\EXn}$ is given by
\begin{align}
 \rmd s^2 = \tfrac{4}{\rho^2}\,\bigl(\rmd \rho^2 + \eta_{mn}\,\rmd x^m\,\rmd x^n\bigr) \,.
\end{align}
In this case, we do not find an inequivalent subgroup $H'$ and do not perform a generalized duality.

\subsubsection{Example 2: \texorpdfstring{$T^{1,1}$}{T(1,1)} compactification of type IIB and its duals}\label{sec:ex4-ExFT}
Here we take $\GDM=\Edd[7]$ and consider a type IIB EGA with the following structure constants:
\begin{align}
\begin{split}
 f_{12}{}^3 &= f_{23}{}^1 = f_{31}{}^2 = f_{45}{}^6 = f_{56}{}^4 = f_{64}{}^5 = 1\,,
\\
 f^{\bm{\alpha}}_{234}&=f^{\bm{\alpha}}_{456}= -f^{\bm{\alpha}}_{123}=-f^{\bm{\alpha}}_{156}=q^{\bm{\alpha}}\,,\quad
 f_{23456} = - f_{12356} = q\,,
\\
 f_2{}_{\bm{\alpha}}^{34} &= - f_3{}_{\bm{\alpha}}^{24} = f_5{}_{\bm{\alpha}}^{16} = - f_6{}_{\bm{\alpha}}^{15} = r_{\bm{\alpha}}\,.
\end{split}
\label{eq:T11-fabc}
\end{align}
The Leibniz identity (also called quadratic constraints) requires that $q^{\bm{\alpha}}\,r_{\bm{\alpha}} = 0$. A simple example is $q^{\bm{\alpha}} = r_{\bm{\alpha}} = 0$. Then the ideal $\cI$ has dimension 27, and the corresponding Lie group $G$ has dimension 29. To understand the structure of $G$, we perform a Levi decomposition of its Lie algebra. This gives rise to a six-dimensional semisimple factor with Lie algebra $\mathfrak{su}(2) \oplus \mathfrak{su}(2)$ and a 23-dimensional solvable radical $\mathfrak{r}$. The $\mathfrak{su}(2) \oplus \mathfrak{su}(2)$ Levi factor is spanned by the following generators:
\begin{align}
\begin{split}
 \mathring{T}_1 &= \hat{T}_1 + \tfrac{1}{2} q \, \hat{T}_{28} + \tfrac{1}{2} q \, \hat{T}_{29} \,, 
\\
 \mathring{T}_2 &= \hat{T}_2 - \tfrac{1}{2} q \, \hat{T}_{23} + \tfrac{1}{2} q \, \hat{T}_{34} \,, 
\\
 \mathring{T}_3 &= \hat{T}_3 + \tfrac{1}{2} q \, \hat{T}_{20} + \tfrac{1}{2} q \, \hat{T}_{37} \,, 
\\
 \mathring{T}_4 &= \hat{T}_4 - \tfrac{1}{2} q \, \hat{T}_{28} - \tfrac{1}{2} q \, \hat{T}_{29} \,, 
\\
 \mathring{T}_5 &= \hat{T}_5 + \tfrac{1}{2} q \, \hat{T}_{27} - \tfrac{1}{2} q \, \hat{T}_{30} \,, 
\\
 \mathring{T}_6 &= \hat{T}_6 - \tfrac{1}{2} q \, \hat{T}_{26} - \tfrac{1}{2} q \, \hat{T}_{31} \,.
\end{split}
\end{align}
The generators $\{\mathring{T}_1,\, \mathring{T}_2,\, \mathring{T}_3\}$ span the first $\mathfrak{su}(2)$ factor, while $\{\mathring{T}_4,\, \mathring{T}_5,\, \mathring{T}_6\}$ span the second $\mathfrak{su}(2)$ factor. The corresponding Lie group takes the form $G = (\text{SU}(2) \times \text{SU}(2)) \ltimes R$, where $R$ is the 23-dimensional solvable Lie group associated with $\mathfrak{r}$. The generalized parallelizable space $G/H$ with $H=R$ is then locally diffeomorphic to $\text{SU}(2) \times \text{SU}(2)$. By choosing $\GS = \text{U}(1)$ as the subgroup generated by $\mathring{T}_1 +\mathring{T}_4=\hat{T}_1 + \hat{T}_4$\,, the physical space $\GS \backslash G / H$ becomes locally diffeomorphic to the standard coset space $\text{U}(1) \backslash \bigl(\text{SU}(2) \times \text{SU}(2)\bigr)$. As one might expect, this coset includes the Sasaki--Einstein manifold $T^{1,1}$ as a special configuration. When $q^{\bm{\alpha}}$ are non-zero, the resulting generalized coset $\GS \backslash G / H$ may include the $T^{1,1}$ background with additional Ramond--Ramond fluxes as a particular configuration. The flux components $f_a{}_{\bm{\alpha}}^{bc}$ correspond to non-geometric $Q$- or $P$-fluxes, and hence the introduction of $r_{\bm{\alpha}}$ leads to a deformation of the $T^{1,1}$ metric.

Before constructing the generalized frame field, let us first study the truncated five-dimensional theory. By changing the basis according to
\begin{align}
  \hat{T}_{\Ah} \to \check{T}_{\Ah} = C_{\Ah}{}^{\Bh}\, \hat{T}_{\Bh}\,,\qquad \text{with}\qquad
 C = \Exp{\frac{\pi}{4}\,(K^1{}_4 - K^4{}_1)}\,,
\label{eq:T11-PS-basis}
\end{align}
the $\text{U}(1)$ generator can be expressed as $\check{T}_{1}$\,. In this new basis, the structure constants become
\begin{align}
\begin{split}
 &\check{f}_{12}{}^3 = - \check{f}_{13}{}^2 = \check{f}_{15}{}^6 = - \check{f}_{16}{}^5 = \check{f}_{23}{}^1 = - \check{f}_{23}{}^4 = \check{f}_{24}{}^3 = -\check{f}_{34}{}^2
\\
 &= \check{f}_{45}{}^6 = -\check{f}_{46}{}^5 = \check{f}_{56}{}^1 = \check{f}_{56}{}^4 =\tfrac{1}{\sqrt{2}}\,,\qquad
 \check{f}_{23456} = \sqrt{2}\,q\,,
\\
 &\check{f}_2{}_{\bm{\alpha}}^{34} = - \check{f}_3{}_{\bm{\alpha}}^{24} = \check{f}_5{}_{\bm{\alpha}}^{16} = - \check{f}_6{}_{\bm{\alpha}}^{15}
 = -\check{f}_2{}_{\bm{\alpha}}^{13} = \check{f}_3{}_{\bm{\alpha}}^{12} = -\check{f}_5{}_{\bm{\alpha}}^{46} = \check{f}_6{}_{\bm{\alpha}}^{45}
 = \tfrac{1}{\sqrt{2}}\,r_{\bm{\alpha}}\,.
\end{split}
\end{align}
We furthermore decompose the generators as
\begin{align}
 \check{T}_{\Ah} = \begin{pmatrix} \check{T}_{\exA}&\check{T}_{A}&\check{T}^{\exA \AA}\end{pmatrix} ,
\end{align}
and the matrix representation of the $\text{U}(1)$ structure group generator is then given by
\begin{align}
 (\mathfrak{t}_{1})_{A}{}^{B} = - \tfrac{1}{\sqrt{2}}\,(K^2{}_3-K^3{}_2 + K^5{}_6 - K^6{}_5)_{A}{}^{B}\,,
\end{align}
At this point, we can confirm that the commutant group is $\mathrm{C}_{\Edd[6]}(\text{U}(1)) = \mathbb{R}^+\times \OO(5,3)$. As expected, there are nine null eigenvectors $K_{\cA}{}^{B}$ ($\cA=1,\dotsc,8,*$), which can be compactly written as
\begin{align}
 K_{\cA}{}^{B}\,\hat{T}_{B} = \begin{pmatrix}
 \hat{T}_{16}&\hat{T}_8& \hat{T}_{13}& 
 \frac{\hat{T}_{19} + \hat{T}_{23}}{\sqrt{2}}&
 \frac{\hat{T}_{20} - \hat{T}_{22}}{\sqrt{2}}&
 -\hat{T}_{25}& \hat{T}_{27}& -\hat{T}_{26}& \hat{T}_{3} \end{pmatrix}.
\end{align}
We also find null eigenvectors $K_{\BB}{}^{\overline{\cA}}$ ($\cA=1,\dotsc,8,\overline{*}$) in the $R_2$-representation, given by
\begin{align}
 K_{\BB}{}^{\overline{\cA}}\,\overline{T}^{\BB} = \begin{pmatrix}
 -\overline{T}^{12}&-\overline{T}^{20}&\overline{T}^{15}&-\frac{\overline{T}^{5}+\overline{T}^{9}}{\sqrt{2}}&-\frac{\overline{T}^{6}-\overline{T}^{8}}{\sqrt{2}}&
 \overline{T}^{3}&\overline{T}^{1}&\overline{T}^{2}&-\overline{T}^{25} \end{pmatrix}.
\end{align}
By decomposing the indices as $\cA=\{\mathtt{A},\,*\}$ and $\overline{\cA}=\{\overline{\mathtt{A}},\,\overline{*}\}$, one finds that the structures expected from the commutant group $\mathbb{R}^+\times\OO(5,3)$ arise from the $\eta$-symbols through
\begin{align}
 \eta_{\cA\cB:\overline{*}} = \begin{pmatrix}
 \eta_{\mathtt{A}\mathtt{B}} & 0 \\
 0 & 0 \end{pmatrix} ,\qquad 
 \eta_{\cA\cB:\overline{\mathtt{C}}} = \begin{pmatrix}
 0 & \eta_{\mathtt{A}\overline{\mathtt{C}}} \\
 \eta_{\mathtt{B}\overline{\mathtt{C}}} & 0 \end{pmatrix} ,\qquad
 \eta_{\mathtt{A}\mathtt{B}} = \scalebox{0.8}{$\begin{pmatrix}
 0 & 0 & 0 & 0 & 0 & 1 & 0 & 0 \\
 0 & 0 & 0 & 0 & 0 & 0 & 1 & 0 \\
 0 & 0 & 0 & 0 & 0 & 0 & 0 & 1 \\
 0 & 0 & 0 & 1 & 0 & 0 & 0 & 0 \\
 0 & 0 & 0 & 0 & 1 & 0 & 0 & 0 \\
 1 & 0 & 0 & 0 & 0 & 0 & 0 & 0 \\
 0 & 1 & 0 & 0 & 0 & 0 & 0 & 0 \\
 0 & 0 & 1 & 0 & 0 & 0 & 0 & 0 
\end{pmatrix}$}
\end{align}
(in analogy with \eqref{eq:etaFromEtaSymbol} from the previous example). Finally, the generalized Ricci tensor reads
\begin{align}
 R_{AB} = \begin{pmatrix} R_{ab} & R_a{}^{b}_{\bm{\beta}} & 0 & 0 \\
 R^a_{\bm{\alpha}}{}_{b} & 0 & 0 & 0 \\
 0 & 0 & 0 & 0 \\
 0 & 0 & 0 & 0 
\end{pmatrix},
\end{align}
where we decompose the indices as $V_{A}=\begin{pmatrix} V_{a} & V^{a}_{\bm{\alpha}} & V^{a_1a_2a_3} & V^{a_1\cdots a_5}_{\bm{\alpha}} \end{pmatrix}$ and define
\begin{align}
 R_{ab} = \tfrac{1}{2}\,\diag(1,1,0,1,1)_{ab}\,,\qquad R_a{}^{b}_{\bm{\beta}} = \tfrac{1}{2}\,r_{\bm{\beta}}\,\diag(1,1,0,-1,-1)_{a}{}^{b} = R^{b}_{\bm{\beta}}{}_a\,.
\end{align}
The generalized torsion $X_{AB}{}^C$ contains many components, but here we present only the intrinsic torsion $X_{\cA\cB}{}^{\cC}$, given by
\begin{align}
 X_{\mathtt{A}\mathtt{B}}{}^{\mathtt{C}} = F_{\mathtt{A}\mathtt{B}}{}^{\mathtt{C}}\,,\qquad 
 X_{*\mathtt{B}}{}^{\mathtt{C}} = \xi_{\mathtt{B}}{}^{\mathtt{C}}\,,
\end{align}
where the non-zero components of $F_{\mathtt{A}\mathtt{B}\mathtt{C}}=F_{[\mathtt{A}\mathtt{B}\mathtt{C}]}=F_{\mathtt{A}\mathtt{B}}{}^{\mathtt{D}}\,\eta_{\mathtt{D}\mathtt{C}}$ and $\xi_{\mathtt{A}\mathtt{B}}=\xi_{[\mathtt{A}\mathtt{B}]}=\xi_{\mathtt{A}}{}^{\mathtt{C}}\,\eta_{\mathtt{C}\mathtt{B}}$ are
\begin{align}
\begin{split}
 F_{236} &= F_{123} = - \tfrac{1}{\sqrt{2}}\,,\qquad \xi_{45} = \sqrt{2} \,,\qquad \xi_{23} = -\sqrt{2}\,q\,,
\\
 \xi_{13} &= \xi_{36} = \sqrt{2}\,q^{\bm{1}}\,,\qquad
 \xi_{12} = \xi_{26} = -\sqrt{2}\,q^{\bm{2}}\,.
\end{split}
\label{eq:T11-ET}
\end{align}
In the following, we consider the case where $q \neq 0$ and $q^{\bm{\alpha}} = r_{\bm{\alpha}} = 0$. In this situation, the generators $K_{\cA}{}^{B}\,\check{T}_{B}$ ($\cA=4,5,7,8$) span the ideal. Upon taking the quotient, we find that the gauge group is $\cG = H_3 \times \text{U}(1)^2$, where $H_3$ is the three-dimensional Heisenberg group.

We note that our embedding tensor \eqref{eq:T11-ET} agrees with results known in the literature \cite{Cassani:2010uw,Cassani:2010na,Bena:2010pr}. One of the advantages of our approach is that the embedding tensor of the gauged supergravity can be derived automatically and algebraically, even before specifying an explicit reduction ansatz. We also emphasize that the embedding tensor \eqref{eq:T11-ET} is non-geometric. As discussed in \cite{Sakatani:2021eqo,Hassler:2023nht}, if the generalized frame field $E_{\mathtt{A}}{}^{\mathtt{I}} \in \mathbb{R}^+\times \OO(d,d+n)$ is assumed to satisfy the section condition, then it is not possible to introduce both $F_{\mathtt{A}\mathtt{B}\mathtt{C}}$ and $\xi_{\mathtt{A}\mathtt{B}}$ simultaneously. But here, the generalized frame field $K_{\cA}{}^I$ is treated as a generalized vector field in the $\Edd[6]$ ExFT, and within this framework, the section condition of $\Edd[6]$ ExFT is not violated. 

In the following, we explicitly construct the generalized frame field $E_A{}^I$ for the case $q \neq 0$ and $q^{\bm{\alpha}} = r_{\bm{\alpha}} = 0$, by selecting a specific subgroup $H$. 

\paragraph{\underline{$T^{1,1}$ compactification:}}
We start with the structure constants \eqref{eq:T11-fabc} and use the parameterization
\begin{align}
\begin{split}
 l &= f\,m\,h\,,\qquad f=\Exp{x\,(\hat{T}_1+\hat{T}_4)} \in \GS\,,\qquad h\in H\,,
\\
 m &=\Exp{\frac{\psi}{2}\,(\hat{T}_1-\hat{T}_4)} \Exp{(\frac{\pi}{2} - \theta_1)\,\hat{T}_2} \Exp{\phi_1\,\hat{T}_3} \Exp{(\theta_2-\frac{\pi}{2})\,\hat{T}_5} \Exp{\phi_2\,\hat{T}_6} \in \GS\backslash G/H\,.
\end{split}
\end{align}
Then, in the basis \eqref{eq:T11-PS-basis}, we can straightforwardly compute the generalized frame field as
\begin{align}
 E_{A}{}^I &= D_{A}{}^J\,N_J{}^I\,,\qquad N_I{}^J = \Exp{q\,\psi\sin\theta_1\sin\theta_2\,R^{\psi\theta_1\phi_1\theta_2\phi_2}}\,,
\\
 D_{A}{}^I&= \diag \begin{pmatrix} e_{\sfa}^{\sfi} & e_{\sfi}^{\sfa} & e_{\sfi}^{\sfa} & e_{[\sfi_1}^{[\sfa_1}e_{\sfi_2}^{\sfa_2}e_{\sfi_3]}^{\sfa_3]} & e_{[\sfi_1}^{[\sfa_1}\cdots e_{\sfi_5]}^{\sfa_5]} & e_{[\sfi_1}^{[\sfa_1}\cdots e_{\sfi_5]}^{\sfa_5]} & e_{[\sfi_1}^{[\sfa_1}\cdots e_{\sfi_6]}^{\sfa_6]}e_{\sfi'}^{\sfa'} \end{pmatrix} ,
\\
 e_{\sfa}^{\sfi} &= e_{\sfa}{}^{\sfi} = \begin{pmatrix}
 -\cos \frac{\psi}{2} & - \frac{\sin \frac{\psi}{2}}{\sin \theta_1} & \frac{\sin \frac{\psi}{2}}{\tan\theta_1} & 0 & 0 \\
 -\sin \frac{\psi}{2} & \frac{\cos \frac{\psi}{2}}{\sin \theta_1} & -\frac{\cos \frac{\psi}{2}}{\tan\theta_1} & 0 & 0 \\
 0 & 0 & -\sqrt{2} & 0 & 0 \\
 0 & 0 & -\frac{\sin \frac{\psi}{2}}{\tan \theta_2} & \cos \frac{\psi}{2} & \frac{\sin \frac{\psi}{2}}{\sin \theta_2} \\
 0 & 0 & -\frac{\cos \frac{\psi}{2}}{\tan \theta_2} & -\sin \frac{\psi}{2} & \frac{\cos \frac{\psi}{2}}{\sin \theta_2}
\end{pmatrix},
\\
 E_{\check{\exA}}{}^{\mu} &= \tfrac{1}{\sqrt{2}}\,,\qquad
 \Omega_{A}^{\check{\exA}} = -\tfrac{1}{\sqrt{2}}\begin{pmatrix}
 \frac{\sin \frac{\psi}{2}}{\tan\theta_1} &
 -\frac{\cos \frac{\psi}{2}}{\tan\theta_1} & 0 &
 \frac{\sin \frac{\psi}{2}}{\tan\theta_2} &
 \frac{\cos \frac{\psi}{2}}{\tan\theta_2} & 0 & \cdots & 0\end{pmatrix} ,
\end{align}
in the coordinates $\mathsf{y}^{\sfi}=(\theta_1,\,\phi_1,\,\psi,\,\theta_2,\,\phi_2)$. We find that the $\text{U}(1)$ action has no fixed points and does not introduce any singularities. Using the generalized frame field, we can explicitly express the nine generalized vector fields $K_{\cA} = K_{\cA}{}^I\,\partial_I$ as
\begin{align}
 K_1&= \tfrac{1}{\sqrt{2}} \sin\theta_1\, \bigl(\partial_{16} + \cos\theta_2\,\partial_{18}\bigr)\,,
\\
 K_2&= -\tfrac{1}{\sqrt{2}} \,\bigl(\partial_{8} -q\,\psi \sin\theta_1 \sin\theta_2\,\partial_{26} +\cos\theta_1\,\partial_{7} +\cos\theta_2\,\partial_{10} \bigr)\,,
\\
 K_3&= -\tfrac{1}{\sqrt{2}} \,\bigl(\partial_{13} -q\,\psi \sin\theta_1 \sin\theta_2\,\partial_{27} +\cos\theta_1\,\partial_{12} +\cos\theta_2\,\partial_{15} \bigr)\,,
\\
 K_4+i\,K_5&= -\tfrac{1}{2}\,\Exp{-i\psi}\,\bigl[ \sin\theta_1 \sin\theta_2\,\partial_{23} - \cos\theta_1 \partial_{17}  +\cos\theta_2\,\partial_{21} -\partial_{19} 
\nn\\
 &\qquad\qquad\quad -i (\sin\theta_1\,(\partial_{22}-\cos\theta_2\,\partial_{24}) + \sin\theta_2\,(\cos\theta_1\,\partial_{18} +\partial_{20})) \bigr]\,,
\\
 K_6&= \tfrac{1}{\sqrt{2}}\sin\theta_2\,\bigl(\partial_{25} + \cos\theta_1\,\partial_{24}\bigr)\,,
\\
 K_7&= \tfrac{1}{\sqrt{2}}\sin\theta_1 \sin\theta_2\,\partial_{27}\,,
\\
 K_8&= -\tfrac{1}{\sqrt{2}}\sin\theta_1 \sin\theta_2\,\partial_{26}\,,
\\
 K_*&= -\sqrt{2}\,\partial_{3} \,.
\end{align}
They can be used to write down the reduction ansatz for the 1-form field $A_1{}^I = A_1{}^{\cA}(x)\,K_{\cA}{}^I$. Taking also into account the $\GS$ singlet, $\cM_{AB}\in \mathbb{R}^+ \times \OO(5,3)\in \Edd[6]$, we also construct the generalized metric
\begin{align}
 \cM_{IJ} = \Exp{-4\Cbeta \Delta} E_I{}^{A}\,E_J{}^{B}\,\cM_{AB}\,,\qquad
 \Exp{-2 \Delta} = \tfrac{1}{\sqrt{2}}\,\sin \theta_1\sin \theta_2\,.
\end{align}
The ansatz for the external metric then becomes
\begin{align}
 \mathfrak{g}_{\EXm\EXn} = \Exp{-4\Cbeta\Delta} g_{\EXm\EXn}(x)= 2^{-\frac{1}{3}}\,(\sin\theta_1\sin\theta_2)^{\frac{2}{3}} \,g_{\EXm\EXn}(x)\,.
\end{align}

In particular, if we choose $\cM_{AB}$ to be a specific element of $\GL(5)\subset \Edd[6]$, namely
\begin{align}
\begin{split}
 \cM_{AB} &= \lvert\det m_{ab}\rvert^{\Cbeta}\diag \begin{pmatrix} m_{ab} & m^{ab} & m^{ab} & \cdots \end{pmatrix},
\\
 m_{ab} &= \diag \begin{pmatrix} \frac{1}{6} & \frac{1}{6} & \frac{2}{9} & \frac{1}{6} & \frac{1}{6} \end{pmatrix},
\end{split}
\label{eq:T11-MAB}
\end{align}
we find that the generalized metric contains only the internal metric $g_{\sfi\sfj}= g_{\sfi\sfj}^{T^{1,1}}$, where \(g_{\sfi\sfj}^{T^{1,1}}\) is the standard \(T^{1,1}\) metric
\begin{align}
\begin{split}
 \rmd s^2_{T^{1,1}} &= \tfrac{1}{6}\,\bigl( \rmd\theta_1^2 + \sin^2\theta_1\, \rmd\phi_1^2 + \rmd\theta_2^2 + \sin^2\theta_2\, \rmd\phi_2^2 \bigr)
\\
 &\quad + \tfrac{1}{9}\,\bigl(\rmd\psi + \cos\theta_1\, \rmd\phi_1 + \cos\theta_2\, \rmd\phi_2 \bigr)^2\,,
\end{split}
\end{align}
complemented by the Ramond--Ramond 4-form potential
\begin{align}
 C_{\theta_1\phi_1\theta_2\phi_2} = q\,\psi\sin\theta_1\sin\theta_2\,.
\end{align}
Combining this result with the parameterization $\mathfrak{g}_{\EXm\EXn}= \lvert\det \sfg_{\sfi\sfj}\rvert^{\Cbeta}\, \sfg_{\EXm\EXn}(x)$ of the type IIB Einstein-frame metric, we find
\begin{align}
 \sfg_{\EXm\EXn} = \frac{\Exp{-4\Cbeta\Delta}}{\lvert\det \sfg_{\sfi\sfj}\rvert^{\Cbeta}}\,\gD_{\EXm\EXn}(x) = 18 \,\gD_{\EXm\EXn}(x)\,. 
\end{align}
If we moreover choose $q=-\frac{1}{27}$ and assume that the external metric $\sfg_{\EXm\EXn}$ is given by the AdS$_5$ metric
\begin{align}
 \rmd s^2_{\text{AdS}_5} = z^2 \,\bigl[-\rmd t^2 + (\rmd x^1)+ (\rmd x^2)+ (\rmd x^3)\bigr] + \tfrac{\rmd z^2}{z^2}\,,
\end{align}
the AdS$_5 \times T^{1,1}$ solution of type IIB supergravity, known as the Klebanov--Witten solution \cite{Klebanov:1998hh}, is reproduced. Here, we do not compute the basis of 4-form potentials, but the external part of the 5-form field strength can be found by the self-duality relation $F_5=*F_5$ as
\begin{align}
 F_5 = 4 \,\text{vol}_{\text{AdS}_5}\,,\qquad \text{vol}_{\text{AdS}_5} = \tfrac{z^4}{4}\,\rmd t\wedge \rmd x^1\wedge \rmd x^2\wedge \rmd x^3\wedge \rmd z\,.
\end{align}

\paragraph{\underline{NATD:}}
Various generalized dualities can be performed in this setup. To demonstrate consistency with previous studies in the literature, we now show that the non-abelian T-duality studied in \cite{Itsios:2013wd} can be reproduced by choosing a certain subgroup $H'$. By performing a change of basis $\hat{T}_{\Ah}\to \hat{T}'_{\Ah}=C_{\Ah}{}^{\Bh}\,\hat{T}_{\Bh}$, with a specific $56\times 56$ matrix $C_{\Ah}{}^{\Bh}$, the EGA is mapped to an M-theory EDA with structure constants
\begin{align}
 f'_{45}{}^6 = f'_{64}{}^5 = f'_{56}{}^4 = 1\,,\qquad
 f'_{56}{}^7 = -q\,,\qquad
 f'_1{}^{237} = f'_2{}^{317} = f'_3{}^{127} = 1\,.
\label{eq:t11-natd-f}
\end{align}
Decomposing its generators as $\hat{T}'_{\Ah} = \begin{pmatrix} \hat{T}'_{\hat{a}} & \hat{T}'_{\tilde{\alpha}}\end{pmatrix}$, the generators $\hat{T}'_{\tilde{\alpha}}$ span the Leibniz subalgebra that contains $\mathrm{Lie}(H')$. Since the coset $G/H'$ has seven dimensions, $H'$ is clearly distinct from the original subgroup $H$.

We employ the parameterization
\begin{align}
\begin{split}
 l &= f\,m'\,h'\,,\qquad f=\Exp{x\,(\hat{T}'_4 + \hat{T}'_{13} + q\,\hat{T}'_{47} + q\,\hat{T}'_{56})} \in \GS\,, 
\\
 m' &=\Exp{\psi\,\hat{T}'_4} \Exp{\theta\,\hat{T}'_6} \Exp{\phi\,\hat{T}'_4} \Exp{y_1\,\hat{T}'_1} \Exp{y_2\,\hat{T}'_2} \Exp{z\,\hat{T}'_7} \in \GS\backslash G/H\,,\qquad h'\in H'\,,
\end{split}
\end{align}
where $f$ remains unchanged due to the relation
\begin{align}
 \hat{T}'_4 + \hat{T}'_{13} + q\,\bigl(\hat{T}'_{47} + \hat{T}'_{56}\bigr) = \hat{T}_1+\hat{T}_4\,.
\end{align}
Following the procedure outlined in preceding subsections, we compute the generalized frame field. While the full matrix is too large to display here, the generalized vector fields $K'_{\cA} = K_{\cA}{}^{B}\,E'_{B}{}^I\,\partial_I$ in coordinates $y^i=(\psi,\, \theta,\, \phi,\,x_1,\,x_2,\,z)$ are given by
\begin{align}
 K'_{1}&=-\tfrac{1}{\sqrt{2}}\,\bigl(\partial_{6}+\partial_{10}+\cos \theta\,\partial_{17}\bigr)\,,
\\
 K'_{2}&=\tfrac{1}{\sqrt{2}}\,\bigl[-\partial_{5} + q\,\partial_{8} +\partial_{11} + \cos \theta\,(\partial_{18}-q\, \partial_{8}) \bigr]\,,
\\
 K'_{3}&=\tfrac{1}{\sqrt{2}}\,\bigl[x_1\,\partial_{9} + x_2\,(\partial_{6}+\partial_{19}) + \cos \theta\,(x_1\,\partial_{16}+ x_2\, \partial_{17}) \bigr]\,,
\\
 K'_{4}+i\,K'_{5}&= \tfrac{1}{2}\,\Exp{i\,\psi}\,\bigl[ \cos\theta\,(q\,\partial_{22} +\partial_{27}) -q\,\partial_{22} -x_1 \sin\theta\,\partial_{8} +\partial_{13}-\partial_{25} 
\nonumber\\
 &\qquad\qquad -i\, \bigl(\sin\theta\, (\partial_{16}-\partial_{26}) -x_1 \cos\theta\,\partial_{12} + x_1\,\partial_{7}\bigr)\bigr]\,,
\\
 K'_{6}&=\tfrac{1}{\sqrt{2}}\sin \theta\,\bigl[ -x_1\,\partial_{23} + x_2\,(\partial_{12}-\partial_{24}) \bigr]\,, 
\\
 K'_{7}&= -\tfrac{1}{\sqrt{2}}\,x_1 \sin \theta\,\partial_{22}\,,
\\
 K'_{8}&=\tfrac{1}{\sqrt{2}}\sin \theta\,\bigl(\partial_{12}-\partial_{24}\bigr)\,, 
\\
 K'_{*}&=\sqrt{2}\, \bigl(\partial_{1} + q\,x_2 \sin \theta\,\partial_{12}\bigr)\,.
\end{align}
The generalized metric takes the form
\begin{align}
 \cM'_{IJ} = \Exp{-4\Cbeta \Delta'} E'_I{}^{A}\,E'_J{}^{B}\,\cM_{AB}\,,\qquad
 \Exp{-2 \Delta'} = \tfrac{1}{\sqrt{2}}\,x_1\sin\theta\,.
\end{align}
Again the $\text{U}(1)$ action has no fixed point, hence no singularity appears.

For the specific configuration \eqref{eq:T11-MAB} corresponding to the Klebanov--Witten solution in the original duality frame, the generalized metric encodes the internal part of the 11D metric and the 3-form potential. The internal metric reads
\begin{align}
 \rmd s'^2 &= \bm{\Delta}^{-\frac{2}{3}}\,\bigl[ (x_1^2+\lambda^2\,\lambda_1^2)\,\rmd x_1^2 + (x_2^2+\lambda_1^4)\,\rmd x_2^2 + 2\,x_1\,x_2\,\rmd x_1\,\rmd x_2
 + \lambda^2\,\lambda_1^2\,x_1^2\,\sigma_3^2
 + \bigl(\rmd x_\sharp - q\,\sigma_3 \bigr)^2\bigr]
\nn\\
 &\quad + \bm{\Delta}^{\frac{1}{3}}\,\lambda_1^2\,\bigl(\rmd \theta^2+\sin^2\theta\,\rmd \phi^2\bigr)\,,
\end{align}
where
\begin{align}
\begin{split}
 \sigma_3&=\rmd \psi + \cos \theta\,\rmd\phi\,,\quad 
 x_\sharp =z + q\,(\phi+\psi)\,,\quad 
 \bm{\Delta} = \lambda_2^2\,x_1^2 + \lambda^2 \bigl(x_2^2+\lambda_2^4\bigr)\,,
\\
 \lambda_1 &= \lambda_2 = \tfrac{1}{\sqrt{6}}\,,\quad
 \lambda=\tfrac{1}{3} \,,
\end{split}
\end{align}
while the resulting 3-form potential reads
\begin{align}
 C'_3 = q\,x_2\sin\theta\,\sigma_3 \wedge \rmd \theta\wedge\rmd\phi - \lambda^2\,\bm{\Delta}^{-1} \bigl[x_1\,x_2\,\rmd x_1 +(x_2^2+\lambda_4^4)\,\rmd x_2\bigr]\wedge \sigma_3\wedge \rmd x_\sharp\,.
\end{align}
From the ansatz for the external metric, we furthermore find
\begin{align}
 \mathfrak{g}'_{\EXm\EXn} = \Exp{-4\,\Cbeta\,\Delta'} \gD_{\EXm\EXn}(x)= 2^{-\frac{1}{3}}\,(x_1\sin\theta)^{\frac{2}{3}} \,\gD_{\EXm\EXn}(x)\,,
\end{align}
and using the parameterization $\mathfrak{g}'_{\EXm\EXn}= \lvert\det g'_{ij}\rvert^{\Cbeta}\, g'_{\EXm\EXn}$, the external part of the standard 11D supergravity metric becomes
\begin{align}
 g'_{\EXm\EXn} = 18\,\bm{\Delta}^{\frac{1}{3}}\,\gD_{\EXm\EXn}(x) = \bm{\Delta}^{\frac{1}{3}}\,g_{\EXm\EXn}^{\text{AdS$_5$}}(x)\,. 
\end{align}
These results agree perfectly with eqs.~(3.31)--(3.32) of \cite{Itsios:2013wd}.

It is worth emphasizing the methodological distinction from \cite{Itsios:2013wd}. That work employs the traditional non-abelian T-duality procedure derived from the string $\sigma$-model, proceeding in two steps: first computing NS--NS sector fields, then obtaining Ramond--Ramond fields via the corresponding duality transformation. In contrast, our approach constructs all bosonic fields directly from the algebraic data encoded in \eqref{eq:t11-natd-f}. Moreover, our duality map applies not only to the Klebanov--Witten solution but to arbitrary configurations within five-dimensional gauged supergravity.

\paragraph{\underline{Generalized duality:}}
We also find a basis where the algebra takes the form of a type IIB EDA, with structure constants
\begin{align}
 f''_{1}{}_{\bm{1}}^{23} 
 = -f''_{2}{}_{\bm{1}}^{13} 
 = f''_{3}{}_{\bm{1}}^{12} 
 = f''_{4}{}_{\bm{1}}^{56} 
 = -f''_{5}{}_{\bm{1}}^{46} 
 = f''_{6}{}_{\bm{1}}^{45} = 1\,,\qquad
 f''_{4\bm{1}}{}^{\bm{2}} = -q\,.
\end{align}
The Killing form of the subgroup $H''$ has six negative eigenvalues, indicating that $H''$ is inequivalent to $H$. Moreover, $\GS$ is contained within $H''$. Upon decomposing
\begin{align}
 l = f\,m''\,h''\,,\qquad m'' \in \GS\backslash G/H''\,, \qquad h''\in H''\,, 
\end{align}
a fixed point or singularity appears at the identity element $m''=1$. While it is in principle possible to construct the generalized frame field and extract the corresponding supergravity fields through a more explicit parameterization, the resulting expressions become rather cumbersome. We therefore omit these details here.

\subsection{Quarter-maximal theories}
To construct truncations to quarter-maximal theories, we consult the classification of \cite{Josse:2021put}, where consistent truncations of $\Edd[6]$ ExFT to quarter-maximal theories have been systematically studied. To keep the extended duality group $\Edd[p]$ finite dimensional, we focus on examples with $\GS = \text{U}(1)$ and $\GDM = \Edd[7]$. Specifically, we consider two cases from Table 1 of \cite{Josse:2021put} with $(n_{\text{VT}},\,n_{\text{H}}) = (8,0)$ and $(n_{\text{VT}},\,n_{\text{H}}) = (4,1)$, where $n_{\text{VT}}$ and $n_{\text{H}}$ denote the numbers of vector and hyper multiplets in the resulting $\cN = 2$ supergravity.

To identify the structure group $\GS = \text{U}(1)$, consider the branching of the maximal compact subgroup $\text{USp}(8) \subset \Edd[6]$ as
\begin{align}
 \text{USp}(8) \supset \text{USp}(6)\times \text{SU}(2)_R\,.
\end{align}
In our convention, the Cartan generator $h$ of $\text{SU}(2)_R$ takes the form
\begin{align}
 \omega_0 = K^1{}_2 - K^2{}_1 + K^3{}_4 - K^4{}_3 + K^5{}_6 - K^6{}_5 - \bigl(R_{123456} + R^{123456}\bigr)\,,
\end{align}
and $\text{USp}(6)$ is defined as the commutant of $\omega_0$ in $\text{USp}(8)$. Under this branching, the spinorial representation $\bm{8}$ of $\text{USp}(8)$ decomposes as
\begin{align}
 \bm{8} = (\bm{6},\bm{1}) \oplus (\bm{1},\bm{2})\,,
\end{align}
and retaining only the second representation, which is a $\text{USp}(6)$ singlet, preserves only $\cN=2$ supersymmetry. Furthermore, $\text{USp}(6)$ contains the subgroup
\begin{align}
 \text{USp}(6) \supset \text{SU}(3)\times \text{U}(1)'\,,
\end{align}
with the generator of $\text{U}(1)'$ given by
\begin{align}
 \omega' = K^1{}_2 - K^2{}_1 + K^3{}_4 - K^4{}_3 + K^5{}_6 - K^6{}_5 + 3\,\bigl(R_{123456} + R^{123456}\bigr)\,.
\end{align}
Then $\text{SU}(3)$ is the commutant of $\omega'$ in $\text{USp}(6)$. We further consider the branching
\begin{align}
 \text{SU}(3) \supset \text{SU}(2)\times \text{U}(1) 
\end{align}
with the $\text{U}(1)$ generator given by
\begin{align}
 \omega'' = K^1{}_2 - K^2{}_1 + K^3{}_4 - K^4{}_3 -2\,\bigl( K^5{}_6 - K^6{}_5\bigr)\,.
\end{align}
According to \cite{Josse:2021put}, choosing $\omega'$ as the generator of the structure group $\GS$ leads to the spectrum $(n_{\text{VT}},\,n_{\text{H}}) = (8,0)$, while choosing $\omega''$ gives $(n_{\text{VT}},\,n_{\text{H}}) = (4,1)$. We now confirm these results and explicitly construct the corresponding generalized frame fields and associated invariant tensors.

In this subsection, we consider an M-theory EGA characterized by the structure constants
\begin{align}
\begin{split}
 f_{1}{}^{234567}&=\lambda\,,\quad
 f_{2}{}^{134567}=-1\,,\quad
 f_{3}{}^{124567}=1\,,\quad
 f_{4}{}^{123567}=-\sigma\,,\quad
 f_{5}{}^{123467}=\sigma\,,
\\
 f_{6}{}^{123457}&=-\sigma\,,\quad
 f_{7}{}^{123456}=\sigma\,,\quad
 f_{1234567} =\lambda\,,
\end{split}
\label{eq:CSO-f}
\end{align}
where $\lambda=0,1$ and $\sigma=\pm 1$. The case $(n_{\text{VT}},\,n_{\text{H}}) = (8,0)$ is realized by $\lambda=1$, while $(n_{\text{VT}},\,n_{\text{H}}) = (4,1)$ corresponds to $\lambda=0$. Before proceeding, we identify the Lie group $G$ associated with this EGA.

For $\lambda=1$, the subspace $\cI$ is spanned by the 28 generators
\begin{align}
 \{\hat{T}_8,\,\dotsc,\,\hat{T}_{28},\,\hat{T}_{50},\,\dotsc,\,\hat{T}_{56}\}\,,
\end{align}
and the Lie group $G$ is 28-dimensional:
\begin{align}
 G= \begin{cases} \SO(8) & (\sigma =1) \\ \SO(4,4) & (\sigma=-1) 
\end{cases}.
\end{align}
For $\lambda=0$, the subspace $\cI$ is spanned by 27 generators,
\begin{align}
 \{\hat{T}_8,\,\dotsc,\,\hat{T}_{28},\,\hat{T}_{51},\,\dotsc,\,\hat{T}_{56}\}\,,
\end{align}
and the Lie group $G$ is 29-dimensional:
\begin{align}
 G= \begin{cases} \text{CSO(6,0,2)}\times \text{U}(1) & (\sigma =1) \\ \text{CSO(4,2,2)}\times \text{U}(1) & (\sigma=-1). 
\end{cases}
\end{align}

To obtain quarter-maximal supergravities, we must choose the structure group $\GS = \text{U}(1)$ appropriately. We now identify a suitable $\text{U}(1)$ subgroup within these Lie groups and construct the corresponding generalized frame fields by choosing the subgroup $H$.

\subsubsection{Example 1: \texorpdfstring{$(n_{\text{VT}},\,n_{\text{H}})=(8,0)$}{(nVT,nH)=(8,0)}}\label{sec:ex5-ExFT}
We perform the change of basis 
\begin{align}
 \check{T}_{\Ah} = C_{\Ah}{}^{\Bh}\,\hat{T}_{\Bh}\,,\qquad
 C_{\Ah}{}^{\Bh} = \Exp{\frac{\sigma}{\sqrt{3}}\,R^{123}}\Exp{\frac{1}{\sqrt{3}}\,R^{145}}\Exp{\frac{1}{\sqrt{3}}\,R^{167}}\,,
\end{align}
under which the algebra is mapped to an M-theory EDA with structure constants
\begin{align}
\begin{split}
 -f_{12}{}^{3}&= 
 f_{13}{}^{2}= 
 -f_{14}{}^{5}= 
 f_{15}{}^{4}=
 -f_{16}{}^{7}=
 f_{17}{}^{6}=\tfrac{1}{3} \,,
\\
 f_{2}{}^{345}&= 
 f_{2}{}^{367}= 
 -f_{3}{}^{245}= 
 -f_{3}{}^{267}= 
 f_{4}{}^{567}= 
 -f_{5}{}^{467}= 
 f_{6}{}^{457}= 
 -f_{7}{}^{456}=\tfrac{1}{\sqrt{3}}\,,
\\
 f_{4}{}^{235}&= 
 -f_{5}{}^{234}= 
 f_{6}{}^{237}= 
 -f_{7}{}^{236}=\tfrac{\sigma}{\sqrt{3}}\,,
\\
 f_{1}{}^{234567}&= 
 -f_{2}{}^{134567}= 
 f_{3}{}^{124567}= 
 f_{1234567} =1\,,
\\
 -f_{4}{}^{123567}&= 
 f_{5}{}^{123467}= 
 -f_{6}{}^{123457}= 
 f_{7}{}^{123456}=\sigma\,.
\end{split}
\end{align}
In this basis, the $\text{U}(1)$ generator of the structure group $\GS$ is $\check{T}_{\exA}$ ($\exA=1$). Explicitly,
\begin{align}
 \check{T}_1 = \hat{T}_1
 + \tfrac{1}{\sqrt{3}}\,\bigl(\sigma\, \hat{T}_{14} + \hat{T}_{23}+\hat{T}_{28}\bigr)
 -\tfrac{\sigma}{3} \, \bigl(\hat{T}_{29}+\hat{T}_{34}\bigr)
 -\tfrac{1}{3}\,\hat{T}_{43}
 +\tfrac{\sigma}{3 \sqrt{3}}\, \hat{T}_{50}\,.
\end{align}
Decomposing the generators as
\begin{align}
 \check{T}_{\Ah} = \begin{pmatrix} \check{T}_{\exA}&\check{T}_{A}&\check{T}^{\exA \AA}\end{pmatrix},
\end{align}
we identify the representation matrix
\begin{align}
 (\check{T}_{1})_A{}^B = \tfrac{1}{3}\,\omega'\,.
\end{align}
Consequently, the truncated theory corresponds to the $(n_{\text{VT}},\,n_{\text{H}})=(8,0)$ class. 

The generalized Ricci tensor takes the form
\begin{align}
 R_{AB} = \begin{pmatrix}
 \bm{1}_2 &  & & & & & -\frac{1}{3}\,\bm{1}_2 \\
 & \sigma\,\bm{1}_2 &  & & & -\frac{\sigma}{3}\,\bm{1}_2 & \\
 & & \sigma\,\bm{1}_2 &  & -\frac{\sigma}{3}\,\bm{1}_2 & & \\
 & & & \bm{0}_{15} & & & \\
 &  & -\frac{\sigma}{3}\,\bm{1}_2& & \sigma\,\bm{1}_2& & \\
 & -\frac{\sigma}{3}\,\bm{1}_2& & & & \sigma\,\bm{1}_2 & \\
 -\frac{1}{3}\,\bm{1}_2 & & & & & & \bm{1}_2
\end{pmatrix},
\end{align}
where zero entries are omitted, $\bm{1}_2$ denotes the $2\times 2$ identity matrix, and $\bm{0}_{15}$ is the $15\times 15$ zero matrix. The generalized torsion $X_{AB}{}^C$ contains many non-zero components, which we omit.

The commutant of the structure group within $\Edd[6]$ is $\mathrm{C}_{\Edd[6]}(\text{U}(1)) = \SL(3,\mathbb{C}) \times \text{SU}(2)_R$, and the truncated theory contains eight scalar fields parameterizing the coset
\begin{align}
 \cM_{AB} \in \SL(3,\mathbb{C})/\text{SU}(3)\,.
\end{align}
There are nine null vectors $K_{\cA}{}^B$ ($\cA=1,\dotsc,9$) given by
\begin{alignat}{5}
 K_1 &= \tfrac{\check{T}_{14} - \check{T}_{11}}{\sqrt{2}}\,,&\quad
 K_2 &= \tfrac{\check{T}_{10} + \check{T}_{15}}{\sqrt{2}}\,,&\quad
 K_3 &= \tfrac{\check{T}_{7} - \sigma \check{T}_{21}}{\sqrt{2}}\,,&\quad
 K_4 &= \tfrac{\check{T}_{17} + \check{T}_{20}}{\sqrt{2}}\,,&\quad
 K_5 &= \tfrac{\check{T}_{18} - \check{T}_{19}}{\sqrt{2}}\,, 
\nn\\
 K_6 &= -\tfrac{\check{T}_{8}+\check{T}_{13}}{\sqrt{2}}\,,&
 K_7 &= \tfrac{\check{T}_{9} - \check{T}_{12}}{\sqrt{2}}\,,&
 K_8 &=\check{T}_{16}\,,&
 K_9 &=\tfrac{\sigma \check{T}_{7} + \check{T}_{21}}{\sqrt{2}}\,,& &
\end{alignat}
where $K_{\cA} = K_{\cA}{}^B\,\check{T}_B$. These vectors generate the nine vector fields in the truncated theory.

We now compute the components of the intrinsic torsion $\cT_{\cA\cB}{}^{\cC}$. Defining $\kappa_{\cA\cB} = \cT_{\cA\cC}{}^{\cD}\,\cT_{\cB\cD}{}^{\cC}$, we obtain a totally antisymmetric tensor $\cT_{\cA\cB\cC} = \cT_{\cA\cB}{}^{\cD}\,\kappa_{\cD\cC}$ with components
\begin{align}
\begin{split}
 \cT_{123} &= - \tfrac{2\sqrt{2}}{\sqrt{3}}\,,\qquad
 \cT_{458} = 
 \sigma\,\cT_{678} = - \tfrac{2}{\sqrt{3}}\,,
\\
 - \cT_{147} &=
 \cT_{156} =
 -\cT_{246} =
 -\cT_{257} =
 -\cT_{345} =
 \sigma\,\cT_{367} =
 \sigma\,\cT_{459} =
 \cT_{679} = \tfrac{\sigma \sqrt{2}}{\sqrt{3}}\,,
\end{split}
\\
 \kappa_{\cA\cB}&=
\begin{pmatrix}
 -2\sigma  & 0 & 0 & 0 & 0 & 0 & 0 & 0 & 0 \\
 0 & -2\sigma & 0 & 0 & 0 & 0 & 0 & 0 & 0 \\
 0 & 0 & -2 & 0 & 0 & 0 & 0 & 0 & 0 \\
 0 & 0 & 0 & -2 & 0 & 0 & 0 & 0 & 0 \\
 0 & 0 & 0 & 0 & -2 & 0 & 0 & 0 & 0 \\
 0 & 0 & 0 & 0 & 0 & -2\sigma & 0 & 0 & 0 \\
 0 & 0 & 0 & 0 & 0 & 0 & -2\sigma & 0 & 0 \\
 0 & 0 & 0 & 0 & 0 & 0 & 0 & -\frac{4}{3} & \frac{2 \sqrt{2}}{3} \\
 0 & 0 & 0 & 0 & 0 & 0 & 0 & \frac{2 \sqrt{2}}{3} & -\frac{2}{3}\end{pmatrix}.
\end{align}
The corresponding gauge group is
\begin{align}
\cG = \begin{cases}
 \text{SU}(3) \times \text{U}(1) & (\sigma = +1) \\
 \text{SU}(2,1) \times \text{U}(1) & (\sigma = -1) ,
\end{cases}
\end{align}
where the $\text{U}(1)$ factor is generated by a specific linear combination of a generator in $\text{SU}(2)_R$ and the generator of the structure group $\GS = \text{U}(1)$. This gauge group matches the classification in Table 3 of \cite{Josse:2021put}, providing a concrete realization through exceptional generalized cosets.

In the $\Edd[6]$ case, the $R_2$ representation is isomorphic to the $R_1$-representation, and the $\eta$-symbol $\eta_{AB;\CC}$ can be redefined as a totally symmetric invariant tensor $d_{ABC}=\eta_{AB;\CC}\,\chi_C^{\CC}$. Contracting with the null eigenvectors yields the non-vanishing components
\begin{align}
\begin{split}
 d_{118} &= d_{228} = -1\,, \qquad d_{899} = -d_{338} = \sigma\,,
\\
 d_{146} &= d_{157} = -d_{247} = d_{256} = d_{344} = d_{355} = d_{669} = d_{779} = -\tfrac{1}{\sqrt{2}}\,,
\\
 d_{366} &= d_{377} = -d_{449} = -d_{559} = \tfrac{\sigma}{\sqrt{2}}\,.
\end{split}
\end{align}

We now construct the generalized frame field. Working in the basis that realizes \eqref{eq:CSO-f}, we adopt the parameterization
\begin{align}
\begin{split}
 l &= f\,m\,h\,,\qquad f=\Exp{s\,\check{T}_1}\in \GS\,,
\\
 m &=\Exp{x\,\hat{T}_2} \Exp{y\,\hat{T}_3}\Exp{z\,\hat{T}_4} \Exp{p\,\hat{T}_5}\Exp{q\,\hat{T}_6} \Exp{r\,\hat{T}_7} \in \GS\backslash G/H\,,\qquad
 h\in H\,.
\end{split}
\end{align}
Computing the matrix $M_{\Ah}{}^{\Bh}$, we find no fixed point. The explicit form of the generalized frame field depends on $\sigma$; here we present the case $\sigma=+1$, corresponding to $G=\SO(8)$ and $H=\SO(7)$. We obtain
\begin{align}\label{eq:EformQuadMax}
 E_A{}^I = \Exp{-2\,\Cbeta\,\Delta} L_A{}^B\,\bar{\cE}_B{}^I\,,
\end{align}
where
\begin{align}
  L_A{}^B &\equiv \begin{pmatrix}
 \delta_a^b & 0 & 0 \\
 \Pi^{ba_1a_2} & 2\,\delta_{a_1a_2}^{b_1b_2} & 0 \\
 -5\,\Pi^{[a_1a_2a_3}\,\Pi^{a_4a_5]b}  & -20\,\delta_{b_1b_2}^{[a_1a_2}\,\Pi^{a_3a_4a_5]} & 5!\,\delta_{a_1\cdots a_5}^{b_1\cdots b_5} 
 \end{pmatrix},\qquad
 \Exp{-2\, \Delta} = c_xc_y^2 c_z^3 c_p^4 c_q^5 c_r^6 \,,
\nn\\
 \bar{\cE}_A{}^I&\equiv (\det e_a{}^i)^{\Cbeta}\begin{pmatrix}
 \bar{e}_a{}^i & 0 & 0 \\
 0 & 2\,\bar{e}_{[i_1}{}^{a_1}\,\bar{e}_{i_2]}{}^{a_2} & 0 \\
 0 & 0 & 5!\,\bar{e}_{[i_1}{}^{a_1}\cdots \bar{e}_{i_5]}{}^{a_5}
 \end{pmatrix},
\nn\\
 \bar{e}_a{}^i&=(c_pc_qc_rc_xc_yc_z)^{-\frac{2}{3}}\begin{pmatrix}
 1 & 0 & 0 & 0 & 0 & 0 \\
 s_x t_y & c_x & 0 & 0 & 0 & 0 \\
 \frac{s_x t_z}{c_y} & c_x s_y t_z & c_x c_y & 0 & 0 & 0 \\
 \frac{s_x t_p}{c_y c_z} & \frac{c_x s_y t_p}{c_z} & c_x c_y s_z t_p & c_x c_y c_z & 0 & 0 \\
 \frac{s_x t_q}{c_y c_z c_p} & \frac{c_x s_y t_q}{c_z c_p} & \frac{c_x c_y s_z t_q}{c_p} & c_x c_y c_z s_p t_q & c_x c_y c_z c_p & 0 \\
 \frac{s_x t_r}{c_y c_z c_p c_q} & \frac{c_x s_y t_r}{c_z c_p c_q} & \frac{c_x c_y s_z t_r}{c_p c_q} & \frac{c_x c_y c_z s_p t_r}{c_q} & c_x c_y c_z c_p s_q t_r & c_x c_y c_z c_p c_q\end{pmatrix},
\end{align}
with $a=2,\dotsc,7$\,, $i=2,\dotsc,7$\,, and the non-vanishing components of the 3-vector $\Pi^{abc}$
\begin{align}
 \Pi^{234}&= \tfrac{t_p}{\sqrt{3}\,c_x c_y c_z}\,,\quad
 \Pi^{235}= -\tfrac{t_z}{\sqrt{3}\,c_x c_y}\,,\quad
 \Pi^{236}= \tfrac{t_r}{\sqrt{3}\,c_p c_q c_x c_y c_z}\,,\quad
 \Pi^{237}= -\tfrac{t_q}{\sqrt{3}\,c_p c_x c_y c_z}\,,
\nn\\
 \Pi^{245}&= \tfrac{t_y}{\sqrt{3}\,c_x}\,,\quad
 \Pi^{267}= \tfrac{t_y}{\sqrt{3}\,c_x}\,,\quad
 \Pi^{345}= -\tfrac{t_x}{\sqrt{3}}\,,\quad
 \Pi^{367}= -\tfrac{t_x}{\sqrt{3}}\,,\quad
 \Pi^{456}= \tfrac{t_r}{\sqrt{3}\,c_p c_q c_x c_y c_z}\,,
\nn\\
 \Pi^{457}&= -\tfrac{t_q}{\sqrt{3}\,c_p c_x c_y c_z}\,,\quad
 \Pi^{467}= \tfrac{t_p}{\sqrt{3}\,c_x c_y c_z}\,,\quad
 \Pi^{567}= -\tfrac{t_z}{\sqrt{3}\,c_x c_y}\,.
\end{align}
We use the shorthand notation $\sin x = s_x$, $\cos x = c_x$, and $\tan x = t_x$.

In principle, one can compute the generalized metric by introducing an explicit parameterization of $\cM_{AB}$ and extracting the corresponding supergravity fields. While a more suitable parameterization of the group element $l$ may lead to simpler expressions, we do not pursue this here and proceed to the next example.

\subsubsection{Example 2: \texorpdfstring{$(n_{\text{VT}},\,n_{\text{H}})=(4,1)$}{(nVT,nH)=(4,1)}}\label{sec:ex6-ExFT}
We now consider the case $\lambda=0$. We begin with the change of basis
\begin{align}
 \check{T}_{\Ah} = C_{\Ah}{}^{\Bh}\,\hat{T}_{\Bh}\,,\qquad
 C_{\Ah}{}^{\Bh} = \Exp{\frac{\sqrt{2}}{\sigma}\,R^{123}}\Exp{\sqrt{2}\,R^{145}}\Exp{-\frac{1}{\sqrt{2}}\,R^{167}}\,,
\end{align}
which produces an M-theory EDA with structure constants
\begin{align}
\begin{split}
 f_{12}{}^{3}&= 
 -f_{13}{}^{2}= 
 f_{14}{}^{5}= 
 -f_{15}{}^{4}=1\,,\quad 
 -f_{16}{}^{7}=
 f_{17}{}^{6}=2\,,
\\
 -f_{2}{}^{345}&= 
 f_{3}{}^{245}= 
 -\sigma\,f_{4}{}^{235}= 
 \sigma\,f_{5}{}^{234}= \tfrac{1}{\sqrt{2}}\,,
\\
 f_{2}{}^{367}&= 
 -f_{3}{}^{267}= 
  f_{4}{}^{567}=
 -f_{5}{}^{467}= 
 \sigma\,f_{6}{}^{237}= 
 f_{6}{}^{457}= 
 -\sigma\,f_{7}{}^{236} =
 -f_{7}{}^{456} = \sqrt{2}\,,
\\
 -f_{2}{}^{134567}&= 
 f_{3}{}^{124567}= 
 -\sigma\, f_{4}{}^{123567}= 
 \sigma\,f_{5}{}^{123467}= 
 -\sigma\,f_{6}{}^{123457}= 
 \sigma\,f_{7}{}^{123456}=1\,.
\end{split}
\end{align}
We here take the $\text{U}(1)$ generator $\check{T}_{\exA}$ ($\exA=1$) as the generator of the structure group $\GS$. It is embedded into the full Leibniz algebra by
\begin{align}
 \check{T}_{\Ah} = \begin{pmatrix} \check{T}_{\exA}&\check{T}_{A}&\check{T}^{\exA \AA}\end{pmatrix}.
\end{align}
The representation matrix is
\begin{align}
 (\check{T}_{1})_A{}^B = -\omega''\,,
\end{align}
and the truncated theory is characterized by $(n_{\text{VT}},\,n_{\text{H}})=(4,1)$.

Following the previous examples, we keep the analysis brief. The commutant group is $\mathrm{C}_{\Edd[6]}(\text{U}(1)) = \SO(3,1)\times \mathbb{R}^+\times \text{SU}(2,1)$ and there are $3+1+7$ scalar fields $\cM_{AB}$ in the truncated theory. The generalized Ricci tensor is
\begin{align}
 R_{AB} = \begin{pmatrix}
 &  & & & & & \bm{1}_2 \\
 & &  & & & \sigma \,\bm{1}_2 & \\
 & & &  & -2\,\sigma \,\bm{1}_2 & & \\
 & & & \bm{0}_{15} & & & \\
 &  & -2\,\sigma\,\bm{1}_2& & & & \\
 & \sigma \,\bm{1}_2& & & & & \\
 \bm{1}_2 & & & & & & 
\end{pmatrix},
\end{align}
where zeros are omitted, $\bm{1}_2$ is a $2\times 2$ identity matrix, and $\bm{0}_{15}$ is a $15\times 15$ zero matrix. The generalized Ricci scalar depends on the concrete parameterization of the scalar field $\cM_{AB}$; for example, restricting to $\cM_{AB}\in \frac{\SO(3,1)}{\SO(3)}\times \mathbb{R}^+$ leads to $R=0$. There are five null eigenvectors:
\begin{align}
\begin{split}
 K_{1}{}^A\,\check{T}_A &= \frac{\check{T}_{9}-\check{T}_{12}}{\sqrt{2}} \,,
\qquad
 K_{2}{}^A\,\check{T}_A = \tfrac{\check{T}_{8}+\check{T}_{13}}{\sqrt{2}} \,,
\qquad
 K_{3}{}^A\,\check{T}_A = \tfrac{\sigma\,\check{T}_{7}-\check{T}_{16}}{\sqrt{2}} \,,
\\
 K_{4}{}^A\,\check{T}_A &= \tfrac{\check{T}_{7}+\sigma\,\check{T}_{16}}{\sqrt{2}} \,,
\qquad
 K_{5}{}^A\,\check{T}_A = \check{T}_{21}\,,
\end{split}
\end{align}
where $K_{\cA}=K_{\cA}{}^B\,\check{T}_B$\,. For the embedding tensor, we obtain
\begin{align}
 \cT_{12}{}^3 = \sigma\,,\qquad \cT_{23}{}^1 = \cT_{31}{}^2 = 1\,,
\end{align}
and the gauge group is either $G=\SO(3)\times \mathbb{R}^+\times \text{U}(1)^2$ for $\sigma=1$, or $G=\SO(2,1)\times \mathbb{R}^+\times \text{U}(1)^2$ for $\sigma=-1$\,. The abelian subgroup $\text{U}(1)^2$ is generated by a compact $\text{U}(1)$ generator of $\text{SU}(2,1)$ and the generator of the structure group $\GS=\text{U}(1)$. The totally symmetric tensor $d_{\cA\cB\cC}$ has non-vanishing components
\begin{align}
 d_{115} = -1\,, \qquad d_{225} = 1\,,\qquad
 d_{335} = -\sigma\,,\qquad d_{445} = \sigma \,.
\end{align}

We now construct the generalized frame field. Working with the basis that realizes the algebra \eqref{eq:CSO-f}, we choose the parameterization
\begin{align}
\begin{split}
 l &= f\,m\,h\,,\qquad f=\Exp{s\,\check{T}_1}\in \GS\,,
\\
 m &=\Exp{x\,\hat{T}_2} \Exp{y\,\hat{T}_3}\Exp{z\,\hat{T}_4} \Exp{p\,\hat{T}_5}\Exp{q\,\hat{T}_6} \Exp{r\,\hat{T}_7} \in \GS\backslash G/H\,,\qquad
 h\in H,
\end{split}
\end{align}
where
\begin{align}
 \check{T}_1 = 
 \hat{T}_1 + \sqrt{2}\,\sigma\, \hat{T}_{14} + \sqrt{2}\,\hat{T}_{23}
 - \tfrac{1}{\sqrt{2}}\,\hat{T}_{28}
 - 2\,\sigma\, \hat{T}_{29} + \sigma\,\hat{T}_{34}
 + \hat{T}_{43} - \sqrt{2}\,\sigma\, \hat{T}_{50}\,.
\end{align}
This exceptional generalized coset has a fixed point, and with it a singularity, at $x=y=z=p=q=r=0$ where $m=1$. The generalized frame field is found as
\begin{align}
 E_A{}^I &\equiv \begin{pmatrix}
 \delta_a^i & 0 & 0 \\
 \Pi^{ia_1a_2} & 2\,\delta_{a_1a_2}^{i_1i_2} & 0 \\
 \Pi^{ia_1\cdots a_5}-5\,\Pi^{[a_1a_2a_3}\,\Pi^{a_4a_5]i}  & -20\,\delta_{i_1i_2}^{[a_1a_2}\,\Pi^{a_3a_4a_5]} & 5!\,\delta_{a_1\cdots a_5}^{i_1\cdots i_5} 
 \end{pmatrix},
\end{align}
where 3-vector $\Pi^{abc}$ and 6-vector $\Pi^{a_1\cdots a_6}$ ($a,b=2,\dotsc,7$) are
\begin{align}
\begin{split}
 \Pi^{234} &= \tfrac{\sigma\,p}{\sqrt{2}}\,,\quad
 \Pi^{235} = -\tfrac{\sigma\,z}{\sqrt{2}}\,,\quad
 \Pi^{236} = -\sigma\sqrt{2}\,r\,,\quad
 \Pi^{237} = \sigma\sqrt{2}\,q\,,\quad
 \Pi^{245} = \tfrac{y}{\sqrt{2}}\,,
\\
 \Pi^{267} &= -\sqrt{2}\,y\,,\quad
 \Pi^{345} = -\tfrac{x}{\sqrt{2}}\,,\quad
 \Pi^{367} = \sqrt{2}\,x\,,\quad
 \Pi^{456} = -\sqrt{2}\,r\,,\quad
 \Pi^{457} = \sqrt{2}\,q\,,
\\
 \Pi^{467} &= -\sqrt{2}\,p\,,\quad
 \Pi^{567} = \sqrt{2}\,z\,,\quad
 \Pi^{234567} = x\,y \,.
\end{split}
\end{align}
The generalized metric is
\begin{align}
 \cM_{IJ} = E_I{}^{A}\,E_J{}^{B}\,\cM_{AB}\,,\qquad
 \Exp{-2 \Delta} = 1 \,,
\end{align}
and, in principle, the corresponding supergravity fields can be extracted. As in the previous example, the computation is rather involved, and we do not pursue it further here

\section{Summary and Discussion}\label{sec:conclusion}
In this article, we have developed a comprehensive framework for generalized cosets, establishing their relation to consistent truncations and dualities in extended field theories. Our starting point was to gain full control over duality covariant torsion and curvature tensors for the respective duality groups $\GD$, inspired by the algebraic approach to differential geometry. In the simplest case $\GD=\GL(d)$, as detailed in section~\ref{sec:GL}, this led us to trivialize the tangent bundle of the space-time manifold $M$ by introducing the auxiliary mega-space $\hat{M}$, which is parallelizable and thus admits a globally defined frame and a flat connection. In this setting, the torsion of the flat connection encodes both the Riemann tensor and the torsion tensor of $M$, unifying them into a single algebraic structure. The analysis of section~\ref{sec:GL} provides the unifying perspective from which more sophisticated duality groups, such as $\GD=\OO(d,d)$ or $\GD=\Edd$, can be systematically treated, as developed in section~\ref{sec:algebras-decom}. Although the explicit results are quite involved, especially for higher-rank exceptional duality groups, the computation can always be reduced to a level decomposition of a mega-space group $\GM$ which unifies the generalized structure group $\GS$ with the duality group. Starting from this structure, the corresponding covariant torsion and curvature tensors are derived in section~\ref{sec:frames-curvatures}.

With a well-defined notion of curvature and torsion established in section~\ref{sec:frames-curvatures}, section~\ref{sec:gen-coset} demonstrates how spaces with covariantly constant generalized curvature and torsion can be constructed. This condition is equivalent to demanding that their flat-index components be constant. These spaces necessarily take the form of a double coset
\begin{equation}
  M = \GS \backslash G / H\,.
\end{equation}
This construction generalizes the standard homogeneous space characterization given by Theorem~\ref{theorem:AS} on page \pageref{theorem:AS}: while ordinary cosets $\GS\backslash G$ are characterized by covariantly constant Riemann curvature and torsion, generalized cosets require all generalized curvature and torsion tensors to be covariantly constant. In analogy with the standard case for $\GD=\GL(d)$ where $G$ is the isometry group and $\GS$ the isotropy group, the additional subgroup $H$ is a genuinely new ingredient of generalized geometry. In section~\ref{sec:gen-dualities}, we have established that this group allows us to switch between different generalized duality frames for a given generalized coset. Similar behavior is known for the special case of generalized parallelizable spaces, which are distinguished by a trivial structure group. Here we have shown it also holds for arbitrary $\GS$. At least at the level of supergravity, this allows us to give a precise definition of generalized dualities by employing the framework of consistent truncations:
\begin{quote}
 Generalized dualities relate different realizations of the target-space metric and gauge fields obtained from generalized cosets by choosing different admissible subgroups $H$. A consistent truncation on any such realization results in the same lower-dimensional theory, whose spectrum, couplings and (super)symmetries are determined solely by $\GS$ and $G$, and remain independent of $H$.
\end{quote}

With this new perspective, some puzzling aspects of dualities can be revisited. In particular, we gain a better understanding of:
\begin{enumerate}[label=\arabic*)]
  \item How singularities arise (see section~\ref{sec:singularity}) in certain duality frames due to an overlap of the actions of the generalized structure group $\GS$ and $H$. It may be possible to resolve these singularities by appropriately extending the duality group, because there is, of course, no singularity in the mega-space itself. The singularity arises only after taking the quotient that defines the generalized coset. From a physical point of view, extending the duality group would couple the theory to additional fields, which may originate from excitations of extra degrees of freedom localized at the singularity.
  \item Another immediate consequence of our construction is that generalized dualities do not affect the amount of supersymmetry preserved in the truncated theory, because the result of the truncation is independent of $H$. At first glance, this seems to contradict the established view that supersymmetry can, in general, be broken by dualities. Our results instead suggest that supersymmetry is simply realized differently. Although we employ the language of a generalized $\GS$-structure, explicit supersymmetry transformations can be constructed \cite{Coimbra:2011nw,Coimbra:2012af}. It would be interesting to study them further.
\end{enumerate}

A challenge for our construction arises when $p = d+n$, the sum of the internal space dimension $d$ and the dimension of $\GS$, exceeds $8$, as the mega-space duality group $\GM=\Edd[p]$ then becomes infinite-dimensional. One possible way to address this is to consider a smaller mega-space duality group, such as $\GM = \GD \ltimes \GS$. While this minimalistic approach has been discussed in \cite{Hassler:2025rag}, it does not leave room to accommodate the automorphisms $C_{\Ah}{}^{\Bh}$ that underlie the generalized dualities described in section~\ref{sec:differentGS}. Thus, a compromise between the two approaches would be ideal: $\GM$ should be sufficiently large to realize all relevant generalized dualities, yet small enough to allow tractable computations.

Another strategy is to employ a level decomposition, a technique commonly used to handle infinite-dimensional algebras such as $E_{11}$. This method allows one to extract a finite, controllable subset from an otherwise infinite structure. In the present context, although it remains an open question whether all necessary frame fields, connections, and curvatures can be explicitly computed when $p>8$, our current study demonstrates that the construction is well-controlled under a level decomposition. This gives good reason to expect that practical calculations can be carried out successfully within this framework.

Another limitation of our approach is that, for example, when truncating to half-maximal supergravity, it is applicable only to cases with a relatively small number of vector multiplets. This issue has also been noted in \cite{Malek:2017cjn}. However, recent work \cite{Hassler:2024yis} shows that the duality group can be reduced from $\OO(p,p)$ to $\OO(d,d+n)$ with arbitrary $n$. If this method can be generalized to the exceptional groups $\Edd[p]$, it may allow for the realization of $\OO(d,d+n)$ with arbitrary $n$ even after truncating to half-maximal supergravity. Building on this idea, a natural future direction is to develop an extension of our framework that enables the discussion of more general consistent truncations, beyond the half-maximal case, thus broadening the scope of generalized coset constructions.

Even without further modifications, the wide range of explicit examples -- from two-spheres and gauged WZW models to M-theory lifts and Sasaki--Einstein reductions -- demonstrates that generalized cosets provide a powerful framework for consistent truncations and generalized dualities. All examples align with the properties of generalized dualities discussed above, emphasizing that, although dual backgrounds often appear very different geometrically, they are nonetheless physically equivalent. Beyond consolidating known results, our construction suggests several avenues for further research. Important open problems include the full classification of exceptional generalized cosets, the systematic study of supersymmetry breaking in these reductions, and the role of singular loci in duality webs. Extending the formalism to incorporate more general duality groups and non-geometric backgrounds may reveal new patterns relevant for string and M-theory compactifications. Taken together, these results indicate that generalized cosets are not merely technical tools but rather central objects bridging geometry, dualities, and field theory. By demonstrating how consistent truncations and dualities can be encoded in a purely algebraic language, they provide both a conceptual and practical foundation for future developments in extended field theories.
\label{end}

\section*{Acknowledgements}
We would like to thank Gianluca Inverso, David Osten, and Alex Swash for helpful discussions. 
The work of FH is supported by the SONATA BIS grant 2021/42/E/ST2/00304 from the National Science Centre (NCN), Poland. 
The work of YS is supported by JSPS KAKENHI Grant Number JP23K03391.

\newpage
\appendix

\section{Formulas and identities}
\label{app:formulas}
\newcommand{\dprime}{^{\prime\!\prime}}

Throughout this appendix, we adopt the notation
\begin{align}
 \hat{\delta}^{a_1\cdots a_p}_{b_1\cdots b_p} = p!\,\delta^{a_1\cdots a_p}_{b_1\cdots b_p}\,,
\end{align}
and employ the short-hand notation $R^{\bar{a}_p}=R^{a_1\cdots a_p}$ to denote multiple antisymmetric indices. When contracting the indices in $\bar{a}_p$, we use the convention of dividing by $p!$, so that for example $v^{\bar{a}_p}\,w_{\bar{a}_p} = \frac{1}{p!}\,v^{a_1\cdots a_p}\,w_{a_1\cdots a_p}$.

\subsection{Decompositions of \texorpdfstring{$\Edd[d]$}{Ed(d)} tensors}

The U-duality group $\Edd[d]$ admits two important decompositions with respect to subgroups, each corresponding to a different aspect of string theory. The first decomposition involves the subgroup $\GL(d)$, which provides the natural framework for describing M-theory. The second decomposition involves $\GL(d)\times \SL(2)$, which is particularly suited for type IIB supergravity. We refer to these as the M-theory parameterization and the type IIB parameterization, respectively. By explicitly performing these decompositions, we construct various $\Edd[d]$ tensors.

\paragraph{M-theory parameterization:}
In the M-theory parameterization, the generators of $\Edd[d]$ (with $d\leq 8$) decompose into representations of $\GL(d)$ as follows \cite{West:2001as}:
\begin{align}
\begin{split}
 t_{\adja} &\equiv \{ R_{\bar{a}_8,a'},\, R_{\bar{a}_6},\,R_{\bar{a}_3},\,K^a{}_b,\,R^{\bar{a}_3},\,R^{\bar{a}_6},\,R^{\bar{a}_8,a'} \}\,,
\\
 t^{\adja} &\equiv \{ R^{\bar{a}_8,a'},\, R^{\bar{a}_6},\,R^{\bar{a}_3},\,K_a{}^b,\,R_{\bar{a}_3},\,R_{\bar{a}_6},\,R_{\bar{a}_8,a'} \}\,,
\end{split}
\end{align}
where $K_a{}^b=-K^b{}_a$, multiple indices are antisymmetric, and $R_{\bar{a}_8,a'}$ carries a Hook-type structure. The matrix representations in the $R_1$-representation can be explicitly constructed as follows:
\begin{align}
 K^c{}_d &\equiv \tilde{K}^c{}_d - \Cbeta\, \delta_d^c\,t_0 \,,
\\
 \tilde{K}^c{}_d &\equiv {\arraycolsep=1mm \begin{pmatrix}
 \delta_a^c\,\delta_d^b & 0 & 0 & 0 & 0 & 0 & 0 \\
 0 & K_2 & 0 & 0 & 0 & 0 & 0\\
 0 & 0 & K_5 & 0 & 0 & 0 & 0 \\
 0 & 0 & 0 & K_{7,1} & 0 & 0 & 0 \\
 0 & 0 & 0 & 0 & K_{8,3} & 0 & 0 \\
 0 & 0 & 0 & 0 & 0 & K_{8,6} & 0 \\
 0 & 0 & 0 & 0 & 0 & 0 & K_{8,8,1}
 \end{pmatrix} } 
\\
 &\left[\begin{array}{l}
 K_p \equiv - \hat{\delta}_{\bar{d}\bar{e}_{p-1}}^{\bar{a}_p}\,\hat{\delta}_{\bar{b}_p}^{\bar{c}\bar{e}_{p-1}}\,,\quad
 K_{s,t} \equiv -\hat{\delta}^{\bar{a}_s}_{d \bar{e}_{s-1}} \hat{\delta}^{c \bar{e}_{s-1}}_{\bar{b}_s}\hat{\delta}^{\bar{a}'_t}_{\bar{b}'_t} - \hat{\delta}^{\bar{a}_s}_{\bar{b}_s} \hat{\delta}^{\bar{a}'_t}_{d\bar{e}_{t-1}}\hat{\delta}^{c\bar{e}_{t-1}}_{\bar{b}'_t}\,,
\\
 K_{8,8,1} \equiv - \hat{\delta}^{\bar{a}_8}_{d\bar{e}_7} \hat{\delta}^{c\bar{e}_7}_{\bar{b}_8}\hat{\delta}^{\bar{a}'_8}_{\bar{b}'_8}\hat{\delta}^{a\dprime}_{b\dprime} - \hat{\delta}^{\bar{a}_8}_{\bar{b}_8} \hat{\delta}^{\bar{a}'_8}_{d\bar{e}_7}\hat{\delta}^{c\bar{e}_7}_{\bar{b}'_8}\hat{\delta}^{a\dprime}_{b\dprime} - \hat{\delta}^{\bar{a}_8}_{\bar{b}_8} \hat{\delta}^{\bar{a}'_8}_{\bar{b}'_8} \hat{\delta}_{d}^{a\dprime} \hat{\delta}^{c}_{b\dprime} \end{array}\right]\,, 
\\
 R_{\bar{c}_3} &\equiv {\footnotesize 
 {\arraycolsep=0.5mm \begin{pmatrix}
 0 & 0 & 0 & 0 & 0 & 0 & 0 \\
 \hat{\delta}^{b \bar{a}_2}_{\bar{c}_3} & 0 & 0 & 0 & 0 & 0 & 0 \\
 0 & -\hat{\delta}_{\bar{b}_2\bar{c}_3}^{\bar{a}_5} & 0 & 0 & 0 & 0 & 0 \\
 0 & 0 & \!\!\! -\hat{\delta}_{\bar{b}_5 \bar{d}_2}^{\bar{a}_7} \hat{\delta}_{\bar{c}_3}^{\bar{d}_2 a'} + \frac{1}{4} \hat{\delta}_{\bar{b}_5 \bar{c}_3}^{\bar{a}_7 a'} \!\!\! & 0 & 0 & 0 & 0 \\
 0 & 0 & 0 & \!\!\! -\hat{\delta}_{\bar{b}_7d}^{\bar{a}_8} \hat{\delta}_{\bar{c}_3}^{d\bar{e}_2} \hat{\delta}_{\bar{e}_2b'}^{\bar{a}'_3} + \frac{1}{4}\hat{\delta}_{\bar{b}_7 b'}^{\bar{a}_8} \hat{\delta}_{\bar{c}_3}^{\bar{a}'_3} \!\!\! & 0 & 0 & 0 \\
 0 & 0 & 0 & 0 & -\hat{\delta}_{\bar{b}_8}^{\bar{a}_8} \hat{\delta}_{\bar{b}'_3 \bar{c}_3}^{\bar{a}'_6} & 0 & 0 \\
 0 & 0 & 0 & 0 & 0 & -\hat{\delta}_{\bar{b}_8}^{\bar{a}_8} \hat{\delta}_{\bar{b}'_6 \bar{d}_2}^{\bar{a}'_8} \hat{\delta}_{\bar{c}_3}^{a\dprime \bar{d}_2} & 0 
 \end{pmatrix} } } , 
\\
 R^{\bar{c}_3} &\equiv {\footnotesize 
 {\arraycolsep=0.5mm \begin{pmatrix}
 0 & -\hat{\delta}_{a\bar{b}_2}^{\bar{c}_3} & 0 & 0 & 0 & 0 & 0 \\
 0 & 0 & \hat{\delta}^{\bar{a}_2\bar{c}_3}_{\bar{b}_5} & 0 & 0 & 0 & 0 \\
 0 & 0 & 0 & \hat{\delta}^{\bar{a}_5 \bar{d}_2}_{\bar{b}_7} \hat{\delta}^{\bar{c}_3}_{\bar{d}_2b'} -\frac{1}{4} \hat{\delta}^{\bar{a}_5 \bar{c}_3}_{\bar{b}_7 b'} & 0 & 0 & 0 \\
 0 & 0 & 0 & 0& \hat{\delta}^{\bar{a}_7 d}_{\bar{b}_8} \hat{\delta}^{\bar{c}_3}_{d \bar{e}_2} \hat{\delta}^{\bar{e}_2a'}_{\bar{b}_3'}-\frac{1}{4}\hat{\delta}^{\bar{a}_7 a'}_{\bar{b}_8} \hat{\delta}^{\bar{c}_3}_{\bar{b}'_3} & 0 & 0 \\
 0 & 0 & 0 & 0& 0 & \hat{\delta}^{\bar{a}_8}_{\bar{b}_8} \hat{\delta}^{\bar{a}'_3\bar{c}_3}_{\bar{b}'_6} & 0 \\
 0 & 0 & 0 & 0& 0 & 0 & \hat{\delta}^{\bar{a}_8}_{\bar{b}_8} \hat{\delta}^{\bar{a}'_6 \bar{d}_2}_{\bar{b}'_8} \hat{\delta}^{\bar{c}_3}_{b\dprime \bar{d}_2} \\
 0 & 0 & 0 & 0& 0 & 0 & 0 
 \end{pmatrix} } } , 
\\
 R_{\bar{c}_6} &\equiv {\footnotesize {\arraycolsep=1mm \begin{pmatrix}
 0 & 0 & 0 & 0 & 0 & 0 & 0 \\
 0 & 0 & 0 & 0 & 0 & 0 & 0 \\
 \hat{\delta}^{b\bar{a}_5}_{\bar{c}_6} & 0 & 0 & 0 & 0 & 0 & 0 \\
 0 & -\hat{\delta}_{\bar{b}_2}^{a'd} \hat{\delta}_{\bar{c}_6 d}^{\bar{a}_7} - \frac{1}{2} \hat{\delta}_{\bar{c}_6 \bar{b}_2}^{\bar{a}_7 a'} & 0 & 0 & 0 & 0 & 0 \\
 0 & 0 & -\hat{\delta}_{\bar{b}_5 \bar{d}_3}^{\bar{a}_8} \hat{\delta}_{\bar{c}_6}^{\bar{a}'_3\bar{d}_3} & 0 & 0 & 0 & 0 \\
 0 & 0 & 0 & -\hat{\delta}_{d\bar{b}_7}^{\bar{a}_8} \hat{\delta}_{\bar{c}_6b'}^{\bar{a}'_6 d} - \frac{1}{2} \hat{\delta}_{\bar{b}_7b'}^{\bar{a}_8} \hat{\delta}_{\bar{c}_6}^{\bar{a}'_6} & 0 & 0 & 0 \\
 0 & 0 & 0 & 0 & \hat{\delta}_{\bar{b}_8}^{\bar{a}_8} \hat{\delta}_{\bar{c}_6 \bar{d}_2}^{\bar{a}'_8} \hat{\delta}_{\bar{b}'_3}^{a\dprime \bar{d}_2} & 0 & 0 \end{pmatrix} } } ,
\\
 R^{\bar{c}_6} &\equiv {\footnotesize \arraycolsep=1mm \begin{pmatrix}
 0 & 0 & -\hat{\delta}_{a\bar{b}_5}^{\bar{c}_6} & 0 & 0 & 0 & 0 \\
 0 & 0 & 0 & \hat{\delta}^{\bar{a}_2}_{b'd} \hat{\delta}^{\bar{c}_6d}_{\bar{b}_7} + \frac{1}{2} \hat{\delta}^{\bar{c}_6\bar{a}_2}_{\bar{b}_7b'} & 0 & 0 & 0 \\
 0 & 0 & 0 & 0 & \hat{\delta}^{\bar{a}_5\bar{d}_3}_{\bar{b}_8} \hat{\delta}^{\bar{c}_6}_{\bar{b}'_2 \bar{d}_3} & 0 & 0 \\
 0 & 0 & 0 & 0& 0 & \hat{\delta}^{d \bar{a}_7}_{\bar{b}_8} \hat{\delta}^{\bar{c}_6 a'}_{\bar{b}'_6 d} + \frac{1}{2} \hat{\delta}^{\bar{a}_7a'}_{\bar{b}_8} \hat{\delta}^{\bar{c}_6}_{\bar{b}'_6} & 0 \\
 0 & 0 & 0 & 0& 0 & 0 & -\hat{\delta}^{\bar{a}_8}_{\bar{b}_8} \hat{\delta}^{\bar{c}_6\bar{d}_2}_{\bar{b}'_8} \hat{\delta}^{\bar{a}'_3}_{b\dprime\bar{d}_2} \\
 0 & 0 & 0 & 0& 0 & 0 & 0 \\
 0 & 0 & 0 & 0& 0 & 0 & 0 \end{pmatrix}} , 
\\
 R_{\bar{c}_8,c'} &\equiv {\footnotesize \arraycolsep=1mm \begin{pmatrix}
 0 & 0 & 0 & 0 & 0 & 0 & 0 \\
 0 & 0 & 0 & 0 & 0 & 0 & 0 \\
 0 & 0 & 0 & 0 & 0 & 0 & 0 \\
 \hat{\delta}^{b\bar{a}_7}_{\bar{c}_8}\hat{\delta}^{a'}_{c'}-\frac{1}{4} \hat{\delta}_{\bar{c}_8}^{\bar{a}_7a'}\hat{\delta}_{c'}^b & 0 & 0 & 0 & 0 & 0 & 0 \\
 0 & \hat{\delta}_{\bar{c}_8}^{\bar{a}_8} \hat{\delta}^{\bar{a}'_3}_{c' \bar{b}_2} & 0 & 0 & 0 & 0 & 0 \\
 0 & 0 & \hat{\delta}_{\bar{c}_8}^{\bar{a}_8} \hat{\delta}_{c' \bar{b}_5}^{\bar{a}'_6} & 0 & 0 & 0 & 0 \\
 0 & 0 & 0 & \hat{\delta}_{\bar{c}_8}^{\bar{a}_8} \hat{\delta}_{c'\bar{b}_7}^{\bar{a}'_8}\hat{\delta}_{b'}^{a\dprime} - \frac{1}{4}\hat{\delta}_{\bar{c}_8}^{\bar{a}_8}\hat{\delta}_{\bar{b}_7b'}^{\bar{a}'_8}\hat{\delta}_{c'}^{a\dprime} & 0 & 0 & 0 \end{pmatrix}} ,
\\
 R^{\bar{c}_8,c'} &\equiv {\footnotesize \arraycolsep=0.5mm \begin{pmatrix}
 0 & 0 & 0 & -\hat{\delta}_{a\bar{b}_7}^{\bar{c}_8}\hat{\delta}_{b'}^{c'}+\frac{1}{4} \hat{\delta}^{\bar{c}_8}_{\bar{b}_7b'}\hat{\delta}^{c'}_{a} & 0 & 0 & 0 \\
 0 & 0 & 0 & 0 & - \hat{\delta}^{\bar{c}_8}_{\bar{b}_8} \hat{\delta}_{\bar{b}'_3}^{c'\bar{a}_2} & 0 & 0 \\
 0 & 0 & 0 & 0 & 0 & -\hat{\delta}^{\bar{c}_8}_{\bar{b}_8} \hat{\delta}^{c' \bar{a}_5}_{\bar{b}'_6} & 0 \\
 0 & 0 & 0 & 0& 0 & 0 & -\hat{\delta}^{\bar{c}_8}_{\bar{b}_8}\hat{\delta}^{c'\bar{a}_7}_{\bar{b}'_8}\hat{\delta}^{a'}_{b\dprime} + \frac{1}{4}\hat{\delta}^{\bar{c}_8}_{\bar{b}_8}\hat{\delta}^{\bar{a}_7a'}_{\bar{b}'_8}\hat{\delta}^{c'}_{b\dprime} \\
 0 & 0 & 0 & 0& 0 & 0 & 0 \\
 0 & 0 & 0 & 0& 0 & 0 & 0 \\
 0 & 0 & 0 & 0& 0 & 0 & 0 \end{pmatrix}} .
\end{align}
Their explicit matrix forms are remarkably simple combinations of antisymmetrized Kronecker deltas that realize the exceptional Lie algebra, although the terms involving Hook-type tensors appear somewhat less transparent.

\paragraph{Type IIB parameterization:}
The type IIB parameterization provides an alternative decomposition that is particularly natural for type IIB string theory. Here, the generators decompose under $\GL(d-1)\times \SL(2)$ as \cite{Tumanov:2014pfa}
\begin{align}
 \!\!\!
\begin{split}
 t_{\adja} &\equiv \bigl\{ R_{\sfa_1\cdots \sfa_7,\sfa'} ,\, R^{\bm{\alpha}}_{\sfa_1\cdots \sfa_6} ,\, R_{\sfa_1\cdots \sfa_4} ,\, R^{\bm{\alpha}}_{\sfa_1\sfa_2} ,\,K^{\sfa}{}_{\sfb},\,R^{\bm{\alpha}}{}_{\bm{\beta}},\, R_{\bm{\alpha}}^{\sfa_1\sfa_2} ,\, R^{\sfa_1\cdots \sfa_4} ,\, R_{\bm{\alpha}}^{\sfa_1\cdots \sfa_6} \,, R^{\sfa_1\cdots \sfa_7,\sfa'} \bigr\}\,,
\\
 t^{\adji} &\equiv \bigl\{ R^{\sfa_1\cdots \sfa_7,\sfa'} ,\, R_{\bm{\alpha}}^{\sfa_1\cdots \sfa_6} ,\, R_{\sfa_1\cdots \sfa_4} ,\, R^{\bm{\alpha}}_{\sfa_1\sfa_2} ,\,K^{\sfa}{}_{\sfb},\,R^{\bm{\alpha}}{}_{\bm{\beta}},\, R_{\bm{\alpha}}^{\sfa_1\sfa_2} ,\, R^{\sfa_1\cdots \sfa_4} ,\, R_{\bm{\alpha}}^{\sfa_1\cdots \sfa_6} \,, R^{\sfa_1\cdots \sfa_7,\sfa'} \bigr\}\,,
\end{split}
\end{align}
where $\sfa=1,\dotsc,d-1$ labels the $\GL(d-1)$ indices and $\bm{\alpha}=1,2$ labels the $\SL(2)$ doublet indices. The structure constants in this parameterization naturally incorporate the $\SL(2)$ invariant tensor $\epsilon^{\bm{\alpha}\bm{\beta}}$.

The $\GL(d-1)$ generator $K^{\mathsf{c}}{}_{\mathsf{d}}$ again takes a block-diagonal form, now with additional $\SL(2)$ tensor product structure in certain blocks. The $\SL(2)$ generator $R^{\bm{\gamma}}{}_{\bm{\delta}}$ acts only on generators carrying $\SL(2)$ indices. The matrix representations of all generators are summarized as follows:
\begin{align}
 K^{\sfc}{}_{\sfd} &\equiv \tilde{K}^{\sfc}{}_{\sfd} - \Cbeta\, \delta_{\sfd}^{\sfc}\,t_0 \,,
\\
 \tilde{K}^{\sfc}{}_{\sfd} &\equiv 
 {\small{\arraycolsep=1mm \begin{pmatrix}
 \delta_{\sfa}^{\sfc}\, \delta_{\sfd}^{\sfb} & 0 & 0 & 0 & 0 & 0 & 0 & 0 & 0 & 0 \\
 0 & \delta_{\bm{\alpha}}^{\bm{\beta}}\,K_1 & 0 & 0 & 0 & 0 & 0 & 0 & 0 & 0 \\
 0 & 0 & K_3 & 0 & 0 & 0 & 0 & 0 & 0 & 0 \\
 0 & 0 & 0 & \delta_{\bm{\alpha}}^{\bm{\beta}}\,K_5 & 0 & 0 & 0 & 0 & 0 & 0 \\
 0 & 0 & 0 & 0 & K_{6,1} & 0 & 0 & 0 & 0 & 0 \\
 0 & 0 & 0 & 0 & 0 & \delta^{({\bm{\alpha}}_{1}}_{({\bm{\beta}}_{1}}\,\delta^{{\bm{\alpha}}_{2})}_{{\bm{\beta}}_{2})}\, K_7 & 0 & 0 & 0 & 0 \\
 0 & 0 & 0 & 0 & 0 & 0 & \delta_{\bm{\alpha}}^{\bm{\beta}}\,K_{7,2} & 0 & 0 & 0 \\
 0 & 0 & 0 & 0 & 0 & 0 & 0 & K_{7,4} & 0 & 0 \\
 0 & 0 & 0 & 0 & 0 & 0 & 0 & 0 & \delta_{\bm{\alpha}}^{\bm{\beta}}\,K_{7,6} & 0 \\
 0 & 0 & 0 & 0 & 0 & 0 & 0 & 0 & 0 & K_{7,7,1} 
\end{pmatrix}}} 
\\
 &\left[\begin{array}{l}
 K_{p} \equiv -\hat{\delta}_{\sfd\bar{\sfe}_{p-1}}^{\bar{\sfa}_p}\,\hat{\delta}_{\bar{\sfb}_p}^{\sfc \bar{\sfe}_{p-1}} \,, \quad
 K_{s,t} \equiv -\hat{\delta}_{\bar{\sfb}_s}^{\sfc\bar{\sfe}_{s-1}}\,\hat{\delta}_{\sfd\bar{\sfe}_{s-1}}^{\bar{\sfa}_s}\,\hat{\delta}_{\bar{\sfb}'_t}^{\bar{\sfa}'_t} -\hat{\delta}_{\bar{\sfb}_s}^{\bar{\sfa}_s}\,\hat{\delta}_{\sfd\bar{d}_{t-1}}^{\bar{\sfa}'_t}\,\hat{\delta}_{\bar{\sfb}'_t}^{\sfc\bar{d}_{t-1}}\,,
\\
 K_{7,7,1}\equiv -\hat{\delta}_{\bar{\sfa}_7}^{\sfc\bar{\sfe}_6} \hat{\delta}_{\sfd\bar{\sfe}_6}^{\bar{\sfb}_7}\hat{\delta}_{\bar{\sfa}'_7}^{\bar{\sfb}'_7}\hat{\delta}_{\sfa\dprime}^{\sfb\dprime} - \hat{\delta}_{\bar{\sfa}_7}^{\bar{\sfb}_7} \hat{\delta}_{\bar{\sfa}'_7}^{\sfc\bar{\sfe}_6}\hat{\delta}_{\sfd\bar{\sfe}_6}^{\bar{\sfb}'_7}\hat{\delta}_{\sfa\dprime}^{\sfb\dprime} - \hat{\delta}_{\bar{\sfa}_7}^{\bar{\sfb}_7} \hat{\delta}_{\bar{\sfa}'_7}^{\bar{\sfb}'_7} \hat{\delta}^{\sfc}_{\sfa\dprime} \hat{\delta}_{\sfd}^{\sfb\dprime}
 \end{array}\right]\,,
\\
 R^{\bm{\gamma}}{}_{\delta} &\equiv 
 {\small{\arraycolsep=1mm \begin{pmatrix}
 0 & 0 & 0 & 0 & 0 & 0 & 0 & 0 & 0 & 0 \\
 0 & R_1\,\hat{\delta}_{\sfb}^{\sfa} & 0 & 0 & 0 & 0 & 0 & 0 & 0 & 0 \\
 0 & 0 & 0 & 0 & 0 & 0 & 0 & 0 & 0 & 0 \\
 0 & 0 & 0 & R_1 \,\hat{\delta}^{\bar{\sfa}_5}_{\bar{\sfb}_5} & 0 & 0 & 0 & 0 & 0 & 0 \\
 0 & 0 & 0 & 0 & 0 & 0 & 0 & 0 & 0 & 0 \\
 0 & 0 & 0 & 0 & 0 & R_2\,\hat{\delta}^{\bar{\sfa}_7}_{\bar{\sfb}_7} & 0 & 0 & 0 & 0 \\
 0 & 0 & 0 & 0 & 0 & 0 & R_1 \,\hat{\delta}^{\bar{\sfa}_7}_{\bar{\sfb}_7}\,\hat{\delta}^{\bar{\sfa}'_2}_{\bar{\sfb}'_2} & 0 & 0 & 0 \\
 0 & 0 & 0 & 0 & 0 & 0 & 0 & 0 & 0 & 0 \\
 0 & 0 & 0 & 0 & 0 & 0 & 0 & 0 & R_1\,\hat{\delta}^{\bar{\sfa}_7}_{\bar{\sfb}_7}\,\hat{\delta}^{\bar{\sfa}'_6}_{\bar{\sfb}'_6} & 0 \\
 0 & 0 & 0 & 0 & 0 & 0 & 0 & 0 & 0 & 0 
\end{pmatrix}}} ,
\\
 &\Bigl[R_1 \equiv \delta_{\bm{\alpha}}^{\bm{\gamma}}\,\delta^{\bm{\beta}}_{\bm{\delta}}-\tfrac{1}{2}\delta_{\bm{\alpha}}^{\bm{\beta}}\delta^{\bm{\gamma}}_{\bm{\delta}}\,,\qquad 
 R_2 \equiv \delta_{({\bm{\alpha}}_1}^{\bm{\gamma}}\,\delta_{{\bm{\alpha}}_2)}^{\bm{\epsilon}}\,\delta^{({\bm{\beta}}_1}_{\bm{\delta}}\,\delta^{{\bm{\beta}}_2)}_{\bm{\epsilon}} - \tfrac{1}{2}\,\delta_{({\bm{\alpha}}_1}^{({\bm{\beta}}_1}\,\delta_{{\bm{\alpha}}_2)}^{{\bm{\beta}}_2)}\,\delta^{\bm{\gamma}}_{\bm{\delta}}\Bigr]\,,
\\
 R^{\bm{\gamma}}_{\bar{\sfc}_2}
 &\equiv {\footnotesize{\arraycolsep=0.2mm 
 \begin{pmatrix}
 0 & 0 & 0 & 0 & 0 & 0 & 0 & 0 & 0 & 0 \\
 -\delta^{\bm{\gamma}}_{\bm{\alpha}}\hat{\delta}^{\sfa\sfb}_{\bar{\sfc}_2} & 0 & 0 & 0 & 0 & 0 & 0 & 0 & 0 & 0 \\
 0 & \!\!\! -\epsilon^{{\bm{\beta}}{\bm{\gamma}}}\hat{\delta}^{\bar{\sfa}_3}_{\sfb\bar{\sfc}_2} \!\!\! & 0 & 0 & 0 & 0 & 0 & 0 & 0 & 0 \\
 0 & 0 & \!\!\! -\delta^{\bm{\gamma}}_{\bm{\alpha}}\hat{\delta}^{\bar{\sfa}_5}_{\bar{\sfb}_3\bar{\sfc}_2} \!\!\!\!\!\! & 0 & 0 & 0 & 0 & 0 & 0 & 0 \\
 0 & 0 & 0 & \!\!\! \epsilon^{{\bm{\beta}}{\bm{\gamma}}} \biggl[\genfrac{}{}{0pt}{1}{-\hat{\delta}^{\bar{\sfa}_6}_{\bar{\sfb}_5\sfd} \hat{\delta}^{\sfa'\sfd}_{\bar{\sfc}_2}}{-c_2\,\hat{\delta}_{\bar{\sfb}_5\bar{\sfc}_2}^{\bar{\sfa}_6\sfa'}}\biggr] \!\!\!\!\!\! & 0 & 0 & 0 & 0 & 0 & 0 \\
 0 & 0 & 0 & \!\!\!\!\!\! -\delta_{({\bm{\alpha}}_1}^{{\bm{\beta}}}\delta_{{\bm{\alpha}}_2)}^{{\bm{\gamma}}} \hat{\delta}_{\bar{\sfb}_5\bar{\sfc}_2}^{\bar{\sfa}_7} \!\!\!\!\!\! & 0 & 0 & 0 & 0 & 0 & 0 \\
 0 & 0 & 0 & 0 & \!\!\!\!\!\! \delta^{\bm{\gamma}}_{\bm{\alpha}} \biggl[\genfrac{}{}{0pt}{1}{-\hat{\delta}^{\bar{\sfa}_7}_{\bar{\sfb}_6\sfd} \hat{\delta}^{\sfd\sfe}_{\bar{\sfc}_2} \hat{\delta}^{\bar{\sfa}'_2}_{\sfe\sfb'}}{-c_2\,\hat{\delta}^{\bar{\sfa}_7}_{\bar{\sfb}_6\sfb'} \hat{\delta}^{\bar{\sfa}'_2}_{\bar{\sfc}_2}}\biggr] & -\delta^{({\bm{\beta}}_1}_{\bm{\alpha}} \epsilon^{{\bm{\beta}}_2){\bm{\gamma}}} \hat{\delta}^{\bar{\sfa}_7}_{\bar{\sfb}_7}\hat{\delta}^{\bar{\sfa}'_2}_{\bar{\sfc}_2} \!\!\!\!\!\! & 0 & 0 & 0 & 0 \\
 0 & 0 & 0 & 0 & 0 & 0 & \!\!\!\!\!\! -\epsilon^{{\bm{\beta}}{\bm{\gamma}}}\hat{\delta}_{\bar{\sfb}_7}^{\bar{\sfa}_7}\hat{\delta}_{\bar{\sfb}'_2\bar{\sfc}_2}^{\bar{\sfa}'_4} \!\!\! & 0 & 0 & 0 \\
 0 & 0 & 0 & 0 & 0 & 0 & 0 & \!\!\!\!\!\! -\delta^{\bm{\gamma}}_{\bm{\alpha}}\hat{\delta}_{\bar{\sfb}_7}^{\bar{\sfa}_7}\hat{\delta}_{\bar{\sfb}'_4\bar{\sfc}_2}^{\bar{\sfa}'_6} \!\!\! & 0 & 0 \\
 0 & 0 & 0 & 0 & 0 & 0 & 0 & 0 & \!\!\! -\epsilon^{{\bm{\beta}}{\bm{\gamma}}}\hat{\delta}_{\bar{\sfb}_7}^{\bar{\sfa}_7}\hat{\delta}_{\bar{\sfb}'_6\sfd}^{\bar{\sfa}'_7}\hat{\delta}_{\bar{\sfc}_2}^{\sfa\dprime\sfd} & 0 
\end{pmatrix}}}
\nn\\
 &\quad \bigl[c_2 \equiv \tfrac{2}{7}+\tfrac{1}{7\sqrt{2}}\bigr]\,,
\\
 R_{\bm{\gamma}}^{\bar{\sfc}_2}
 &\equiv {\footnotesize{\arraycolsep=0.2mm 
 \begin{pmatrix}
 0 & \delta_{\bm{\gamma}}^{\bm{\beta}}\hat{\delta}_{\sfb\sfa}^{\bar{\sfc}_2} & 0 & 0 & 0 & 0 & 0 & 0 & 0 & 0 \\
 0 & 0 & \!\!\!\epsilon_{{\bm{\alpha}}{\bm{\gamma}}}\hat{\delta}_{\bar{\sfb}_3}^{\sfa\bar{\sfc}_2}\!\!\! & 0 & 0 & 0 & 0 & 0 & 0 & 0 \\
 0 & 0 & 0 & \delta_{\bm{\gamma}}^{\bm{\beta}}\hat{\delta}_{\bar{\sfb}_5}^{\bar{\sfa}_3\bar{\sfc}_2} & 0 & 0 & 0 & 0 & 0 & 0 \\
 0 & 0 & 0 & 0 & \!\!\!\!\!\! \epsilon_{{\bm{\alpha}}{\bm{\gamma}}} \biggl[\genfrac{}{}{0pt}{1}{\hat{\delta}_{\bar{\sfb}_6}^{\bar{\sfa}_5\sfd} \hat{\delta}_{\sfb'\sfd}^{\bar{\sfc}_2}}{+c_2\,\hat{\delta}^{\bar{\sfa}_5\bar{\sfc}_2}_{\bar{\sfb}_6\sfb'}}\biggr] & \delta^{({\bm{\beta}}_1}_{({\bm{\alpha}}} \delta^{{\bm{\beta}}_2)}_{{\bm{\gamma}})} \hat{\delta}^{\bar{\sfa}_5\bar{\sfc}_2}_{\bar{\sfb}_7} \!\!\! & 0 & 0 & 0 & 0 \\
 0 & 0 & 0 & 0 & 0 & 0 & \!\!\!\!\!\!\delta_{\bm{\gamma}}^{\bm{\beta}} \biggl[\genfrac{}{}{0pt}{1}{\hat{\delta}_{\bar{\sfb}_7}^{\bar{\sfa}_6\sfd} \hat{\delta}_{\sfd\sfe}^{\bar{\sfc}_2} \hat{\delta}_{\bar{\sfb}'_2}^{\sfe\sfa'}}{+c_2\,\hat{\delta}_{\bar{\sfb}_7}^{\bar{\sfa}_6\sfa'} \hat{\delta}_{\bar{\sfb}'_2}^{\bar{\sfc}_2}}\biggr]\!\!\!\!\!\! & 0 & 0 & 0 \\
 0 & 0 & 0 & 0 & 0 & 0 & \!\!\!\!\!\! \delta_{({\bm{\alpha}}_1}^{\bm{\beta}} \epsilon_{{\bm{\alpha}}_2){\bm{\gamma}}} \hat{\delta}_{\bar{\sfb}_7}^{\bar{\sfa}_7}\hat{\delta}_{\bar{\sfb}'_2}^{\bar{\sfc}_2}\!\!\!\!\!\! & 0 & 0 & 0 \\
 0 & 0 & 0 & 0 & 0 & 0 & 0 & \epsilon_{{\bm{\alpha}}{\bm{\gamma}}}\hat{\delta}^{\bar{\sfa}_7}_{\bar{\sfb}_7}\hat{\delta}^{\bar{\sfa}'_2\bar{\sfc}_2}_{\bar{\sfb}'_4} & 0 & 0 \\
 0 & 0 & 0 & 0 & 0 & 0 & 0 & 0 & \!\!\!\!\!\! \delta_{\bm{\gamma}}^{\bm{\beta}}\hat{\delta}^{\bar{\sfa}_7}_{\bar{\sfb}_7}\hat{\delta}^{\bar{\sfa}'_4\bar{\sfc}_2}_{\bar{\sfb}'_6} \!\!\!\!\!\! & 0 \\
 0 & 0 & 0 & 0 & 0 & 0 & 0 & 0 & 0 & \epsilon_{{\bm{\alpha}}{\bm{\gamma}}}\hat{\delta}^{\bar{\sfa}_7}_{\bar{\sfb}_7}\hat{\delta}^{\bar{\sfa}'_6\sfd}_{\bar{\sfb}'_7}\hat{\delta}^{\bar{\sfc}_2}_{\sfb\dprime\sfd} \\
 0 & 0 & 0 & 0 & 0 & 0 & 0 & 0 & 0 & 0 
\end{pmatrix}}} ,
\\
 R_{\bar{\sfc}_4} &\equiv {\footnotesize {\arraycolsep=1mm 
 \begin{pmatrix}
 0 & 0 & 0 & 0 & 0 & 0 & 0 & 0 & 0 & 0 \\
 0 & 0 & 0 & 0 & 0 & 0 & 0 & 0 & 0 & 0 \\
 -\hat{\delta}_{\bar{\sfc}_4}^{\bar{\sfa}_3\sfb} & 0 & 0 & 0 & 0 & 0 & 0 & 0 & 0 & 0 \\
 0 & \delta^{\bm{\beta}}_{\bm{\alpha}} \hat{\delta}_{\sfb\bar{\sfc}_4}^{\bar{\sfa}_5} & 0 & 0 & 0 & 0 & 0 & 0 & 0 & 0 \\
 0 & 0 & -\biggl[\genfrac{}{}{0pt}{1}{ \hat{\delta}_{\bar{\sfb}_3\bar{\sfd}_3}^{\bar{\sfa}_6}\hat{\delta}_{\bar{\sfc}_4}^{\sfa'\bar{\sfd}_3}}{+c_4\,\hat{\delta}_{\bar{\sfb}_3\bar{\sfc}_4}^{\bar{\sfa}_6\sfa'}}\biggr] & 0 & 0 & 0 & 0 & 0 & 0 & 0 \\
 0 & 0 & 0 & 0 & 0 & 0 & 0 & 0 & 0 & 0 \\
 0 & 0 & 0 & \!\!\!\!\!\! -\delta^{\bm{\beta}}_{\bm{\alpha}}\hat{\delta}_{\bar{\sfb}_5\bar{\sfd}_2}^{\bar{\sfa}_7}\hat{\delta}_{\bar{\sfc}_4}^{\bar{\sfd}_2\bar{\sfa}'_2} \!\!\!\!\!\! & 0 & 0 & 0 & 0 & 0 & 0 \\
 0 & 0 & 0 & 0 & \!\!\! -\biggl[\genfrac{}{}{0pt}{1}{ \hat{\delta}_{\bar{\sfb}_6\sfd}^{\bar{\sfa}_7}\hat{\delta}_{\bar{\sfe}_3\sfb'}^{\bar{\sfa}'_4}\hat{\delta}_{\bar{\sfc}_4}^{\sfd\bar{\sfe}_3}}{+c_4\,\hat{\delta}_{\bar{\sfb}_6\sfb'}^{\bar{\sfa}_7}\hat{\delta}_{\bar{\sfc}_4}^{\bar{\sfa}'_4}}\biggr] \!\!\! & 0 & 0 & 0 & 0 & 0 \\
 0 & 0 & 0 & 0 & 0 & 0 & \delta^{\bm{\beta}}_{\bm{\alpha}}\hat{\delta}^{\bar{\sfa}_7}_{\bar{\sfb}_7}\hat{\delta}^{\bar{\sfa}'_6}_{\bar{\sfb}'_2\bar{\sfc}_4} & 0 & 0 & 0 \\
 0 & 0 & 0 & 0 & 0 & 0 & 0 & -\hat{\delta}^{\bar{\sfa}_7}_{\bar{\sfb}_7}\hat{\delta}^{\sfa'\bar{\sfd}_3}_{\bar{\sfb}'_4}\hat{\delta}^{\bar{\sfa}'_7}_{\bar{\sfd}_3\bar{\sfc}_4} & 0 & 0 
\end{pmatrix}}}
\nn\\
 &\quad \bigl[c_4\equiv \tfrac{4}{7} + \tfrac{2}{7\sqrt{2}}\bigr]\,,
\\
 R^{\bar{\sfc}_4} &\equiv {\footnotesize {\arraycolsep=1mm 
 \begin{pmatrix}
 0 & 0 & \hat{\delta}^{\bar{\sfc}_4}_{\bar{\sfb}_3\sfa} & 0 & 0 & 0 & 0 & 0 & 0 & 0 \\
 0 & 0 & 0 & -\delta_{\bm{\alpha}}^{\bm{\beta}} \hat{\delta}^{\sfa\bar{\sfc}_4}_{\bar{\sfb}_5} & 0 & 0 & 0 & 0 & 0 & 0 \\
 0 & 0 & 0 & 0 & \!\!\! \biggl[\genfrac{}{}{0pt}{1}{ \hat{\delta}^{\bar{\sfa}_3\bar{\sfd}_3}_{\bar{\sfb}_6}\hat{\delta}^{\bar{\sfc}_4}_{\sfb'\bar{\sfd}_3}}{+c_4\,\hat{\delta}^{\bar{\sfa}_3\bar{\sfc}_4}_{\bar{\sfb}_6\sfb'}}\biggr] & 0 & 0 & 0 & 0 & 0 \\
 0 & 0 & 0 & 0 & 0 & 0 & \delta_{\bm{\alpha}}^{\bm{\beta}}\hat{\delta}^{\bar{\sfa}_5\bar{\sfd}_2}_{\bar{\sfb}_7}\hat{\delta}^{\bar{\sfc}_4}_{\bar{\sfd}_2\bar{\sfb}'_2} & 0 & 0 & 0 \\
 0 & 0 & 0 & 0 & 0 & 0 & 0 & \!\!\! \biggl[\genfrac{}{}{0pt}{1}{ \hat{\delta}^{\bar{\sfa}_6\sfd}_{\bar{\sfb}_7}\hat{\delta}^{\bar{\sfe}_3\sfa'}_{\bar{\sfb}'_4}\hat{\delta}^{\bar{\sfc}_4}_{\sfd\bar{\sfe}_3}}{+c_4\,\hat{\delta}^{\bar{\sfa}_6\sfa'}_{\bar{\sfb}_7}\hat{\delta}^{\bar{\sfc}_4}_{\bar{\sfb}'_4}}\biggr] \!\!\! & 0 & 0 \\
 0 & 0 & 0 & 0 & 0 & 0 & 0 & 0 & 0 & 0 \\
 0 & 0 & 0 & 0 & 0 & 0 & 0 & 0 & \!\!\! -\delta_{\bm{\alpha}}^{\bm{\beta}}\hat{\delta}_{\bar{\sfb}_7}^{\bar{\sfa}_7}\hat{\delta}_{\bar{\sfb}'_6}^{\bar{\sfa}'_2\bar{\sfc}_4} \!\!\! & 0 \\
 0 & 0 & 0 & 0 & 0 & 0 & 0 & 0 & 0 & \hat{\delta}_{\bar{\sfb}_7}^{\bar{\sfa}_7}\hat{\delta}_{\sfb'\bar{\sfd}_3}^{\bar{\sfa}'_4}\hat{\delta}_{\bar{\sfb}'_7}^{\bar{\sfd}_3\bar{\sfc}_4} \\
 0 & 0 & 0 & 0 & 0 & 0 & 0 & 0 & 0 & 0 \\
 0 & 0 & 0 & 0 & 0 & 0 & 0 & 0 & 0 & 0 
\end{pmatrix}}} ,
\\
 R^{\bm{\gamma}}_{\bar{\sfc}_6} &\equiv {\footnotesize {\arraycolsep=0.4mm 
 \begin{pmatrix}
 0 & 0 & 0 & 0 & 0 & 0 & 0 & 0 & 0 & 0 \\[-1mm]
 0 & 0 & 0 & 0 & 0 & 0 & 0 & 0 & 0 & 0 \\[-1mm]
 0 & 0 & 0 & 0 & 0 & 0 & 0 & 0 & 0 & 0 \\[-1mm]
 -\delta^{\bm{\gamma}}_{\bm{\alpha}}\hat{\delta}_{\bar{\sfc}_6}^{\bar{\sfa}_5\sfb} & 0 & 0 & 0 & 0 & 0 & 0 & 0 & 0 & 0 \\[-0.5mm]
 0 & \!\!\! \epsilon^{{\bm{\beta}}{\bm{\gamma}}}\biggl[\genfrac{}{}{0pt}{1}{\hat{\delta}_{\bar{\sfc}_6}^{\bar{\sfa}_6}\hat{\delta}^{\sfa'}_\sfb}{-c_6\,\hat{\delta}_{\sfb\bar{\sfc}_6}^{\bar{\sfa}_6\sfa'}}\biggr] \!\!\! & 0 & 0 & 0 & 0 & 0 & 0 & 0 & 0 \\[-0.5mm]
 0 & \!\!\! \delta_{({\bm{\alpha}}_1}^{({\bm{\beta}}} \delta_{{\bm{\alpha}}_2)}^{{\bm{\gamma}})}\hat{\delta}_{\sfb\bar{\sfc}_6}^{\bar{\sfa}_7} \!\!\! & 0 & 0 & 0 & 0 & 0 & 0 & 0 & 0 \\[-0.5mm]
 0 & 0 & \delta^{\bm{\gamma}}_{\bm{\alpha}}\hat{\delta}_{\bar{\sfc}_6\sfd}^{\bar{\sfa}_7}\hat{\delta}_{\bar{\sfb}_3}^{\bar{\sfa}'_2\sfd} & 0 & 0 & 0 & 0 & 0 & 0 & 0 \\[-0.5mm]
 0 & 0 & 0 & \!\!\! -\epsilon^{{\bm{\beta}}{\bm{\gamma}}}\hat{\delta}_{\bar{\sfc}_6\sfd}^{\bar{\sfa}_7}\hat{\delta}_{\bar{\sfb}_5}^{\bar{\sfa}'_4\sfd} \!\!\! & 0 & 0 & 0 & 0 & 0 & 0 \\[-0.5mm]
 0 & 0 & 0 & 0 & \!\!\! \delta^{\bm{\gamma}}_{\bm{\alpha}}\biggl[\genfrac{}{}{0pt}{1}{-\hat{\delta}_{\bar{\sfb}_6}^{\bar{\sfa}'_6}\hat{\delta}_{\sfb'\bar{\sfc}_6}^{\bar{\sfa}_7}}{+c_6\,\hat{\delta}_{\bar{\sfb}_6\sfb'}^{\bar{\sfa}_7}\hat{\delta}_{\bar{\sfc}_6}^{\bar{\sfa}'_6}}\biggr] & -\delta_{\bm{\alpha}}^{({\bm{\beta}}_1} \epsilon^{{\bm{\beta}}_2){\bm{\gamma}}} \hat{\delta}_{\bar{\sfb}_7}^{\bar{\sfa}_7}\hat{\delta}_{\bar{\sfc}_6}^{\bar{\sfa}'_6} \!\!\! & 0 & 0 & 0 & 0 \\
 0 & 0 & 0 & 0 & 0 & 0 & \!\!\!\!\!\! -\epsilon^{{\bm{\beta}}{\bm{\gamma}}}\hat{\delta}_{\bar{\sfb}_7}^{\bar{\sfa}_7}\hat{\delta}^{\bar{\sfa}'_7}_{\bar{\sfc}_6\sfd}\hat{\delta}_{\bar{\sfb}'_2}^{\sfa\dprime\sfd} & 0 & 0 & 0 
\end{pmatrix}}} ,
\\
 R_{\bm{\gamma}}^{\bar{\sfc}_6} &\equiv {\footnotesize {\arraycolsep=0.4mm 
 \begin{pmatrix}
 0 & 0 & 0 & \delta_{\bm{\gamma}}^{\bm{\beta}} \hat{\delta}^{\bar{\sfc}_6}_{\bar{\sfb}_5\sfa} & 0 & 0 & 0 & 0 & 0 & 0 \\[-0.5mm]
 0 & 0 & 0 & 0 & \!\!\! \epsilon_{{\bm{\alpha}}{\bm{\gamma}}}\biggl[\genfrac{}{}{0pt}{1}{-\hat{\delta}^{\bar{\sfc}_6}_{\bar{\sfb}_6}\hat{\delta}_{\sfb'}^\sfa}{+c_6\,\hat{\delta}^{\sfa\bar{\sfc}_6}_{\bar{\sfb}_6\sfb'}}\biggr] & -\delta_{({\bm{\alpha}}}^{({\bm{\beta}}_1}\delta^{{\bm{\beta}}_2)}_{{\bm{\gamma}})}\hat{\delta}^{\sfa\bar{\sfc}_6}_{\bar{\sfb}_7} \!\!\! & 0 & 0 & 0 & 0 \\[-0.5mm]
 0 & 0 & 0 & 0 & 0 & 0 & \!\!\!\!\!\! -\delta_{\bm{\gamma}}^{\bm{\beta}}\hat{\delta}^{\bar{\sfc}_6\sfd}_{\bar{\sfb}_7}\hat{\delta}^{\bar{\sfa}_3}_{\bar{\sfb}'_2\sfd} \!\!\! & 0 & 0 & 0 \\[-0.5mm]
 0 & 0 & 0 & 0 & 0 & 0 & 0 & \epsilon_{{\bm{\alpha}}{\bm{\gamma}}}\hat{\delta}^{\bar{\sfc}_6\sfd}_{\bar{\sfb}_7}\hat{\delta}^{\bar{\sfa}_5}_{\bar{\sfb}'_4\sfd} & 0 & 0 \\[-0.5mm]
 0 & 0 & 0 & 0 & 0 & 0 & 0 & 0 & \!\!\! \delta_{\bm{\gamma}}^{\bm{\beta}}\biggl[\genfrac{}{}{0pt}{1}{\hat{\delta}^{\bar{\sfa}_6}_{\bar{\sfb}'_6}\hat{\delta}^{\sfa'\bar{\sfc}_6}_{\bar{\sfb}_7}}{-c_6\,\hat{\delta}^{\bar{\sfa}_6\sfa'}_{\bar{\sfb}_7}\hat{\delta}^{\bar{\sfc}_6}_{\bar{\sfb}'_6}}\biggr] \!\!\! & 0 \\[-0.5mm]
 0 & 0 & 0 & 0 & 0 & 0 & 0 & 0 & \!\!\! \delta^{\bm{\beta}}_{({\bm{\alpha}}_1} \epsilon_{{\bm{\alpha}}_2){\bm{\gamma}}} \hat{\delta}^{\bar{\sfa}_7}_{\bar{\sfb}_7}\hat{\delta}^{\bar{\sfc}_6}_{\bar{\sfb}'_6} \!\!\! & 0 \\[-0.5mm]
 0 & 0 & 0 & 0 & 0 & 0 & 0 & 0 & 0 & \epsilon_{{\bm{\alpha}}{\bm{\gamma}}}\hat{\delta}^{\bar{\sfa}_7}_{\bar{\sfb}_7}\hat{\delta}_{\bar{\sfb}'_7}^{\bar{\sfc}_6\sfd}\hat{\delta}^{\bar{\sfa}'_2}_{\sfb\dprime\sfd} \\[-1mm]
 0 & 0 & 0 & 0 & 0 & 0 & 0 & 0 & 0 & 0 \\[-1mm]
 0 & 0 & 0 & 0 & 0 & 0 & 0 & 0 & 0 & 0 \\[-1mm]
 0 & 0 & 0 & 0 & 0 & 0 & 0 & 0 & 0 & 0 
\end{pmatrix}}} ,
\\
 R_{\bar{\sfc}_7,\sfc'} &\equiv {\footnotesize {\arraycolsep=1mm 
 \begin{pmatrix}
 0 & 0 & 0 & 0 & 0 & 0 & 0 & 0 & 0 & 0 \\[-1mm]
 0 & 0 & 0 & 0 & 0 & 0 & 0 & 0 & 0 & 0 \\[-1mm]
 0 & 0 & 0 & 0 & 0 & 0 & 0 & 0 & 0 & 0 \\[-1mm]
 0 & 0 & 0 & 0 & 0 & 0 & 0 & 0 & 0 & 0 \\[-1mm]
 \Bigl[\genfrac{}{}{0pt}{1}{\hat{\delta}_{\bar{\sfc}_7}^{\sfb\bar{\sfa}_6}\hat{\delta}_{\sfc'}^{\sfa'}}{-c_{7,1}\,\hat{\delta}_{\bar{\sfc}_7}^{\bar{\sfa}_6\sfa'}\hat{\delta}^\sfb_{\sfc'}}\Bigr] & 0 & 0 & 0 & 0 & 0 & 0 & 0 & 0 & 0 \\[-1mm]
 0 & 0 & 0 & 0 & 0 & 0 & 0 & 0 & 0 & 0 \\[-1mm]
 0 & -\delta^{\bm{\beta}}_{\bm{\alpha}} \hat{\delta}_{\bar{\sfc}_7}^{\bar{\sfa}_7} \hat{\delta}_{\sfb\sfc'}^{\bar{\sfa}'_2} & 0 & 0 & 0 & 0 & 0 & 0 & 0 & 0 \\[-1mm]
 0 & 0 & - \hat{\delta}_{\bar{\sfc}_7}^{\bar{\sfa}_7}\hat{\delta}_{\bar{\sfb}_3\sfc'}^{\bar{\sfa}'_4} & 0 & 0 & 0 & 0 & 0 & 0 & 0 \\[-1mm]
 0 & 0 & 0 & -\delta^{\bm{\beta}}_{\bm{\alpha}} \hat{\delta}_{\bar{\sfc}_7}^{\bar{\sfa}_7} \hat{\delta}_{\bar{\sfb}_5\sfc'}^{\bar{\sfa}'_6} & 0 & 0 & 0 & 0 & 0 & 0 \\[-1mm]
 0 & 0 & 0 & 0 & \biggl[\genfrac{}{}{0pt}{1}{\hat{\delta}_{\bar{\sfc}_7}^{\bar{\sfa}_7}\hat{\delta}_{\bar{\sfb}_6\sfc'}^{\bar{\sfa}'_7}\hat{\delta}_{\sfb'}^{\sfa\dprime}}{-c_{7,1}\,\hat{\delta}_{\bar{\sfc}_7}^{\bar{\sfa}_7}\hat{\delta}_{\bar{\sfb}_6\sfb'}^{\bar{\sfa}'_7}\hat{\delta}^{\sfa\dprime}_{\sfc'}}
\biggr] & 0 & 0 & 0 & 0 & 0 
\end{pmatrix}}} ,
\\
 R^{\bar{\sfc}_7,\sfc'} &\equiv {\footnotesize {\arraycolsep=1mm 
 \begin{pmatrix}
 0 & 0 & 0 & 0 & \biggl[\genfrac{}{}{0pt}{1}{-\hat{\delta}^{\bar{\sfc}_7}_{\sfa\bar{\sfb}_6}\hat{\delta}^{\sfc'}_{\sfb'}}{+c_{7,1}\,\hat{\delta}^{\bar{\sfc}_7}_{\bar{\sfb}_6\sfb'}\hat{\delta}_\sfa^{\sfc'}}\biggr] & 0 & 0 & 0 & 0 & 0 \\[-1mm]
 0 & 0 & 0 & 0 & 0 & 0 & \delta_{\bm{\alpha}}^{\bm{\beta}} \hat{\delta}^{\bar{\sfc}_7}_{\bar{\sfb}_7}\hat{\delta}^{\sfa\sfc'}_{\bar{\sfb}'_2} & 0 & 0 & 0 \\[-1mm]
 0 & 0 & 0 & 0 & 0 & 0 & 0 & \hat{\delta}^{\bar{\sfc}_7}_{\bar{\sfb}_7}\hat{\delta}^{\bar{\sfa}_3\sfc'}_{\bar{\sfb}'_4} & 0 & 0 \\[-1mm]
 0 & 0 & 0 & 0 & 0 & 0 & 0 & 0 & \delta_{\bm{\alpha}}^{\bm{\beta}} \hat{\delta}^{\bar{\sfc}_7}_{\bar{\sfb}_7}\hat{\delta}^{\bar{\sfa}_5\sfc'}_{\bar{\sfb}'_6} & 0 \\[-1mm]
 0 & 0 & 0 & 0 & 0 & 0 & 0 & 0 & 0 & \biggl[\genfrac{}{}{0pt}{1}{-\hat{\delta}^{\bar{\sfc}_7}_{\bar{\sfb}_7}\hat{\delta}^{\bar{\sfa}_6\sfc'}_{\bar{\sfb}'_7}\hat{\delta}^{\sfa'}_{\sfb\dprime}}{+c_{7,1}\,\hat{\delta}^{\bar{\sfc}_7}_{\bar{\sfb}_7}\hat{\delta}^{\bar{\sfa}_6\sfa'}_{\bar{\sfb}'_7}\hat{\delta}_{\sfb\dprime}^{\sfc'}}\biggr] \\[-1mm]
 0 & 0 & 0 & 0 & 0 & 0 & 0 & 0 & 0 & 0 \\[-1mm]
 0 & 0 & 0 & 0 & 0 & 0 & 0 & 0 & 0 & 0 \\[-1mm]
 0 & 0 & 0 & 0 & 0 & 0 & 0 & 0 & 0 & 0 \\[-1mm]
 0 & 0 & 0 & 0 & 0 & 0 & 0 & 0 & 0 & 0 \\[-1mm]
 0 & 0 & 0 & 0 & 0 & 0 & 0 & 0 & 0 & 0 
\end{pmatrix}}} 
\nn\\
 &\quad \Bigl[ c_6 \equiv \tfrac{1}{7}-\tfrac{3}{7\sqrt{2}}\qquad c_{7,1}\equiv \tfrac{1}{7}+\tfrac{4}{7\sqrt{2}}\Bigr]\,.
\end{align}

\subsection{Invariant tensors of $\Edd$}
Having established the explicit generator decompositions, we now turn to the construction of invariant tensors of $\Edd$, such as $\eta_{AB;\CC}$ and $Z_A{}^{\BB}{}_{\CCC}$\,, which play an important role in this paper.

Our strategy proceeds systematically through level decomposition. We begin with the $\mathfrak{e}_{d+n(d+n)}$ generators and perform level decompositions appropriate to the dimension under consideration. From the commutation relations established in section~\ref{sec:commutation}, we can then extract the explicit forms of the invariant tensors.

Since we have constructed the matrix form of the $\mathfrak{e}_{8(8)}$ generators, choosing $n=2$ allows us to determine $\eta_{AB;\CC}$ for $\Edd[6]$ from its defining relation. This reproduces the known results of \cite{Sakatani:2017xcn} up to conventions. To determine $Z_A{}^{\BB}{}_{\CCC}$\,, which requires $n>2$, we set $n=3$ and extract the tensor for $\Edd[5]$. Fortunately, in $\Edd[6]$, the $R_2$-representation is isomorphic to $R_1$ and the $R_3$-representation coincides with the adjoint representation, allowing us to construct $Z_A{}^{\BB}{}_{\CCC}$ directly from the generators $(t_{\adja})_A{}^B$. Through this procedure, we obtain explicit forms for $Z_A{}^{\BB}{}_{\CCC}$ up to $\Edd[6]$.

\paragraph{Level decomposition:}
We organize the generators at each level in the M-theory parameterization as follows:
\begin{align}
 K^\exA{}_{\exB}&= K^\exA{}_{\exB}\,,
\\
 R_{\exA_1\exA_2\exA_3;\AAA} &= \{R_{\exA_1\exA_2\exA_3} ,\,R_{\exA_1\exA_2\exA_3 \bar{a}_3} \}\,,
\\
 R_{\exA_1\exA_2;\AA} &= \{R_{\exA_1\exA_2 a} ,\,R_{\exA_1\exA_2 \bar{a}_4} \}\,,
\\
 R_\exA^A &= \{K^a{}_\exA,\,R_{\exA \bar{a}_2} ,\,R_{\exA \bar{a}_5} \}\,,
\\
 t_{\adja} &= \{R^{\bar{a}_3},\,\check{K}^a{}_b,\,R_{\bar{a}_3}\}\,,
\\
 t^{\adja} &= \{R_{\bar{a}_3},\,\check{k}_a{}^b,\,R^{\bar{a}_3}\}\,,
\\
 R^\exA_A &= \{-K^\exA{}_a,\,R^{\exA \bar{a}_2} ,\,R^{\exA \bar{a}_5} \}\,,
\\
 R^{\exA_1\exA_2;\AA} &= \{R^{\exA_1\exA_2 a} ,\,R^{\exA_1\exA_2 \bar{a}_4} \}\,,
\\
 R^{\exA_1\exA_2\exA_3;\AAA} &= \{R^{\exA_1\exA_2\exA_3} ,\,R^{\exA_1\exA_2\exA_3 \bar{a}_3} \}\,,
\end{align}
and so forth, with modified trace parts $\check{K}^{\hat{a}}{}_{\hat{b}}\equiv K^{\hat{a}}{}_{\hat{b}} - \Cbeta \,\delta^{\hat{a}}_{\hat{b}}\,K^\exC{}_\exC$\,. The structure truncates appropriately for smaller values of $n$: for $n=2$, the level-$3$ generators are absent, while for $n=1$, both level-$2$ and level-$3$ generators disappear.

\paragraph{The $\eta$-symbol}

The invariant tensor $\eta^{AB;\CC}$ in $\Edd$ with $d\leq 6$ decomposes as
\begin{align}
 \eta^{AB;\CC} = \begin{pmatrix} \eta^{AB;c} & \eta^{AB;\bar{c}_4} & \eta^{AB;\bar{c}_6,c'} \end{pmatrix},
\end{align}
where all multiple indices are antisymmetric. The explicit forms are:
\begin{align}
 (\eta^c)^{AB} 
 &\equiv \eta^{AB;c}\equiv \begin{pmatrix}
 0 & \hat{\delta}^{ac}_{\bar{b}_2} & 0 \\
 \hat{\delta}^{bc}_{\bar{a}_2} & 0 & 0 \\
 0 & 0 & 0 
 \end{pmatrix} , 
\\
 \eta^{\bar{c}_4} &\equiv 
 \eta^{AB;\bar{c}_4} 
 \equiv \begin{pmatrix}
 0 & 0 & \hat{\delta}^{a\bar{c}_4}_{\bar{b}_5} \\
 0 & \hat{\delta}^{\bar{c}_4}_{\bar{a}_2\bar{b}_2} & 0 \\
 \hat{\delta}^{b\bar{c}_4}_{\bar{a}_5} & 0 & 0 
 \end{pmatrix} , 
\\
 (\eta^{\bar{c}_6,c'})^{AB} &\equiv 
 \eta^{AB;c_1\cdots c_6,c'} 
\nn\\
 &\equiv \begin{pmatrix}
 0 & 0 & 0 \\
 0 & 0 & -\hat{\delta}^{\bar{c}_6}_{\bar{b}_5d}\,\hat{\delta}^{dc'}_{\bar{a}_2} \\
 0 & -\hat{\delta}^{\bar{c}_6}_{\bar{a}_5d}\,\hat{\delta}^{dc'}_{\bar{b}_2} & 0 \end{pmatrix} .
\end{align}
These tensors satisfy the defining property that they intertwine between different representations and play an important role in formulating the section condition in ExFT. The extension of this $\eta$-symbol to $\Edd[7]$ has been constructed in \cite{Sakatani:2017xcn}. Note that for smaller $d$, we need only truncate the rows and columns where the antisymmetrization trivially vanishes.

\paragraph{The $Z$-tensor}
The invariant tensor $Z_{A}{}^{\BB}{}_{\CCC}$ decomposes as
\begin{align}
 Z_{A}{}^{\BB}{}_{\CCC}= \begin{pmatrix} Z_{A}{}^{\BB}{}_{*} & Z_{A}{}^{\BB}{}_{\bar{c}_3} & Z_{A}{}^{\BB}{}_{\bar{c}_5,c'} & Z_{A}{}^{\BB}{}_{\bar{c}_6,\bar{c}'_3} & Z_{A}{}^{\BB}{}_{\bar{c}_6,\bar{c}'_6}\end{pmatrix} \,.
\end{align}
The components are given explicitly by:
\begin{align}
 Z_{A}{}^{\BB}{}_{*} &= \begin{pmatrix} -\delta_a^b & 0 & 0 \\
 0 & 0 & 0 \\
 0 & 0 & 0 \end{pmatrix},\qquad
 Z_{A}{}^{\BB}{}_{\bar{c}_3} = \begin{pmatrix} 0 & -\hat{\delta}^{\bar{b}_4}_{a\bar{c}_3} & 0 \\ \hat{\delta}^{\bar{a}_2b}_{\bar{c}_3} & 0 & 0 \\ 0 & 0 & 0 \end{pmatrix},
\\
 Z_{A}{}^{\BB}{}_{\bar{c}_5,c'} &= \begin{pmatrix} 0 & 0 & - \hat{\delta}_{a\bar{c}_5}^{\bar{b}_6}\,\delta_{c'}^{b'} \\ 0 & \hat{\delta}_{c'd}^{\bar{a}_2}\,\hat{\delta}^{\bar{b}_4d}_{\bar{c}_5} & 0 \\ \hat{\delta}_{\bar{c}_5}^{\bar{a}_5}\,\delta^b_{c'} & 0 & 0 \end{pmatrix},
\\
 Z_{A}{}^{\BB}{}_{\bar{c}_6,\bar{c}'_3} &= \begin{pmatrix} 0 & 0 & 0 \\ 0 & 0 & -\hat{\delta}_{\bar{c}_6}^{\bar{b}_6}\,\hat{\delta}^{\bar{a}_2b'}_{\bar{c}'_3} \\ 0 & \hat{\delta}_{\bar{c}_6}^{\bar{b}_4\bar{d}_2}\,\hat{\delta}^{\bar{a}_5}_{\bar{d}_2\bar{c}'_3} & 0 \end{pmatrix},
\\
 Z_{A}{}^{\BB}{}_{\bar{c}_6,\bar{c}'_3} &= \begin{pmatrix} 0 & 0 & 0 \\ 0 & 0 & 0 \\ 0 & 0 & \hat{\delta}_{\bar{c}_6}^{\bar{b}_6}\,\hat{\delta}^{\bar{a}_5b'}_{\bar{c}'_6} \end{pmatrix},
\end{align}

\paragraph{Special cases: $\Edd[6]$, $\Edd[7]$, and $\Edd[8]$}
For $\Edd[6]$, we determine $Z_{A}{}^{\BB}{}_{\CCC}$ through the relation
\begin{align}
 Z_{A}{}^{\BB}{}_{\CCC} = (t_{\adja})_A{}^E\,\chi_E^{\BB}\,\chi^{\adja}_{\CCC}\,,
\end{align}
where the intertwiners $\chi_A^{\BB}$ and $\chi^{\adja}_B$ connect isomorphic representations. The explicit forms involve the totally antisymmetric tensor $\epsilon^{123456}=1=\epsilon_{123456}$\,:
\begin{align}
 \chi_A^{\BB} &\equiv \begin{pmatrix} 0 & 0 & \epsilon^{\bar{b}_6}\delta_a^{b'} \\ 0 & \epsilon^{\bar{a}_2\bar{b}_4} & 0 \\ \epsilon^{\bar{a}_5b} & 0 & 0 \end{pmatrix}, 
\\
 \chi^{\adja}_{\BB} &\equiv \begin{pmatrix}
 0 & 0 & 0 & 0 & -\epsilon_{\bar{b}_6}\,\hat{\delta}^{\bar{a}_6}_{\bar{b}'_6} \\
 0 & 0 & 0 & -\epsilon_{\bar{b}_6}\,\hat{\delta}^{\bar{a}_3}_{\bar{b}'_3} & 0 \\
 0 & 0 & - \bigl(\epsilon_{a\bar{b}_5}\, \delta_{b'}^{a'} + \tfrac{1+\sqrt{3}}{6\sqrt{3}}\,\delta_a^{a'}\,\epsilon_{\bar{b}_5b'} \bigr) & 0 & 0 \\
 0 & \epsilon_{\bar{a}_3\bar{b}_3} & 0 & 0 & 0 \\
 -\epsilon_{\bar{a}_6} & 0 & 0 & 0 & 0 \end{pmatrix}.
\end{align}
Note that $\chi_A^{\BB}$ is the intertwiner introduced in the main text.

For completeness, we also provide the intertwiner $\chi_{\adja B}$ in $\Edd[8]$ and the invariant tensor $\omega_{AB}$ in $\Edd[7]$, which find applications in formulating the section condition and gauge transformations in exceptional field theory. Their explicit forms are
\begin{align}
 \chi_{\adja B} &= \begin{pmatrix}
 0 & 0 & 0 & 0 & 0 & 0 & \delta_a^b \\
 0 & 0 & 0 & 0 & 0 & \hat{\delta}_{\bar{a}_6}^{\bar{b}_6} & 0 \\
 0 & 0 & 0 & 0 & \hat{\delta}_{\bar{a}_3}^{\bar{b}_3} & 0 & 0 \\
 0 & 0 & 0 & \epsilon^{a\bar{b}_7}\delta_c^b - \tfrac{1}{4} \,\epsilon^{\bar{b}_7b'}\,\delta_c^a & 0 & 0 & 0 \\
 0 & 0 & -\epsilon^{\bar{a}_3\bar{b}_5} & 0 & 0 & 0 & 0 \\
 0 & \epsilon^{\bar{a}_6\bar{b}_2} & 0 & 0 & 0 & 0 & 0 \\
 \delta^a_b & 0 & 0 & 0 & 0 & 0 & 0
\end{pmatrix},
\\
 \omega_{AB} &= \begin{pmatrix} 0 & 0 & 0 & -\epsilon^{\bar{b}_6}\,\delta_a^{b'} \\
 0 & 0 & -\epsilon^{\bar{a}_2\bar{b}_5} & 0 \\
 0 & \epsilon^{\bar{a}_5\bar{b}_2} & 0 & 0 \\
 \epsilon^{\bar{a}_7}\,\delta^{a'}_b & 0 & 0 & 0
\end{pmatrix}.
\end{align}

\section{Exceptional geometric algebra for $d\leq 7$}\label{app:algebra}
When we construct exceptional generalized cosets, we begin by constructing a mega-space as a generalized parallelizable space $G/H$. This space is constructed purely from a Leibniz algebra, called EGA, introduced in \cite{Hassler:2022egz}. Depending on the choice of the section condition in ExFT, there are two EGAs: the M-theory EGA and the type IIB EGA. These two EGAs are generically related by a redefinition of generators $T_A\to C_A{}^B T_B$ with $C_A{}^B\in \Edd$, which corresponds to a generalized U-duality, though such a transformation is not always possible.

Here we present the explicit forms of these algebras for $\Edd$ with $d\leq 7$. The case of $\Edd[8]$ can be found in \cite{Hassler:2022egz}. The M-theory EGA is given by:
\begin{align}
 T_a \circ T_b &= f_{ab}{}^c\,T_c + f_{ab\bar{c}_2}\,T^{\bar{c}_2} + f_{ab\bar{c}_5}\,T^{\bar{c}_5} \,,
\nn\\
 T_a \circ T^{\bar{b}_2} &= -f_a{}^{\bar{b}_2c}\,T_c + \hat{\delta}^{\bar{b}_2}_{de}\,f_{ac}{}^{d}\,T^{ec}
 + 3\,Z_a\,T^{\bar{b}_2}
 - f_{a \bar{c}_3}\,T^{\bar{b}_2 \bar{c}_3}
 + \hat{\delta}^{\bar{b}_2}_{ad}\,f_{\bar{c}_7}\,T^{\bar{c}_7,d} \,,
\nn\\
 T_a \circ T^{\bar{b}_5} &= f_a{}^{\bar{b}_5c}\,T_c + \hat{\delta}^{\bar{b}_5}_{\bar{c}_3\bar{d}_2}\,f_{a}{}^{\bar{c}_3}\,T^{\bar{d}_2} - \hat{\delta}^{\bar{b}_5}_{d\bar{e}_4}\,f_{ac}{}^{d}\,T^{\bar{e}_4c} 
 +6\,Z_a\,T^{\bar{b}_5}
 + \hat{\delta}^{\bar{b}_5}_{\bar{d}_4e}\,f_{a\bar{c}_3}\,T^{\bar{c}_3\bar{d}_4,e} \,,
\nn\\
 T_a \circ T^{\bar{b}_7,b'} &= -\hat{\delta}^{\bar{b}_7}_{\bar{c}_6 d}\,f_a{}^{\bar{c}_6}\,T^{d b'} 
 + \hat{\delta}^{\bar{b}_7}_{\bar{c}_2\bar{d}_5}\, f_a{}^{b' \bar{c}_2}\,T^{\bar{d}_5}
\nn\\
 &\quad - \hat{\delta}^{\bar{b}_7}_{d\bar{e}_6}\,f_{ac}{}^{d}\,T^{\bar{e}_6 c,b'}
 - f_{ac}{}^{b'}\,T^{\bar{b}_7,c} 
 + 9\,Z_a\,T^{\bar{b}_7,b'}\,,
\nn\\
 T^{\bar{a}_2} \circ T_b &= f_b{}^{\bar{a}_2c}\,T_c + \hat{\delta}^{\bar{a}_2}_{de}\,\hat{\delta}_{\bar{c}_2b}^{\bar{f}_2e}\,f_{\bar{f}_2}{}^{d}\,T^{\bar{c}_2}
 -3\,Z_c\,\hat{\delta}^{c\bar{a}_2}_{b\bar{e}_2}\,T^{\bar{e}_2}
 + \hat{\delta}^{\bar{a}_2\bar{c}_4}_{b\bar{d}_5}\,f_{\bar{c}_4}\,T^{\bar{d}_5} \,,
\nn\\
 T^{\bar{a}_2} \circ T^{\bar{b}_2} &= \hat{\delta}^{\bar{b}_2}_{de}\, f_c{}^{\bar{a}_2d}\, T^{ec} - \hat{\delta}^{\bar{a}_2}_{de}\,f_{\bar{c}_2}{}^{d}\,T^{e\bar{b}_2\bar{c}_2}
 +3\,Z_c\,T^{\bar{a}_2\bar{b}_2c}
 +\hat{\delta}^{\bar{b}_2}_{de}\,f_{\bar{c}_4}\,T^{\bar{a}_2\bar{c}_4d,e} \,,
\nn\\
 T^{\bar{a}_2} \circ T^{\bar{b}_5} &= -\hat{\delta}^{\bar{b}_5}_{d\bar{e}_4}\,f_c{}^{\bar{a}_2 d}\, T^{\bar{e}_4c} - \hat{\delta}^{\bar{a}_2}_{de}\,\hat{\delta}^{\bar{c}_2e}_{\bar{f}_2g}\,f_{\bar{c}_2}{}^d\,T^{\bar{b}_5\bar{f}_2,g} 
 + 3\,\hat{\delta}^{\bar{a}_2c}_{\bar{d}_2e}\,Z_c\,T^{\bar{b}_5\bar{d}_2,e}\,,
\nn\\
 T^{\bar{a}_2} \circ T^{\bar{b}_7,b'} &= -\hat{\delta}^{\bar{b}_7}_{d\bar{e}_6}\,f_c{}^{\bar{a}_2d}\, T^{\bar{e}_6 c,b'} - f_c{}^{\bar{a}_2b'}\, T^{\bar{b}_7,c} \,,
\nn\\
 T^{\bar{a}_5} \circ T_b &= - f_b{}^{\bar{a}_5c}\,T_c - \hat{\delta}^{\bar{a}_5}_{\bar{c}_3\bar{d}_2}\,f_b{}^{\bar{c}_3}\,T^{\bar{d}_2}
 - \hat{\delta}^{\bar{a}_5}_{\bar{d}_3bf}\,f_c{}^{\bar{d}_3}\,T^{fc} 
\\
 &\quad + \hat{\delta}^{\bar{a}_5}_{d\bar{e}_4}\,f_{bc}{}^{d}\,T^{\bar{e}_4c}
 + \hat{\delta}^{\bar{a}_5}_{db\bar{f}_3}\,f_{\bar{c}_2}{}^{d}\, T^{\bar{f}_3 \bar{c}_2} 
 - 6\,\hat{\delta}^{c\bar{a}_5}_{b\bar{d}_5}\,Z_c\, T^{\bar{d}_5}\,,
\nn\\
 T^{\bar{a}_5} \circ T^{\bar{b}_2} &= - \hat{\delta}^{\bar{b}_2}_{de}\,f_c{}^{\bar{a}_5 d}\,T^{ec}
  + \hat{\delta}^{\bar{a}_5}_{\bar{d}_3\bar{e}_2}\,f_c{}^{\bar{d}_3}\, T^{\bar{e}_2 \bar{b}_2c}
\nn\\
 &\quad + \hat{\delta}^{\bar{a}_5}_{d\bar{e}_4}\,\hat{\delta}^{\bar{b}_2}_{fg}\,f_{\bar{c}_2}{}^{d}\,T^{\bar{e}_4 \bar{c}_2f,g} 
 -6\,\hat{\delta}^{\bar{b}_2}_{de}\,Z_c\,T^{c \bar{a}_5d,e}\,,
\nn\\
 T^{\bar{a}_5} \circ T^{\bar{b}_5} &= \hat{\delta}^{\bar{b}_5}_{d\bar{e}_4}\,f_c{}^{\bar{a}_5 d}\, T^{\bar{e}_4 c}
 + \hat{\delta}^{\bar{a}_5}_{\bar{f}_3\bar{g}_2}\,\hat{\delta}^{\bar{g}_2c}_{\bar{d}_2e}\, f_c{}^{\bar{f}_3}\,T^{\bar{b}_5 \bar{d}_2,e} \,,
\nn\\
 T^{\bar{a}_5} \circ T^{\bar{b}_7,b'} &= \hat{\delta}^{\bar{b}_7}_{d\bar{e}_6}\,f_c{}^{\bar{a}_5 d}\,T^{\bar{e}_6 c,b'} + f_c{}^{\bar{a}_5 b'}\,T^{\bar{b}_7,c} \,,
\nn\\
 T^{\bar{a}_7,a'} \circ T_b &= 
 \hat{\delta}^{\bar{a}_7}_{\bar{e}_6f}\,\hat{\delta}^{fa'c}_{b \bar{d}_2} \,f_c{}^{\bar{e}_6}\, T^{\bar{d}_2}
 + \hat{\delta}^{\bar{a}_7}_{\bar{e}_2\bar{f}_5}\,f_c{}^{a' \bar{e}_2}\, \hat{\delta}_{b \bar{d}_5}^{\bar{f}_5 c}\,T^{\bar{d}_5}\,,
\nn\\
 T^{\bar{a}_7,a'} \circ T^{\bar{b}_2} &= - \hat{\delta}^{\bar{a}_7}_{\bar{d}_6e}\, f_c{}^{\bar{d}_6}\,T^{e a' c \bar{b}_2} 
 + \hat{\delta}^{\bar{a}_7}_{\bar{d}_2\bar{e}_5}\,\hat{\delta}^{\bar{b}_2}_{fg}\,f_c{}^{a' \bar{d}_2}\,T^{\bar{e}_5 c f,g}\,,
\nn\\
 T^{\bar{a}_7,a'} \circ T^{\bar{b}_5} &= - \hat{\delta}^{\bar{a}_7}_{\bar{f}_6g}\,\hat{\delta}^{ga'c}_{\bar{d}_2e}\,f_c{}^{\bar{f}_6}\,T^{\bar{b}_5 \bar{d}_2,e} \,,
\nn\\
 T^{\bar{a}_7,a'} \circ T^{\bar{b}_7,b'} &= 0 \,,
\nn
\end{align}
with structure constants
\begin{align}
 f_{\bar{a}_7}\,,\quad f_{\bar{a}_4}\,,\quad f_{ab}{}^c\,,\quad f_a{}^{\bar{b}_3}\,,\quad f_a{}^{\bar{b}_6}\,,\quad Z_a\,.
\label{eq:EGA-M-f}
\end{align}
If we compactly write this algebra as $T_A\circ T_B = X_{AB}{}^C\,T_C$\,, the Leibniz identity takes the form
\begin{align}
 [X_{A},\,X_{B}] = - X_{AB}{}^{C}\,X_{C}\,.
\end{align}
By decomposing this identity in terms of the structure constants in \eqref{eq:EGA-M-f}, we obtain the constraints that these structure constants must satisfy.

The type IIB EGA is given by:
{\small
\begin{align}
\begin{split}
 T_{\sfa}\circ T_{\sfb} &=f_{\sfa\sfb}{}^{\sfc}\,T_{\sfc}
 + f^{\bm{\gamma}}_{\sfa\sfb\sfc}\,T_{\bm{\gamma}}^{\sfc}
 + f_{\sfa\sfb\bar{\sfc}_3}\,T^{\bar{\sfc}_3} \,,
\\
 T_{\sfa}\circ T^{\sfb}_{\bm{\beta}} &= f_{\sfa}{}_{{\bm{\beta}}}^{\sfb\sfc}\,T_{\sfc}
 - f_{\sfa{\bm{\beta}}}{}^{{\bm{\gamma}}}\,T_{\bm{\gamma}}^{\sfb} - f_{\sfa\sfc}{}^{\sfb}\,T_{\bm{\beta}}^{\sfc} 
 + 2\,Z_{\sfa}\,T^{\sfb}_{\bm{\beta}} 
 - \epsilon_{{\bm{\beta}}{\bm{\gamma}}}\,f^{{\bm{\gamma}}}_{\sfa\bar{\sfc}_2}\, T^{\sfb\bar{\sfc}_2} 
 + f_{\sfa\bar{\sfc}_4}\,T_{\bm{\beta}}^{\sfb\bar{\sfc}_4}\,, 
\\
 T_{\sfa}\circ T^{\bar{\sfb}_3} &=
 f_{\sfa}{}^{\bar{\sfb}_3\sfc}\, T_{\sfc}
 - \hat{\delta}^{\bar{\sfb}_3}_{\bar{\sfc}_2\sfd}\,\epsilon^{{\bm{\gamma}}{\bm{\delta}}}\,f_{\sfa}{}_{{\bm{\gamma}}}^{\bar{\sfc}_2}\, T_{{\bm{\delta}}}^{\sfd}
 - \hat{\delta}^{\bar{\sfb}_3}_{\sfd\bar{\sfe}_2}\,f_{\sfa\sfc}{}^{\sfd}\, T^{\bar{\sfe}_2 \sfc} 
 +4\,Z_{\sfa}\,T^{\bar{\sfb}_3}
 - f^{\bm{\gamma}}_{\sfa\bar{\sfc}_2} T_{\bm{\gamma}}^{\bar{\sfb}_3\bar{\sfc}_2}
 - \hat{\delta}^{\bar{\sfb}_3}_{\bar{\sfc}_2\sfd}\,f_{\sfa\bar{\sfe}_4}\,T^{\bar{\sfc}_2\bar{\sfe}_4,\sfd} \,,
\\
 T_{\sfa}\circ T_{\bm{\beta}}^{\bar{\sfb}_5} &=
 f_{\sfa}{}_{{\bm{\beta}}}^{\bar{\sfb}_5\sfc}\,T_{\sfc}
 - \hat{\delta}^{\bar{\sfb}_5}_{\bar{\sfc}_4\sfd}\, f_{\sfa}{}^{\bar{\sfc}_4}\,T_{\bm{\beta}}^{\sfd}
 + \hat{\delta}^{\bar{\sfb}_5}_{\bar{\sfc}_2\bar{\sfd}_3}\, f_{\sfa}{}_{{\bm{\beta}}}^{\bar{\sfc}_2}\,T^{\bar{\sfd}_3} 
\\
 &\quad 
 - f_{\sfa{\bm{\beta}}}{}^{{\bm{\gamma}}}\, T_{\bm{\gamma}}^{\bar{\sfb}_5} 
 - \hat{\delta}^{\bar{\sfb}_5}_{\sfd\bar{\sfe}_4}\,f_{\sfa\sfc}{}^{\sfd}\,T_{\bm{\beta}}^{\bar{\sfe}_4\sfc} 
 + 6\,Z_{\sfa}\,T^{\bar{\sfb}_5}_{\bm{\beta}}
 + \epsilon_{{\bm{\beta}}{\bm{\gamma}}}\,f^{{\bm{\gamma}}}_{\sfa\sfc\sfd} T^{\bar{\sfb}_5\sfc,\sfd} \,,
\\
 T_{\sfa}\circ T^{\bar{\sfb}_6,\sfb'} &=
 \epsilon^{{\bm{\gamma}}\bm{\delta}}\,f_{\sfa}{}_{{\bm{\gamma}}}^{\bar{\sfb}_6}\, T_{\bm{\delta}}^{\sfb'}
 - \hat{\delta}^{\bar{\sfb}_6}_{\bar{\sfc}_3\bar{\sfd}_3}\, f_{\sfa}{}^{\sfb' \bar{\sfc}_3}\,T^{\bar{\sfd}_3}
 + \hat{\delta}^{\bar{\sfb}_6}_{\sfc\bar{\sfd}_5}\,\epsilon^{{\bm{\gamma}}{\bm{\delta}}}\,f_{\sfa}{}_{{\bm{\gamma}}}^{\sfb' \sfc}\,T_{\bm{\delta}}^{\bar{\sfd}_5}
\\
 &\quad 
 - \hat{\delta}^{\bar{\sfb}_6}_{\sfd\bar{\sfe}_5}\,f_{\sfa\sfc}{}^{\sfd}\,T^{\sfc \bar{\sfe}_5,\sfb'}
 - f_{\sfa\sfc}{}^{\sfb'}\,T^{\bar{\sfb}_6,\sfc}
 + 8\,Z_{\sfa}\,T^{\bar{\sfb}_6,\sfb'}\,,
\\
 T^{\sfa}_{\bm{\alpha}}\circ T_{\sfb} &= 
 - f_{\sfb}{}_{{\bm{\alpha}}}^{\sfa\sfc} \, T_{\sfc} 
 - \hat{\delta}^{\sfa\sfd}_{\sfb\sfc}\,f_{\sfd{\bm{\alpha}}}{}^{{\bm{\gamma}}}\, T_{\bm{\gamma}}^{\sfc}
 + f_{\sfb\sfc}{}^{\sfa}\,T_{\bm{\alpha}}^{\sfc} 
 +2\,\hat{\delta}^{\sfa\sfc}_{\sfb\sfd}\,Z_{\sfc}\, T^{\sfd}_{\bm{\alpha}}
 - \hat{\delta}^{\sfa\bar{\sfd}_3}_{\sfb\bar{\sfc}_3}\,\epsilon_{{\bm{\alpha}}{\bm{\gamma}}}\,f^{{\bm{\gamma}}}_{\bar{\sfd}_3}\,T^{\bar{\sfc}_3}
 + \hat{\delta}^{\sfa\bar{\sfc}_5}_{\sfb\bar{\sfd}_5}\,f_{\bar{\sfc}_5}\,T_{{\bm{\alpha}}}^{\bar{\sfd}_5}\,,
\\
 T^{\sfa}_{\bm{\alpha}}\circ T^{\sfb}_{\bm{\beta}} &= f_{\sfc}{}_{{\bm{\alpha}}}^{\sfa\sfb}\,T_{\bm{\beta}}^{\sfc} 
 + f_{\sfc{\bm{\alpha}}}{}^{\bm{\gamma}}\,\epsilon_{{\bm{\gamma}}{\bm{\beta}}}\,T^{\sfc \sfa\sfb} 
 + \epsilon_{{\bm{\alpha}}{\bm{\beta}}}\,f_{\bar{\sfc}_2}{}^{\sfa}\,T^{\bar{\sfc}_2\sfb}
 -2\,\epsilon_{{\bm{\alpha}}{\bm{\beta}}}\,Z_{\sfc}\,T^{\sfa\sfb\sfc} 
 + \epsilon_{{\bm{\alpha}}{\bm{\gamma}}}\,f^{{\bm{\gamma}}}_{\bar{\sfc}_3}\,T_{\bm{\beta}}^{\sfa\sfb\bar{\sfc}_3} 
 - \epsilon_{{\bm{\alpha}}{\bm{\beta}}}\,f_{\bar{\sfc}_5}\,T^{\sfa\bar{\sfc}_5,\sfb}\,,
\\
 T^{\sfa}_{\bm{\alpha}}\circ T^{\bar{\sfb}_3}
 &= \hat{\delta}^{\bar{\sfb}_3}_{\sfd\bar{\sfe}_2}\,f_{\sfc}{}_{{\bm{\alpha}}}^{\sfa \sfd}\,T^{\bar{\sfe}_2 \sfc} 
  + f_{\sfc{\bm{\alpha}}}{}^{\bm{\gamma}}\,T_{\bm{\gamma}}^{\sfa\sfc \bar{\sfb}_3} 
  - f_{\bar{\sfc}_2}{}^{\sfa}\,T_{\bm{\alpha}}^{\bar{\sfc}_2 \bar{\sfb}_3}
 + 2\,Z_{\sfc}\,T^{\sfa \bar{\sfb}_3\sfc}_{\bm{\alpha}} 
 + \hat{\delta}^{\sfa\bar{\sfd}_3}_{\sfe\bar{\sfc}_3}\,\epsilon_{{\bm{\alpha}}{\bm{\gamma}}}\,f^{{\bm{\gamma}}}_{\bar{\sfd}_3} \,T^{\bar{\sfb}_3\bar{\sfc}_3,\sfe}\,,
\\
 T^{\sfa}_{\bm{\alpha}}\circ T_{\bm{\beta}}^{\bar{\sfb}_5}
 &= \hat{\delta}^{\bar{\sfb}_5}_{\sfd\bar{\sfe}_4}\, f_{\sfc}{}_{{\bm{\alpha}}}^{\sfa \sfd}\,T_{\bm{\beta}}^{\bar{\sfe}_4\sfc}
 + \hat{\delta}^{\sfa\sfc}_{\sfd\sfe}\,f_{\sfc{\bm{\alpha}}}{}^{\bm{\gamma}}\,\epsilon_{{\bm{\gamma}}{\bm{\beta}}}\,T^{\bar{\sfb}_5\sfd,\sfe}
 - \epsilon_{{\bm{\alpha}}{\bm{\beta}}}\,f_{\sfc\sfd}{}^{\sfa}\,T^{\bar{\sfb}_5 \sfc,\sfd}
 +4\,\epsilon_{{\bm{\alpha}}{\bm{\beta}}}\,Z_{\sfc}\,T^{\sfa \bar{\sfb}_5,\sfc}\,,
\\
 T^{\sfa}_{\bm{\alpha}}\circ T^{\bar{\sfb}_6,\sfb'} &=
 \hat{\delta}^{\bar{\sfb}_6}_{\sfd\bar{\sfe}_5}\, f_{\sfc}{}_{{\bm{\alpha}}}^{\sfa \sfd}\,T^{\sfc \bar{\sfe}_5,\sfb'}
 + f_{\sfc}{}_{{\bm{\alpha}}}^{\sfa\sfb'}\,T^{\bar{\sfb}_6,\sfc}\,,
\\
 T^{\bar{\sfa}_3} \circ T_{\sfb} 
 &= - f_{\sfb}{}^{\bar{\sfa}_3\sfc}\,T_{\sfc} 
 - \hat{\delta}^{\bar{\sfa}_3}_{\bar{\sfd}_2\sfe}\,\hat{\delta}_{\sfb\sfc}^{\sfe\sff}\,\epsilon^{{\bm{\gamma}}{\bm{\delta}}}\,f_{\sff}{}_{{\bm{\gamma}}}^{\bar{\sfd}_2}\,T_{\bm{\delta}}^{\sfc}
 + \hat{\delta}^{\bar{\sfa}_3}_{\sfd\bar{\sfe}_2}\,f_{\sfb\sfc}{}^{\sfd}\, T^{\bar{\sfe}_2\sfc}
 - \hat{\delta}^{\bar{\sfa}_3}_{\sfb\sfd\sfe}\, f_{\bar{\sfc}_2}{}^{\sfd}\, T^{\sfe\bar{\sfc}_2}
\nn\\
 &\quad +4\,\hat{\delta}^{\bar{\sfa}_3\sfc}_{\sfb\bar{\sfd}_3}\,Z_{\sfc}\, T^{\bar{\sfd}_3} 
 - \hat{\delta}^{\bar{\sfa}_3\bar{\sfd}_3}_{\sfb\bar{\sfc}_5}\,f^{\bm{\gamma}}_{\bar{\sfd}_3}\,T_{\bm{\gamma}}^{\bar{\sfc}_5}\,,
\\
 T^{\bar{\sfa}_3} \circ T_{\bm{\beta}}^{\sfb} 
 &= f_{\sfc}{}^{\bar{\sfa}_3 \sfb}\,T_{\bm{\beta}}^{\sfc}
  - \hat{\delta}^{\bar{\sfa}_3}_{\bar{\sfd}_2\sfe}\, f_{\sfc}{}_{{\bm{\beta}}}^{\bar{\sfd}_2}\,T^{\sfe \sfb\sfc}
  + \hat{\delta}^{\bar{\sfa}_3}_{\sfd\bar{\sfe}_2}\, f_{\bar{\sfc}_2}{}^{\sfd}\,T_{\bm{\beta}}^{\bar{\sfe}_2 \sfb \bar{\sfc}_2}
 - 4\,Z_{\sfc}\,T_{\bm{\beta}}^{\bar{\sfa}_3\sfb\sfc}
 - \hat{\delta}^{\bar{\sfa}_3\bar{\sfd}_3}_{\bar{\sfc}_6}\, \epsilon_{{\bm{\beta}}{\bm{\gamma}}}\,f^{\bm{\gamma}}_{\bar{\sfd}_3}\,T^{\bar{\sfc}_6,\sfb}\,,
\\
 T^{\bar{\sfa}_3} \circ T^{\bar{\sfb}_3} 
 &= \hat{\delta}^{\bar{\sfb}_3}_{\sfd\bar{\sfe}_2}\, f_{\sfc}{}^{\bar{\sfa}_3 \sfd}\, T^{\bar{\sfe}_2 \sfc}
   - \hat{\delta}^{\bar{\sfa}_3}_{\bar{\sfd}_2\sfe}\, \epsilon^{{\bm{\gamma}}{\bm{\delta}}}\,f_{\sfc}{}_{{\bm{\gamma}}}^{\bar{\sfd}_2}\,T_{\bm{\delta}}^{\sfe\bar{\sfb}_3\sfc}
\\
 &\quad + \hat{\delta}^{\bar{\sfa}_3}_{\sfd\sfe\sff}\,f_{\bar{\sfc}_2}{}^{\sfd}\, T^{\bar{\sfb}_3\bar{\sfc}_2\sfe, \sff}
   + \hat{\delta}^{\bar{\sfa}_3}_{\sfe\bar{\sff}_2}\,f_{\sfc\sfd}{}^{\sfe}\,T^{\bar{\sff}_2 \bar{\sfb}_3 \sfc, \sfd} 
 +4\,\hat{\delta}^{\bar{\sfa}_3\sfc}_{\bar{\sfd}_3\sfe}\,Z_{\sfc}\,T^{\bar{\sfb}_3 \bar{\sfd}_3,\sfe} \,,
\\
 T^{\bar{\sfa}_3} \circ T_{\bm{\beta}}^{\bar{\sfb}_5} 
 &= \hat{\delta}^{\bar{\sfb}_5}_{\sfd\bar{\sfe}_4}\,f_{\sfc}{}^{\bar{\sfa}_3 \sfd}\,T_{\bm{\beta}}^{\bar{\sfe}_4 \sfc} 
 - \hat{\delta}^{\bar{\sfa}_3}_{\bar{\sff}_2\sfg}\,\hat{\delta}^{\sfg\sfc}_{\sfd\sfe}\,f_{\sfc}{}_{{\bm{\beta}}}^{\bar{\sff}_2}\,T^{\bar{\sfb}_5 \sfd,\sfe} \,,
\\
 T^{\bar{\sfa}_3} \circ T^{\bar{\sfb}_6,\sfb'} &=
  \hat{\delta}^{\bar{\sfb}_6}_{\sfd\bar{\sfe}_5}\,f_{\sfc}{}^{\bar{\sfa}_3 \sfd}\,T^{\sfc \bar{\sfe}_5, \sfb'}
  + f_{\sfc}{}^{\bar{\sfa}_3 \sfb'}\,T^{\bar{\sfb}_6, \sfc} \,,
\\
 T_{\bm{\alpha}}^{\bar{\sfa}_5} \circ T_{\sfb} 
 &= - f_{\sfb}{}_{{\bm{\alpha}}}^{\bar{\sfa}_5 \sfc}\,T_{\sfc}
 - \hat{\delta}^{\bar{\sfa}_5}_{\bar{\sfd}_4\sfe}\,\hat{\delta}^{\sfe\sff}_{\sfb\sfc}\,f_{\sff}{}^{\bar{\sfd}_4}\, T_{\bm{\alpha}}^{\sfc}
 + \hat{\delta}^{\bar{\sfa}_5}_{\sfb\bar{\sfd}_2\bar{\sfe}_2}\,f_{\sfc}{}_{{\bm{\alpha}}}^{\bar{\sfd}_2}\, T^{\bar{\sfe}_2\sfc}
 - \hat{\delta}^{\bar{\sfa}_5}_{\bar{\sfc}_2\bar{\sfd}_3}\,f_{\sfb}{}_{{\bm{\alpha}}}^{\bar{\sfc}_2}\,T^{\bar{\sfd}_3}
 - \hat{\delta}^{\bar{\sfa}_5}_{\sfb\bar{\sfd}_4}\,f_{\sfc{\bm{\alpha}}}{}^{{\bm{\gamma}}}\, T_{\bm{\gamma}}^{\bar{\sfd}_4\sfc}
\\
 &\quad
 + f_{\sfb{\bm{\alpha}}}{}^{{\bm{\gamma}}}\,T_{\bm{\gamma}}^{\bar{\sfa}_5}
 + \hat{\delta}^{\bar{\sfa}_5}_{\sfd\bar{\sfe}_4}\,f_{\sfb\sfc}{}^{\sfd}\, T_{\bm{\alpha}}^{\bar{\sfe}_4\sfc}
 + \hat{\delta}^{\bar{\sfa}_5}_{\sfd\sfb\bar{\sfe}_3}\,f_{\bar{\sfc}_2}{}^{\sfd}\, T_{\bm{\alpha}}^{\bar{\sfe}_3 \bar{\sfc}_2}
 + 6\,\hat{\delta}^{\bar{\sfa}_5\sfc}_{\sfb\bar{\sfd}_5}\,Z_{\sfc}\, T_{\bm{\alpha}}^{\bar{\sfd}_5}\,,
\\
 T_{\bm{\alpha}}^{\bar{\sfa}_5} \circ T_{\bm{\beta}}^{\sfb} 
 &= f_{\sfc}{}_{{\bm{\alpha}}}^{\bar{\sfa}_5\sfb}\,T_{\bm{\beta}}^{\sfc}
 + \hat{\delta}^{\bar{\sfa}_5}_{\bar{\sfd}_4\sfe}\,\epsilon_{{\bm{\alpha}}{\bm{\beta}}}\,f_{\sfc}{}^{\bar{\sfd}_4}\,T^{\sfe \sfb\sfc}
 - \hat{\delta}^{\bar{\sfa}_5}_{\bar{\sfd}_2\bar{\sfe}_3}\,f_{\sfc}{}_{{\bm{\alpha}}}^{\bar{\sfd}_2}\,T_{\bm{\beta}}^{\bar{\sfe}_3 \sfb\sfc} 
\\
 &\quad - f_{\sfc{\bm{\alpha}}}{}^{\bm{\gamma}}\,\epsilon_{{\bm{\gamma}}{\bm{\beta}}}\,T^{\sfc \bar{\sfa}_6,\sfb}
 - \hat{\delta}^{\bar{\sfa}_5}_{\sfd\bar{\sfe}_4}\,\epsilon_{{\bm{\alpha}}{\bm{\beta}}}\,f_{\bar{\sfc}_2}{}^{\sfd}\,T^{\bar{\sfe}_4 \bar{\sfc}_2,\sfb}
 -6\,\epsilon_{{\bm{\alpha}}{\bm{\beta}}}\,Z_{\sfc}\,T^{\bar{\sfa}_5\sfc,\sfb}\,,
\\
 T_{\bm{\alpha}}^{\bar{\sfa}_5} \circ T^{\bar{\sfb}_3} 
 &= \hat{\delta}^{\bar{\sfb}_3}_{\sfd\bar{\sfe}_2}\,f_{\sfc}{}_{{\bm{\alpha}}}^{\bar{\sfa}_5 \sfd}\, T^{\bar{\sfe}_2\sfc}
 - \hat{\delta}^{\bar{\sfa}_5}_{\bar{\sfd}_4\sfe}\,f_{\sfc}{}^{\bar{\sfd}_4}\, T_{\bm{\alpha}}^{\sfe \bar{\sfb}_3\sfc} 
 + \hat{\delta}^{\bar{\sfa}_5}_{\bar{\sff}_2\bar{\sfg}_3}\,\hat{\delta}^{\bar{\sfg}_3\sfc}_{\bar{\sfd}_3\sfe}\,f_{\sfc}{}_{{\bm{\alpha}}}^{\bar{\sff}_2}\,T^{\bar{\sfb}_3 \bar{\sfd}_3,\sfe}\,,
\\
 T_{\bm{\alpha}}^{\bar{\sfa}_5} \circ T_{\bm{\beta}}^{\bar{\sfb}_5} 
 &= \hat{\delta}^{\bar{\sfb}_5}_{\sfd\bar{\sfe}_4}\,f_{\sfc}{}_{{\bm{\alpha}}}^{\bar{\sfa}_5 \sfd}\, T_{\bm{\beta}}^{\bar{\sfe}_4\sfc}
 + \hat{\delta}^{\bar{\sfa}_5}_{\bar{\sff}_4\sfg}\,\hat{\delta}^{\sfg\sfc}_{\sfd\sfe}\,\epsilon_{{\bm{\alpha}}{\bm{\beta}}}\,f_{\sfc}{}^{\bar{\sff}_4}\,T^{\bar{\sfb}_5\sfd,\sfe} \,,
\\
 T^{\bar{\sfa}_5} \circ T^{\bar{\sfb}_6,\sfb'} &= \hat{\delta}^{\bar{\sfb}_6}_{\sfd\bar{\sfe}_5}\,f_{\sfc}{}_{{\bm{\alpha}}}^{\bar{\sfa}_5 \sfd}\, T^{\sfc \bar{\sfe}_5,\sfb'}
 + f_{\sfc}{}_{{\bm{\alpha}}}^{\bar{\sfa}_5 \sfb'}\, T^{\bar{\sfb}_6,\sfc} \,,
\\
 T^{\bar{\sfa}_6,\sfa'} \circ T_{\sfb} &= 
 \hat{\delta}^{\sfa'\sfc}_{\sfb\sfd}\,\epsilon^{{\bm{\gamma}}{\bm{\delta}}}\,f_{\sfc}{}_{{\bm{\gamma}}}^{\bar{\sfa}_6}\, T^{\sfd}_{\bm{\delta}}
 + \hat{\delta}^{\bar{\sfa}_6}_{\bar{\sfc}_3\bar{\sfd}_3}\,f_{\sfb}{}^{\sfa' \bar{\sfc}_3}\,T^{\bar{\sfd}_3}
 + \hat{\delta}^{\bar{\sfa}_6}_{\sfb\bar{\sfd}_3\bar{\sfe}_2}\,f_{\sfc}{}^{\sfa' \bar{\sfd}_3}\, T^{\bar{\sfe}_2\sfc} 
\\
 &\quad - \hat{\delta}^{\bar{\sfa}_6}_{\sfc\bar{\sfd}_5}\,\epsilon^{{\bm{\gamma}}{\bm{\delta}}}\,f_{\sfb}{}_{{\bm{\gamma}}}^{\sfa' \sfc}\,T^{\bar{\sfd}_5}_{\bm{\delta}}
 - \hat{\delta}^{\bar{\sfa}_6}_{\sfb\sfd\bar{\sfe}_4}\,\epsilon^{{\bm{\gamma}}{\bm{\delta}}}\,f_{\sfc}{}_{{\bm{\gamma}}}^{\sfa' \sfd}\, T^{\bar{\sfe}_4\sfc}_{\bm{\delta}} \,,
\\
 T^{\bar{\sfa}_6,\sfa'} \circ T_{\bm{\beta}}^{\sfb} &= f_{\sfc}{}_{{\bm{\beta}}}^{\bar{\sfa}_6}\,T^{\sfa'\sfb\sfc} 
 + \hat{\delta}^{\bar{\sfa}_6}_{\bar{\sfd}_3\bar{\sfe}_3}\,f_{\sfc}{}^{\sfa' \bar{\sfd}_3}\,T_{\bm{\beta}}^{\bar{\sfe}_3\sfb\sfc}
 + \hat{\delta}^{\bar{\sfa}_6}_{\sfd\bar{\sfe}_5}\,f_{\sfc}{}_{{\bm{\beta}}}^{\sfa' \sfd}\,T^{\bar{\sfe}_5\sfc,\sfb} \,,
\\
 T^{\bar{\sfa}_6,\sfa'} \circ T^{\bar{\sfb}_3} &= 
 \epsilon^{{\bm{\gamma}}{\bm{\delta}}}\,f_{\sfc}{}_{{\bm{\gamma}}}^{\bar{\sfa}_6}\,T_{\bm{\delta}}^{\sfa'\bar{\sfb}_3\sfc} 
 - \hat{\delta}^{\bar{\sfa}_6}_{\bar{\sff}_3\bar{\sfg}_3}\,\hat{\delta}^{\bar{\sfg}_3\sfc}_{\bar{\sfd}_3\sfe}\,f_{\sfc}{}^{\sfa' \bar{\sff}_3}\,T^{\bar{\sfb}_3\bar{\sfd}_3,\sfe} \,,
\\
 T^{\bar{\sfa}_6,\sfa'} \circ T_{\bm{\beta}}^{\bar{\sfb}_5} &= \hat{\delta}^{\sfa'\sfc}_{\sfd\sfe}\,f_{\sfc}{}_{{\bm{\beta}}}^{\bar{\sfa}_6}\,T^{\bar{\sfb}_5 \sfd,\sfe} \,,
\\
 T^{\bar{\sfa}_6,\sfa'} \circ T^{\bar{\sfb}_6,\sfb'} &= 0 \,.
\end{split}
\end{align}}
The structure constants here are
\begin{align}
 f_{\bar{\sfa}_5}\,,\quad f_{\bar{\sfa}_3}^{\bm{\alpha}}\,,\quad f_{\sfa\sfb}{}^{\sfc}\,,\quad f_{\sfa\bm{\beta}}{}^{\bm{\gamma}}\,,\quad f_{\sfa}{}^{\bar{\sfb}_2}_{\bm{\alpha}}\,,\quad f_{\sfa}{}^{\bar{\sfb}_4}\,,\quad f_{\sfa}{}^{\bar{\sfb}_6}_{\bm{\alpha}}\,,\quad Z_{\sfa}\,.
\end{align}

\bibliography{literature}

\bibliographystyle{JHEP}

\end{document}